\documentclass[preprint]{aastex}
\usepackage[usenames,dvipsnames]{color}
\usepackage{natbib,graphicx,latexsym,lscape}

\newcommand{\Spitzer}{{\sl Spitzer}}
\newcommand{\Hipparcos}{{\sl Hipparcos}}
\newcommand{\HST}{{\sl HST}}
\newcommand{\JWST}{{\sl JWST}}
\newcommand{\WISE}{{\sl WISE}}
\newcommand{\ROSAT}{{\sl ROSAT}}
\newcommand{\Msun}{\mbox{$M_{\sun}$}}
\newcommand{\Mearth}{\mbox{$M_{\oplus}$}}
\newcommand{\Lsun}{\mbox{$L_{\sun}$}}
\newcommand{\Rsun}{\mbox{$R_{\sun}$}}
\newcommand{\Mjup}{\mbox{$M_{\rm Jup}$}}
\newcommand{\Rjup}{\mbox{$R_{\rm Jup}$}}
\newcommand{\degree}{\mbox{$^{\circ}$}}
\newcommand{\perpix}{\mbox{pixel$^{-1}$}}
\newcommand{\cms}{\hbox{cm~s$^{-1}$}}
\newcommand{\kms}{\hbox{km~s$^{-1}$}}
\newcommand{\masyr}{\hbox{mas~yr$^{-1}$}}
\newcommand{\asyr}{\hbox{$\arcsec$~yr$^{-1}$}}
\newcommand{\Jc}{\mbox{$J_{\rm cont}$}}
\newcommand{\Hc}{\mbox{$H_{\rm cont}$}}
\newcommand{\Kc}{\mbox{$K_{\rm cont}$}}
\newcommand{\Kp}{\mbox{$K^{\prime}$}}
\newcommand{\Ks}{\mbox{$K_S$}}
\newcommand{\CHs}{\mbox{$CH_4s$}}
\newcommand{\Kn}{\mbox{$K_{\rm H2}$}}
\newcommand{\Mtot}{\mbox{$M_{\rm tot}$}}
\newcommand{\Lbol}{\mbox{$L_{\rm bol}$}}
\newcommand{\Rstar}{\mbox{$R_{\star}$}}
\newcommand{\Mstar}{\mbox{$M_{\star}$}}
\newcommand{\Mbol}{\mbox{$M_{\rm bol}$}}
\newcommand{\Teff}{\mbox{$T_{\rm eff}$}}
\newcommand{\logg}{\mbox{$\log(g)$}}
\newcommand{\Vtan}{\mbox{$V_{\rm tan}$}}
\newcommand{\Rc}{\mbox{${R}$}}
\newcommand{\Ic}{\mbox{${I}$}}
\newcommand{\zp}{\mbox{${z^\prime}$}}
\newcommand{\Lp}{\mbox{${L^\prime}$}}
\newcommand{\Mp}{\mbox{${M^\prime}$}}

\shorttitle{Hawaii IR Parallaxes: Binaries and the L/T Transition}
\shortauthors{Dupuy \& Liu}

\begin{document}

\title{The Hawaii Infrared Parallax Program. \\ I.~Ultracool Binaries
  and the L/T Transition\altaffilmark{*,\dag}}

\author{Trent J.\ Dupuy\altaffilmark{1,2,3} and
        Michael C.\ Liu\altaffilmark{2}}

      \altaffiltext{*}{Based on observations obtained with WIRCam,
        a joint project of CFHT, Taiwan, Korea, Canada, France, at the
        Canada-France-Hawaii Telescope (CFHT) which is operated by the
        National Research Council (NRC) of Canada, the Institute
        National des Sciences de l'Univers of the Centre National de
        la Recherche Scientifique of France, and the University of
        Hawaii.}

      \altaffiltext{\dag}{Some of the data presented herein were
        obtained at the W.M.\ Keck Observatory, which is operated as a
        scientific partnership among the California Institute of
        Technology, the University of California, and the National
        Aeronautics and Space Administration. The Observatory was made
        possible by the generous financial support of the W.M.\ Keck
        Foundation.}

      \altaffiltext{1}{Harvard-Smithsonian Center for Astrophysics,
        60 Garden Street, Cambridge, MA 02138}

      \altaffiltext{2}{Institute for Astronomy, University of Hawai`i,
        2680 Woodlawn Drive, Honolulu, HI 96822}

      \altaffiltext{3}{Hubble Fellow}

\begin{abstract}

  We present the first results from our high-precision infrared (IR)
  astrometry program at the Canada-France-Hawaii Telescope.  We
  measure parallaxes for 83 ultracool dwarfs (spectral types M6--T9)
  in 49 systems, with a median uncertainty of 1.1~mas (2.3\%) and as
  good as 0.7~mas (0.8\%).
  We provide the first parallaxes for 48 objects in 29 systems, and
  for another 27 objects in 17 systems, we significantly improve upon
  published results, with a median (best) improvement of 1.7$\times$
  (5$\times$).
  Three systems show astrometric perturbations indicative of orbital
  motion; two are known binaries (2MASS~J0518$-$2828AB and
  2MASS~J1404$-$3159AB) and one is spectrally peculiar
  (SDSS~J0805+4812).  In addition, we present here a large set of Keck
  adaptive optics imaging that more than triples the number of
  binaries with L6--T5 components that have both multi-band photometry
  and distances.  Our data enable an unprecedented look at the
  photometric properties of brown dwarfs as they cool through the L/T
  transition.  Going from $\approx$L8 to $\approx$T4.5, flux in the
  $Y$ and $J$ bands increases by $\approx$0.7~mag and
  $\approx$0.5~mag, respectively (the $Y$- and $J$-band ``bumps''),
  while flux in the $H$, $K$, and \Lp\ bands declines monotonically.
  This wavelength dependence is consistent with cloud clearing over a
  narrow range of temperature, since condensate opacity is expected to
  dominate at 1.0--1.3~\micron.  Interestingly, despite more than
  doubling the near-IR census of L/T transition objects, we find a
  conspicuous paucity of objects on the color--magnitude diagram just
  blueward of the late-L/early-T sequence.  This ``L/T gap'' occurs at
  $(J-H)_{\rm MKO} = 0.1$--0.3~mag, $(J-K)_{\rm MKO} = 0.0$--0.4~mag,
  and implies that the last phases of cloud evolution occur rapidly.
  Finally, we provide a comprehensive update to the absolute
  magnitudes of ultracool dwarfs as a function of spectral type using
  a combined sample of 314 objects.

\end{abstract}



\section{Introduction}

Few astronomical measurements are as direct and model-independent as
trigonometric parallaxes, as they rely solely on geometry and an
accurate ephemeris of the Earth's orbit.  Distances determined by
parallaxes form the foundation of much of modern astrophysics, e.g.,
enabling the creation of the Hertzsprung-Russell diagram and
establishing a key rung in the cosmological distance ladder.  Since
the first stellar parallax measurement
\citep[61~Cyg;][]{1838AN.....16...65B}, astrometry programs have
continuously evolved using new technology to achieve ever-expanding
science objectives.  Photographic plates dominated parallax work for
many decades, but the need to reach fainter stars eventually required
the use CCDs with their low noise, high quantum efficiency, and
capacity for large dynamic range.  Pioneering work in this area
demonstrated that precise astrometry was in fact possible with such
devices \citep[e.g., see][]{1983AJ.....88.1489M}, even though the
field-of-view of early detectors was small by today's standards.  As
CCDs have grown in size they have become the dominant tool for
high-precision astrometry.  With the advent of large-format infrared
(IR) arrays it is now possible to extend parallax measurements to
large samples of the coldest known objects outside the solar system:
brown dwarfs.

Over the past decade, several ground-based astrometry programs have
laid the foundation for understanding the basic evolution of brown
dwarfs on the color--magnitude diagram. Infrared parallax programs
account for about two thirds of parallaxes for brown dwarfs with
spectral types $\geq$L4 \citep[e.g.,][]{2003AJ....126..975T,
  2004AJ....127.2948V, 2010A&A...524A..38M}, with red optical programs
providing the remaining one third, mostly at earlier types
\citep[e.g.,][]{2002AJ....124.1170D, 2009A&A...493L..27S,
  2011AJ....141...54A}.  Companions to stars with \Hipparcos\
parallaxes also make up a significant fraction of the current sample
of $\geq$L4 dwarfs with parallaxes, roughly half as many as have been
measured directly in infrared astrometry programs.  Parallax
measurements for very low-mass stars and brown dwarfs of earlier
spectral types (M6--L4) are dominated by red optical astrometry
programs at the USNO \citep{1992AJ....103..638M, 2002AJ....124.1170D}
and elsewhere \citep[e.g.,][]{1995AJ....110.3014T,
  1996MNRAS.281..644T, 2006AJ....132.1234C, 2009AJ....137..402G,
  2009AJ....137.4109L, 2009A&A...493L..27S, 2011AJ....141...54A}.

There is a pressing demand for the highest possible precision in
ultracool dwarf distance measurements.  This is because dynamical mass
studies are now providing the strongest tests of substellar models
\citep[e.g.,][]{2004A&A...423..341B, 2008ApJ...689..436L,
  2009ApJ...692..729D, 2009ApJ...699..168D, 2010ApJ...721.1725D,
  2010ApJ...711.1087K}, and precise parallaxes are crucial for such
work.  Dynamical mass uncertainties from visual binary orbits are
almost always dominated by the error in the distance since mass
$\propto d^3$. Thus, to achieve a 10\% mass uncertainty requires
parallax errors of $\approx$3\%. Among ground-based measurements for
$\geq$L4~dwarfs such precision is not common (only 26\% of parallaxes)
and has previously been achieved \emph{only} for relatively nearby
objects ($\leq$13~pc).

Furthermore, despite the past successes of the parallax programs
described above, there are still important aspects of brown dwarf
evolution that would benefit from a larger set of distance
measurements: young field brown dwarfs
\citep[e.g.,][]{2006ApJ...639.1120K, 2009ApJ...697..824A,
  2009AJ....137.3345C}, the coldest brown dwarfs
\citep[$\lesssim$500~K, e.g.,][]{2010MNRAS.408L..56L,
  2011ApJ...743...50C}, and the L/T transition
\citep[e.g.,][]{2006ApJ...647.1393L, 2008ApJ...689.1327S}. Samples
pertaining to the first two subjects have only recently begun to be
uncovered, and parallax measurements are underway by multiple groups
for both young field dwarfs (e.g., \citealp{2008A&A...489..825T}; Liu,
Dupuy \& Allers, submitted) and the latest-type T~dwarfs
\citep[e.g.,][]{2010A&A...511A..30S, 2011ApJ...740..108L}.  In
contrast, objects with properties intermediate between red L~dwarfs
and blue T~dwarfs have been known since some of the earliest surveys
to yield brown dwarfs \citep{2000ApJ...536L..35L,
  2002ApJ...564..466G}. However, to date only 6 single objects in this
range (L9--T4) have parallaxes, compared to 33 parallaxes for single
T4.5--T9 dwarfs and 29 parallaxes for single L4--L8.5 dwarfs. (There
are an additional $\approx$4 components of binaries in the L9--T4
range with parallaxes, but this exact number is subject to the
somewhat uncertain spectral classification of most of these
components.)  There is a present deficiency in the number of L/T
transition objects with parallaxes and thus in our ability to
characterize one of the most important phases of brown dwarf
evolution.

To address the need for high precision parallaxes of ultracool
binaries, we initiated an infrared parallax program at the
Canada-France-Hawaii Telescope (CFHT) in 2007.  We concentrated our
observations on a sample of ultracool binaries with a wide range of
component spectral types (M6--T9) that includes all systems observable
with CFHT that are likely to yield dynamical masses in the next
$\approx$decade. This dynamical mass sample also forms the basis of
our ongoing Keck adaptive optics (AO) orbital monitoring program,
which to date has tripled the number of ultracool binaries with
dynamical masses sufficiently precise for model testing (see
\citealp{me-cs16} and references therein). The primary goals of this
first phase of our CFHT program are to expand the sample of dynamical
mass measurements for brown dwarfs and enable more precise masses from
the existing sample of orbits by reducing distance errors. In addition
to the dynamical mass sample, we included in our original parallax
program several other binaries that are not necessarily amenable to
orbit determination in the near future but that have components
bridging the L/T transition. This L/T sample is motivated by the
deficit of parallaxes for objects with spectral types L9--T4 and by
the inherent utility of binaries for substellar model tests given
their identical age and composition
\citep[e.g.,][]{2005astro.ph..8082L, 2010ApJ...722..311L}. This supplemental
sample of L/T binaries provides the context needed for comparisons to
the field population as our orbital monitoring program yields
dynamical mass measurements for L/T transition objects
\citep[e.g.,][]{2009ApJ...699..168D}.  Finally, we have also been targeting
binaries with the coldest known components ($\gtrsim$T8), and this has
resulted in a parallax for CFBDS~J1458+1013AB, which has component
types of T9 and $>$T10 \citep[][updated parallax given in this
paper]{2011ApJ...740..108L}.

We present here the first large set of results of our CFHT infrared
parallax program along with a complete description of our astrometric
methods (Section~\ref{sec:cfht}).  This sample includes 34 binaries
and 15 single objects that have been chosen because they will be
useful for measuring dynamical masses in the future, studying the L/T
transition, and increasing the number of parallaxes for mid- to late-T
dwarfs.  We also present supporting observations from other
telescopes, including a large collection of resolved photometry for
tight binaries from Keck, \HST, and VLT (Section~\ref{sec:keck}) and
integrated-light near-infrared spectroscopy (Section~\ref{sec:irtf}).
The ensemble of these new measurements provides an unprecedented view
of the L/T transition.


\section{CFHT/WIRCam Astrometric Monitoring \label{sec:cfht}}

Since 2007, we have been using the facility near-IR camera WIRCam at
CFHT to conduct an astrometric monitoring program with the goal of
measuring parallaxes for ultracool dwarfs. WIRCam comprises a mosaic
of four 2048$\times$2048 Hawaii-2RG infrared arrays, each with a
field-of-view of $10.4\arcmin \times 10.4\arcmin$ and pixelscale of
$0\farcs3$~\perpix\ \citep{2004SPIE.5492..978P}. At each epoch, we
obtained $\approx$20--30 dithered images of our targets, which were
always centered on the northeast array of WIRCam. All images were
first processed at CFHT using the WIRCam pipeline `I`iwi, which
performs a non-linearity correction, dark subtraction, flat fielding,
bad pixel masking, sky subtraction, and cross-talk removal for each
image.\footnote{\url{http://cfht.hawaii.edu/Instruments/Imaging/WIRCam/IiwiVersion1Doc.html}}
We obtained data in $J$ band for most targets, as this filter afforded
the lowest sky background and thus the most reference stars. Targets
brighter than $J < 13.3$~mag were at risk of saturating in the
5-second minimum integration time of WIRCam, so for these targets we
used the narrow $K$-band filter (\Kn) centered at 2.122~\micron\ with
a bandwidth of 0.032~\micron\ (1.5\%).  Table~\ref{tbl:sample}
summarizes our target list and the details of our observations.

The CFHT data presented herein were mostly collected from the fall
semester of 2007 to spring 2010, with 89~hours of queue-scheduled CFHT
time allocated over 6 semesters. We have continued monitoring some
targets in later semesters to improve their parallax errors, and the
most recent data presented here comes from early 2012. The median
seeing for all the CFHT data presented here is $0\farcs63$, as judged
by the target full-width half maximum (FWHM), and 85\% of the data
were taken in $<0\farcs80$ seeing. Our goal is to obtain a minimum of
$\approx$10 epochs spread over three or more observing seasons for
each target, and in this paper we include targets with 6--24
observations obtained over 2--5 seasons.

\begin{figure}
\centerline{\includegraphics[width=6.0in,angle=0]{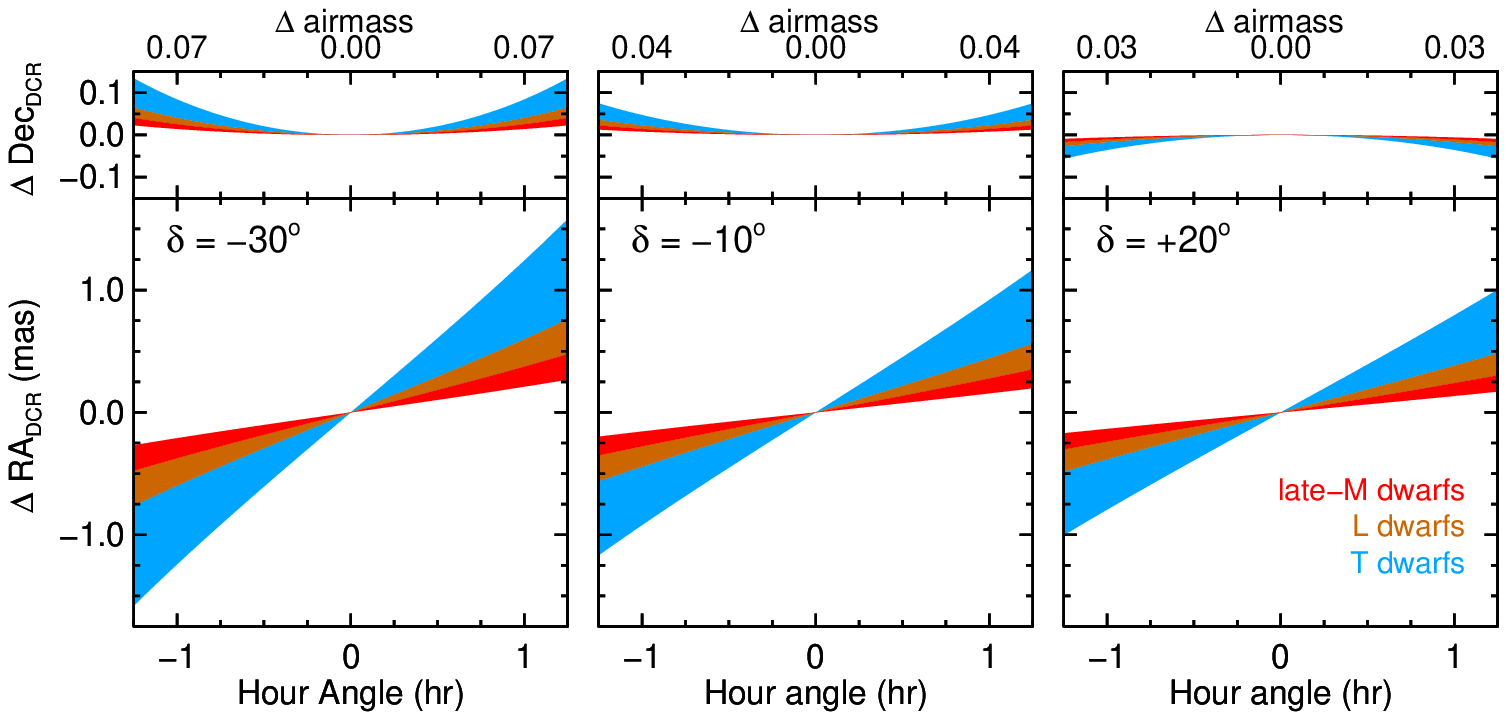}}

\caption{\normalsize Astrometric offsets in $J$ band due to the
  differential chromatic refraction (DCR) between our targets and
  background stars computed for Mauna Kea.  These offsets result from
  our targets' $J$-band spectra being dissimilar from those of
  background stars, and thus the offsets increase at later spectral
  types because the differences are more pronounced.  Each colored
  swath shows the range of offsets predicted for the variety of
  subtypes within each spectral classification (e.g., T0--T8 for the T
  dwarfs).  The offsets increase with airmass, so our observations
  were constrained to be as close to transit as possible, and the
  effects are expected to be worse for targets farther from zenith
  ($\delta = 19\fdg8$ at Mauna Kea).  By always obtaining data within
  1~hr of transit (and typically within 30~min), we have ensured that
  the effects of DCR on our astrometry are negligible
  ($\lesssim1$~mas).  \label{fig:dcr}}

\end{figure}

CFHT is operated in queue mode, providing significant advantages for
astrometric monitoring. Foremost is the ability to virtually eliminate
the systematic effects of differential chromatic refraction (DCR)
between observation epochs for every target. This is accomplished by
obtaining data only within a narrow specified range of airmass, which
can be done automatically within the CFHT queue software. Our $J$-band
targets were typically observed within $\Delta$airmass of 0.03 and
never more than 1~hour from transit (Table~\ref{tbl:sample}). (DCR is
completely negligible for the \Kn-band targets because of the narrow
bandpass.) Figure~\ref{fig:dcr} shows the expected DCR offsets in $J$
band between our targets and background reference stars, as determined
using the method described in Section~2.2 of \citet{2009ApJ...692..729D}. We
computed the effective wavelength at $J$ band for late-M, L, and T
dwarf spectral standards given in the SpeX Prism
Library\footnote{\url{http://www.browndwarfs.org/spexprism},
  maintained by Adam Burgasser.} and then determined the DCR offsets
using equations from \citet{1984A&A...138..275S} and
\citet{1992AJ....103..638M}. We found that GKM stars all have
virtually the same effective wavelength at $J$ band, and for our
calculations we used the value derived from the M0 spectral standard
(HD~19305; $\lambda_{\rm eff} = 1.2462$~\micron).  Systematic
astrometric offsets due to DCR result from the fact that atmospheric
refraction shifts the grid of reference stars by a different amount
than our ultracool target.  Our calculations show that even at our
most extreme deviation in airmass, DCR is only a $\approx$1~mas effect
for T dwarfs, $\approx$0.5~mas effect for L dwarfs, and
$\approx$0.3~mas effect for late-M dwarfs. As will be shown in
Section~\ref{sec:plx-fit}, such DCR offsets have a negligible effect
on our resulting parallaxes and uncertainties.

The other major advantage afforded by queue service mode is the
ability to obtain excellent parallax phase coverage for targets widely
distributed on the sky with minimal impact from poor weather or
seeing. Note however that WIRCam is bolted onto the telescope when in
use and must be removed to use other instruments, so there are
discrete WIRCam runs of $\approx$1--2 weeks each undertaken
$\approx$4--5 times per semester. These runs could be at irregular
intervals, depending on the queue pressure each semester. This,
combined with the fact that a string of very poor weather could
cripple a given run, means that targets at some right ascensions
received much better phase coverage in our program than others, with
targets at 12h--01h generally getting the most coverage and targets at
04h--10h getting somewhat less.

\subsection{Creating an Astrometric Catalog at each Epoch
  \label{sec:dither-match}}

\subsubsection{Position Measurements}

At each epoch we obtained $\approx$20--30 individual dithered frames
of our target fields. We obtained positional measurements for all of
the sources in each field from SExtractor \citep{1996A&AS..117..393B}
using the ``windowed'' parameters (e.g., {\tt XWIN\_IMAGE}) rather
than the classic isophotal parameters (e.g., {\tt X\_IMAGE}). Windowed
parameters have the advantage of being less noisy, because they are
computed with a Gaussian weight function that decreases the impact of
pixels far from the PSF core on the measured positions. We used flag
maps within SExtractor to track sources that were either saturated or
located near bad pixels, as identified by the CFHT data processing
pipeline. These flagged sources were excluded from subsequent
analysis. We also used the S/N estimates from
SExtractor\footnote{S/N~$\equiv {\rm \tt FLUX\_AUTO/FLUXERR\_AUTO}$.}
to exclude sources with S/N~$< 10$. We did not attempt to exclude
galaxies based on SExtractor shape parameters at this stage, but in a
later step (Section \ref{sec:epoch-match}) non-stellar sources
typically ended up being excised because of their large positional
rms.

\subsubsection{Cross-identifying Detections}

The first step in creating an astrometric catalog was to associate all
of the detections across multiple frames as belonging to a common set
of objects. We found that the most robust method for cross-identifying
stars was to first match detections in a given frame to an astrometric
reference catalog, either the Sloan Digital Sky Survey Data Release 7
\citep[SDSS-DR7;][]{2009ApJS..182..543A}, the Two Micron All Sky
Survey Point Source Catalog \citep[2MASS-PSC;][]{2006AJ....131.1163S},
or the USNO-B1.0 Catalog \citep{2003AJ....125..984M}. We used the
information in the CFHT FITS headers to obtain an initial guess for
the image coordinates, and we refined this initial guess by using the
catalog matching software SCAMP \citep{2006ASPC..351..112B}. We
thereby determined approximate source positions in celestial
coordinates, adequate for cross-identifying detections that have
corresponding entries in the reference catalog; we used whichever
catalog gave the most matches for a target field. We then determined a
more precise astrometric solution for the given frame that included
second order terms (i.e., $x^2$, $y^2$, $xy$) since these distortion
terms are significant at the $\approx$1$\arcsec$ level. This fit was
performed using the MPFIT implementation of the Levenberg--Marquardt
least-squares minimization routine in IDL \citep{2009ASPC..411..251M}.
This temporary best-fit astrometric solution was then applied to all
the detections in the frame so that we could crossmatch them between
frames.

We constructed our catalog of associated detections by starting with
the list of detections in the first image and then adding detections
from the next image by either finding a match in the existing catalog
or creating a new entry if no match was found. After adding a new
image, the catalog position of each object was recomputed as the
median of currently associated measurements. This procedure was
repeated for each image until all positional measurements from that
epoch were included in the catalog. We then discarded objects from the
catalog that were detected fewer than ten times in order to focus on
stars that will have the most robust astrometry. This cut excludes
stars on the periphery of the field that were only captured in a
subset of dithers as well as bright sources near the saturation limit
and image artifacts (e.g., cosmic rays, persistence spots, and array
defects).  Note that because we created a separate catalog for each
epoch, sources with large proper motion would not be discarded at this
step.

\subsubsection{Registering Dithers}

We next optimally registered the positions of stars cross-associated
in individual images at a given epoch. The only information we used
from the initial pass of reference catalog matching were the
coordinates of the tangent point and linear terms for the first frame,
and these were only a temporary guess because later in our analysis we
solve for all of these parameters directly. Our optimization operates
in spherical rather than $(x,y)$ coordinates in order to properly
account for the fact that our measurements are actually tangent
projections of celestial positions. For example, our largest dithers
of 1$\arcmin$ can cause the relative positions of stars at the edges
of our 10$\arcmin$ field to appear to move by $\sim$10~mas due to
tangent projection effects. The best-fit registration solution was
found using MPFIT to jointly minimize unweighted residuals in right
ascension, $(\alpha - {\rm mean}(\alpha))\cos{\delta}$, and
declination, $\delta - {\rm mean}(\delta)$. The only parameters
allowed to vary between frames in this fit were the $(\alpha, \delta)$
coordinates of the tangent point (i.e., only a shift). After
performing the fit the first time, we clipped any positional
measurements that were more than 3.5$\sigma$ discrepant with the
median catalog position to eliminate corrupted measurements (e.g.,
affected by a cosmic ray hit) or image artifacts that were erroneously
associated with real sources. This cut was chosen because it would
eliminate $\lesssim1$ true measurement even in our richest data sets
of a few $\times 10^3$ detections, and typically $\lesssim10$
detections were actually clipped. After clipping, the fit was then
repeated a second and final time.

\subsubsection{Accounting for Distortion and Linear Terms}

\begin{figure}
\centerline{
\includegraphics[width=1.6in,angle=0]{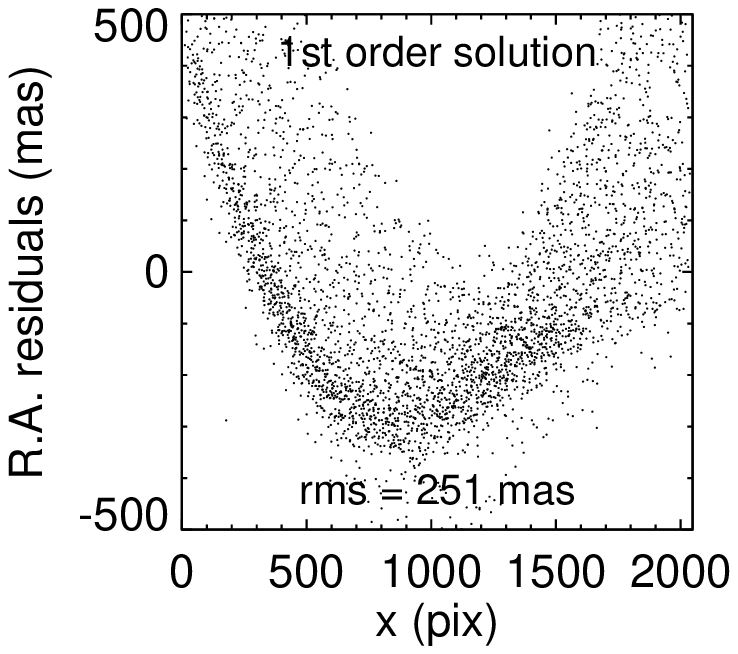}
\hskip -0.1in
\includegraphics[width=1.6in,angle=0]{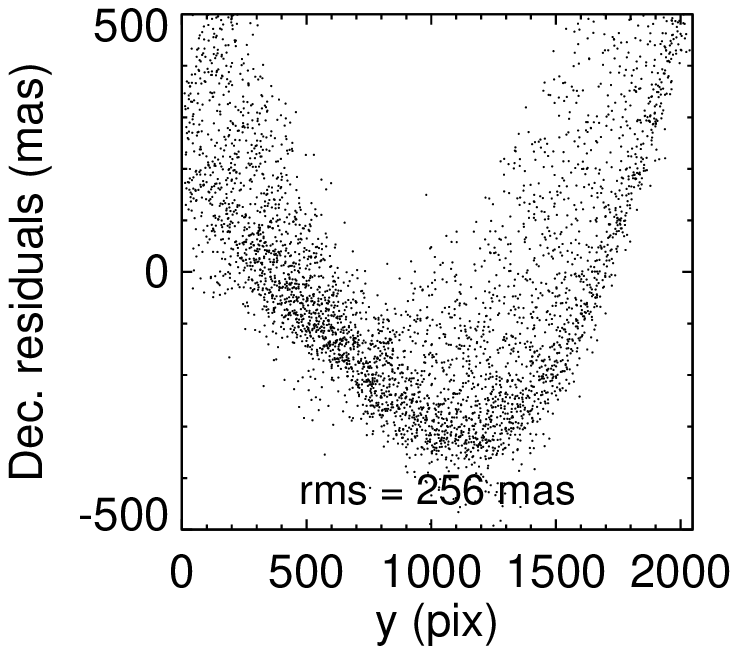}
\hskip -0.1in
\includegraphics[width=1.6in,angle=0]{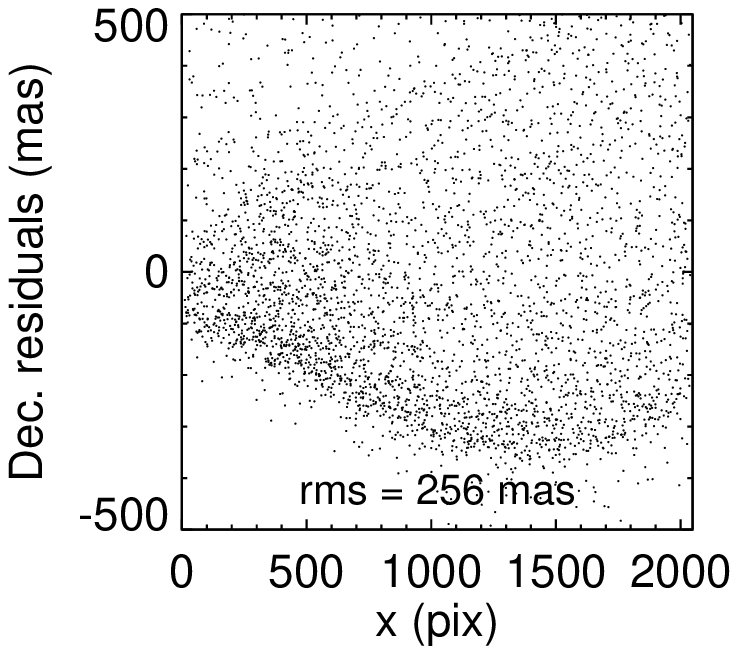}
\hskip -0.1in
\includegraphics[width=1.6in,angle=0]{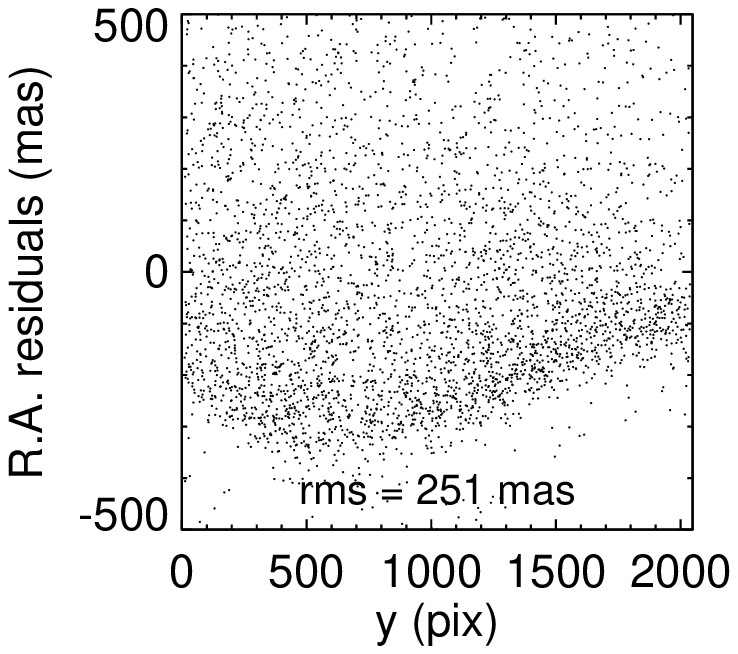}}
\vskip 0.1 in
\centerline{
\includegraphics[width=1.6in,angle=0]{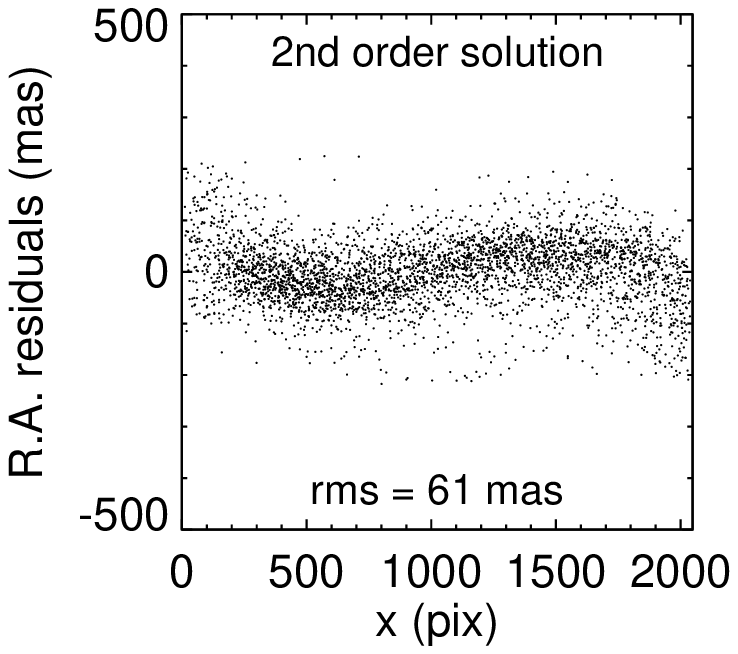}
\hskip -0.1in
\includegraphics[width=1.6in,angle=0]{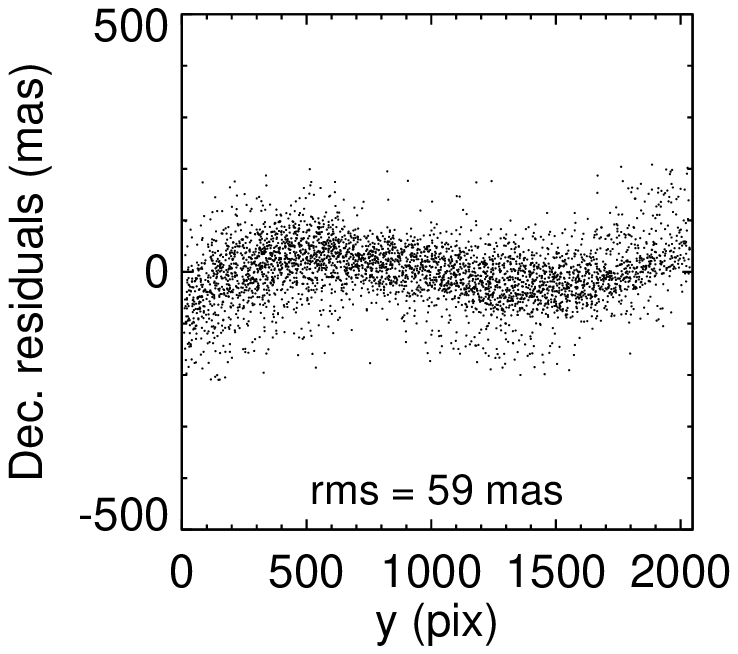}
\hskip -0.1in
\includegraphics[width=1.6in,angle=0]{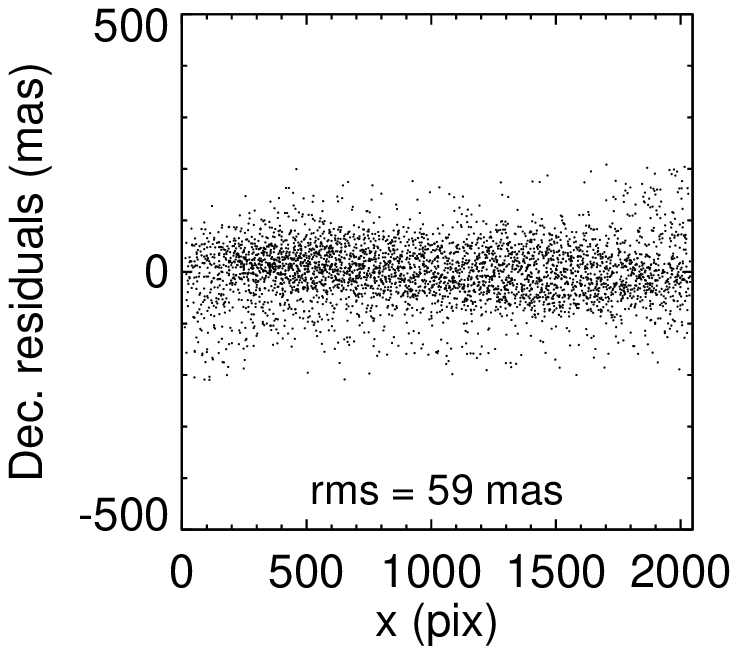}
\hskip -0.1in
\includegraphics[width=1.6in,angle=0]{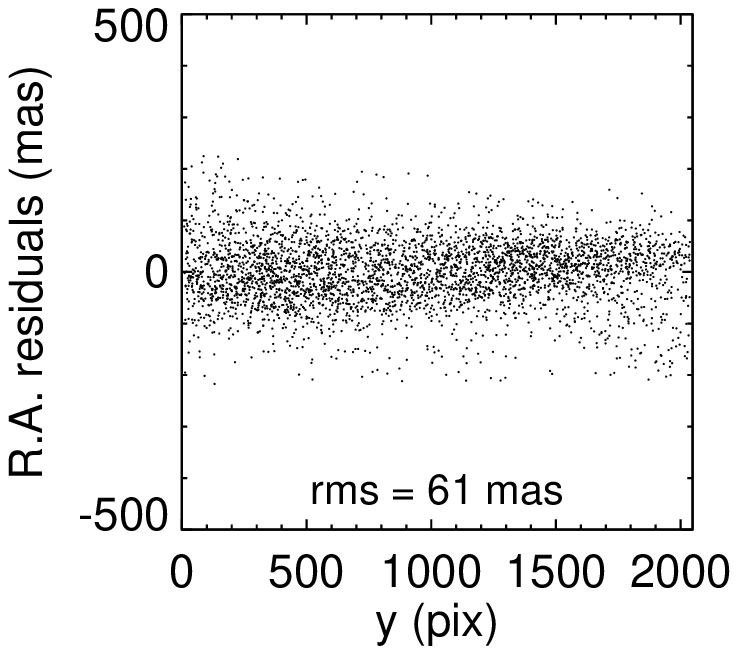}}
\vskip 0.1 in
\centerline{
\includegraphics[width=1.6in,angle=0]{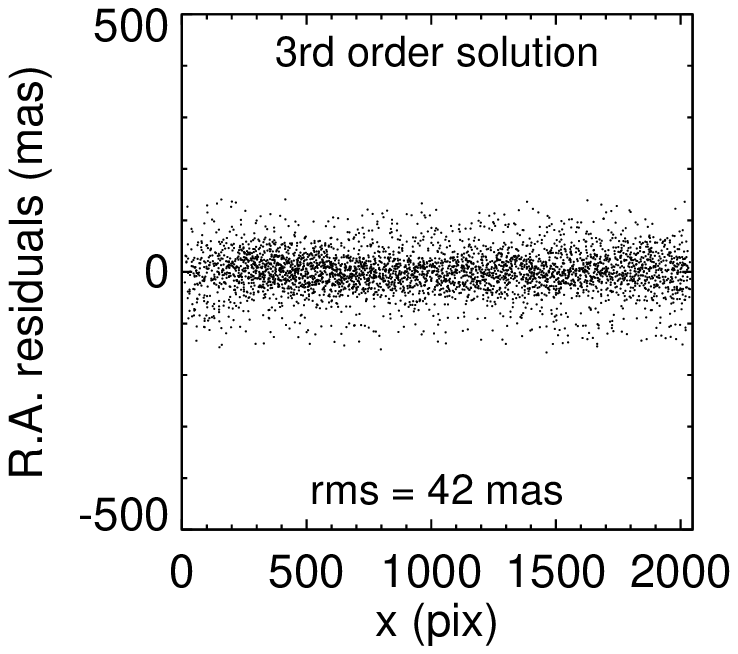}
\hskip -0.1in
\includegraphics[width=1.6in,angle=0]{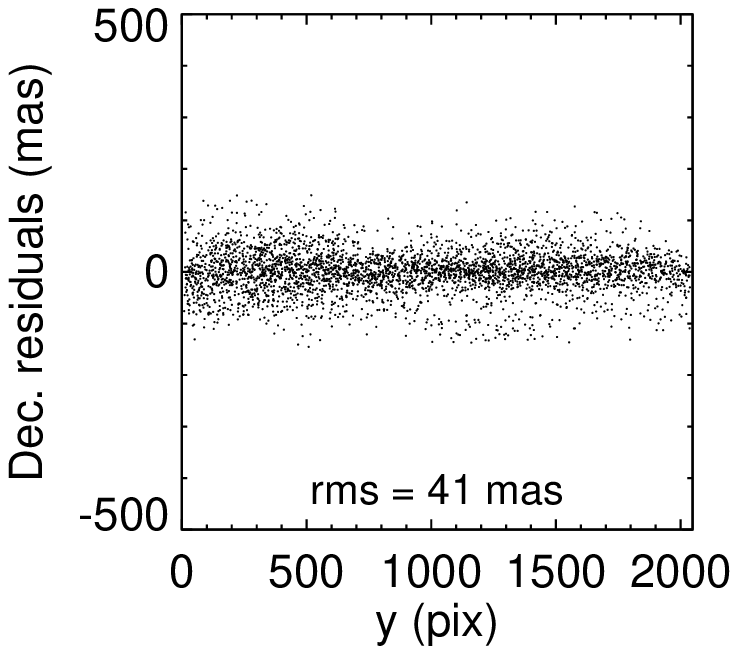}
\hskip -0.1in
\includegraphics[width=1.6in,angle=0]{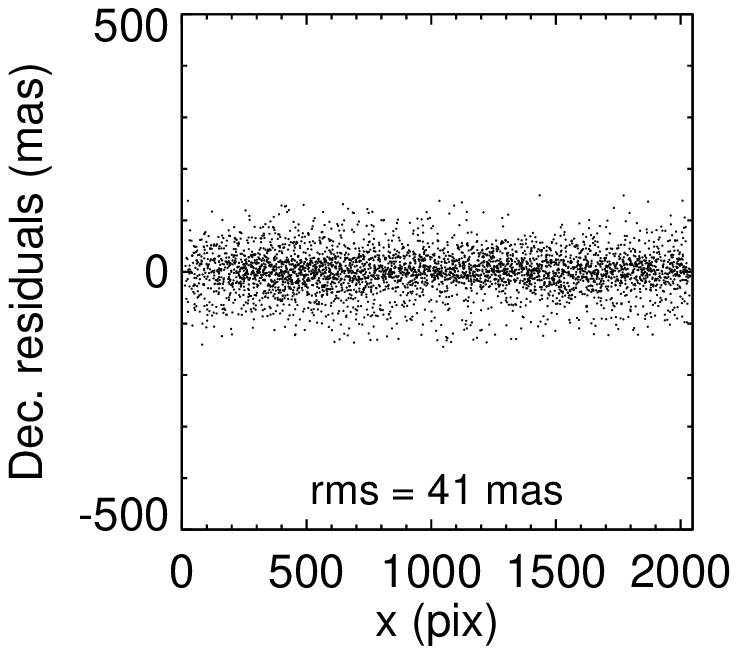}
\hskip -0.1in
\includegraphics[width=1.6in,angle=0]{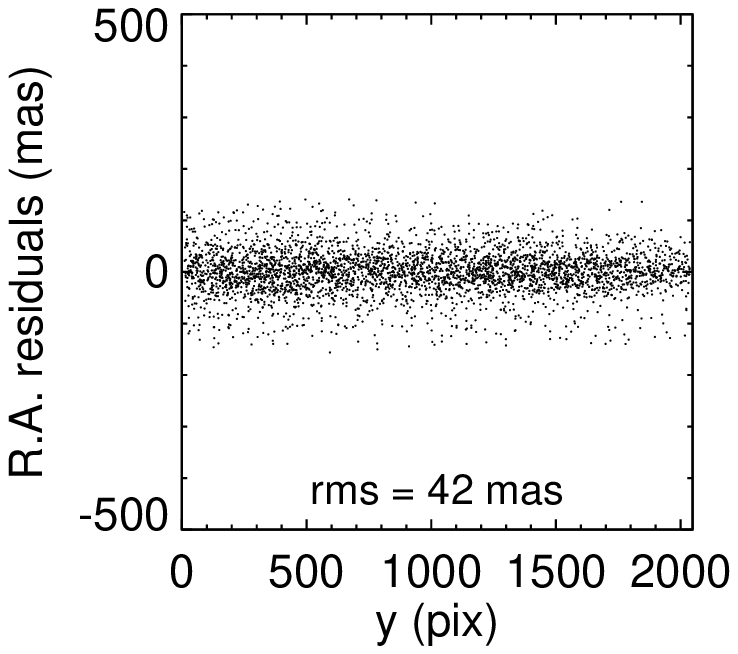}}

\caption{\normalsize Residuals in the fit of measured WIRCam star
  positions to the SDSS-DR7 catalog, using linear and higher order
  distortion terms, as a function of $x$ and $y$ position. The data
  set shown here is for $\approx$200 stars in the 2MASS~J0850+1057
  field observed over 21 dithers with offsets of 1$\arcmin$. Both
  second- and third-order terms are needed in the distortion solution,
  and the resulting residual rms is $\approx$40~mas, dominated by SDSS
  positional errors. There is no obvious remaining structure in the
  residuals, indicating that a third-order solution is sufficient for
  WIRCam. \label{fig:distort-resid}}

\end{figure}

\begin{figure}
\centerline{
\includegraphics[width=3.1in,angle=0]{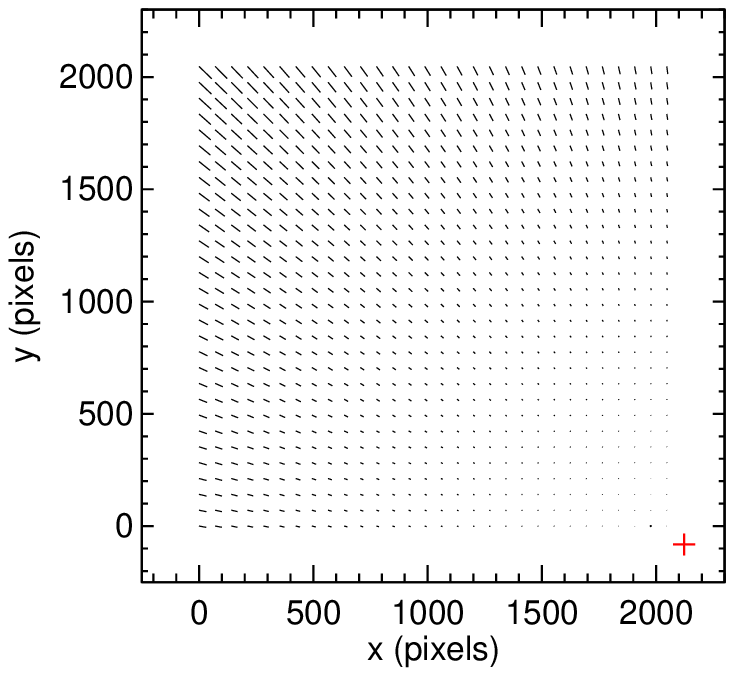}
\includegraphics[width=3.1in,angle=0]{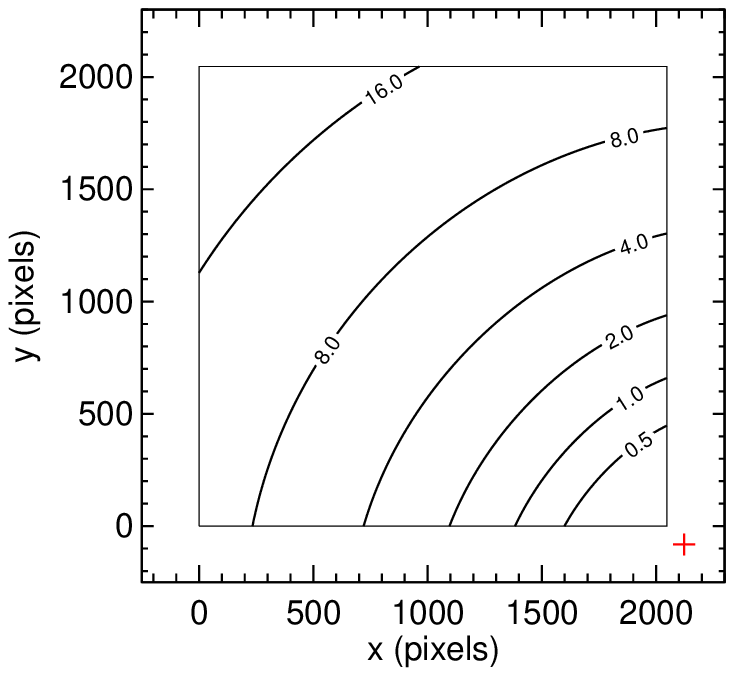}}

\caption{\normalsize {\em Left:} Map of the optical distortion present
  in the northeast array of the WIRCam mosaic (the only array we use).
  Positional offsets due to distortion are multiplied by 3 to make
  them more easily visible. The largest distortion offset has an
  amplitude of 27~pixels (i.e., from corner to corner), but the actual
  shifts induced in our dithered data sets are at most 1--2~pixels
  because our largest dithers are $\approx$200~pixels. {\em Right:}
  Contour plot showing how the distortion amplitude increases radially
  from the optical axis (red cross), which is roughly the midpoint of
  the four-array WIRCam mosaic.  Contours are labeled with the
  amplitude of the offset in pixels. \label{fig:distort}}

\end{figure}

In optimizing the registration of dithers, we allowed for optical
distortion as a 3$^{\rm rd}$-order polynomial function in $x$ and $y$,
which was applied before the tangent projection. These distortion
terms were derived from several data sets of the densest target field
that lies within the Sloan footprint (2MASS~0850+1057) by fitting our
measured $(x,y)$ positions to SDSS-DR7 reference catalog coordinates.
SDSS provides the best combination of source density on the sky and
positional accuracy ($\approx$40~mas as judged from the rms of our
fits) among astrometric reference catalogs currently available. The
residuals of our fits using first, second, and third order terms are
shown in Figure~\ref{fig:distort-resid}. There was no discernable
improvement by including fourth order terms, so we adopted the
best-fit terms up to third order for our distortion solution, shown in
Figure~\ref{fig:distort} with coefficients given in
Table~\ref{tbl:distort}. We note that we also tried fitting for the
distortion from our data alone, since dithered images can in principle
constrain any nonlinear terms \citep[e.g.,][]{2003PASP..115..113A}.
However, the largest observed offset of any given star between two of
our 1$\arcmin$ dithers is only $\approx$2--3~pixels, even though the
largest absolute offsets due to distortion are $\approx$10--20~pixels.
Thus, we found that we have more leverage for determining the
distortion by using a comparison to an absolute reference catalog. The
scatter in the best-fit distortion terms determined from different
data sets of 2MASS~0850+1057 reflects this fact as it is much lower
for the catalog matching approach compared to using the internal
position residuals alone. We also tested the stability of the
distortion pattern by both fixing and fitting for it in dense fields
observed throughout our program. The astrometric residuals of star
positions did not change significantly, validating our approach of
using a single distortion solution for all images.

We also accounted for differential aberration and refraction offsets
in the process of registering dithered images. Both effects are
essentially a linear transformation of star positions, since stars on
one side of our 10$\arcmin$ field experience slightly different
positional offsets due to annual stellar aberration and atmospheric
refraction than the opposite side of the field. Differential
refraction can cause up to a few~$\times 10^{-4}$ expansion of the
scale along the elevation axis, and differential aberration can cause
up to a $\pm2\times10^{-4}$ seasonal change in the scale. Thus, it is
important to account for these effects in order to monitor the
stability of WIRCam's linear terms over time and between targets. We
computed the appropriate offsets from equations in \citet[][p.\
121--141]{2004fuas.book.....K} and applied the differential values
(i.e., with the median offset subtracted) to the celestial coordinates
in our minimization routine.

\subsubsection{Resulting Positional Errors}

\begin{figure}
\centerline{
\includegraphics[width=2.0in,angle=0]{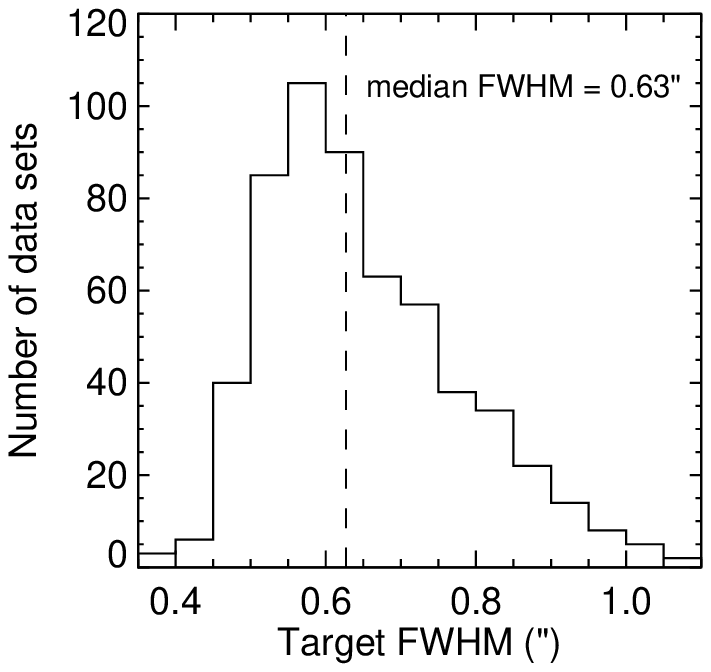}
\includegraphics[width=2.0in,angle=0]{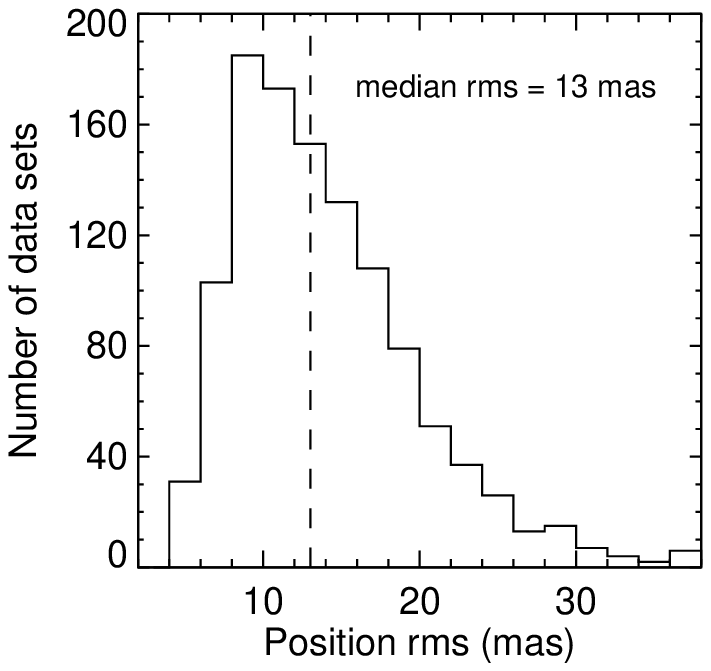}
\includegraphics[width=2.0in,angle=0]{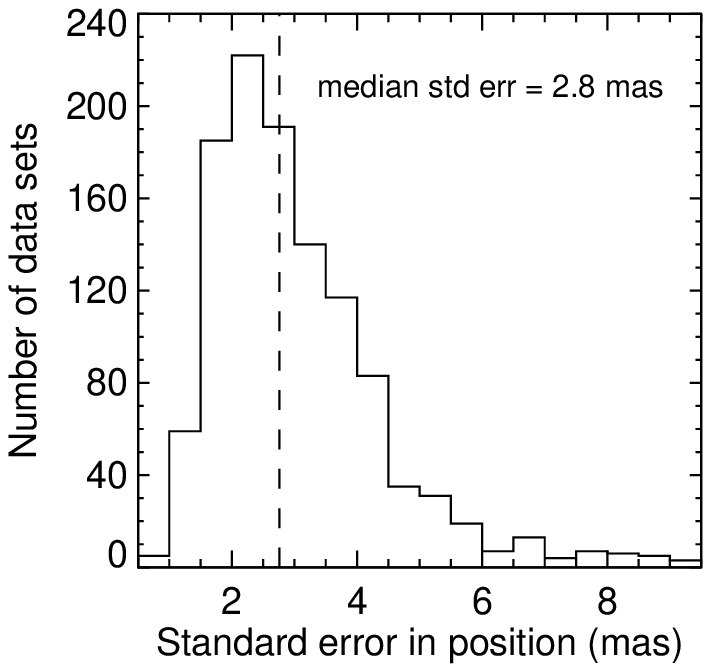}}

\caption{\normalsize {\em Left:} Distribution of the FWHM of our
  observations.  The median FWHM for our target within each dithered
  data set at each epoch is plotted, so the total number of frames we
  obtained is actually 20--30$\times$ the number of measurements shown
  here. {\em Middle:} Distribution of the rms of measured positions
  among each dithered data set.  {\em Right:} Distribution of the
  standard error (i.e., rms/$\sqrt{N_{\rm frame}}$) of our position
  measurements at each observation epoch. \label{fig:dither-rms}}

\end{figure}

The end product of combining measurements from each dithered data set
was a catalog of median positions in celestial coordinates\footnote{We
  emphasize that while our dither registration is done in celestial
  coordinates, the output \emph{relative} positions are deliberately
  not tied to an absolute reference frame. This is because most
  astrometric catalogs have a lower precision than our measurements
  and would unnecessarily introduce systematic errors into the
  relative measurements at this stage. Our initial catalog match
  provides only the approximate field coordinates, to within
  $\approx$1$\arcsec$, which is needed to achieve an accurate tangent
  projection and match our measured positions to an absolute reference
  catalog at a later step (Section~\ref{sec:catalog-match}).} and the
and the rms for each source as determined from $\geq10$ dithered
measurements. These rms values correspond to the often quoted
astrometric quality metric of the ``mean error for a single
observation of unit weight'' (m.e.1). \citet{1992AJ....103..638M}
quote m.e.1 values of 3--5~mas for the highest S/N stars in the USNO
CCD program, \citet{2004AJ....127.2948V} quote 8--10~mas for the
brightest reference stars in the USNO infrared astrometry program, and
\citet{2003AJ....126..975T} report a median rms of 12~mas for the NTT
infrared astrometry program. The ultimate astrometric precision at
each epoch may be expected to scale as $1/\sqrt{N_{\rm frames}}$, and
the USNO CCD, USNO IR, and NTT programs obtained 1--2, 3, and 8 frames
per epoch, respectively. Therefore, their precisions per epoch are
2--4~mas, 5--6~mas, and 4~mas, respectively. For our program, the rms
of the position measurements for our targets were typically 6--18~mas
(13~mas median; Figure~\ref{fig:dither-rms}). Because we obtained
20--30 frames for each data set, our astrometric precision per epoch
is 1.5--3.0~mas (2.8~mas median). Thus, the quality of our astrometry
is comparable to or better than previous ground-based parallax
programs targeting ultracool dwarfs in the optical or infrared.

\subsection{Registering Astrometry between Epochs
  \label{sec:epoch-match}}

In order to obtain multi-epoch astrometry for all objects in each of
our target fields, we next associated the sources measured in
different dithered data sets. We excluded the noisiest measurements
from this analysis, typically applying an rms threshold of 30--60~mas
(0.1--0.2~pixels). The positional shifts between epochs were estimated
using a two-dimensional histogram approach as follows: all $n_1$
objects from the first image were each paired with all $n_2$ objects
from a second image; the $\alpha$ and $\delta$ offsets between all
$n_1 \times n_2$ possible pairings were computed and binned in a
two-dimensional histogram; the peak bin in $(\Delta{\alpha},
\Delta{\delta})$ space, which contains the min($n_1, n_2$) true
pairings, was found; the shift was computed by taking the median of
the offsets contained in the peak bin.  The bin size used was
initially set to be arbitrarily large and then iteratively decreased
until the number of pairs in the peak bin was $<2\times$ the expected
number (i.e., until true pairs dominated the peak bin).  The
crossmatching of positions was then performed in similar fashion as
for the individual dithered images: a match radius of $2\farcs0$ was
employed to associate objects detected at different epochs. Such a
large match radius is needed if the proper motion is large
($\gtrsim1\arcsec$ yr$^{-1}$), as is the case for some targets. We
excluded sources from the multi-epoch astrometry catalog if they were
detected at fewer than half of the epochs. This excludes faint sources
that were only well-detected in the best conditions, bright sources
that were only below the saturation limit in poor conditions, and any
other transient sources or long-lived artifacts that may be in the
data set.

Because the initial association of object positions was based only on
rough estimates of the position offsets between epochs, we optimized
this registration by fitting for the offsets as well as allowing for
relative changes in linear terms across different data sets. Thus, we
replaced the initial guesses of the linear terms generated by the
reference catalog matching (except for the first epoch, which we solve
for later). We parameterized the linear terms as a rotation, $x$-axis
pixel scale, ratio of $y/x$-axis pixel scales, and a shear term
($\Delta{y} \propto x$). We used MPFIT to perform an unweighted least
squares minimization in a similar fashion as described for the dither
matching in the previous section. After the first optimization, we fit
every object in the field for proper motion and parallax and
temporarily excluded objects that displayed significant parallax
($>3\sigma$) or proper motion ($>30$~\masyr). This automated procedure
typically excluded no more than 5\% of the reference stars, and it
always excluded the science target. We then determined the optimal
registration solution a second and final time after excluding these
objects.

\subsection{Absolute Astrometric Calibration \label{sec:catalog-match}}

We have performed as much of our analysis as possible using relative
astrometry in order to preserve the fidelity of our position
measurements.  However, we must ultimately tie our astrometry to an
absolute reference frame in order to determine, e.g., the actual pixel
scale and orientation of our images. The most suitable catalogs for
this purpose are 2MASS, which provides positions for infrared sources
over the entire sky, and SDSS, which has a higher sky density of
sources and higher astrometric precision but more limited sky
coverage.  In our shallowest images taken with the \Kn-band filter, we
found that shallower reference catalogs were usually more appropriate
(USNO-B1, \citealp{2003AJ....125..984M}; and UCAC-3,
\citealp{2010AJ....139.2184Z}).  For each field, we constructed a
reference frame from the catalog that had the most sources in common
with our images. We required reference catalog sources to have
absolute position errors $\leq150$~mas (e.g., for 2MASS:
\texttt{ERRMAJ}~$<$~0.15\arcsec). We found the rough offset between
our own astrometric catalog and the reference sources by using our
aforementioned two-dimensional histogram approach. We then matched
reference sources to our own using a match radius of 2$\farcs$0. We
excluded from this analysis any sources in our astrometric catalog
that displayed significant proper motion ($>30$~\masyr), as these
would have introduced substantial scatter ($\gtrsim$0$\farcs$3) to our
comparison with reference catalog position measurements from typically
$\approx$5--10~years ago.

Using the sources in common between our science images and the
reference catalog (typically $\gtrsim30$ stars; see
Table~\ref{tbl:sample}), we determined the absolute astrometric frame
for our CFHT images. We registered our positions to the reference
catalog allowing for an offset (i.e., to determine the absolute
coordinates of our astrometry) and the linear terms. This solution
allows us to compute the pixel scale and orientation in an absolute
sense, completely replacing the temporary guess from the initial
catalog crossmatching. In the final best-fit registration, the rms of
all stars about their catalog positions was typically 60--80~mas for
2MASS and 30--50~mas for SDSS. This scatter is dominated by the
reference catalog positional errors. (Thus, the actual relative
astrometric uncertainties of 2MASS positions over our 10$\arcmin$
field are a factor of $\sim$2 smaller than the nominal catalog errors
of 100--150~mas.) After this final absolute calibration we found that
our input guess for the absolute coordinates from image headers were
accurate to within $\lesssim1$\arcsec.

\subsubsection{Astrometric Stability of WIRCam \label{sec:stability}}

\begin{figure}
\centerline{
\hskip -0.2in
\includegraphics[width=3.8in,angle=0]{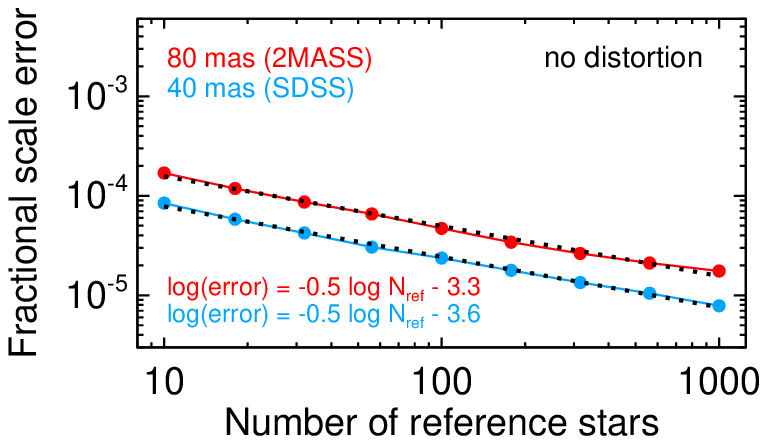}
\hskip -0.8in
\includegraphics[width=3.8in,angle=0]{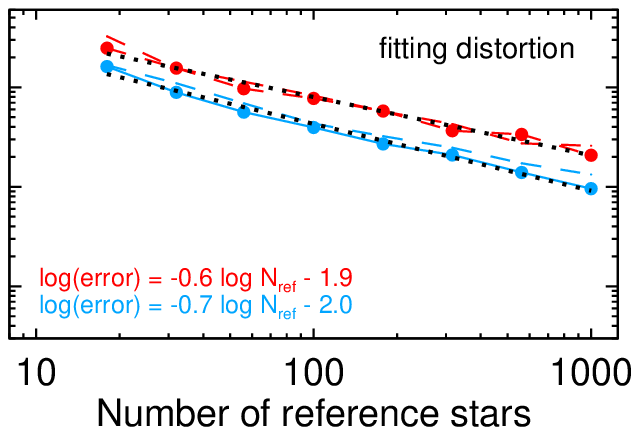}}

\caption{\normalsize {\em Left:} Error in the derived pixel scale of
  WIRCam due only to random errors in the catalog positions as
  determined from our Monte Carlo simulations.  If the distortion of
  WIRCam were known perfectly, this would set the limit on how well
  the pixel scale is known, i.e., a fractional uncertainty of
  $2\times10^{-5}$ for our calibration field containing $\approx$200
  SDSS stars.  (Dashed lines show first order polynomial fits to the
  simulation results.) {\em Right:} Same as the left except that the
  third order distortion terms have been treated as free parameters.
  Because of the strong degeneracy between linear and higher order
  terms in the fit, the precision in the pixel scale more than an
  order of magnitude worse than in the case of no (or known)
  distortion.  This fundamentally limits our absolute calibration of
  WIRCam to a precision of $3\times10^{-4}$. \label{fig:sclsim}}

\end{figure}

The best-fit parameters from the registration of multi-epoch data sets
to an astrometric reference catalog enables us to assess the long-term
astrometric stability of WIRCam. The level of precision with which we
are able to monitor the changes in linear terms such as scale and
rotation is fundamentally limited in two ways: (1)~positional errors
both in our data and reference catalogs introduce random and
systematic errors in the derived terms; and (2)~the uncertainty in the
higher order distortion terms is a source of systematic error in the
derived linear terms. We have assessed the level of uncertainty in the
scale introduced by both of these effects through Monte Carlo
simulations. To test the contribution of random errors alone (i.e.,
case 1), we simulated many star fields with random positions
distributed uniformly over a $10\arcmin \times 10\arcmin$ field and
found the best-fit scale to match them to a reference catalog that had
normally distributed noise added to it. For a reference catalog
accurate to 80~mas (i.e., akin to 2MASS), $\approx$30 reference stars
were needed to achieve a fractional precision in the scale of
$1\times10^{-4}$ (Figure~\ref{fig:sclsim}). This situation is typical
of about half of our targets. For a higher fidelity reference catalog
accurate to 40~mas (e.g., like SDSS), 30 reference stars give a much
better scale precision of $5\times10^{-5}$, and the very best case
among our targets of 190 SDSS reference stars would give a precision
of $2\times10^{-5}$.

The second source of error present in our determinations of linear
terms is the uncertainty in the distortion solution. This is because
the linear and higher order terms are partially degenerate when
fitting polynomials for the distortion. In the reduction procedure
described above, we used data sets containing $\approx$200 SDSS
reference stars to determine the WIRCam distortion, and the catalog
errors were estimated to be 40~mas from the rms of the fit residuals.
Thus, we simulated many random star fields each containing 200 stars
with normally distributed noise of 40~mas and found that fitting
freely for both linear and distortion terms resulted in a scale
uncertainty of $3 \times 10^{-4}$ (Figure~\ref{fig:sclsim}). This
result is effectively independent of the assumed centroiding error in
the star positions even up to our worst errors of 0.1~pixel because
the reference catalog scatter dominates. This source of error is a few
times larger than the uncertainty due simply to random errors in the
reference catalog, and thus it is the limiting factor in our ability
to measure the scale of WIRCam. From these simulations, the limiting
systematic uncertainties in shear and rotation are $3 \times 10^{-4}$
and 0$\fdg$02 (i.e., $3 \times 10^{-4}$ radians), respectively.

\begin{figure}
\centerline{
\includegraphics[width=3.3in,angle=0]{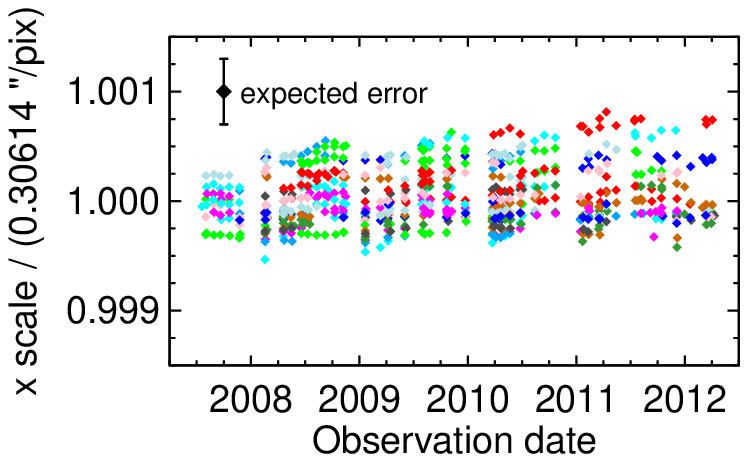}
\includegraphics[width=3.3in,angle=0]{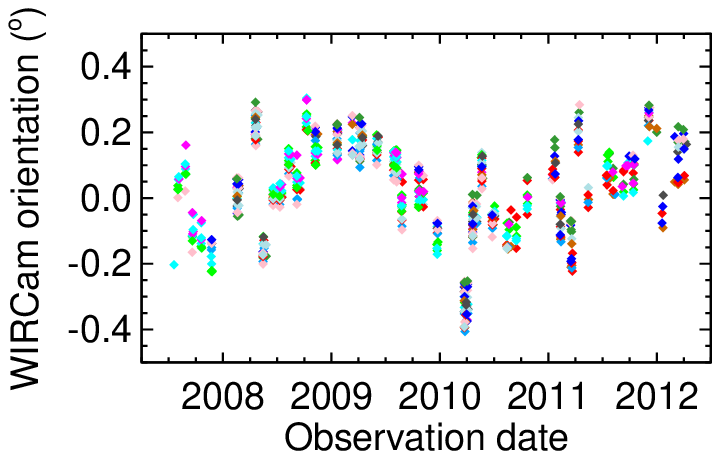}}

\caption{\normalsize {\em Top:} Relative pixel scale of the $x$-axis
  of WIRCam over the duration of our observing program, with diamonds
  of different colors indicating different targets (not uniquely since
  there are 49 targets and only 11 colors). The scatter is consistent
  with the expected error in linear terms due to the uncertainty in
  the distortion solution ($3\times10^{-4}$, illustrated by black
  diamond and error bar).  {\em Bottom:} Orientation of WIRCam over
  the course of our observing program (0\degree\ corresponds to the
  $y$-axis aligned with north).  Changes in the orientation are
  clearly evident and are correlated with the observing run.  (Runs
  can be seen as groupings of points very close together in time.)
  This is expected as WIRCam is taken off the telescope between
  runs.  \label{fig:sclrot}}

\end{figure}

With these results in mind, we can now assess the stability of WIRCam
from our astrometric monitoring data (Figure~\ref{fig:sclrot}). (1)~We
are most sensitive to changes in the orientation of WIRCam, and we
found a highly significant scatter of $\pm0\fdg14$ among data sets
taken over our program. This scatter is clearly not Gaussian but
rather is highly correlated with the observation date; the orientation
of data sets taken on the same WIRCam observing run were nearly
identical.  This is consistent with the fact that instrument is taken
off of the telescope between observing runs. (2)~We found the $x$
pixel scale to be $0\farcs30614 \pm 0\farcs00008$~\perpix\ (i.e., a
fractional error of $3\times10^{-4}$). Given the errors estimated
above, this scatter is consistent with the pixel scale being constant
over the duration of our program. This stability is impressive given
that WIRCam is taken on and off the telescope for $\sim$8 observing
runs per year. (3)~We found the ratio between $y$ and $x$ pixel scales
to be consistent with unity ($0.9997\pm0.0003$), and the scatter in
this value is consistent with the uncertainty given by our Monte Carlo
simulations.  (4)~Finally, we found a significant shear term (which we
have defined as $\Delta{y} \propto x$) of $-0.0013 \pm 0.0004$. If the
angle between WIRCam's $x$ and $y$ axes were different from the
90\degree\ angle between north and east only by a rotation, this term
would be zero. Instead, this shear term implies that the angle between
WIRCam's $x$ and $y$ axes is actually $89\fdg93 \pm 0\fdg02$ when
projected onto the sky.

\subsection{Parallax and Proper Motion Determination \label{sec:plx-fit}}

Using our final astrometric catalog of WIRCam position measurements
calibrated against an absolute reference frame, we fit for the proper
motion and parallax of all sources in each target field. For each
source, we used MPFIT to perform a least-squares minimization weighted
by the standard errors of the position measurements. We fitted three
parameters to the combined $(\alpha, \delta)$ data: proper motion in
right ascension ($\mu_{\alpha}$), proper motion in declination
($\mu_{\delta}$), and parallax ($\pi$). This is notably different from
one standard approach taken in the literature of fitting two separate
values of the parallax in $\alpha$ and $\delta$. The parallax offsets
were computed as follows:
\begin{equation}
  \Delta{\alpha} = \pi(X\sin{\alpha} - Y\cos{\alpha})/\cos{\delta}
\end{equation}
\begin{equation}
  \Delta{\delta} = \pi(X\cos{\alpha}\sin{\delta} +
                       Y\sin{\alpha}\sin{\delta} - Z\cos{\delta})
\end{equation}
where $X$, $Y$, and $Z$ are the coordinates of the Earth relative to
the barycenter of the solar system as given by the JPL ephemeris
DE405.  MPFIT minimized the residuals in $(\alpha, \delta)$ after
subtracting the relative parallax and proper motion offsets (3
parameters) and the mean $(\alpha, \delta)$ position (effectively
removing 2 additional degrees of freedom).  Thus, each fit to $2
\times N_{\rm epoch}$ measurements had $2 \times N_{\rm epoch}-5$
degrees of freedom (dof).

For each target, we then performed a Markov Chain Monte Carlo (MCMC)
analysis on the astrometry in order to accurately determine the
posterior distributions of all parameters.  We adopted the formalism
described by \citet{2005AJ....129.1706F}, which uses a
Metropolis-Hastings jump acceptance criterion with Gibbs sampling that
chooses only one parameter (at random) to be altered at each step in
the chain. Before running our science chains, we first ran a test
chain to determine the optimal step size ($\beta$) for each of our
parameters in order to ensure efficient convergence. This initial
chain was run according to the procedure outlined by
\citet{2006ApJ...642..505F} in which each value of $\beta$ is
periodically adjusted until the acceptance rate for that parameter
comes within some tolerance (we chose 5\%) of the target rate (we
chose 0.25). We then ran 30 chains of $10^4$ steps each that started
at different points in parameter space drawn at random by adding
Gaussian noise, with $\sigma$ equal to the step size, to the best-fit
parameters from the MPFIT results. We computed the Gelman-Rubin
statistic for our set of 30 chains, which \citet{2005AJ....129.1706F}
suggests should be $<1.2$ to ensure that the results are converged and
well-mixed. The Gelman-Rubin statistic was always $<$1.03 for all
parameters, and typically $<$1.01. Finally, we discarded the first
10\% of each chain as the ``burn in'' portion, using only the latter
90\% for deriving the probability distributions of parameters.

At this stage we investigated the impact of DCR on our resulting
parallaxes for targets observed at $J$ band. We assumed an effective
wavelength of 1.2462~\micron\ for the background star reference frame,
based on the typical values for GKM stars as discussed earlier, and
computed individual DCR offsets for the measured positions at each
epoch using the method described in the introduction to this section
(also see Figure~\ref{fig:dcr}). We added these offsets to the
measured astrometry and performed our MCMC analysis a second time. We
found that the change in the resulting parallax was almost always
$\leq$0.15$\sigma$.  As a source of systematic error this is
completely negligible as it would boost the final error by $\leq$1\%
when added in quadrature. In a few special cases that are most
sensitive to DCR shifts (i.e., three T7--T8 dwarfs with fewer than 10
epochs) the change in parallax was as large as 0.2--0.4$\sigma$.  This
would give a slightly larger boost of 2--7\% to their errors, but this
is still negligible.  In examining the ensemble of the 33 $J$-band
targets for which we computed DCR parallax offsets, we found a
mean$\pm$rms offset of $-0.10\pm0.19$~mas ($-0.06\pm0.10$~mas when
excluding objects with parallax errors $>$2~mas), indicating that
there is also no systematic offset in our parallaxes due to DCR.

We also performed tests on our data to determine when to consider a
parallax measurement ``done''.  Even though our MCMC analysis fully
captures any uncertainty due to the degeneracy between proper motion
and parallax over datasets spanning modest time baselines
($\lesssim2$~years), we wanted to confirm that the parallaxes we
present here will not change substantially with the addition of future
data.  To check this, for each object we determined the best-fit
parallax using subsets of the data starting with the first 3 epochs
(the minimum needed to constrain the 5-parameter fit) and then adding
one data point at a time for each successive epoch.  As expected, the
most important criterion for reaching a stable parallax solution was
the time baseline.  For all of our targets we found that a time
baseline of $\approx$1.2~years was sufficient to reach a best-fit
parallax value that remained stable with the addition of new data up
to the last observation epoch (our longest time baseline to date is
4.3~years).  Therefore, all of the parallaxes presented here are
expected to have a reached a stable, final value (median baseline of
2.4~years, minimum 1.8~years).  We note that this minimum needed time
baseline of 1.2~years will necessarily be longer for cases where the
astrometric errors are significantly larger than ours or when the
target parallax is smaller.

The results from our MCMC analysis are given in Table~\ref{tbl:plx},
and the astrometric data are shown in Figures~\ref{fig:plx1},
\ref{fig:plx2}, \ref{fig:plx3}, and \ref{fig:plx4}.  The minimum
$\chi^2$ value for each chain is commensurate with the degrees of
freedom, which verifies that our adopted positional errors are
accurate. There are three exceptions, the known binaries
2MASS~J0518$-$2828AB \citep{2006ApJS..166..585B} and
2MASS~J1404$-$3159AB \citep{2008ApJ...685.1183L} and the candidate
unresolved binary SDSS~J0805+4812 \citep{2007AJ....134.1330B}. Their
large $\chi^2$/dof values can be attributed to the large perturbations
present in the residuals after fitting for parallax and proper motion
due to orbital motion.

\subsubsection{Correction for Orbital Motion \label{sec:orbcorr}}

For the binaries in our sample that have orbit determinations, we
apply a correction to the best-fit parallax and proper motion
parameters to account for photocenter shifts due to orbital motion.
All binaries in our sample only have relative astrometric orbit
determinations, so the center of mass is not known.  Thus, we use the
relative orbital offsets and an assumed mass ratio (derived from
evolutionary models) to compute the motion of the center of mass.
This motion is then modified by a factor that depends on the binary's
flux ratio in the bandpass used in our CFHT imaging in order to
determine the actual motion of the photocenter.  The coefficient by
which to multiply the relative orbital motion is thus $(\xi-\gamma)$,
where
\begin{equation}
  \xi \equiv l / (1+l),
\end{equation}
\begin{equation}
  \gamma \equiv q / (1+q),
\end{equation}
$l$ is the flux ratio ($\equiv f_2/f_1 = 10^{-0.4\Delta{m}}$), and $q$
is the mass ratio $(\equiv M_2/M_1)$. The coefficient $(\xi-\gamma)$
is typically negative because flux is a steeper-than-linear function
of mass, and thus the photocenter motion is opposite in sign from that
of the orbit of the secondary relative to the primary.

We compute the astrometric offsets in a Monte Carlo fashion, using the
Markov chain from the published orbit determination to draw the
binary's orbital elements \citep[e.g., see][]{2008ApJ...689..436L}. We
also draw random: (1)~flux ratios corresponding to the error in the
measured $J$-band resolved photometry of the target binaries (or
$K$-band photometry for targets using the CFHT \Kn\ filter); and
(2)~mass ratios from evolutionary models are derived using the method
described in Section~5.4 of \citet{2009ApJ...692..729D}. Our Monte
Carlo approach enables us to appropriately track the correlation
between the different parameters (e.g., the derived mass ratio depends
on the $K$-band flux ratio and also the system mass via the orbital
elements). For each Monte Carlo trial we subtracted the orbital motion
offsets from our CFHT astrometry and recomputed the best-fit parallax
and proper motion. This enabled us to derive systematic offsets and
corresponding errors due to the uncertainties in the various input
parameters.

Our corrections to the parallax and proper motion are given in
Table~\ref{tbl:orbcorr} along with the predicted semimajor axis of the
photocenter motion for each binary ($a_{\rm phot}$). We quote $a_{\rm
  phot}$ as positive for photocenter motion has the same sign as the
primary motion.  For each target we added the randomly drawn orbital
motion offsets to our MCMC chains in a Monte Carlo fashion. We note
that the minimum $\chi^2$ of the parallax fit either improved or did
not change significantly for all binaries. The latter cases correspond
to orbital motion that is nearly linear (e.g., for very long period
binaries) and thus easily compensated by a slightly different proper
motion than the pre-corrected fit. The parallax offsets are quite
small (always $<$0.5$\sigma_{\pi}$, median 0.08$\sigma_{\pi}$), which
is not surprising since the orbital periods of these binaries are very
long ($\approx$10--20~years). The proper motion offsets are much more
significant, since the orbital motion over 2--4~years of monitoring
can largely be expressed as a linear term. (We note that even though
these proper motions are corrected for orbital motion, they are still
not ``absolute'' since the bulk proper motion of the background stars
that define the astrometric reference frame is not known.)

The three binaries showing large perturbations due to orbital motion
(2MASS~J0518$-$2828AB, SDSS~J0805+4812AB, and 2MASS~J1404$-$3159AB)
unfortunately do not have orbit determinations, and thus we are unable
to correct their astrometry as described in this section. Additional
astrometric monitoring is needed before the orbits of these binaries
can be determined from CFHT data alone.

\subsubsection{Correction from Relative to Absolute Parallax \label{sec:abscorr}}

The final step in determining the parallaxes of our targets is the
conversion from relative to absolute parallax. Because our reference
objects are almost all stars, not galaxies,\footnote{The small number
  of galaxies among our reference sources is both supported by the
  lack of extended sources and expected from galaxy count
  measurements.  For our typical exposure time of 5~s, our high S/N
  reference sources are brighter than $J$ of 17~mag.
  \citet{2009ApJ...696.1554C} predict 500~galaxies~deg$^{-2}$ with $J
  < 17$~mag, and for the 0.028~deg$^2$ field of view of one WIRCam
  detector, this implies 14 galaxies. This is much smaller than the
  $\approx$40--200 reference stars in our fields (see
  Table~\ref{tbl:plx}).  In fact, given that galaxies have larger FWHM
  they will have a larger rms in their position measurement, and thus
  most will be excluded by our rms cuts at the beginning of the
  epoch-matching step.} their finite distances will result in a small
parallax motion of the reference frame that erases part of the true
parallax motion of our target. This introduces a systematic error in
the parallax measurement that varies in amplitude depending on the
distance distribution of the reference stars in each target field.

We have computed a correction to account for this effect using the
Besan\c{c}on model of the Galaxy \citep{2003A&A...409..523R}. Given
the celestial coordinates of each field, the Besan\c{c}on model
generates a list of simulated stars with distances, magnitudes, and
proper motions. We oversampled the model output for our WIRCam fields
by a factor of 40 in order to ensure that our derived corrections are
not dominated by small number statistics. For each field, we used the
SExtractor photometry to determine the magnitude range of our
reference stars, and we used only model stars within this range for
our calculations. The distribution of the model star distances for our
fields is typically peaked at 0.5--2~kpc, giving corrections of
0.5--2.0~mas (i.e., $\approx$1--2$\sigma_{\pi}$). We added the
model-predicted parallax offsets to the actual reference stars within
the analysis pipeline in a Monte Carlo fashion to determine the impact
on the final derived target parallax. We found that different
approaches such as applying offsets as a function of star brightness,
applying offsets randomly, or not applying offsets to a subset of our
reference sources (e.g., simulating the fact that some reference
sources may be galaxies) all produced essentially the same systematic
error in the target parallax. The resulting shift was always very
close to the mean of the model-predicted parallax distribution. Thus,
we used the mean Besan\c{c}on parallax for each field as the
correction from relative to absolute parallax. We adopted an
uncertainty in this correction factor based on sampling variance in a
Monte Carlo fashion. For example, if a target field's astrometric
catalog contained 100 stars, we drew random subsets of 100 stars from
the oversampled Besan\c{c}on model output and determined the mean
Besan\c{c}on parallax for each trial. The rms of $10^3$ trials was
adopted as the error in the absolute parallax correction (median error
in the correction was 0.2~mas). In Table~\ref{tbl:sample} we list the
values of these corrections derived for our target fields.


\section{Keck/NIRC2,  \HST, and VLT Resolved Photometry \label{sec:keck}}

We have used the AO system at the Keck~II Telescope on Mauna Kea,
Hawaii, to resolve 17 binaries in our sample and measure relative
photometry. We employed the facility near-infrared camera NIRC2 to
obtain images in the standard Mauna Kea Observatories (MKO)
photometric system \citep{mkofilters1, mkofilters2}. Depending on the
target and observing conditions (see Table~\ref{tbl:keck}), we used
laser guide star (LGS) AO \citep{2006PASP..118..297W,
  2006PASP..118..310V} or natural guide star (NGS) AO
\citep{2000PASP..112..315W, 2004SPIE.5490....1W}. At some epochs we
obtained data using the 9-hole non-redundant aperture mask installed
in the filter wheel of NIRC2 \citep{2006SPIE.6272E.103T}. Our
procedure for obtaining, reducing, and analyzing such imaging and
masking data is described in detail in our previous work
\citep[e.g.,][]{2006ApJ...647.1393L, 2009ApJ...692..729D,
  2009ApJ...699..168D}. Table~\ref{tbl:keck} summarizes the Keck
observations presented here, and Figures~\ref{fig:lgs1},
\ref{fig:lgs2}, and \ref{fig:lgs3} show our imaging data.

We also analyzed \HST/NICMOS and VLT/NACO archival images of eight
ultracool binaries with parallaxes to supplement our sample of
resolved near-IR photometry.  Five of these have had their NICMOS data
published previously \citep{2004AJ....128.1733G, 2006ApJS..166..585B,
  2011AJ....141...70B}, sometimes without errors given
\citep{2006AJ....132..891R}.  Our re-analysis thus provides a check on
the published values and errors.  Two of these binaries are among the
17 that we have observed with Keck/NIRC2.  Table~\ref{tbl:arch}
summarizes the results of our (re)analysis of these archival data.


\section{NASA IRTF/SpeX Spectroscopy \label{sec:irtf}}

We have obtained near-IR spectroscopy for targets in our sample that
did not have previously published data. Spectra were obtained using
SpeX \citep{2003PASP..115..362R} at the NASA Infrared Telescope
Facility (IRTF) either in prism or SXD mode. Prism mode delivers
continuous wavelength coverage from 0.75~\micron\ to 2.5~\micron\ ($R
= 120$ with the 0$\farcs$5 slit), while SXD mode has five separate
orders spanning 0.81--2.42~\micron\ ($R$~=~1200 with the 0$\farcs$5
slit). We calibrated, extracted, and telluric-corrected all data using
the SpeXtool software package \citep{2003PASP..115..389V,
  2004PASP..116..362C}. The data presented herein were obtained on 6
different nights (2008 Jul 6, Aug 15; 2011 Jan 22, 27, 30; 2011 Sep 8)
using either the 0$\farcs$3, 0$\farcs$5, or 0$\farcs$8 slit. We
obtained prism data for 2MASS~J1750+4424AB and SXD data for the
remaining targets (LSPM~J1735+2634AB, 2MASS~J2140+1625AB,
2MASS~J1847+5522AB, Gl~417BC, 2MASS~J1017+1308AB, 2MASS~J1047+4026AB,
2MASS~J0700+3157AB).


\section{Results \label{sec:results}}

\subsection{Comparison to Published Parallaxes \label{sec:compare}}

While our primary goal is to measure new parallaxes with CFHT, we also
monitored several objects with published parallaxes to validate our
methods. In this subsection we provide a detailed comparison of our
parallaxes to published values in order to determine if there are any
unaccounted for sources of systematic error in our data.  As shown
below, we determine that our parallaxes generally agree very well with
published values, with a few exceptions, and thus readers primarily
interested in science results may wish to skip this subsection.

Our ``control'' sample of ultracool dwarfs included (1)~single objects
(e.g., 2MASS~J2224$-$0158); (2)~wide but unresolved binaries that will
have negligible orbital motion over our observations (e.g.,
2MASS~J1146+2230AB); and (3)~binaries that are in our Keck AO
dynamical mass sample for which we can independently check and/or
improve the published parallax measurements (e.g.,
SDSS~J0423$-$0414AB).  In Figure~\ref{fig:vs-pub} and
Table~\ref{tbl:compare} we show our absolute parallaxes compared to
published measurements.  (Note that we do not compare our proper
motion measurements to published values because all such measurements
are relative, not absolute, and we have no way to ascertain the
absolute proper motion of the reference frame for published results
that generally will be many times larger than the relative proper
motion uncertainty.)  Our parallaxes are consistent within
$<1.8\sigma$ in 23 of 27 cases (i.e., 85\% of the time) and this
subset of comparisons has a reasonable $\chi^2$ (19.9 for 23 dof). The
published values largely come from the USNO CCD
\citep[8~objects;][]{1992AJ....103..638M, 2002AJ....124.1170D} and IR
\citep[8~objects;][]{2004AJ....127.2948V} programs. The parallax
values for this subsample range from $25.7\pm0.9$~mas
\citep{2007hnrr.book.....V} to $174.3\pm2.8$~mas
\citep{2004AJ....127.2948V}.  In Table~\ref{tbl:compare}, we also show
how well previously published parallaxes have agreed with each other.
We note that there are several instances of published values that are
discrepant with each other at the $\geq$2$\sigma$ level (9 of the 31
cases listed), whereas only 1--2 would be expected for Gaussian
errors.  This implies that some of the parallax errors for the
published sample are underestimated.  We now consider the 4 objects
for which our parallax is discrepant with the published value at
$>$1.8$\sigma$.
\begin{itemize}

\item SDSS~J0423$-$0414AB disagrees by 3.1$\sigma$ with the parallax
  of \citet{2004AJ....127.2948V}. These authors emphasize the
  preliminary nature of all their parallaxes (see their Section~6) and
  present evidence that their errors may be somewhat
  underestimated. There are 7 objects in common between their IR
  program and the USNO CCD program. The two sets of measurements are
  only consistent to within 0.5--2.7$\sigma$, with an unreasonably
  large $\chi^2$ of 18.8 (7 dof). To achieve the median expected value
  of $\chi^2 = 6.3$ would require multiplying their errors by a factor
  of 2.0. (Alternatively, the parallax errors from \emph{both}
  programs may be underestimated by a factor of 1.72.) If we multiply
  the published parallax error of SDSS~J0423$-$0414AB by 2.0, the
  discrepancy between our two measurements is much more modest
  (1.7$\sigma$).

\item 2MASS~J0700+3157AB and 2MASS~J1534$-$2952AB are 2.0$\sigma$ and
  6.3$\sigma$ discrepant with the measurements of
  \citet{2003PASP..115.1207T} and \citet{2003AJ....126..975T},
  respectively. We find that both published errors may be
  underestimated based on Monte Carlo simulations of the published
  data using an appropriate astrometric precision per epoch. Using the
  actual measurement epochs and precision per epoch of the published
  2MASS~J0700+3157AB data (J.\ Thorstenen 2010, private
  communication), we find an uncertainty in the parallax of 3.8~mas
  that is $\approx$2$\times$ larger than the published error. Adopting
  this error would result in better agreement with our measurement
  (1.2$\sigma$ difference). In the case of 2MASS~J1534$-$2952AB, we
  retrieved the epochs of the observations from the ESO archive and
  assumed a range of astrometric precision based on the values given
  in \citet{2003AJ....126..975T}, namely (7~mas to 20~mas)/$\sqrt{8}$
  added in quadrature to the DCR offset error of 2--6~mas. This
  resulted in a parallax uncertainty of 2.7--3.7~mas, which is
  2.3--3.1$\times$ larger than the published error. At this level, the
  discrepancy with our parallax measurement is significantly
  decreased, though it still disagrees at the 2.9$\sigma$ level.  We
  also checked if orbital motion was significant and found that the
  correction offset for the parallax was negligible for the
  \citet{2003AJ....126..975T} epochs as it is for ours. We note that
  \citet{2003AJ....126..975T} used only 8 reference stars (cf.\ our
  475) and data spanning 6 distinct epochs over 2.0~years (cf.\ our 16
  over 2.4~years), so our solution should be more robust.

\item For LP~349-25AB, our parallax is 3.4$\sigma$ different from the
  published value \citep[$75.8\pm1.6$~mas;][]{2009AJ....137..402G}.
  This is the one case that we cannot readily explain with information
  at hand.  \citet{2009AJ....137..402G} do not discuss their
  astrometric precision per epoch, so we cannot assess their quoted
  error with Monte Carlo simulations. One source of systematic error
  could be their relatively small number of reference stars (12 versus
  our 33).  Another effect could be DCR as \citet{2009AJ....137..402G}
  seem to have observed in a broadband optical filter (not described
  in their paper).  For this object we used the narrow-band \Kn\
  filter, so DCR will be completely negligible. We note that our value
  for the correction from relative to absolute parallax
  ($1.7\pm0.3$~mas) agrees very well with theirs ($1.6\pm0.8$~mas), so
  this cannot be the source of the discrepancy.  Our orbit correction
  is very small ($-0.40\pm0.13$~mas) and \citet{2009AJ....137..402G}
  see no significant perturbations due to orbital motion, so this is
  also unlikely to explain the discrepancy.

\item There is one other published parallax that is discrepant with
  our results, 2MASS~J0850+1057AB \citep[2.5$\sigma$ different
  than][]{2002AJ....124.1170D}. This object has already been discussed
  by \citet{2011AJ....141...71F}.  They found that the
  \citet{2002AJ....124.1170D} parallax was likely biased by a
  background star which was blended with the target at the time of
  those observations but which is now clearly separated from the
  science target at $\approx$4$\arcsec$ in both our data and that of
  \citet{2011AJ....141...71F}. Our parallax of $30.1\pm0.8$~mas for
  2MASS~J0850+1057AB is in good agreement with both the values of
  $35\pm8$~mas and $26.2\pm4.2$~mas determined by
  \citet{2011AJ....141...71F} and \citet{2004AJ....127.2948V},
  respectively, but with 5--10$\times$ smaller error bars.

\end{itemize}

Finally, we note that on average our absolute parallax measurements
are not different from published results in any systematic fashion.
The mean and rms of the differences between published parallax values
and our own is $1.5\pm3.4$~mas (excluding the 4 discrepant cases
discussed above and published values with $>$20~mas parallax errors).
In fact, by using a much larger number of reference stars than
previous parallax programs, we should be less sensitive to systematic
errors in our correction to absolute parallax.  Previous surveys
typically used $\approx$5--10 reference stars, whereas on average we
have $>$100 reference stars per field ($N_{\rm ref} = 20$--475) and so
are much less biased by outliers. This is further supported by the
fact that the one target with a \Hipparcos\ parallax for its stellar
companion (HD~225118; $25.7\pm0.9$~mas) is in excellent agreement with
our CFHT value for the M7.5 dwarf (2MASS~J0003$-$2822;
$25.0\pm1.9$~mas).

\subsection{Spectral Decomposition of Binaries \label{sec:spt}}

The vast majority of our ultracool binary targets do no have any
published resolved spectroscopy providing component spectral type
determinations.  Some binaries have spectral types in the literature
determined by a spectral decomposition, i.e., using integrated-light
spectra and resolved photometry to estimate the deblended component
spectra.  However, methods used in the literature have varied
substantially \citep[e.g.,][]{2006ApJ...647.1393L, 2006ApJ...639.1114R,
  2007AJ....133.2320S} and often rely on an input assumption for the
relationship between spectral type absolute magnitude
\citep[e.g.,][]{2007AJ....134.1330B, 2010ApJ...710.1142B}.  This is
problematic because the binary components cannot then truly be used to
assess such empirical relations, which is a major goal of our work.
Therefore, we have determined component types for all binaries with
parallaxes in a uniform fashion that is completely independent of any
assumptions about how magnitude should depend on spectral type.

Our approach is modeled somewhat after the spectral template matching
method of \citet{2010ApJ...710.1142B}, but we have removed assumption
for the relationship between spectral type absolute magnitude.  A
library of template spectra for single\footnote{All spectral
  decomposition methods effectively assume that the templates being
  used are accurate representations of a single object of that
  spectral type.  This is most important around the L/T transition as
  a blended L+T dwarf spectrum can show significant anomalies relative
  to single objects \citep[e.g.,][]{2004ApJ...604L..61C}.  The
  templates we used have been cleaned of all known binaries, as well
  as the six strong spectral blend binary candidates proposed by
  \citet{2010ApJ...710.1142B}.} objects is compiled, and all possible
pairs of these spectra are added together. Each pairing is allowed to
have an arbitrary component flux ratio. The flux ratio and overall
scale factor are adjusted to minimize the $\chi^2$ of the difference
with the target's measured integrated-light spectrum. For each pairing
we then compute synthetic photometry in the bands for which we have
measured flux ratios. We reject pairings that disagree significantly
with the measured resolved photometry ($p$-value $<0.05$ in a $\chi^2$
test). Thus, our final set of modeled binary spectra is purely
selected on how well they match the measured integrated-light spectrum
and resolved photometry. We then ranked this ensemble based on the
$\chi^2$ of the match to the integrated-light spectrum and computed
weighted averages and errors of the component types and synthesized
flux ratios using the method outlined in Section~4.3 of
\citet{2010ApJ...710.1142B}. When assessing component types, we take
these quantities and their nominal errors into consideration but do
not treat them as absolute truth.

The input library of template spectra we used necessarily varied with
the component spectral types.  For binaries composed wholly of
$\geq$L3 dwarfs, we used the same library of 178 IRTF/SpeX prism
spectra as \citet{2010ApJ...710.1142B}.  Although this library is
somewhat less numerous than the full set of spectra in the SpeX Prism
Library, it has the significant advantage that
\citet{2010ApJ...710.1142B} report infrared spectral types on a
consistent scheme for all templates.  This is in contrast to types
available in the literature, particularly for L~dwarfs, which are
based on a variety of infrared flux indices and sometimes only have
optical types.  Because this library only has a handful of early L
dwarf templates and no late-M dwarfs we had to use a different subset
of spectra for earlier type binaries.  For binaries with at least one
$<$L3 component we simply used the full SpeX Prism Library with
whatever spectral types were available in the literature (i.e., a mix
of optical and infrared types).  For uniformity, we resampled all
spectra to a wavelength grid with 0.004~\micron\ steps ranging from
0.78--2.40~\micron.  To reduce systematic errors due to inaccurate
correction of telluric absorption, we excluded two wavelength regions
($1.34~\micron < \lambda < 1.41~\micron$ and $1.81~\micron < \lambda <
1.94~\micron$) when performing the spectral matching.  In some cases,
we had to use measured integrated-light spectra obtained with SpeX in
SXD mode ($R = 1200$--2000), which we degraded to the standard SpeX
prism resolution of 120 for accurate comparison to library templates.
For such SXD data we exclude the $K$-band portion of the spectrum
since that order does not overlap with the $JH$ orders and thus its
relative normalization would need to account for the uncertainty in
the measured and synthesized integrated-light photometry in $K$ band.

Throughout our analysis, we conservatively assume that infrared types
of late-M and L~dwarfs are uncertain by at least 1~subtype, with some
templates having larger uncertainties of 1.5--2~subtypes, and that
T~dwarf types are uncertain by at least 0.5~subtype. This is based on
the analysis of infrared types done by \citet{2010ApJ...710.1142B} who
compared their types to published values for 189 spectra of 178
sources. These authors found an intrinsic rms scatter of 1.1 and 0.5
subtypes in the ensemble of L and T~dwarfs, respectively.

We assigned component types and uncertainties on a case by case basis,
taking into account various factors such as: larger than average
spectral type uncertainties in the best-match templates; the full
range of properties implied when there were multiple matches giving
equally good fits; and constraints imposed by requiring consistency
with the integrated-light type.  When flux ratios were available from
multiple sources (e.g., our MKO Keck photometry and \HST/NICMOS
medium-band data), we checked for consistency.  We sometimes noted
discrepancies with photometry from the literature when published
errors were rather small.  In these cases we excluded the published
values as their errors are likely underestimated, and it did not
change the derived spectral types significantly within the errors.

Our derived component spectral types and their corresponding
uncertainties are listed in Table~\ref{tbl:spt}, and the single best
template pairing for each binary is shown in Figures~\ref{fig:spec1},
\ref{fig:spec2}, and \ref{fig:spec3}. In Table~\ref{tbl:spt}, we give
references for the literature photometry used and also a list of the
bandpasses utilized in constraining each fit. We also list separately
those binaries for which we do not use component types from our
spectral template matching because their types have been determined
directly from resolved spectroscopy \citep[e.g.,
LHS~1070BC;][]{2000A&A...353..691L} or other analysis \citep[e.g.,
CFBDSIR~J1458+1013AB;][]{2011ApJ...740..108L}.

Finally, we note that there are several binaries with parallaxes for
which we cannot derive component spectral types using this method,
either because they do not have the needed spectral or photometric
data or the data available do not sufficiently constrain the component
types.  Binaries with parallaxes for which we do not have spectra are
2MASS~J0025+4759AB, HD~65216BC, LSPM~J1314+1320AB,
DENIS-P~J1441$-$0945AB, and 2MASS~J2331$-$0406AB.  2MASS~J0856+2235AB,
2MASS~J0952$-$1924AB, 2MASS~J1239+5515AB, and LSR~J1610$-$0040AB have
no near-IR photometry.  And LHS~2397aAB has a spectrum and near-IR
photometry, but the late-L companion is too faint to enable accurate
spectral decomposition.

Our spectral decomposition analysis produces estimates of the flux
ratios for bandpasses without resolved photometry.  We synthesize flux
ratios for every template pairing that agrees with the available
resolved photometry ($p$-value $<0.05$).  To determine the flux ratio
and corresponding uncertainty in a given bandpass, we then use the
weighted average and error given by the aforementioned
\citet{2010ApJ...710.1142B} weighting scheme.  There are several cases
which benefit from these flux ratios, in order of most to least
reliable: (1)~binaries with resolved $K_S$ photometry that we convert
to MKO $K$ (and vice versa, e.g., for $\Delta{J}_{\rm MKO}$ to
$\Delta{J}_{\rm 2MASS}$); (2)~binaries with \HST/NICMOS
0.9--1.8~$\mu$m photometry that we convert to $JH$ photometry;
(3)~binaries with, e.g., only a $K$-band ratio for which we determine
$J$ and $H$ flux ratios; and (4)~one binary without any flux ratios
(SDSS~0805+4812AB).  In the following analysis we account for these
differing levels of reliability when reporting absolute magnitudes and
examining the location of binary components on color--magnitude
diagrams.

\subsection{Absolute Magnitudes of Single and Binary
  Objects \label{sec:absmag}}

We combine our parallaxes with integrated-light photometry (and flux
ratios in the case of binaries) to compute the absolute magnitudes for
our sample. We have also compiled such measurements for all ultracool
objects with published parallaxes. Table~\ref{tbl:plx-all} presents a
list of all ultracool dwarf parallax measurements, including objects
that have parallax determinations by virtue of their companionship to
stellar primaries. We compiled photometry from the literature for this
entire sample, and for objects that have published photometry in only
one system (MKO or 2MASS) we use near-IR spectra, when available, to
synthesize photometric conversions. For such objects we also
synthesize $Y-J$ colors to provide $Y$-band photometry when it is not
measured directly. Using objects in the SpeX Prism Library with both
2MASS and MKO photometry we can test the quality of our synthetic
photometric system offsets.  The $\chi^2$ of our computed 2MASS/MKO
offsets compared to measured values was 52.0 (98 dof), 96.6 (95 dof),
and 51.4 (93 dof) for $J$, $H$, and $K$ bands, respectively.  Since
$\chi^2$ is reasonable in all cases, we find that any systematic error
in our computed offsets must be negligible compared to the uncertainty
in the measured photometry ($\lesssim$0.02~mag).  However, when
computing colors across different bandpasses we found an additional
error of 0.05~mag was needed to explain the scatter in observed minus
computed values. Thus, we treat synthesized photometric system offsets
(e.g., $J_{\rm MKO} - J_{\rm 2MASS}$) has having zero error, while we
add 0.05~mag in quadrature to all synthesized $Y-J$ photometry.

We also include mid-IR photometry from \Spitzer/IRAC
\citep[e.g.,][]{2006ApJ...651..502P, 2007ApJ...655.1079L,
  2010ApJ...710.1627L} and the \WISE\ All-Sky Source
Catalog\footnote{\url{http://wise2.ipac.caltech.edu/docs/release/allsky/expsup/}}
\citep{2010AJ....140.1868W}.  We checked for \WISE\ quality flags
indicating possibly spurious or contaminated detections for all
objects after noting that \citealp{2011ApJS..197...19K} include
sources with nonzero flags in their Table~1. We visually inspected the
\WISE\ image atlas in the worst cases, i.e., ``H'' and ``D'' flags
indicating possible spurious detections, and found that the sources
are in fact likely to be real.  A published example of one such object
is HD~46588B shown in Figure~1 of \citet{2011ApJ...739...81L}, which
is flagged in the \WISE\ catalog as potentially spurious even though
it is real.  After vetting sources against the \WISE\ image atlas, we
do not find the need to exclude any flagged \WISE\ photometry from the
following analysis.

Tables~\ref{tbl:nir-mags} and \ref{tbl:mir-mags} present the resulting
collections of apparent magnitudes in the near- and mid-IR,
respectively.  In total, there are 314 objects in 255 systems that
have parallax measurements.  In Tables~\ref{tbl:nir-absmag} and
\ref{tbl:mir-absmag} we list absolute magnitudes sorted by spectral
type, along with references for any high angular resolution imaging
available.  This encompasses numerous \HST\ imaging programs with
WFPC2, NICMOS, ACS, and STIS; AO surveys at Keck, VLT, Gemini, Subaru,
Palomar, CFHT, and Lick; as well as speckle and lucky imaging surveys.
We note that there are unpublished archival data for many of the
objects compiled in our table, but we only count observations for
which analysis has been published.  The only exception is for the
subset of objects observed by our Keck AO binary survey that we have
determined to be unresolved in FWHM = 0$\farcs$05--0$\farcs$10 imaging
(Liu et al., in preparation).  The intention of this compilation is to
identify a clean sample of likely single objects (i.e., with no
companions outside $\approx$0$\farcs$1) from objects that have not
been surveyed for binarity.  Thus, we assign null entries for the
handful of unresolved spectroscopic and astrometric binaries that have
been imaged at high angular resolution in order to remove them from
the subset of likely single objects.

While our tables give complete compilations of the available data, we
have excluded objects from our plots and analysis if: (1)~their
fractional parallax uncertainty is greater than 24\% (i.e., an error
in the distance modulus $>$0.50~mag) or (2)~their color uncertainty is
$>$0.50~mag. Binary components are sometimes absent from plots if they
have no resolved photometry (e.g., 2MASS~J0856+2235AB, which has only
been resolved by \HST\ in F814W). Note that we retain objects that
lack spectral type determinations, as these objects are still useful
for color--magnitude diagrams (CMDs). For objects with multiple
published parallax measurements, we use the one with the lowest
uncertainty. In the following analysis (e.g., in polynomial fits), we
use optical spectral types for M and L~dwarfs when available (infrared
types otherwise) and infrared types for T~dwarfs.

Figures~\ref{fig:cmd-jhk1-mko}--\ref{fig:cmd-l} show CMDs for the
near-IR and Figures~\ref{fig:cmd-irac1}--\ref{fig:cmd-wise} show CMDs
for the mid-IR. In these plots we have excluded any objects with
subdwarf classifications (i.e., tabulated spectral types denoted as
``d/sd'', ``sd'', or ``esd'') for clarity, as well as
AB~Pic~b.\footnote{Although it formally passes our selection criteria
  for plotting, we exclude the young L~dwarf AB~Pic~b from the CMD
  figures. The exceptionally red $(J-H)_{\rm 2MASS}$ and $(J-K)_{\rm
    2MASS}$ colors (1.37~mag and 2.02~mag, respectively) makes this
  object an unusually prominent outlier in the CMDs, being
  $\approx$0.4--0.5~mag redder than objects of comparable absolute
  magnitude or spectral type. While this may reflect a unique SED for
  this source, another possibility is that the $J$-band photometry
  uncertainty is larger than reported. This speculation is supported
  by two possible pieces of evidence.  (1)~Figure~6 of
  \citet{2011ApJ...729..139W} shows that the $J-H$ color of AB~Pic~b
  is far redder than all other known young companions and field
  ultracool dwarfs, but not its $H-K$ color.
  (2)~\citet{2010A&A...512A..52B} show that the near-IR spectra of
  AB~Pic~b in the individual $J$, $H$, and $K$ bandpasses are
  consistent with previously known young early-L dwarfs. But if the
  published $JHK$ photometry is used to assemble a flux-calibrated
  SED, the resulting near-IR spectrum has a very peculiar broadband
  appearance (B.\ Bowler, 2011, private communication). Thus we
  conservatively choose to exclude AB~Pic~b from the CMD plots.}
However, note that one object (SDSS~J1416+1348) stands out as
significantly blue in all $JHK$ colors compared to the L~dwarf
sequence, as expected if it were a subdwarf; this object is discussed
in detail in Section~\ref{sec:1416}.

Figures~\ref{fig:spt-abs-jhk} and \ref{fig:spt-abs-yl} show near-IR
absolute magnitude as a function of spectral type, and
Figures~\ref{fig:spt-abs-irac} and \ref{fig:spt-abs-wise} show the
same for the mid-IR. For these relations we have excluded subdwarfs,
likely unresolved binaries (2MASS~J0559$-$1404 and SDSS~J1504+1027,
see Section~\ref{sec:overl}), and very low gravity objects. The last
cut excludes planetary mass objects (HR~8799bcde, 2M~1207b, and
$\beta$~Pic~b) as well as very young objects in stellar associations:
2MASS~J1207$-$3932 (TWA), PZ~Tel~B ($\beta$~Pic), HR~7329B
($\beta$~Pic), AB~Pic~b (Tuc-Hor), SSSPM~J1102$-$3431 (TWA), and
2MASS~J2234+4041AB (LkH$\alpha$~233). We fit polynomials to the
remaining field dwarf data, accounting for errors in both spectral
types and absolute magnitudes in a Monte Carlo fashion. This is
necessary as least squares regression algorithms are unable to
properly handle data sets in which the independent variables have
errors, as in the case of our spectral types. We drew $10^4$
realizations of each data point with normally distributed magnitude
errors and uniformly distributed spectral type errors. We then found
the single best-fit polynomial to all $N\times10^4$ simulated points,
using a standard least squares method since all points now have equal
weight. We fit magnitude as a function of spectral type for all bands,
and for bands that were sufficiently monotonic in their decline we
were able to also fit spectral type as a function of magnitude. The
latter fits are applicable in the situation where an observer wishes
to estimate the spectral type of an object based on photometry,
whereas the more often quoted former fits are useful for spectroscopic
distance estimates.

The coefficients of all of our polynomial fits are given in
Table~\ref{tbl:poly} along with the bulk rms of each fit. In
Figures~\ref{fig:spt-abs-jhk}, \ref{fig:spt-abs-yl},
\ref{fig:spt-abs-irac}, and \ref{fig:spt-abs-wise} the rms of the fits
over specific spectral type ranges are also given. These rms values
are useful in diagnosing the intrinsic scatter in absolute magnitude
over different ranges of spectral types. They are also the relevant
numbers to use, e.g., when determining a spectroscopic distance to a
single object of known spectral type since T~dwarfs generally show
more scatter in absolute magnitude than L~dwarfs, as opposed to
adopting the single rms value given in Table~\ref{tbl:poly}.  To test
the impact of our choice of using optical types for M and L~dwarfs
when available, we tried using only infrared types for all objects.
The median absolute value of the difference between polynomial fits
computed these two ways was 0.01--0.02~mag for the MKO photometry and
0.03--0.06~mag for the 2MASS photometry.  This is negligible compared
to the scatter in the data about the fits, which is typically
$\approx$0.4~mag, and thus using optical versus infrared spectral
types does not significantly affect our results.

We have also tabulated the mean absolute magnitude at each spectral
type in Tables~\ref{tbl:spt-abs-mko}--\ref{tbl:spt-abs-wise}, and we
plot the resulting values in Figures~\ref{fig:spt-wavg} and
\ref{fig:cmd-wavg}.  This information enables a more direct way of
assessing the changes in broadband fluxes as a function of spectral
type, since polynomial fits are not guaranteed to be a good
description of the data.  In these tables we give the weighted average
along with the rms for ``normal'' field dwarfs (i.e., those not
flagged as atypical in Table~\ref{tbl:plx-all}).  We also quantify the
level of intrinsic scatter at each spectral type by computing $\chi^2$
for each type's collection of measurements and finding the amount of
additional magnitude error that is needed to make reduced $\chi^2
\approx 1$, i.e., $p(\chi^2) = 0.5$.  When there are small numbers of
objects in a bin, or when the measurement errors are large, the
additional error needed is typically small, but this does not
necessarily mean that the intrinsic scatter is small.  Thus, the value
we find for the needed additional error is only a lower limit to the
intrinsic scatter at a given spectral type.

Some well-known patterns are seen in our intrinsic scatter estimates,
e.g., it is relatively low for late-M dwarfs (0.1--0.3~mag) and high
for mid- to late-L dwarfs (0.3--0.5~mag) at near-IR wavelengths. However,
we also find quite large scatter, previously unappreciated, among mid-
to late-T dwarfs (0.3--0.8~mag) in the near-IR. This highlights the fact
that cloud properties are not the only ``second parameter'' after
\Teff\ that can induce large near-IR flux variations --- metallicity and
surface gravity can produce variations in late-T dwarfs
\citep[e.g.,][]{2006ApJ...639.1095B, 2007ApJ...660.1507L} of similar or
greater amplitude than that seen among dusty L~dwarfs.


\section{Discussion of Individual Objects and Subsamples of Interest \label{sec:indiv}}

\subsection{Astrometric Binaries \label{sec:astrombin}}

As mentioned in Section~\ref{sec:plx-fit}, our CFHT astrometry has
revealed perturbations due to binary orbital motion for some of our
targets.  The strongest cases are SDSS~J0805+4812AB and
2MASS~J1404$-$3159AB, with 2MASS~J0518$-$2828AB showing smaller
residual scatter that we also tentatively attribute to orbital motion.
2MASS~J1404$-$3159AB has previously been resolved by Keck LGS AO
imaging \citep{2008ApJ...685.1183L}, and 2MASS~J0518$-$2828AB was
marginally resolved in \HST/NICMOS imaging
\citep{2006ApJS..166..585B}.  In contrast, SDSS~J0805+4812AB has only
previously been suggested as a binary due to its unusual spectral
morphology \citep{2007AJ....134.1330B}. Thus, our CFHT astrometry is
the first confirmation that SDSS~J0805+4812AB is indeed a binary.

Our spectral decomposition analysis gives spectral types of L5: and T5
for SDSS~J0805+4812AB and flux ratios of $\Delta{J} =
1.46\pm0.05$~mag, $\Delta{H} = 2.43\pm0.13$~mag, and $\Delta{K} =
3.13\pm0.17$~mag in the MKO photometric system. These are almost
identical to the values derived in a similar analysis by
\citet{2007AJ....134.1330B}. Although the orbit of the system is not
readily determined from our CFHT astrometry, the perturbation
amplitude ($\pm$15~mas) can be combined with the $J$-band flux ratio
and an assumed mass ratio to estimate the semimajor axis following the
equations in Section~\ref{sec:orbcorr}. Using evolutionary models,
\citet{2007AJ....134.1330B} estimate $q = 0.55$--0.88 for an age range
of 1--5~Gyr, respectively, which gives a factor of 0.35--0.15 by which
the photocenter motion should be divided. Thus, a rough estimate of
the semimajor axis is 40--100~mas. At a distance of $22.9\pm0.6$~pc
(using a parallax uncorrected for orbital motion) this corresponds to
0.9--2.3~AU and an orbital period of 2.7--9.1~years (again assuming
the masses from \citealp{2007AJ....134.1330B}). The short end of this
orbital period range is broadly consistent with the oscillations in
the astrometric residuals over our 4.0-year time baseline, thereby
suggesting a lower value for the mass ratio (i.e., a younger age and
lower masses of 0.066+0.036~\Msun). This is also consistent with the
binary being unresolved in Keck LGS AO images (Liu et al., in prep.)
since the semimajor axis would be small ($\approx$40~mas). An age much
younger than 1~Gyr becomes problematic since lithium absorption would
be expected in the optical spectrum but is not observed
\citep{2002astro.ph..4065H}, though \citet{2007AJ....134.1330B}
caution that this could simply due to insufficient S/N in the
spectrum.

Using the same approach, we can estimate the properties of
2MASS~J1404$-$3159AB, which displays astrometric residuals of
$\pm$12~mas.  Using the mass ratio estimate of $0.80\pm0.09$ from
\citet{2008ApJ...685.1183L} and our flux ratio of $\Delta{J} =
-0.54\pm0.08$~mag gives a photocenter correction factor of
$0.18\pm0.05$ and thereby semimajor axis estimate of $70\pm20$~mas. At
a distance of $23.8\pm0.6$~pc (again using the parallax without
correcting for orbital motion), this corresponds to $1.7\pm0.5$~AU and
orbital period of $8\pm4$~years for an assumed total mass of
0.07~\Msun\ from \citet{2008ApJ...685.1183L}. This semimajor axis is
somewhat at odds with the projected separation of the binary
($133.6\pm0.6$~mas on 2006~June~3~UT; \citealp{2008ApJ...685.1183L})
unless the orbit is fairly eccentric. An eccentric orbit would however
also be consistent with the short timescale of the astrometric
perturbations ($\approx$2~years) relative to the $8\pm4$ year orbital
period, since the binary could be passing through periastron during
our CFHT observations. We note that eccentric orbits are not very
common for such ultracool binaries \citep{me-ecc}, but they do occur.

2MASS~J0518$-$2828AB does not have a measured $J$-band flux ratio, but
the NICMOS F110W flux ratio ($0.8\pm0.5$~mag) combined with our
spectral deconvolution gives $\Delta{J} = 0.8\pm0.6$~mag.  This very
uncertain flux ratio means we cannot estimate the orbital properties
for this system.  We note that 2MASS~J0518$-$2828AB does not show as
clear a signature of orbital motion in its residuals as the other two
binaries, and its reduced $\chi^2$ is also much lower (3.9 vs.\ 10.9
and 7.0).  Although we are unable to estimate this binary's orbital
properties from our CFHT data, orbital monitoring currently underway
with \HST/ACS will perhaps yield more information.

\subsection{Overluminous Objects: Unresolved Binaries? \label{sec:overl}}

One simple result from measuring the absolute magnitudes for a large
sample is the identification of potential binaries as those objects
that are overluminous.  For L and T~dwarfs, this is complicated by the
large dispersion in colors at a given magnitude and in magnitudes at a
given spectral type.  Perhaps the cleanest sequence seen in any CMD is
that of $\approx$T0--T7 dwarfs in the IRAC $[$3.6$]$ and $[$4.5$]$
channels or similarly the W1 and W2 \WISE\ bands
(Figures~\ref{fig:cmd-irac1} and \ref{fig:cmd-wise}). At earlier types
($\lesssim$T0) these colors do not change at all with magnitude and at
later types there is appears to be a large amount of intrinsic
scatter. Another place that we may be able to look for unresolved
binaries is actually just above the L/T transition in the near-IR CMDs,
because the only way to reach that location is by being an extremely
blue L~dwarf or an overluminous early-T dwarf.  With these
considerations in mind, we identify the following overluminous objects
as candidate binaries:
\begin{itemize}

\item \textit{2MASS~J0559$-$1404:} This T4.5 dwarf has long been
  suspected to be an unresolved binary, because it stands out in both
  CMDs and the spectral type--absolute magnitude relations as being
  very bright compared to both the late-L and early-T dwarfs.
  Alternatively, it could simply represent the most extreme outcome of
  the brightening seen across the L/T transition.  We note that this
  object not only continues to stand out on the CMDs in the near-IR
  but also in the mid-IR with bands 1 and 2 of \WISE\ and IRAC.  This
  greatly favors the unresolved binary hypothesis, since no such
  brightening is seen in the mid-IR CMDs.  However, a companion to
  2MASS~J0559$-$1404 has remained elusive in both direct imaging
  (\citealp{2003ApJ...586..512B}; but also see footnote~15 in
  \citealp{2008ApJ...689..436L}) and radial velocity monitoring
  \citep{2007ApJ...666.1205Z}.

\item \textit{SDSS~J1021$-$0304A:} Our parallax of $29.9\pm1.3$~mas
  for this system is consistent with the \citet{2003AJ....126..975T}
  value $34.4\pm4.6$~mas but 3.5$\times$ more precise.  This has
  revealed that the T0 primary component lies significantly above the
  L/T transition in most near-IR CMDs (note that without resolved
  mid-IR photometry we can only use near-IR magnitudes here).
  SDSS~J1021$-$0304A could instead be described as being bluer than
  other objects of its absolute magnitude, akin to SDSS~J1416+1348 (a
  possible L6 subdwarf).  However, unlike SDSS~J1416+1348, which has
  normal $J$ and $H$ absolute magnitudes for its spectral type but is
  fainter than average in $K$, SDSS~J1021$-$0304A is 0.3--0.5~mag
  brighter in $J$ and $H$ for its spectral type and normal at $K$.
  This suggests that its blueness (or overluminosity) is due to a
  different reason than SDSS~J1416+1348.  Perhaps the simplest
  explanation is that SDSS~J1021$-$0304A is an unresolved binary
  itself -- a hypothesis that can be validated if future orbital
  monitoring determines that the total dynamical mass of
  SDSS~J1021$-$0304AB turns out to be $>$2$\times$ the substellar mass
  limit ($\gtrsim$0.16~\Msun).  Since the location of
  SDSS~J1021$-$0304A in the near-IR CMDs is not shared by any other
  known single objects, it is difficult to come up with another
  explanation without resorting to models.  In the framework of
  \citet{2001ApJ...556..872A}, SDSS~J1021$-$0304A could be a brown
  dwarf with a large value of $f_{\rm sed}$ (i.e., rapid grain growth
  leading to optically thin clouds with a low number density of
  particles).

\item \textit{SDSS~J1504+1047:} We measure the distance to this T7
  dwarf for the first time ($21.7\pm0.7$~pc), and it appears very
  similar to 2MASS~J0559$-$1404 in its location on the \WISE\ and IRAC
  bands 1 and 2 CMDs (i.e., $\approx$0.7~mag brighter than the T~dwarf
  sequence). However, because of its later spectral type,
  SDSS~J1504+1047 is effectively buried in the nearly vertical T~dwarf
  sequence in the near-IR CMDs.  But it does stand out as the
  brightest T7 in all near-IR bands for which it has data ($JHK$) and
  this is even more clear in the spectral type--absolute magnitude
  relations in the mid-IR, owing to their much lower dispersion in
  magnitude as a function of spectral type
  (Figures~\ref{fig:spt-abs-irac} and \ref{fig:spt-abs-wise}). Thus,
  we find that SDSS~J1504+1047 is a strong candidate for being a
  nearly equal magnitude binary. There is no published high-resolution
  imaging for this object to date, and we note that its lack of
  astrometric perturbations in our CFHT data would be consistent with
  this picture (i.e., nearly equal magnitude binaries have
  undetectable photocenter motion).

\end{itemize}

We note that 2MASS~J0939$-$2448 (T8), 2MASS~J0937+2931 (T6p), and to a
lesser extent 2MASS~J1237+6526 (T6.5) show up as brighter than the
T~dwarf sequence in mid-IR CMDs, similar to the candidate binaries
2MASS~J0559$-$1404 and SDSS~J1504+1047 discussed above. However, it
seems more likely that the atypical locations of 2MASS~J0937+2931 and
2MASS~J0939$-$2448 may be explained by unusually low metallicity
and/or high gravity \citep[e.g.,][]{2003ApJ...594..510B}, since they
are not brighter than other objects of similar spectral type in the
near-IR bands (in fact, they are both the faintest objects of their
type at $K$ band). In other words, 2MASS~J0937+2931 and
2MASS~J0939$-$2448 are unusually red in WISE and IRAC bands 1 and 2,
not unusually bright. The very active T6.5 dwarf 2MASS~J1237+6526 also
does not display unusually bright near-IR magnitudes and so is
probably more accurately described as unusually red. 2MASS~J1237+6526
has been discussed extensively by \citet{2006astro.ph..9793L} who
found that it is likely old, high-gravity, and with slightly subsolar
metallicity.  Thus, its location on the mid-IR CMDs may be due to
similar, but somewhat weaker, effects as for 2MASS~J0937+2931 and
2MASS~J0939$-$2448.\footnote{\citet{2008ApJ...689L..53B} determined
  that 2MASS~J0939$-$2448 is overluminous for its model atmosphere
  derived temperature, concluding that it was likely an unresolved,
  nearly equal-flux binary.  This conclusion was also reached by
  \citet{2009ApJ...695.1517L} from analysis based on model
  atmospheres.  Our interpretation does not necessarily require
  unresolved binarity to explain the observations since we find that
  2MASS~J0939$-$2448 is unusual in color rather than in magnitude. If
  single, the model-derived \Teff\ from previous work would be
  systematically offset from the actual \Teff, possibly due to this
  object's subsolar metallicity and/or high gravity.}

\subsection{SDSS~J1416+1348 and ULAS~J1416+1348 \label{sec:1416}}

SDSS~J1416+1348 was identified by \citet{2010ApJ...710...45B} as a
nearby L6 dwarf ($8.4\pm1.9$~pc spectrophotometric distance estimate)
with unusually blue near-IR colors that might normally be indicative
of being a metal-poor subdwarf. However, \citet{2010ApJ...710...45B}
did not find metal-poor features in its optical or near-IR spectra,
thereby suggesting that its color was instead due to unusual cloud
properties for its spectral type.  \citet{2010AJ....139.1045S}
independently discovered this object and found a consistent spectral
type (L5).  \citet{2010MNRAS.404.1952B} assigned an intermediate
classification of d/sdL7 based on an alternative interpretation of its
optical spectrum and identified a late-T companion ULAS~J1416+1348
(T7.5p) at a projected separation of 9$\farcs$81 (also independently
discovered by \citealp{2010A&A...510L...8S}).

Our distance measurement of $9.10\pm0.15$~pc is 15--20$\times$ more
precise than preliminary parallaxes computed by
\citet{2010A&A...510L...8S} and \citet{2010ApJ...710...45B}, enabling
us to robustly assess the absolute magnitudes of both of these unusual
brown dwarfs for the first time. SDSS~J1416+1348 appears to be of
normal brightness for its spectral type in both near- and mid-IR
magnitudes.  This is in contrast with results from
\citet{2008ApJ...672.1159B} for 2MASS~J0532+8246 (sdL7) that showed
this subdwarf to be 1--2~mag brighter in the near-IR than objects of
similar spectral type and slightly brighter at
$[$4.5$]$.\footnote{Note that the updated parallax from
  \citet{2009A&A...493L..27S} for 2MASS~J0532+8246 decreases its
  distance by 2$\sigma$ (13\%), resulting in normal mid-IR magnitudes.}
Enhanced $J$-band flux, such as seen for 2MASS~J0532+8246, would be
expected for SDSS~J1416+1348 if thin clouds or large condensate grains
were responsible for its unusual colors.  It may be that this
enhancement is present but is too small to show up in the comparison
to other objects given the relatively large scatter in $J$-band
absolute magnitude as a function of spectral type ($\gtrsim$0.5~mag
for L6, 2MASS system).  Its offset from typical field $(J-\Ks)_{\rm
  2MASS}$ colors is indeed small in an absolute sense (1.04~mag vs.\
1.75~mag for field L6 dwarfs from \citealp{2009AJ....137....1F}).
Thus, only a small offset in absolute magnitudes is expected,
especially if the color offset is also due in part to $K$ band being
suppressed by stronger-than-average collisionally induced H$_2$
absorption as expected at slightly subsolar metallicity
\citep{1969ApJ...156..989L, 1997A&A...324..185B}.

ULAS~J1416+1348 (T7.5p) on the other hand is much fainter than other
T7--T8 dwarfs. It is $\approx$1~mag fainter than an average T7.5
dwarf; in fact it is the faintest known T7--T8 dwarf in $YJH$ bands
except for the recently discovered T8p dwarf BD~+01~2920B, which has
comparable $YJH$ magnitudes \citep{2012arXiv1201.3243P}. In $K$ band
($M_K = 19.14\pm0.18$~mag) ULAS~J1416+1348 is fainter than \emph{all}
known T dwarfs with parallaxes except for CFBDS~J1458+1013B ($>$T10;
$M_K = 20.4\pm0.5$~mag) and possibly UGPS~J0722$-$0540 (T9; $M_K =
19.0\pm0.3$~mag). This behavior is similar to, but much more extreme
than, the proposed T subdwarf 2MASS~J0937+2931, classified as d/sdT6
by \citet{2007ApJ...657..494B} and \citet{2009A&A...493L..27S}.
ULAS~J1416+1348 also has very red $[3.6]-[4.5]$ colors consistent with
enhanced CH$_4$ absorption at $[3.6]$ and weaker CO absorption at
$[4.5]$, which may occur at subsolar metallicities \citep[e.g.,
see][]{2006astro.ph..9793L}. The \WISE\ All-Sky Source Catalog
photometry is also quite red ($W1-W2 = 3.33\pm0.20$~mag) and, like the
IRAC photometry, shows that ULAS~J1416+1348 is indeed fainter at
3--4~\micron\ by $\approx$0.2~mag and brighter at 4--5~\micron\ by
$\approx$0.4~mag compared to other T7.5 dwarfs.
Thus, we conclude that ULAS~J1416+1348 likely has lower metallicity
than typical field brown dwarfs, and so by extension the unusual
properties of SDSS~J1416+1348 are also affected by subsolar
metallicity. However, we note that this does not exclude unusual cloud
properties or high surface gravity as an explanation for some of the
unusual features observed in these objects.

Finally, our precise distance enables a much better constraint on the
projected separation of this binary system, $89.3\pm1.5$~AU.  To
convert this separation to semimajor axis we use the results from the
Appendix of \citet{me-ecc} for the very low-mass visual binary
eccentricity distribution with no discovery bias, as is appropriate
for such a wide binary.  The median and 68.3\% confidence limits on
the conversion factor is thus $1.16_{-0.31}^{+0.81}$, giving a
semimajor axis of $104_{-72}^{+28}$~AU.  This is the widest known
binary with likely substellar components.\footnote{The only ultracool
  binaries wider than SDSS~J1416+1348AB are pairs with late-M
  primaries: 2MASS~J01303563$-$4445411AB (M9+L6:, $130\pm50$~AU;
  \citealp{2011AJ....141....7D}); DENIS-P~J055146.0$-$443412AB
  (M8.5+L0, $250\pm50$~AU; \citealp{2005A&A...440L..55B});
  Koenigstuhl~1 (M6:+M9.5, $1800\pm170$~AU;
  \citealp{2007A&A...462L..61C}); 2MASS~J01265549$-$5022388AB
  (M6.5+M8, $5100\pm400$~AU; \citealp{2007ApJ...659L..49A}); and
  2MASS~J12583501+4013083AB (M6:+M7:, $6700\pm800$~AU;
  \citealp{2009ApJ...698..405R}). Note that the values listed here are
  projected separations.}

\subsection{Wide Companions \label{sec:comp}}

Some objects in our sample have been proposed to be wide companions to
stars based on common proper motion. We have checked if our improved
proper motions and parallaxes for these objects are still consistent
with companionship. We measure a relative proper motion and absolute
parallax for 2MASS~J0003$-$2822 (M7.5) of $\mu_{\alpha} \cos{\delta} =
280.3\pm1.5$~\masyr, $\mu_{\delta} = -123.3\pm1.7$~\masyr, $\pi =
25.0\pm1.9$~mas. This is in good agreement with the absolute
\Hipparcos\ values for HD~225118 ($\mu_{\alpha} \cos{\delta} =
280.8\pm1.1$~\masyr, $\mu_{\delta} = -141.5\pm0.6$~\masyr, $\pi =
25.7\pm0.9$~mas). Thus, we confirm the result of
\citet{2007AJ....133..439C} that this is a common proper motion pair,
and we show that it is common in parallax as well.

For 2MASS~J0850+1057AB, we measure a proper motion
($144.7\pm0.6$~\masyr) and parallax ($30.1\pm0.8$~mas) 10$\times$ more
precise than \citet{2011AJ....141...71F}, who found that this binary
is a common proper motion companion to NLTT~20346AB.  (Note that the
proper motions for 2MASS~J0850+1057AB and NLTT~20346AB as measured by
\citet{2011AJ....141...71F} are different by 3.3$\sigma$, not
$<$2$\sigma$ as stated in their Section~3.2.) Our improved proper
motion for 2MASS~J0850+1057AB is discrepant with their value for
NLTT~20346AB by $\Delta\mu = 47\pm7$~\masyr\ (6.7$\sigma$) in
two-dimensional proper motion space, where $\Delta\mu \equiv
\sqrt{(\Delta\mu_{\alpha} \cos{\delta})^2 +
  \Delta\mu_{\delta}^2}$. This discrepancy is about a third of the
total proper motion amplitude of the object ($\Delta\mu/\mu = 0.33$),
larger than all other accepted common proper motion pairs in the
literature ($\Delta\mu/\mu$ always $\lesssim0.2$ as discussed below).
We also note that the two proper motions do not satisfy the criterion
of \citet{2007AJ....133..889L} for being a co-moving pair (their
Equation~5), which is specifically valid for the range of proper
motions in the LSPM catalog from which NLTT~20346AB was originally
selected. \citet{2007AJ....133..889L} based their criterion on how
often chance alignments would occur as a function of separation on the
sky and difference in proper motion vectors for LSPM-N. NLTT~20346AB
and 2MASS~J0850+1057AB form a pair with an exceptionally large
separation (248\arcsec), making it very likely that this is only a
chance alignment of marginally consistent proper motions (see Figure~1
of \citealp{2007AJ....133..889L}). Therefore, we conclude that
NLTT~20346AB and 2MASS~J0850+1057AB are not physically associated.

We also searched for previously unrecognized common proper motion
companions to all ultracool dwarfs with parallax measurements
(Table~\ref{tbl:plx-all}), and as a check on our results we included
objects with known companions as well. We queried proper motion
catalogs using a 10\arcmin\ radius around each object, and where
possible for the known companions we used an independent measurement
of the object's proper motion (i.e., not the primary's proper
motion). Our search of \Hipparcos, Tycho, and LSPM-N recovered all
known wide companions present in those catalogs.  We assessed
companionship using both the \citet{2007AJ....133..889L} criterion,
which is valid for proper motions of $\approx$150--450~\masyr, and
also simply the fractional difference in proper motion,
$\Delta\mu/\mu$. We found that all known common proper motion pairs
had $\Delta\mu/\mu \leq 0.21$, and 14 of the 19 pairs (74\%) had
$\Delta\mu/\mu \leq 0.08$. The only exceptions were
2MASS~J0850+1057AB, as discussed above, and 2MASS~J2331$-$0406AB. The
latter inconsistency was simply due to the fact that we used an
apparently erroneous proper motion from Table~4 of
\citet{2009AJ....137....1F}, originally from
\citet{2000AJ....120.1085G}, that gave $\Delta\mu = 225$~\masyr\ and
$\Delta\mu/\mu = 0.49$ for the companion HD~221356. However, both
\citet{2007ApJ...667..520C} and the PPMXL catalog
\citep{2010AJ....139.2440R} give proper motions for
2MASS~J2331$-$0406AB that are consistent with its companion
($\Delta\mu = 4$~\masyr, $\Delta\mu/\mu = 0.01$).

Our search of the \Hipparcos, Tycho, and LSPM-N catalogs revealed only
two previously unrecognized candidate wide companions having
$\Delta\mu/\mu \leq 0.20$:
\begin{itemize}

\item SSSPM~J1102$-$3431 (M8.5) is a member of TWA with a relatively
  small proper motion ($\mu = 68.6\pm0.6$~\masyr;
  \citealp{2008A&A...489..825T}) that appears to be co-moving with the
  Tycho star TYC~7208-592-1 ($\Delta\mu/\mu = 0.07$). With a projected
  separation of 197\arcsec\ this would be an extremely wide pair ($1.1
  \times 10^4$~AU or 0.05~pc). We note that TYC~7208-592-1 is an
  otherwise anonymous star with no X-ray detection in \ROSAT, implying
  it may not be young and thus may not be physically associated with
  SSSPM~J1102$-$3431. Spectroscopy of TYC~7208-592-1 should readily
  determine if it is indeed a young star at the age of TWA, and thus
  whether this is a physically associated pair.  We note that
  SSSPM~J1102$-$3431 has previously been suggested by
  \citet{2005A&A...430L..49S} to be a wide companion of the star
  TW~Hya, and its parallax \citep[$18.1 \pm
  0.5$~mas;][]{2008A&A...489..825T} is consistent with the \Hipparcos\
  value for TW~Hya \citep[$18.6 \pm 2.1$~mas;][]{2007hnrr.book.....V}.
  However, because of the extremely wide projected separation
  ($4\times10^4$~AU or 0.2~pc) \citet{2008A&A...489..825T} point out
  that this is unlikely to survive as a gravitationally bound system.
  From Equation~18 of \citet{2010AJ....139.2566D}, only pairs tighter
  than $\lesssim$0.12~pc are expected to remain bound over 10~Gyr.

\item ULAS~J1315+0826 (T7.5) has a modest proper motion
  ($113\pm10$~\masyr; \citealp{2010A&A...524A..38M}) that is
  marginally consistent with TYC~884-383-1 ($\Delta\mu/\mu = 0.18$).
  If physically associated the projected separation of 383\arcsec\
  would correspond to 9000~AU. A more precise proper motion for this
  late-T dwarf would be useful in determining whether this pair is
  truly associated.

\end{itemize}

\subsection{High Tangential Velocity Objects \label{sec:vtan}}

%
We have computed the tangential velocities (\Vtan) of all ultracool
dwarfs with parallaxes and proper motions (Table~\ref{tbl:plx-all}).
This direct observable is related to an object's kinematic history, as
stars in the halo tend to have larger velocities than stars in the
disk, and likewise the youngest members of the disk are kinematically
colder than old members.  Very high tangential velocity is often used
as an indicator of old age and thereby possibly low metallicity,
especially for faint objects like brown dwarfs where the radial
velocity (and thus full three-dimensional space motion) is not readily
measurable \citep[e.g.,][]{2009AJ....137....1F, 2011ApJ...735...62L,
  2011A&A...532L...5S}.  To put such associations on quantitative
footing, we use a model of the Galaxy to compute the projected motion
on the sky for different kinematic populations and investigate how
this varies along different sight lines.  Since the objects we are
concerned with are all within $\approx$100~pc (median distance of
19~pc), they essentially represent a single point in the Galactic
potential, which simplifies this problem.

We compute probabilities for membership in the thin disk, thick disk,
and halo as a function of \Vtan\ by using the Besan\c{c}on model of
the Galaxy \citep{2003A&A...409..523R}.  We used a custom ``all sky''
simulation, as in our previous kinematic analysis work
\citep[e.g.,][]{2009ApJ...699..168D, 2011ApJ...740L..32L}, that
comprises $8 \times 10^5$ model stars with a thin/thick disk
proportion of 0.977/0.023 and a halo star fraction of $1.5 \times
10^{-4}$.  To simulate observational uncertainties we added Gaussian
noise to the model tangential velocities, and then we computed the
fraction of each population as a function of \Vtan\ to determine the
membership probability for a given combination of \Vtan\ and
$\sigma_V$.  We calculated membership probabilities for a wide range
of observational uncertainties ($\sigma_V = 1$--70~\kms), and the
results are shown in Figure~\ref{fig:vtan}.

We are interested in determining the probability of non-thin disk
membership, and our simulations quantify the degree to which this
membership probability drops as the uncertainty in \Vtan\ increases.
The probability contours in Figure~\ref{fig:vtan} follow very closely
an exponential relationship, so we have fit exponential functions to
the results from our numerical simulations to provide easy-to-use
criteria for determining if an object is likely to be a thin disk
member or a kinematically old thick disk or halo object:
\begin{equation}
  \Vtan > 91 + 56 \exp(0.024 \sigma_V)~~~~(p_{\rm thin} < 0.1)
\end{equation}
\begin{equation}
  \Vtan = 77 + 35\exp(0.028 \sigma_V)~~~~(p_{\rm thin} = 0.5)
\end{equation}
\begin{equation}
  \Vtan < 50 + 26\exp(0.025 \sigma_V)~~~(p_{\rm thin} > 0.9)
\end{equation}
where all velocities are in units of \kms.  At this point we
investigated the effect of observing along different lines of sight of
the local population.  We randomly selected 100 locations uniformly
distributed on the celestial sphere and computed the \Vtan\ cutoffs
for $p_{\rm thin} = 0.1$, 0.5, and 0.9 for zero error in \Vtan.  The
mean values agree with those listed above, and the rms over the sky
was 9\%--15\%, demonstrating that using a single mean value is a
reasonable simplification.  We emphasize that the relations derived
here provide criteria for membership \emph{probability}, which is
always just a statistical argument for any individual object, and if a
radial velocity is available then full three-dimensional space motion
should be used to assess membership probability instead.

We applied the above criteria to all ultracool dwarfs with parallaxes
(Table~\ref{tbl:plx-all}) to check their effectiveness and determine
if any previously unrecognized likely thick disk or halo members are
in this sample. We recovered all objects with $\Vtan > 150$~\kms\ as
likely non-thin disk members ($p_{\rm thin} < 0.1$), except for one
object with a very large error ($253\pm71$~\kms; SDSS~J1256$-$0224).
All 10 of the recovered objects have also been spectroscopically
identified as subdwarfs, and only one known subdwarf in our sample was
not recovered (HD~114762B; $\Vtan = 106\pm3$~\kms). We found 4
additional objects with $\Vtan > 115$~\kms\ as being somewhat unlikely
thin disk members ($0.1 < p_{\rm thin} < 0.5$): LHS~207, LHS~330,
GRH~2208$-$20, and Gl~802B. None of these are known to be subdwarfs,
but some have been suggested as possible thick disk members based on
their space motion (e.g., GRH~2208$-$20 in
\citealp{2002AJ....124.1170D} and Gl~802B in
\citealp{2008ApJ...678..463I}). We did not find any previously
unrecognized kinematically old objects in our sample.

\subsection{Spectral Type ``Flips'' \label{sec:spt-flip}}

It is conventional to assume that if one component of an ultracool
binary is brighter at \emph{all} near-IR bands, then it must either be
of earlier spectral type than the secondary or else an unresolved
binary.  This is largely due to prevailing notion that the parameters
inducing scatter in the absolute magnitude vs.\ spectral type
relations (e.g., metallicity, gravity, and cloud properties) will
always be shared in common between the two components of a
binary. However, as mentioned in Section~\ref{sec:spt}, methods for
determining spectral types of ultracool binaries sometimes assume
\emph{a priori} that absolute magnitude declines monotonically with
spectral type and thus are not well suited to assessing whether that
is actually true.

We find no strong evidence for spectral types ``flips'' for the
binaries in our sample, namely where the brighter primary appears to
be later type than the fainter secondary.  The closest case is
Gl~337CD where we find L$8.5\pm1$ and L$7.5\pm2$ for the two
components.  We did not find any template pairings in which the
primary was earlier type than the secondary.  However, there was
substantial scatter in the spectral types of templates used for the
best pairings, resulting in very uncertain component types.  Thus, we
lack the ability to determine if the primary is indeed later type than
the secondary, and our results are consistent with the secondary being
later type than the primary.  The only other binary with similar
results was 2MASS~J0920+3517AB, but in this case only about half of
the best pairings used a later type primary template.  For this
system, we conservatively assigned types of L$5.5\pm1$ and L$9\pm1.5$
corresponding to the template pairings with earlier type primaries but
marginally consistent with equal type components.

\subsection{2MASS~J0850+1057AB and 2MASS~J1728+3948AB \label{sec:0850}}

\citet{2011AJ....141...70B} recently presented analysis of the two
binaries 2MASS~J0850+1057AB and 2MASS~J1728+3948AB with the main
results that: (1)~2MASS~J0850+1057A is anomalously bright for its
spectral type, implying that it is likely an unresolved binary; and
(2)~2MASS~J1728+3948A is unusually faint in $J$ for its spectral type
(L5 from their analysis), requiring thick condensate clouds.

For 2MASS~J0850+1057AB we find that the best match to the spectrum and
photometry are spectral templates with types of L$6.5\pm1$ and
L$8.5\pm1$, in contrast with the results of
\citet{2011AJ....141...70B} that require an later type secondary
(L7+L6). One reason for this difference is that we find that the
\citet{2011AJ....141...70B} F110W and F110M photometry is highly
discrepant with our own $J$-band photometry.  In addition, we found
essentially no template pairs that both matched the photometry in
these two NICMOS bands and the blended spectra simultaneously,
suggesting that the published flux ratio errors were underestimated.
At L$6.5\pm1$, we find that 2MASS~J0850+1057A is not anomalously
bright compared to other L5.5--L7.5 dwarfs (e.g., it is fainter than
all L6 dwarfs in Table~\ref{tbl:nir-absmag}).  We note that photometry
from \citet{2011AJ....141...70B} in other NICMOS bandpasses (F145M and
F170M) is consistent with our template pair matching and the
F145M$-$F170M colors in fact provide evidence that the secondary
should be later type than the primary. This is because this color is
quite sensitive to the H$_2$O band depths in $H$ band.  From
synthesized F145M$-$F170M colors for field dwarfs we find that the
measured color difference of $0.11\pm0.07$~mag for 2MASS~J0850+1057AB
implies ${\rm SpT}_{\rm B} - {\rm SpT}_{\rm A} =1.9\pm1.2$~subtypes.
This is consistent with our spectral type determination ($\Delta$SpT =
$2.0\pm1.4$~subtypes) and inconsistent with $\Delta$SpT =
$-1.0\pm0.7$~subtypes from \citet{2011AJ....141...70B}.

For 2MASS~J1728+3948AB (L$5\pm1$ and L$7\pm1$), we find essentially
identical spectral types as the L5+L6.5 values of
\citet{2011AJ....141...70B}.  We confirm that 2MASS~J1728+3948A is
quite red for its spectral type, $(J-K)_{\rm MKO} = 2.13\pm0.11$~mag,
and it is the reddest object in the field dwarf sample except for
SDSS~J0107+0041 (L8, $(J-K)_{\rm MKO} = 2.17\pm0.04$~mag) and
2MASS~J1711+2232 (L6.5, $(J-K)_{\rm MKO} = 2.25\pm0.21$~mag). It is
also fainter in $J$ and $H$ than any other L4--L6 dwarf, supporting
the interpretation from \citet{2011AJ....141...70B} that it has
thicker than average dust clouds.\footnote{Note that these comparisons
  assume negligible near-IR variability, which is actually unknown for
  these specific objects but which is generally found to be
  $\lesssim$0.05~mag for objects of similar spectral type
  \citep{2004MNRAS.354..466K, 2005MNRAS.362..727K,
    2008MNRAS.386.2009C, 2009ApJ...701.1534A}. Thus, variability is
  expected to have a negligible impact in our analysis since it is
  comparable to or much smaller than the uncertainties in the colors
  and absolute magnitudes. In addition, Radigan et al.\ (2012,
  submitted) find that the colors of variable ultracool dwarfs stay
  relatively constant while it is their overall flux that increases
  and decreases, so variability should have an even smaller impact on
  our noncontemporaneous color comparisons.}


\section{The L/T Transition \label{sec:lt}}

The transformation of L~dwarfs into T~dwarfs as brown dwarfs cool has
been an long-standing topic of interest.  The dramatic differences
between L~and T~dwarf spectra are generally understood to be due to a
combination of effects as \Teff\ decreases in ultracool objects: the
formation and subsequent removal of condensate clouds from the
photosphere and the change from CO to CH$_4$ being the dominant
carbon-bearing molecule.  One-dimensional models have reproduced the
general properties of the spectra, colors, and magnitudes of late-L to
mid-T dwarfs based on prescriptions for the clouds
\citep{2002ApJ...568..335M, 2002ApJ...575..264T, 2005astro.ph..9066B},
and parameterized models can be successfully fitted to
broad-wavelength observations of individual objects
\citep[e.g.,][]{2008ApJ...678.1372C, 2009ApJ...702..154S,
  2010A&A...510A..99K}.  However, given the difficulty of modeling
clouds \citep[e.g.,][]{2008MNRAS.391.1854H}, a robust physical theory
is still lacking. Consequently no model accurately reproduces the
complete color-magnitude sequence of L~and T~dwarfs (though see
\citealp{2008ApJ...689.1327S} and \citealp{2010arXiv1011.5405A}).

One observational challenge to theory is the fact that the change
between the near-IR SEDs of the late-L~dwarfs and early-T dwarfs (with
very red colors) and those of the mid-T dwarfs (with very blue colors)
occurs over a small range in effective temperature
\citep[$\Teff\approx1100$--1400~K, e.g.,][]{2000AJ....120..447K,
  2004AJ....127.3516G, 2004AJ....127.2948V}.  An additional challenge
is the non-monotonic behavior of the 1.0--1.3~\micron\ fluxes through
the L/T transition region, where the T3--T5 dwarfs can appear brighter
than earlier objects, a phenomenon known as the ``$J$-band bump''
\citep{2002AJ....124.1170D, 2003AJ....126..975T, 2004AJ....127.2948V}.
Both of these effects point to relatively rapid removal of clouds from
the photospheres of the late-L and early-T dwarfs, including
non-equilibrium (dynamic) processes such rapid particle
growth/sedimentation \citep{2004AJ....127.3553K, 2009ApJ...702..154S}
and cloud disruption leading to spatially inhomogeneous photospheres
\citep{2001ApJ...556..872A, 2002ApJ...571L.151B, 2010ApJ...723L.117M}.
The driving role played by cloud evolution is highlighted by the
wavelength-dependence of the brightening. Condensate opacity is
expected to dominate over gas opacity in the 1.0--1.3~\micron\ region
\citep[e.g.,][]{2001ApJ...556..872A, 2005astro.ph..9066B}, and thus
the removal of condensates should be most pronounced at these
wavelengths.

Binarity both enlightens and complicates our understanding. Two
binaries in the L/T region clearly show a reversal in their $J$-band
flux ratios between their two components, indicating that the $J$-band
bump is truly a physical effect that occurs as brown dwarfs cool
\citep{2006ApJ...647.1393L, 2008ApJ...685.1183L} and not solely due to
a spread in the age/surface gravity of the field population
\citep{2003ApJ...585L.151T}. Additional flux-reversal binaries have
been proposed by \citet{2004ApJ...604L..61C},\footnote{The
  decomposition of 2MASS~J0518$-$2828AB by \citet{2004ApJ...604L..61C}
  used the spectrum of SDSS~J1021$-$0304AB as a template, which was
  later found to be a binary \citep{2006ApJS..166..585B}.  The latest
  decomposition presented here (Figure~\ref{fig:spec1}) suggests no
  flux reversal between the components, which is also consistent with
  the results from \citet{2010ApJ...710.1142B}.}
\citet{2006ApJS..166..585B}, and \citet{2010ApJ...710.1142B} based on
decomposition of their integrated-light spectra. Since the near-IR
absolute magnitudes are roughly constant from $\approx$L6--T5 while
the spectra are greatly changing, unresolved binaries can
substantially enhance the dispersion in the absolute magnitudes and
colors, the amplitude of the $J$-band bump, and the binary frequencies
at these spectral types \citep{2006ApJ...647.1393L,
  2006ApJS..166..585B, 2007ApJ...659..655B}.  Further complications
arise from strong photometric variability is which present in at least
some objects \citep{2003AJ....126.1006E, 2008MNRAS.386.2009C,
  2009ApJ...701.1534A} and the age/gravity-dependence of the L/T
transition \citep[e.g.,][]{2006ApJ...651.1166M, 2006astro.ph..9464L,
  2009ApJ...699..168D, 2009ApJ...702..154S, 2010ApJ...723..850B,
  2011ApJ...733...65B}.

Resolved photometry for binaries of known distance offers perhaps the
clearest view of the L/T transition for field objects, since the two
components of each system represent a single isochrone of common
(albeit unknown) metallicity. In addition, most pairs of binary
components have very similar surface gravity, given the nearly
constant radii of all old ($\gtrsim$0.5~Gyr) substellar objects and
the prevalence for brown dwarf binaries to have mass ratios near
unity. Finally, higher order multiplicity is very rare among ultracool
binaries, with DENIS-P~J0205$-$1159 being the only clear example
\citep{2005AJ....129..511B} out of hundreds of objects that have been
imaged with AO and \HST. Thus we can consider each binary component to
be a truly single object, with much less concern about complications
from unresolved binarity, as compared to studying the entire field
sample.

To date, study of the L/T transition with binary components has been
hampered by the small sample available. Previously, only six L/T
binaries with at least one component in the L6--T5 range had both a
measured parallax and resolved multi-band near-IR photometry. Four of
these had \HST/NICMOS photometry covering the $J$ and $H$ bands:
SDSS~J0423$-$0414AB \citep{2005ApJ...634L.177B}; SDSS~J1021$-$0304AB
\citep{2006ApJS..166..585B}; and 2MASS~J0850+1057AB and
2MASS~J1728+3948 \citep{2011AJ....141...70B}. Two had full $JHK$
coverage from ground-based photometry: $\epsilon$~Ind~Bab
\citep{2004A&A...413.1029M, 2010A&A...510A..99K} and
2MASS~J1534$-$2952AB \citep{2008ApJ...689..436L}. By chance, three of
these six also had significant problems with their published parallax
values (i.e., errors underestimated by 2--3$\times$ or contaminated by
an unresolved background star). Our new parallaxes and resolved
photometry greatly expands this sample, resulting in a total of 19
binaries with at least one component in the L/T transition (L6--T5).
We present Keck photometry for 12 of these binaries and high-precision
parallaxes for 15 of them (9 new, 6 significantly improved). Thus, we
have increased the sample of L/T binaries by at least a factor of 3,
or more than a factor of 6 if problems with literature parallaxes are
considered. Note that we have also added two new parallaxes for single
objects in the transition, SDSS~J0000+2554 (T4.5) and 2MASS~J1503+2525
(T5).

\subsection{Magnitudes and Colors in the L/T Transition}

The significant increase in the number of objects with parallaxes and
multi-band infrared photometry provided by our work motivates a new
look at the absolute magnitudes and colors of objects spanning the L/T
transition. We examine two primary diagnostics: (1)~the absolute
magnitude as a function of spectral type and (2)~the color--magnitude
diagram. Our work here almost doubles the number of objects that can
be considered and increases the number of resolved binaries by nearly
a factor of three. Thus a much richer view of the transition's
spectrophotometric behavior is revealed. This is particularly
noteworthy for the peak of the $J$-band flux inversion, which was
previously mapped by only three T3--T4.5 objects with parallaxes (two
of which had 0.3~mag uncertainties in their distance moduli).  Our
compilation (Tables~\ref{tbl:nir-absmag} and \ref{tbl:mir-absmag})
adds 5~more objects with substantially higher precision parallaxes in
this spectral type range.

\subsubsection{Absolute magnitude dependence on spectral type}

We first examine the behavior of absolute magnitude as a function of
spectral type in Figures~\ref{fig:spt-abs-jhk} and
\ref{fig:spt-abs-yl} (all objects) and Figure~\ref{fig:spt-abs-bin}
(binary components only). The plots are consistent, both showing the
increase in $J$-band flux for the early/mid-T~dwarfs relative to the
late-L~dwarfs and the later T~dwarfs. The brightening effect is also
seen in $Y$ band, becoming more of a plateau at $H$ band, and then
showing largely monotonic behavior at $K$ band (see also
\citealp{2010ApJ...710.1627L}). We quantify the amplitude of this
brightening by using the weighted mean of absolute magnitude as a
function of spectral type from Table~\ref{tbl:spt-abs-mko}, which
shows a local flux minimum at $\approx$L8 and a local peak at
$\approx$T4.5. The difference between these extrema is 0.7~mag in the
$Y$ band and 0.5~mag in the $J$ band (MKO).  Fitting a line to the
tabulated fluxes over this spectral type range gives similar results,
with a brightening of 0.8~mag in $Y$ and 0.3~mag in $J$.  In
comparison, the flux decreases over this same range of spectral types
by 0.5~mag in $H$ and 1.4~mag in $K$.

If instead we gauge the brightening effect relative to the brightest
object in the L/T transition, 2MASS~J0559$-$1404 (T4.5, $M_J =
13.49\pm0.06$~mag), these values would be $\approx$0.7~mag larger.
This object is discussed in Section~\ref{sec:overl}, where we find
that its near- and mid-IR magnitudes are unlikely to be consistent
with a pronounced brightening due to unusual cloud properties. Rather,
the simplest explanation is that this object is an unresolved, nearly
equal-flux binary, and thus its photometry should not be used to
assess the $J$-band bump. The next brightest object in the L/T
transition is SDSS~J0000+2554 (T4.5, $M_J = 13.98\pm0.08$~mag).

Note that previous studies have referred to the ``amplitude'' of the
$J$-band brightening, with this term being used loosely
\citep{2002AJ....124.1170D, 2003AJ....126..975T, 2004AJ....127.2948V}.
This lack of specificity was appropriate, given the small sample of
transition objects --- the description of the phenomenon was largely
based on the outstanding object 2MASS~J0559$-$1404, which was
$\sim$1~mag brighter compared to the late-L and mid/late-T~dwarfs in
those earlier studies. With larger parallax samples now available,
some care is warranted when using this description. In particular, the
cited amplitude of the brightening sometimes comes from comparing the
brightest mid-T dwarfs with low-order polynomial fits to the absolute
magnitudes for L~and T~dwarfs \citep{2008ApJ...685.1183L,
  2010ApJ...710.1142B}. Since polynomial fits are a convenient, but
nonphysical, model for the large changes in magnitude as a function of
spectral type, they inevitably do not provide a good match to all the
data and serve to artificially enhance the outlier nature of the
$\approx$T3--T4 objects. Thus, benchmarking the $J$-band behavior
against polynomial fits should now be superseded by a direct
comparison of measured absolute magnitudes as a function of spectral
type (Tables~\ref{tbl:spt-abs-mko} and \ref{tbl:spt-abs-2mass};
Figure~\ref{fig:spt-wavg}).

\subsubsection{Near-infrared color--magnitude diagrams \label{sec:lt-cmd}}

Perhaps the most natural representation of the L/T transition can be
found in near-IR CMDs (Figures~\ref{fig:cmd-mko-zoom} and
\ref{fig:cmd-2mass-zoom}). Here, the view of the transition is much
clearer, as the large change in near-IR colors over a small range in
spectral type is displayed with a long horizontal extent in the CMD.
Objects in the $J$-band bump appear as the brightest objects in the
blue vertical locus of the mid/late-T dwarfs, with 2MASS~J0559$-$1404
being the most protruding object.

Figures~\ref{fig:cmd-mko-zoom} and \ref{fig:cmd-2mass-zoom} shows the
CMDs assembled from resolved binary components, focusing on the L/T
transition region. The distribution of the components is in accord
with the CMD of the entire sample of objects, suggesting that
unresolved binarity is not a significant issue for the latter. With
this much larger sample of objects compared to previous work, one new
feature appears: there is a ``gap'' in the color distribution in the
transition, with many fewer objects seen with $(J-H)_{\rm MKO} \approx
0.1$--0.3~mag and $(J-K)_{\rm MKO} \approx 0.0$--0.4~mag as compared
to redder (early-T and late-L) or bluer (mid-T) objects. There is no
corresponding gap in $H-K$, and thus the above ranges in color appear
to due almost entirely to changes in the $J$-band flux at fixed $H-K$
color. (However, note that there appears to be a separate, much less
prominent gap in $H-K$ color just blueward of the red L~dwarf
sequence.) Since the density of objects in the CMD is related to the
lifetimes of the various evolutionary phases, the natural
interpretation is that the gap reflects the shortest lived-phase of
the L/T transition, shorter than the hotter or cooler stages.  We also
note that this gap appears when simply plotting the weighted averages
of magnitudes binned by spectral type (Figure~\ref{fig:cmd-wavg}).

To highlight the gap, Figure~\ref{fig:col-hist} shows the histogram of
near-IR colors for the range of absolute magnitudes representative of
the L/T transition. In addition to the L/T gap, these plots also
suggest a pileup of objects redward of the gap. This finding is highly
evocative of work by \citet{2008ApJ...689.1327S}. They combine
evolutionary models with dusty model atmospheres to simulate the
distribution of objects in the near-IR CMD. To model the L/T
transition, they build a ``hybrid'' prescription that combines the
hotter dusty atmospheres with the cooler dustless ones, by linearly
interpolating the surface boundary conditions in the model atmospheres
from 1400~K to 1200~K.  Such an approach produces a pileup of objects
in this transition temperature range (see their Figure~13), as the
hotter dusty objects must release more energy to transform into a
cooler dust-free object than compared to objects which do not change
cloud properties.  Their simulated CMD shows a pileup of L/T objects
at $(J-K)_{\rm MKO} \approx 1.0$~mag, which they discuss extensively,
and a relative paucity of objects at $(J-K)_{\rm MKO} \approx
0.2$--0.6~mag, which they do not discuss. While the model-predicted
colors of these features may not exactly match our data, the
qualitative agreement is compelling. Our binary component CMDs suggest
a prolonged stage of brown dwarf color evolution during which
condensate clouds slowly dissipate before rapidly transitioning to
bluer near-IR colors in the last stages of condensate
removal. Although the CMDs most directly probe the \emph{color}
evolution of brown dwarfs (i.e., cloud dispersal), in the theoretical
perspective of \citet{2008ApJ...689.1327S} this pileup and gap are
inextricably tied to \emph{luminosity} evolution as well.

Since our collection of binary components is not a rigorously defined
sample (e.g., volume-limited or magnitude-limited), selection effects
are a natural concern but seem unlikely to fundamentally alter the
outcome.  The target lists for previous high angular resolution
searches for ultracool binaries were derived primarily from three
magnitude-limited searches: the SDSS ultracool dwarf search
\citep[e.g.,][]{2004AJ....127.3553K, 2006AJ....131.2722C}, the 2MASS
L~dwarf search \citep[e.g.,][]{2007AJ....133..439C,
  2008AJ....136.1290R}, and the 2MASS T~dwarf search
\citep[e.g.,][]{2004AJ....127.2856B}. The 2MASS searches were based on
near-IR color criteria that were inevitably incomplete from the latest
L~dwarfs to the mid-T dwarfs ($\approx$L7--T5), while the SDSS search
was based on far-red optical colors and thus sensitive to the full
range of L~and T~dwarfs. (In fact, most of the objects redward of the
$J$-band gap are from SDSS.)  Moreover, it would be highly contrived
to imagine a selection bias whereby integrated-light measurements of
binaries containing a $(J-H)_{\rm MKO}\approx0.2$ component are
avoided, while binaries with somewhat redder or bluer components are
selected, especially as absolute magnitudes are relatively constant as
a function of color across the transition. Thus, we conclude the ``L/T
gap'' is real, though more rigorous samples are needed to quantify the
relative numbers of bluer and redder objects straddling the gap. The
parallax-based census possible with upcoming all-sky surveys like
Pan-STARRS and LSST offer the most robust means to achieve this goal.

\subsection{Individual L/T Binaries of Interest}

A few objects warrant discussion based on comparison of our results
with previous work:
\begin{itemize}

\item SDSS~J0423$-$0414AB (T0 integrated-light near-IR type):
  \citet{2010ApJ...710.1142B} decompose the integrated-light spectra
  based on the \citet{2005ApJ...634L.177B} \HST/NICMOS $F110W$ and
  $F170M$ resolved photometry and find $\Delta{K} = 1.13\pm0.07$~mag,
  in excellent agreement with our observed $\Delta{K} =
  1.18\pm0.08$~mag from Keck LGS AO.

\item SDSS~J1021$-$0304AB (T3 integrated-light near-IR type):
  \citet{2006ApJS..166..585B} resolved this system into a binary with
  \HST/NICMOS and based on spectral decomposition suggested it shows a
  $J$-band flux inversion. This is seen for the first time with our
  Keck LGS AO data, making this the fourth system to show a flux
  inversion after 2MASS~J1728+3948 (see below), SDSS~J1534+1615, and
  2MASS~J1404$-$3159. More recent spectral decomposition by
  \citet{2010ApJ...710.1142B} derive $J$ and $K$-band flux ratios of
  0.16$\pm$0.41~mag (i.e., no brightening) and 1.46$\pm$0.29~mag,
  respectively. Within the large fitting uncertainties, this is
  consistent with our LGS AO measurements of $-$0.10$\pm$0.03~mag
  (i.e., a $J$-band flux reversal) and 1.00$\pm$0.03~mag.

\item 2MASS~J1404$-$3159AB (T3 integrated-light near-IR type):
  \citet{2008ApJ...685.1183L} published this object as a binary using
  the same Keck data as presented here. Our flux ratio measurements
  are consistent with theirs within the quoted errors.  Our
  uncertainties are a factor of 2--4$\times$ smaller, which likely
  stems from the different analysis methods. The key differences are
  that \citet{2008ApJ...685.1183L} manually adjust their image
  subtractions, use aperture photometry, and choose PSF reference
  stars that do not necessarily match the science data.

\item 2MASS~J1728+3948AB (L7 integrated-light optical type):
  \citet{2003AJ....125.3302G} identified this system as a binary from
  \HST/WFPC2 optical far-red imaging. This was the first known
  ultracool binary to show an inversion in its flux ratios with
  wavelength, where the earlier-type component (identified as being
  the optically bluer object) was brighter in $F814W$ but fainter in
  $F1042M$.  Interestingly, our Keck LGS AO $J$-band imaging shows no
  inversion in the $JHK$ flux ratios, indicating the
  wavelength-dependent behavior of the brightening can be rather
  complex.  (This assumes variability effects between the
  non-simultaneous \HST\ and Keck data are negligible.) To date, this
  binary is the only one with direct evidence for the brightening
  phenomenon extending as blue as 1~\micron, though spectral
  decomposition suggests this occurs in other binaries (e.g.,
  Figures~\ref{fig:spec1}--\ref{fig:spec3}, and also see
  \citealp{2010ApJ...710.1142B}).

\item SDSS~J2052$-$1609AB (T1: integrated-light near-IR type): This object
  was identified as a weak candidate for binarity by
  \citet{2010ApJ...710.1142B} based on spectral decomposition and
  subsequently resolved by \citet{2011A&A...525A.123S} with VLT NACO
  in 2009.  We present here an independent identification of this
  binary, obtained almost 4~years earlier in 2005. The flux ratios in
  $J$ and $K$ bands are consistent between VLT and Keck, but the
  $H$-band flux ratio appears to have changed from 0.33$\pm$0.07~mag
  in 2005 to 0.57$\pm$0.01~mag in 2009.

\end{itemize}


\section{Conclusions}

We present here the first results from our ongoing high-precision
infrared astrometry program at CFHT targeting ultracool dwarfs (M6 to
$>$T9).  We have found that CFHT/WIRCam offers an excellent platform
for measuring parallaxes to ultracool objects, given its relatively
large aperture and the excellent seeing on Mauna Kea.  Queue
scheduling at CFHT is also a major advantage, as it enables good
parallax phase coverage for targets widely distributed on the sky with
almost no impact from poor weather.  Queue mode also allows data to be
obtained only during the times of best seeing while also following
rigorous airmass constraints to eliminate the effects of differential
chromatic refraction.  The work we present here is the first to use
CFHT/WIRCam for precision astrometry.

Using CFHT/WIRCam data collected since 2007, we have measured
parallaxes for 34 binaries and 15 single objects (i.e., 83 objects in
49 systems) to a median precision of 1.1~mas (2.3\%), and the best
uncertainties are 0.7~mas (0.8\%).  For 48 objects in 29 systems we
provide the first parallax measurements.  For the 35 objects in 20
systems with published parallaxes we improve the precision in the vast
majority of cases (29 objects in 17 systems).  In these cases the
median improvement in the published parallax error is a factor of 1.7,
and as good as a factor of 5.  Comparison of targets in common between
our program and published samples provides an independent check on our
methods, and we generally find good agreement in parallax values.
However, there are more $>$2$\sigma$ outliers than is statistically
expected, and Monte Carlo simulations for these objects reveal that
this is likely because some published errors are underestimated by a
factor of $\approx$2--3.

To enable detailed analysis of the complete sample of ultracool
binaries with parallaxes, we also present here a large set of resolved
near-IR photometry obtained with Keck AO imaging and aperture masking
and archival \HST\ and VLT data.  Combining this photometry with
near-IR spectroscopy from IRTF/SpeX, we determine component spectral
types using a spectral decomposition technique.  Unlike some previous
studies, our method does not assume any relation between spectral type
and absolute magnitude so that our resulting types may be used to
assess this relationship.  Our full sample comprises 17 M6--L1 dwarfs,
27 L1.5--L8 dwarfs, 22 L8.5--T5 dwarfs, and 17 $\geq$T5.5 dwarfs.
This doubles the number of L/T transition dwarfs with parallaxes and
provides many high-precision distance measurements for ultracool
binaries that will be crucial for future dynamical mass
determinations.

These first results from our ongoing CFHT program provide
high-precision parallaxes for a large sample of ultracool dwarfs,
enabling some basic quantitative tests of brown dwarf evolution. We
combine our sample of new or improved parallaxes for 74 objects with
previously published parallaxes for a total sample of 314 objects that
allows us to form an unprecedented view of the absolute magnitudes of
ultracool dwarfs and provide an update of key empirical relations:
\begin{enumerate}

\item We determine empirical relations between absolute magnitude and
  spectral type for a wide variety of near- and mid-IR photometric
  systems (MKO, 2MASS, \Spitzer/IRAC, and \WISE). We compute simple
  polynomial fits to these relations but suggest that using the actual
  tabulated values of mean and rms absolute magnitude is preferred for
  quantitative analysis.

\item We are able to quantify the intrinsic scatter in absolute
  magnitude at a given spectral type with our high precision
  parallaxes.  As expected, this reveals relatively small intrinsic
  variations in the near-IR among late-M dwarfs (0.1--0.3~mag) that
  increases for L~dwarfs (0.3--0.5~mag) as dust properties become an
  important ``second parameter'' after \Teff.  We also identify a
  large, previously unappreciated amount of intrinsic scatter among
  mid- to late-T dwarfs in the near-IR (0.3--0.8~mag), presumably due
  to metallicity and surface gravity variations in the field
  population.

\item We identify astrometric perturbations due to orbital motion in
  three targets: SDSS~J0805+4812AB, previously suggested to be a
  binary based on its unusual spectrum; and the known binaries
  2MASS~J0518$-$2828AB (L6.5+T5) and 2MASS~J1404$-$3159AB (L9+T5).

\item We find evidence for unresolved, nearly equal-flux binaries
  based on their overluminosity in near- and mid-IR CMDs and spectral
  type--absolute magnitude relations: 2MASS~J0559$-$1404 (T4.5), which
  was previously known to be overluminous; SDSS~J1504+1047 (T7), for
  which we measure a parallax for the first time; and
  SDSS~J1021$-$0304A (T$0\pm1$), which our 3.5$\times$ improved
  parallax precision reveals lies $\approx$0.5~mag above the L/T
  transition in near-IR CMDs and which is unusually bright for its
  spectral type.  If SDSS~J1021$-$0304A is indeed binary, it would be
  a member of a hierarchical triple with SDSS~J1021$-$0304B (T5).
  This idea can be tested with a dynamical mass for the system in the
  near future.

\item Our parallax measurement for the wide pair SDSS~J1416+1348 (L6)
  and ULAS~J1416+1348 (T7.5p) shows that the components occupy unusual
  locations on near- and mid-IR CMDs. We conclude the system has lower
  metallicity than typical field dwarfs, with the possibility
  remaining that unusual cloud properties and high surface gravity
  could also be affecting the components' observed features.

\item We investigate the kinematics of all ultracool dwarfs with
  parallaxes, searching for wide common proper motion companions and
  deriving criteria for identifying likely thick disk or halo members
  based on large tangential velocities.  We identify two new candidate
  wide companions, and find that one previously identified pair is
  likely to be a chance alignment based on our improved proper motion
  (2MASS~J0850+1057AB and NLTT~20346AB).  We do not identify any new
  thick disk or halo members.

\item We find no evidence for a spectral type ``flip'' in the
  components of 2MASS~J0850+1057AB, as recently suggested by
  \citet{2011AJ....141...70B}.  We find types of L$6.5\pm1$ and
  L$8.5\pm1$, in contrast to L7+L6 from their analysis, thereby making
  2MASS~J0850+1057A normal for its spectral type and thus requiring no
  special explanation such as youth or unresolved multiplicity.

\item We have increased the sample of resolved L/T systems having
  multi-band near-IR photometry and a measured parallax by more than a
  factor of 3.  We use these resolved components to provide the
  clearest view to date of the L/T transition.  We find that the
  amplitude of the $J$-band brightening (``bump'') is
  $\approx$0.5~mag, as defined by the mean absolute magnitude as a
  function of spectral type.  As brown dwarfs cool they appear to
  reach a local minimum in $J$-band brightness at $\approx$L8.  In the
  framework of current models, this would correspond to the maximal
  suppression of $J$-band flux due to high condensate opacity.  As
  objects evolve from red to blue near-IR colors, the $J$-band flux
  increases, presumably due to cloud dissipation, reaching a local
  maximum in $J$-band flux at $\approx$T4.5.  A similar pattern is
  seen in $Y$ band, but perhaps with a larger amplitude of
  $\approx$0.7~mag.  Brightening is not seen in the $H$, $K$, and \Lp\
  bands, which instead are consistent with a monotonic decline as a
  function of spectral type.  This behavior is consistent with flux
  ratios measured in near-IR bandpasses for binaries that span the L/T
  transition \citep[e.g.,][]{2006ApJ...647.1393L, 2008ApJ...685.1183L,
    2010ApJ...710.1142B, 2011A&A...525A.123S}.

\item We find an apparent ``gap'' in the evolution of brown dwarfs as
  they traverse the L/T transition in near-IR CMDs at roughly constant
  absolute magnitude.  There is a conspicuous paucity of objects over
  specific color ranges, $(J-H)_{\rm MKO} \approx 0.1$--0.3~mag and
  $(J-K)_{\rm MKO} \approx 0.0$--0.4~mag, with no gap in $(H-K)_{\rm
    MKO}$.  Immediately redward of this gap is an apparent pileup of
  objects in $(J-K)_{\rm MKO}$ color.  This is highly evocative of the
  pileup and gap seen in the ``hybrid'' tracks of
  \citet{2008ApJ...689.1327S}, which self-consistently model brown
  dwarf evolution using a prescription for cloud dissipation at the
  L/T transition.  Regardless of the exact cloud prescription, they
  suggest that there should always be a pileup of some kind because
  hotter dusty objects must release much more energy to become cooler
  dust-free objects compared to objects that do not change dust
  properties.  (They do not discuss the subsequent gap, though it is
  apparent in their model CMDs.)  The features we observe in the
  near-IR CMDs thus indicate a slowing of color evolution at the last
  stages of condensate cloud dissipation (possibly related to a
  slowing of luminosity evolution) before brown dwarfs rapidly
  transform to their final, dust-free, blue near-IR colors
  ($\gtrsim$T4.5).

\end{enumerate}

The capability of measuring $\approx$1~mas parallaxes for faint
infrared sources is novel. We have achieved the highest precision to
date for such faint objects ($J = 13.5$--16.5~mag, and as faint as
19.7~mag at somewhat reduced precision). Although our precision goal
has initially been driven by the need for high-quality dynamical
masses, this new capability opens the door to other previously
inaccessible samples. For example, rare classes of ultracool dwarfs
are on average more distant and thus need high precision for useful
parallaxes. In addition, the faintest ultracool dwarfs known
($J\gtrsim18$~mag) are beyond the reach of previous parallax programs
but can be efficiently monitored with CFHT. Such new samples will be
the subject of our future publications.


\acknowledgments

We are deeply indebted to the CFHT staff for their constant observing
support and dedication to delivering the highest quality data
products, and in particular to Loic Albert.  We also thank Brendan P.\
Bowler, Kimberly Aller, and Mark Pitts for assistance in conducting
our IRTF/SpeX observations.
We are grateful to S.\ K.\ Leggett and Michael J.\ Ireland for
suggestions that significantly improved our analysis. We have
benefited from discussions with Jan Kleyna, Gene Magnier, Dave Monet,
John Thorstenen, Chris Tinney, John Tonry, and Fred Vrba about
astrometry and parallaxes. We are grateful to C\'{e}line Reyl\'{e} for
customized Besan\c{c}on Galaxy models.
It is a pleasure to thank Joel Aycock, Randy Campbell, Al Conrad,
Heather Hershley, Jim Lyke, Jason McIlroy, Gary Punawai, Julie
Riviera, Hien Tran, Cynthia Wilburn,
and the Keck Observatory staff for assistance with the Keck
observations. 
Our research has employed the 2MASS data products; NASA's
Astrophysical Data System; the SIMBAD database operated at CDS,
Strasbourg, France; and the SpeX Prism Spectral Libraries, maintained
by Adam Burgasser at \url{http://www.browndwarfs.org/spexprism}.
This publication makes use of data products from the Wide-field
Infrared Survey Explorer, which is a joint project of the University
of California, Los Angeles, and the Jet Propulsion
Laboratory/California Institute of Technology, funded by the National
Aeronautics and Space Administration.
This research has made use of the NASA/IPAC Infrared Science Archive,
which is operated by the Jet Propulsion Laboratory, California
Institute of Technology, under contract with the National Aeronautics
and Space Administration.
This publication has made use of contour plotting code written by
James R.\ A.\ Davenport.
T.J.D.\ and M.C.L.\ acknowledge support for this work from NSF grants
AST-0507833 and AST-0909222. M.C.L.\ acknowledges support from an
Alfred P.\ Sloan Research Fellowship. T.J.D.\ acknowledges support
from Hubble Fellowship grant HST-HF-51271.01-A awarded by the Space
Telescope Science Institute, which is operated by AURA for NASA, under
contract NAS 5-26555.
Finally, the authors wish to recognize and acknowledge the very
significant cultural role and reverence that the summit of Mauna Kea has
always had within the indigenous Hawaiian community.  We are most
fortunate to have the opportunity to conduct observations from this
mountain.

{\it Facilities:} \facility{Keck:II (LGS AO, NIRC2)}, \facility{CFHT
  (WIRCam)}, \facility{IRTF (SpeX), \facility{Spitzer (IRAC)},
  \facility{WISE}}



\begin{landscape}

\begin{figure}

\centerline{
\includegraphics[width=2.1in,angle=0]{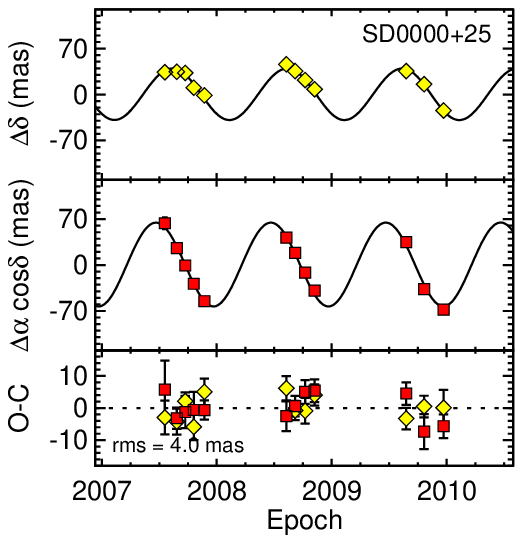}
\includegraphics[width=2.1in,angle=0]{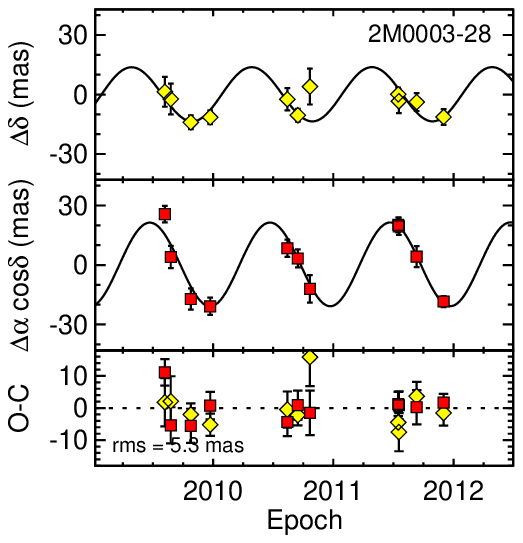}
\includegraphics[width=2.1in,angle=0]{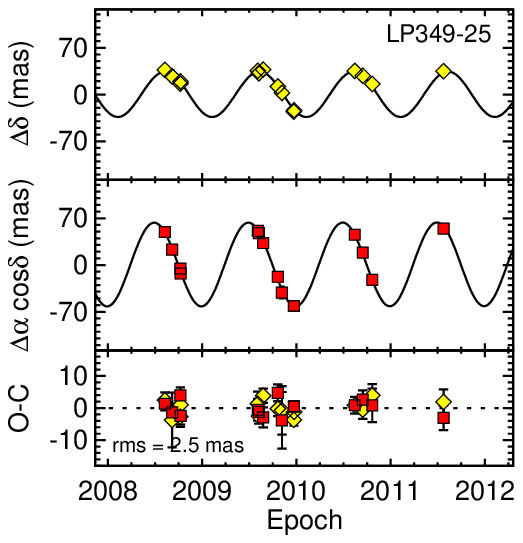}
\includegraphics[width=2.1in,angle=0]{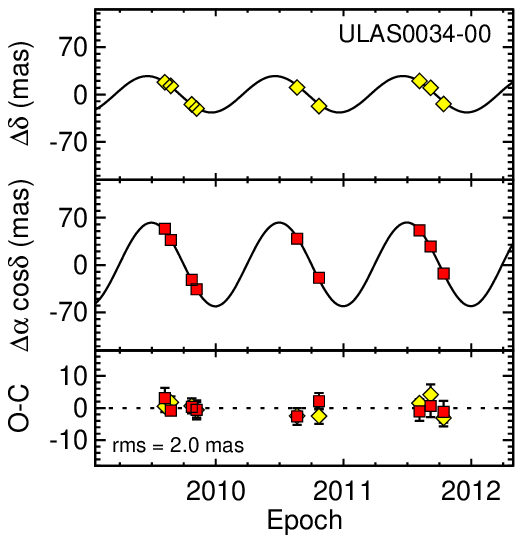}
}
\centerline{
\includegraphics[width=2.1in,angle=0]{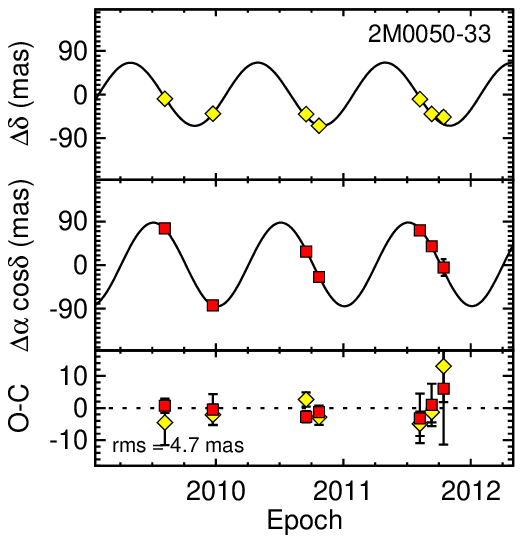}
\includegraphics[width=2.1in,angle=0]{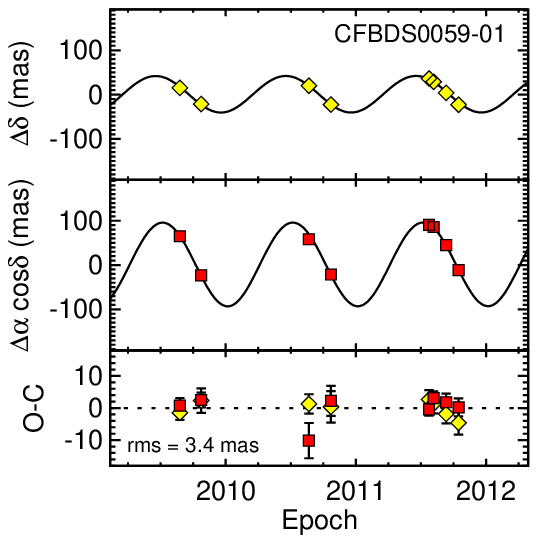}
\includegraphics[width=2.1in,angle=0]{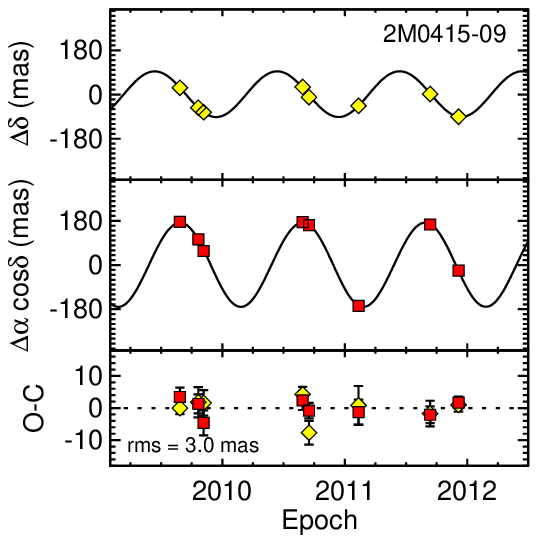}
\includegraphics[width=2.1in,angle=0]{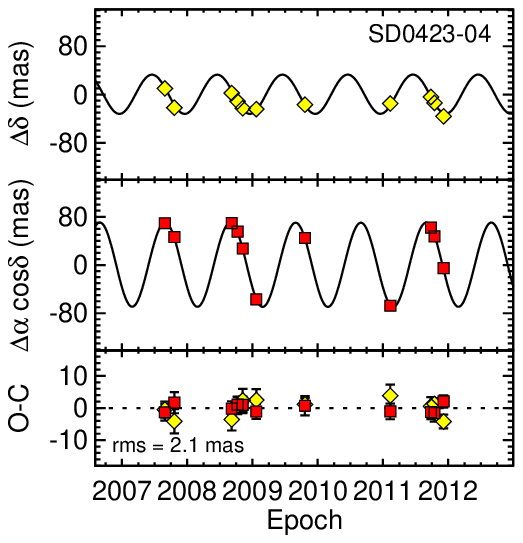}
}

\caption{\normalsize For each object, the top and middle panels show
  relative astrometry in $\delta$ and $\alpha$, respectively, as a
  function of Julian year after subtracting the best-fit proper
  motion.  (This is for display purposes only; in our analysis we fit
  for both the proper motion and parallax simultaneously.)  The bottom
  panels show the residuals after subtracting both the parallax and
  proper motion. \label{fig:plx1}}

\end{figure}

\begin{figure}

\centerline{
\includegraphics[width=2.1in,angle=0]{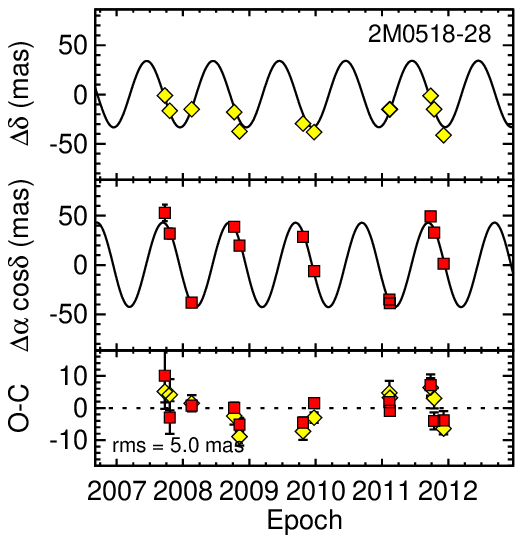}
\includegraphics[width=2.1in,angle=0]{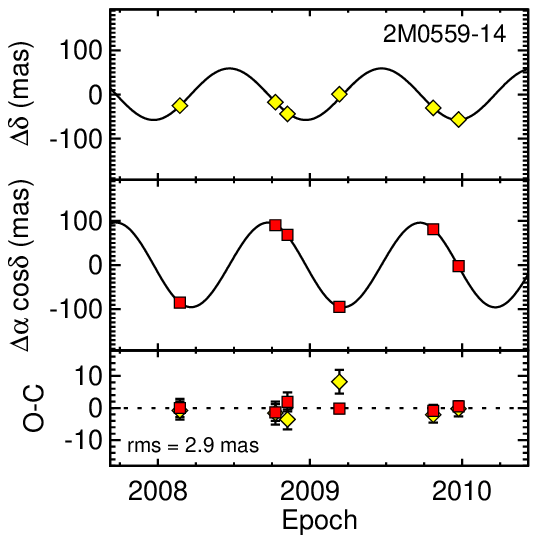}
\includegraphics[width=2.1in,angle=0]{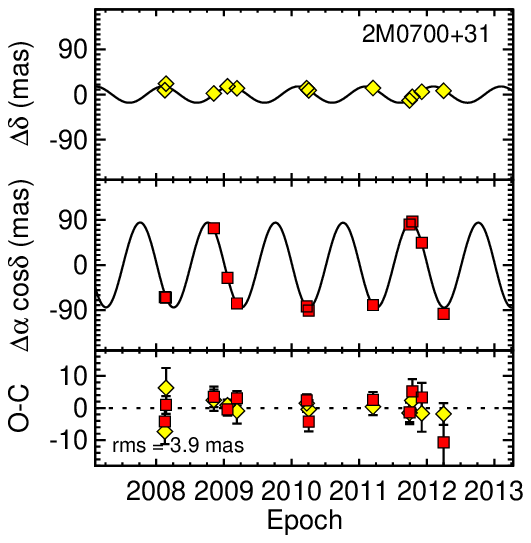}
\includegraphics[width=2.1in,angle=0]{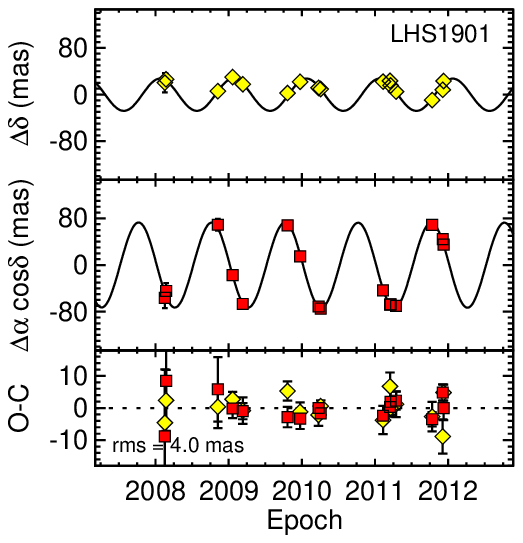}
}
\centerline{
\includegraphics[width=2.1in,angle=0]{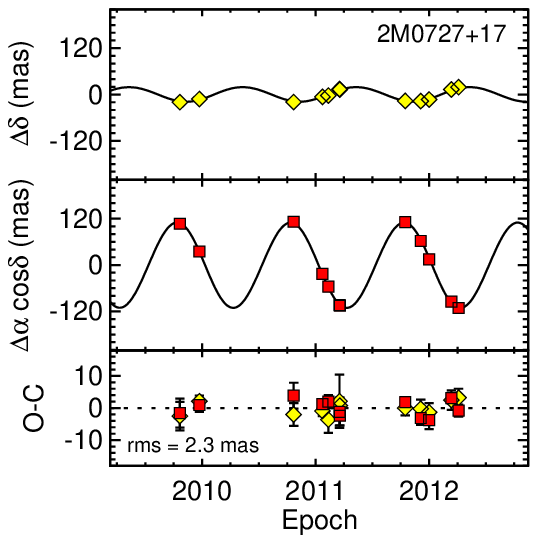}
\includegraphics[width=2.1in,angle=0]{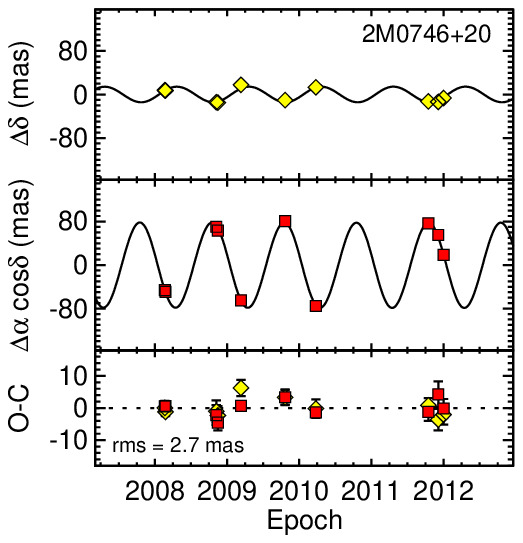}
\includegraphics[width=2.1in,angle=0]{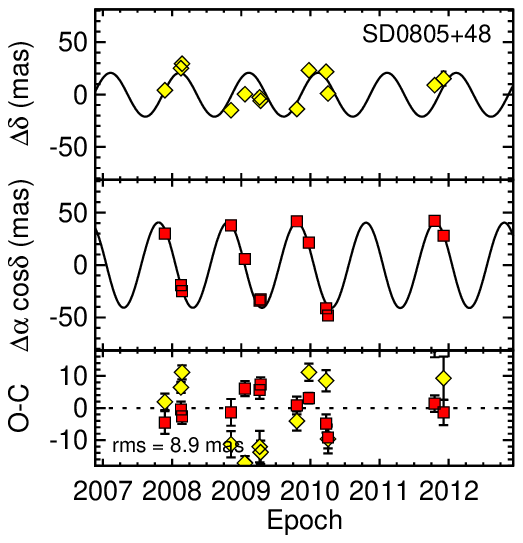}
\includegraphics[width=2.1in,angle=0]{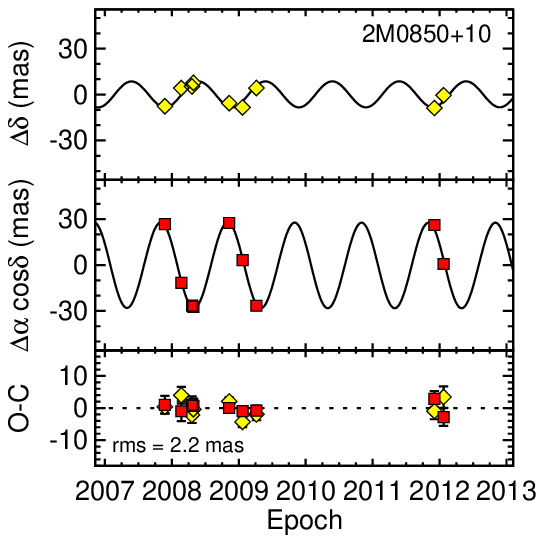}
}
\centerline{
\includegraphics[width=2.1in,angle=0]{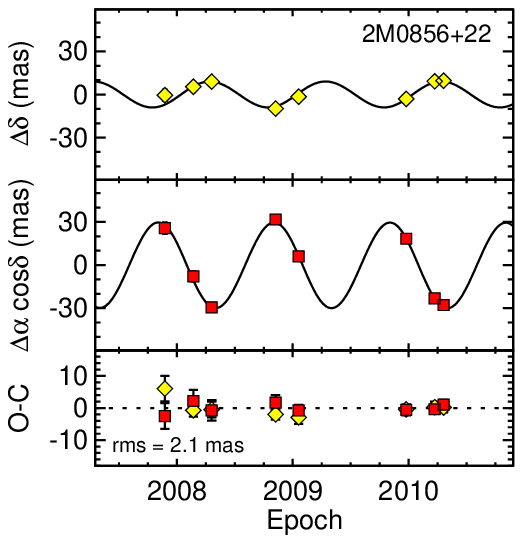}
\includegraphics[width=2.1in,angle=0]{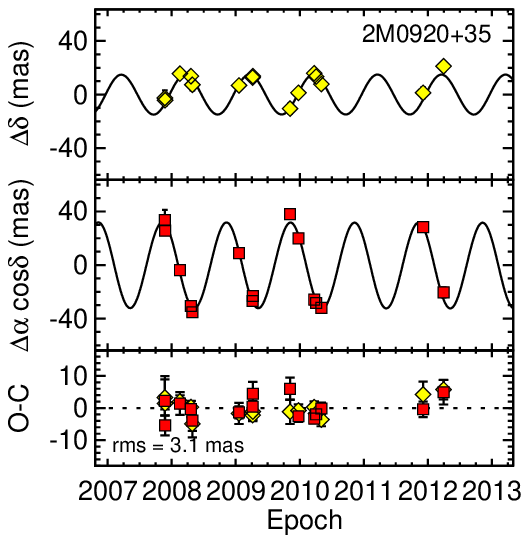}
\includegraphics[width=2.1in,angle=0]{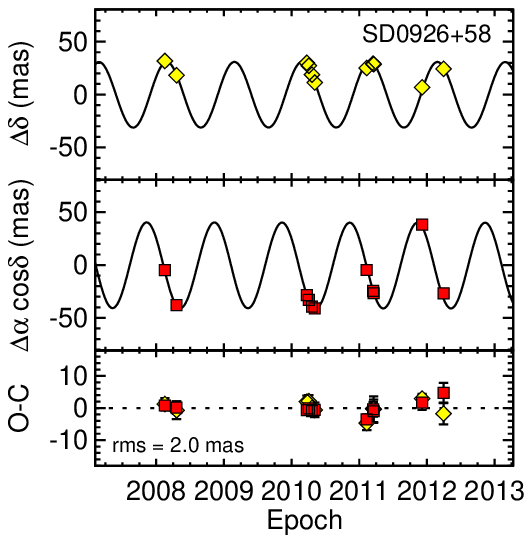}
\includegraphics[width=2.1in,angle=0]{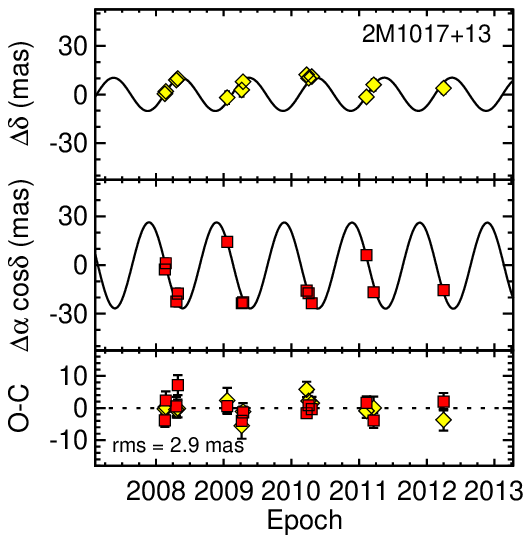}
}

\caption{\normalsize Same as Figure~\ref{fig:plx1}. \label{fig:plx2}}

\end{figure}

\begin{figure}

\centerline{
\includegraphics[width=2.1in,angle=0]{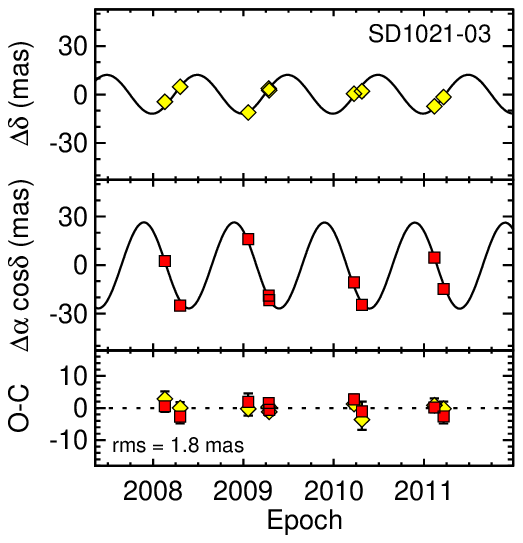}
\includegraphics[width=2.1in,angle=0]{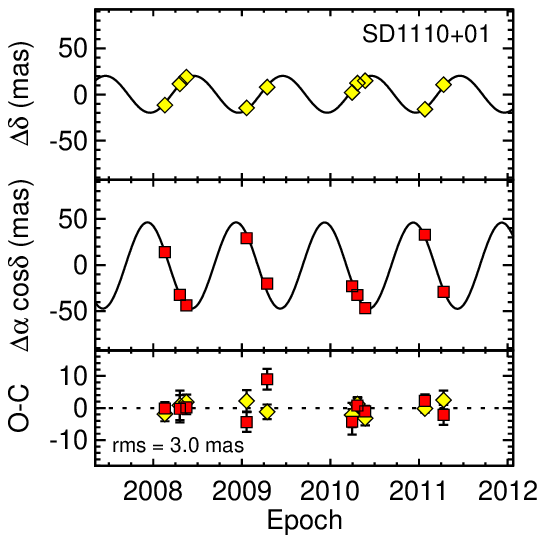}
\includegraphics[width=2.1in,angle=0]{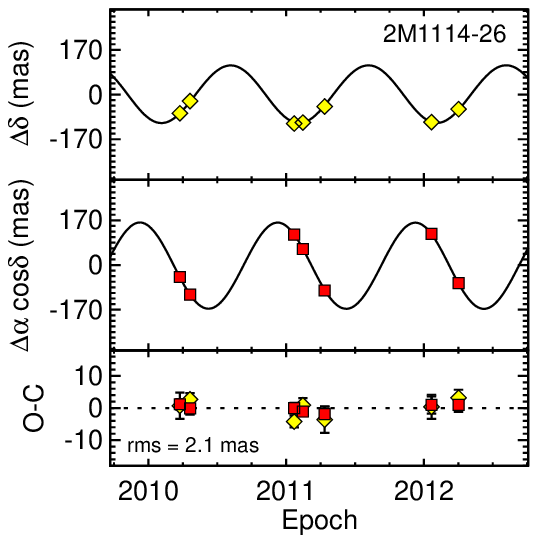}
\includegraphics[width=2.1in,angle=0]{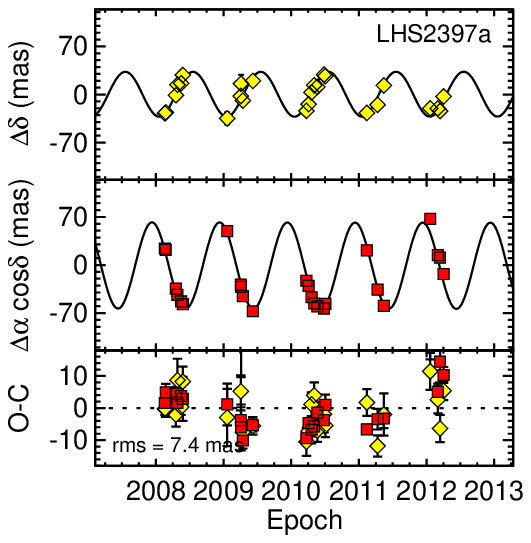}
}
\centerline{
\includegraphics[width=2.1in,angle=0]{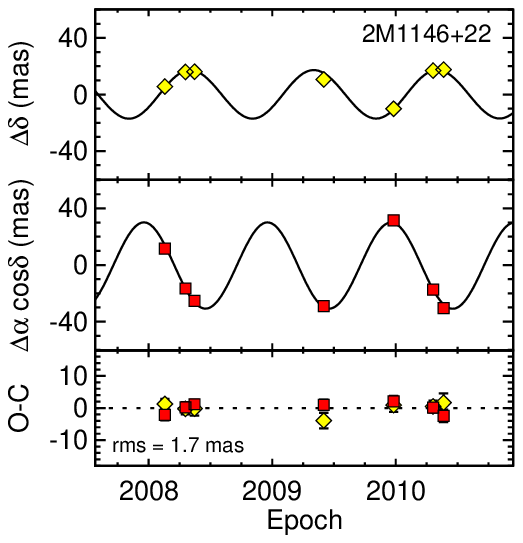}
\includegraphics[width=2.1in,angle=0]{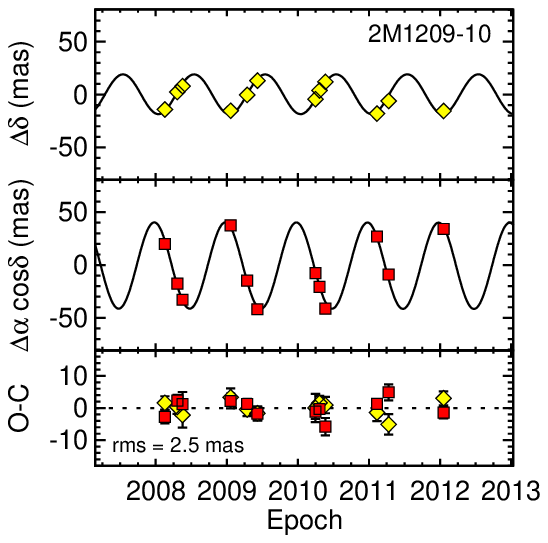}
\includegraphics[width=2.1in,angle=0]{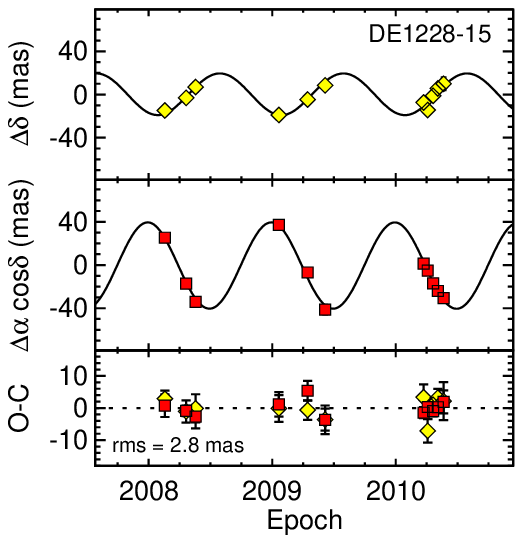}
\includegraphics[width=2.1in,angle=0]{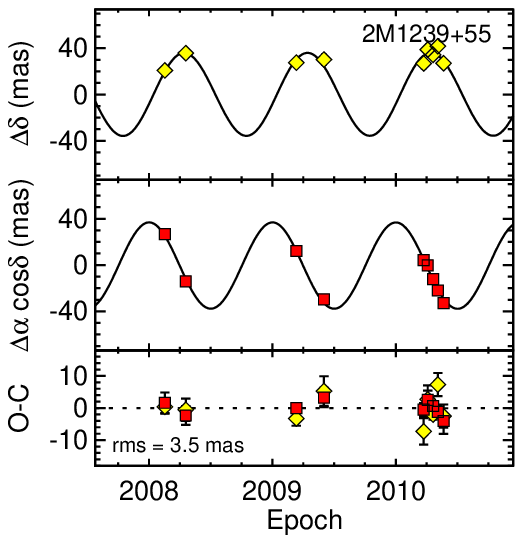}
}
\centerline{
\includegraphics[width=2.1in,angle=0]{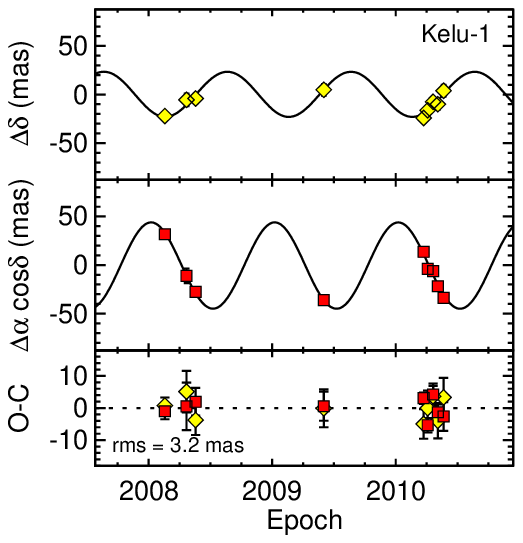}
\includegraphics[width=2.1in,angle=0]{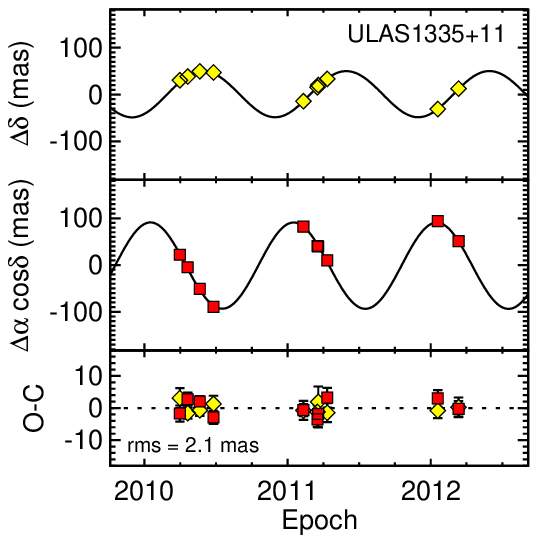}
\includegraphics[width=2.1in,angle=0]{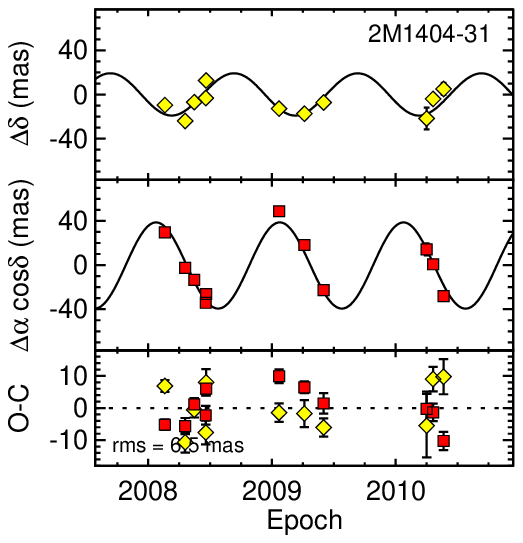}
\includegraphics[width=2.1in,angle=0]{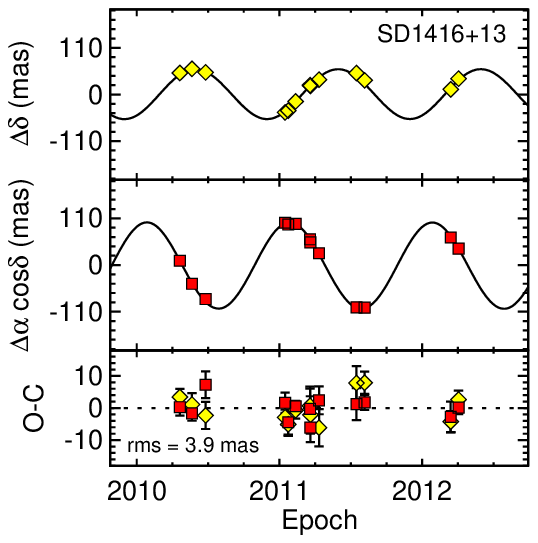}
}

\caption{\normalsize Same as Figure~\ref{fig:plx1}. \label{fig:plx3}}

\end{figure}

\begin{figure}

\centerline{
\includegraphics[width=2.1in,angle=0]{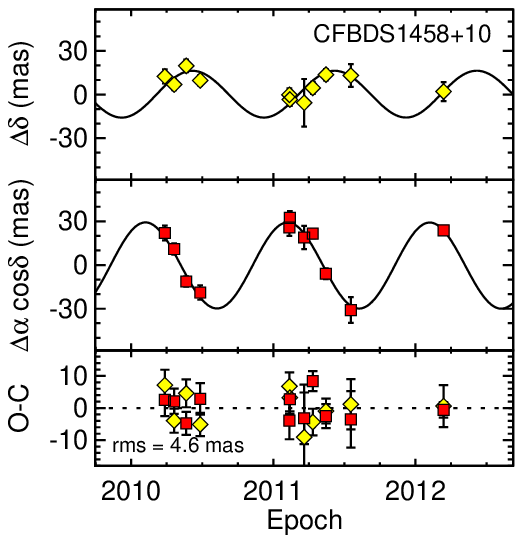}
\includegraphics[width=2.1in,angle=0]{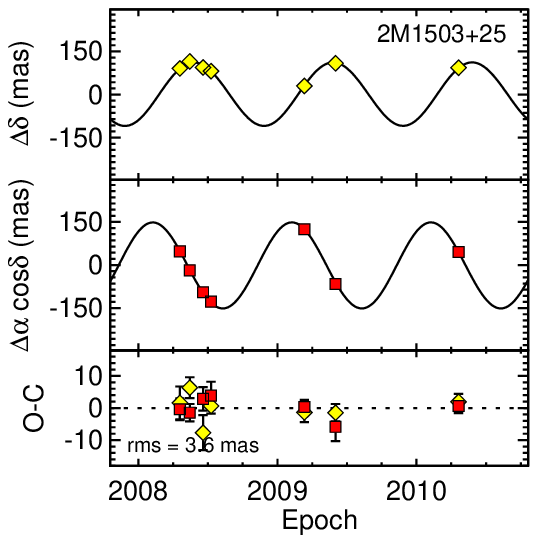}
\includegraphics[width=2.1in,angle=0]{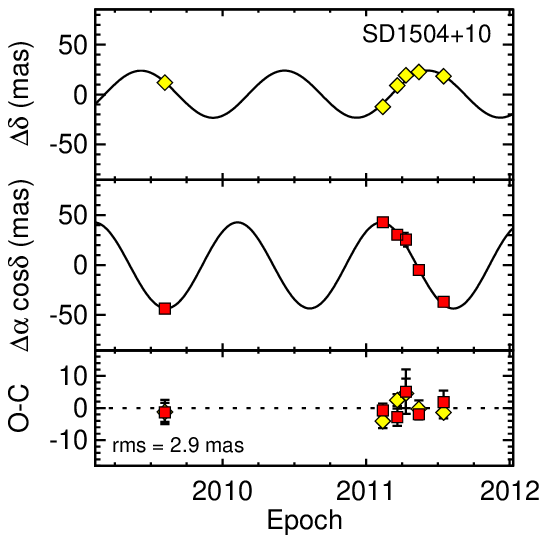}
\includegraphics[width=2.1in,angle=0]{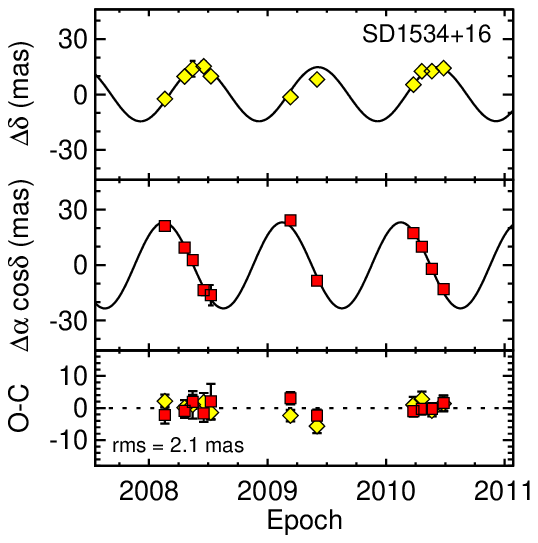}
}
\centerline{
\includegraphics[width=2.1in,angle=0]{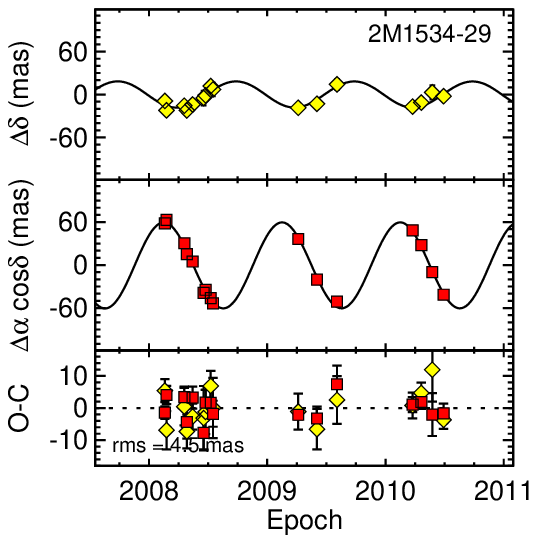}
\includegraphics[width=2.1in,angle=0]{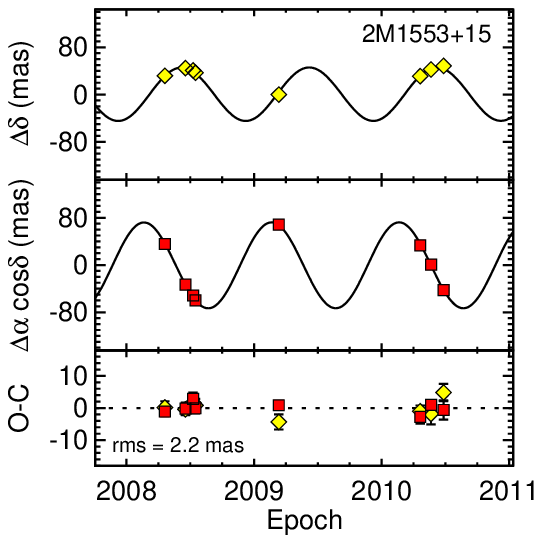}
\includegraphics[width=2.1in,angle=0]{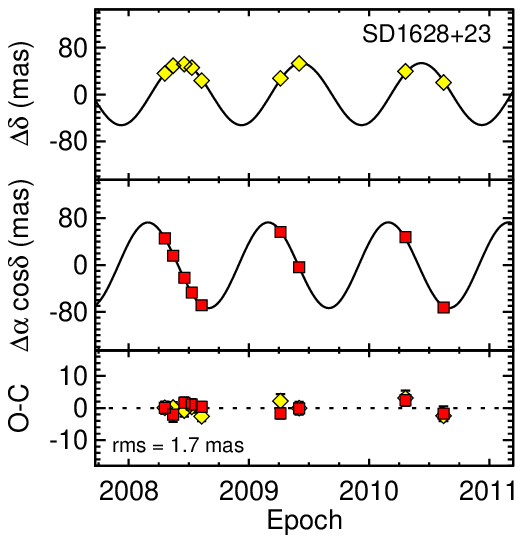}
\includegraphics[width=2.1in,angle=0]{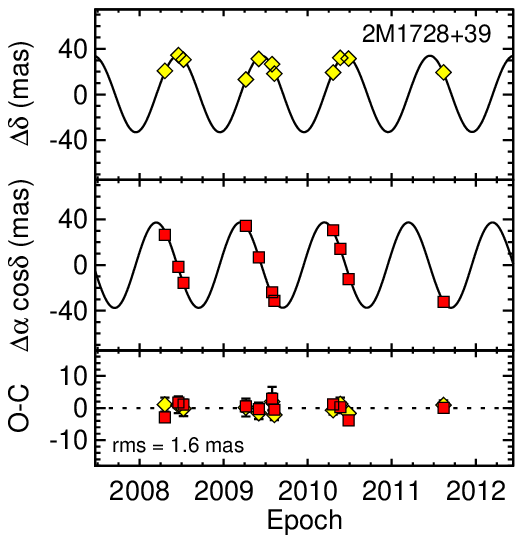}
}
\centerline{
\includegraphics[width=2.1in,angle=0]{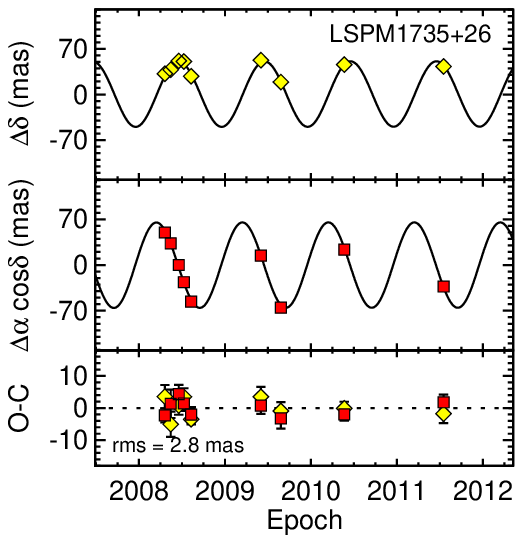}
\includegraphics[width=2.1in,angle=0]{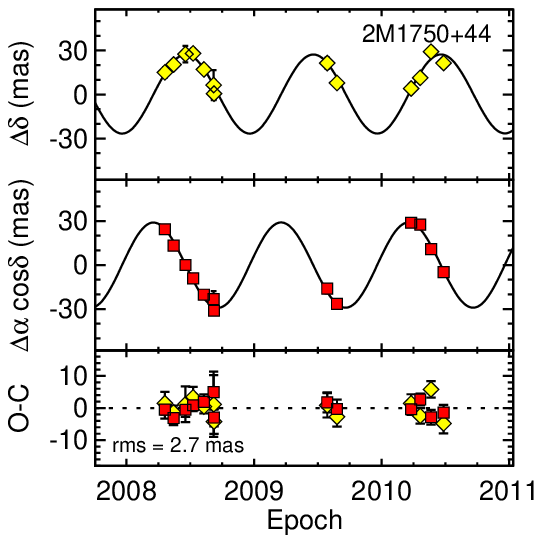}
\includegraphics[width=2.1in,angle=0]{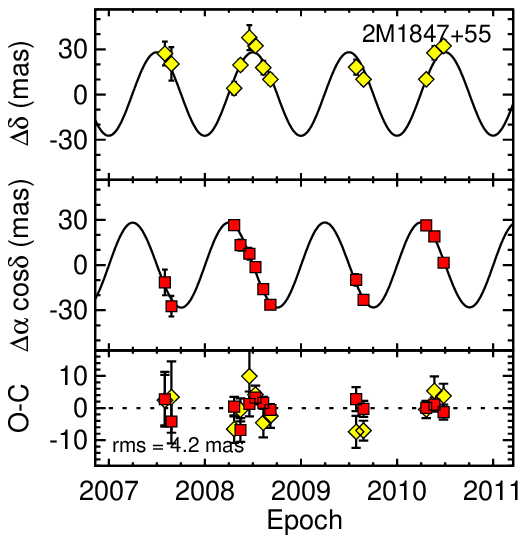}
\includegraphics[width=2.1in,angle=0]{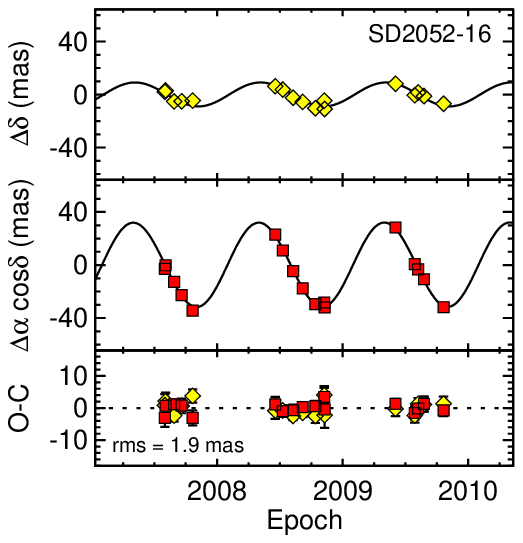}
}

\caption{\normalsize Same as Figure~\ref{fig:plx1}. \label{fig:plx4}}

\end{figure}

\begin{figure}

\centerline{
\includegraphics[width=2.1in,angle=0]{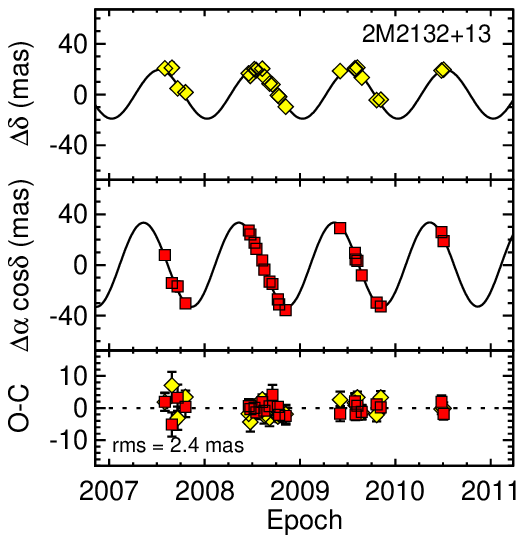}
\includegraphics[width=2.1in,angle=0]{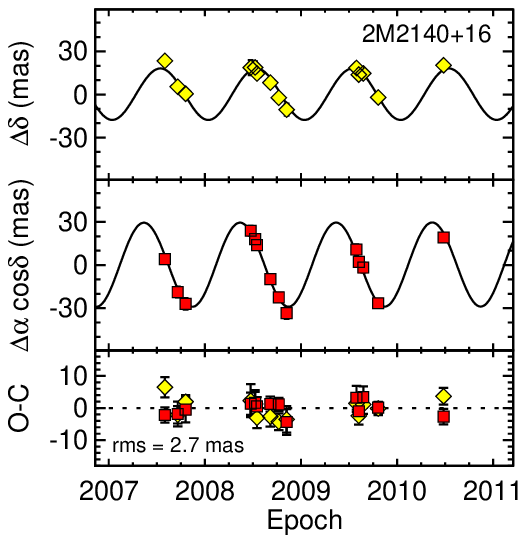}
\includegraphics[width=2.1in,angle=0]{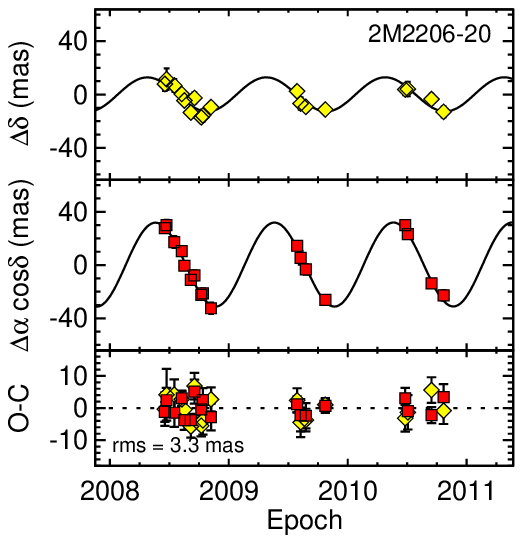}
\includegraphics[width=2.1in,angle=0]{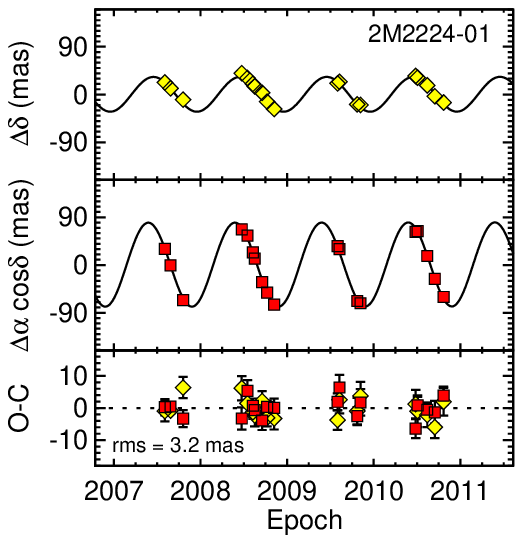}
}
\includegraphics[width=2.1in,angle=0]{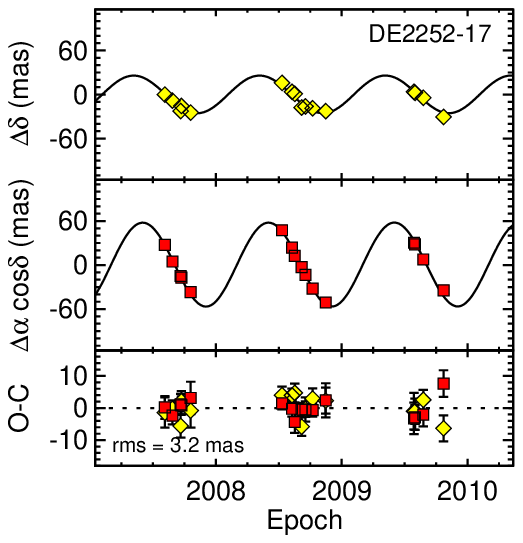}

\caption{\normalsize Same as Figure~\ref{fig:plx1}. \label{fig:plx5}}

\end{figure}

\end{landscape}

\begin{figure}

\centerline{\includegraphics[width=6.4in,angle=0]{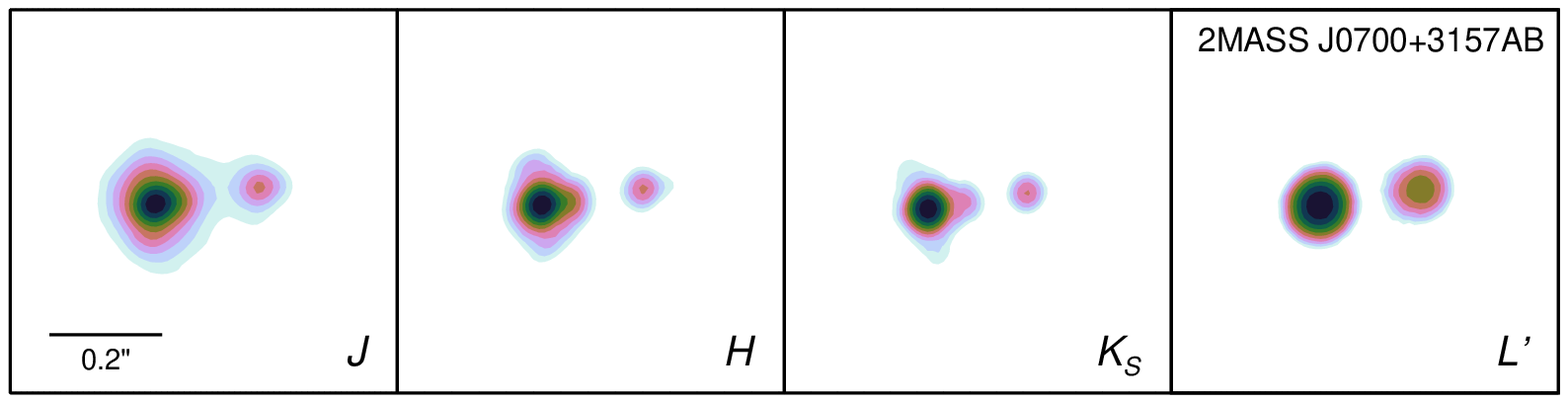}} \vskip 0.10in
\centerline{\includegraphics[width=4.8in,angle=0]{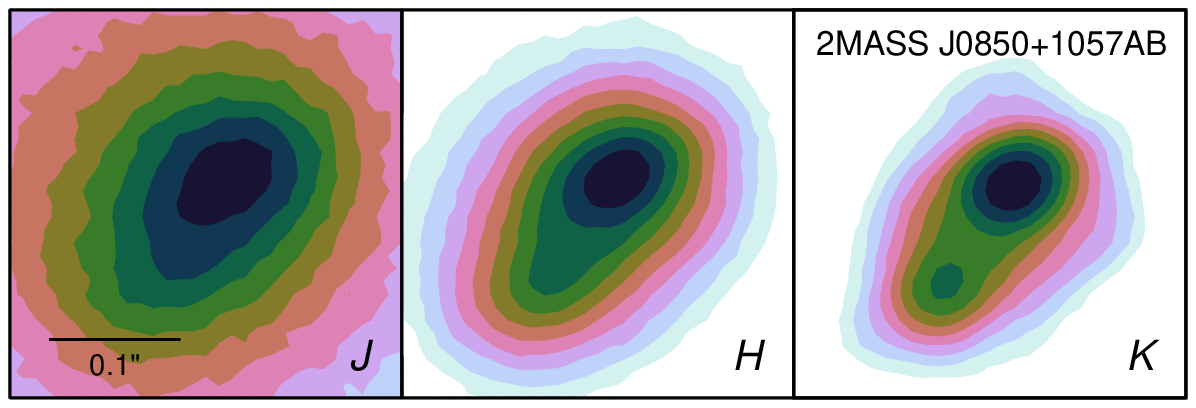}} \vskip 0.10in
\centerline{\includegraphics[width=4.8in,angle=0]{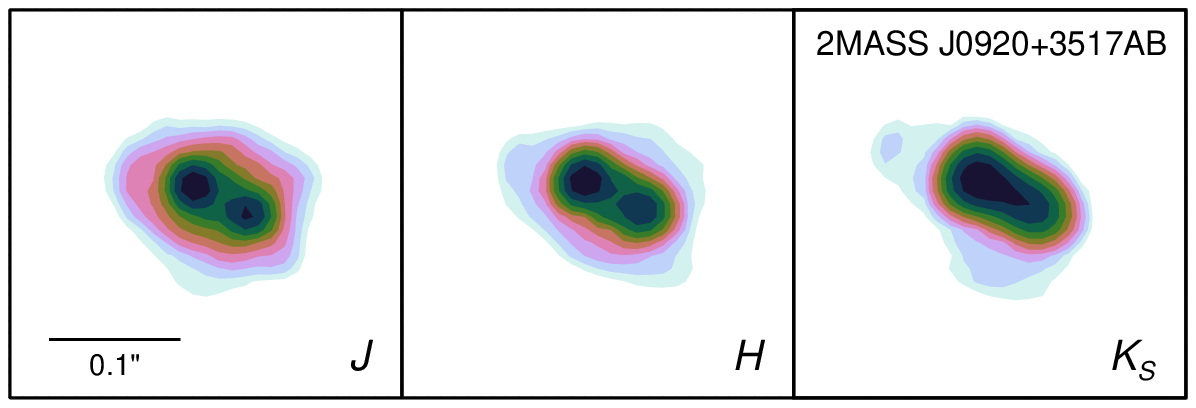}} \vskip 0.10in
\centerline{\includegraphics[width=4.8in,angle=0]{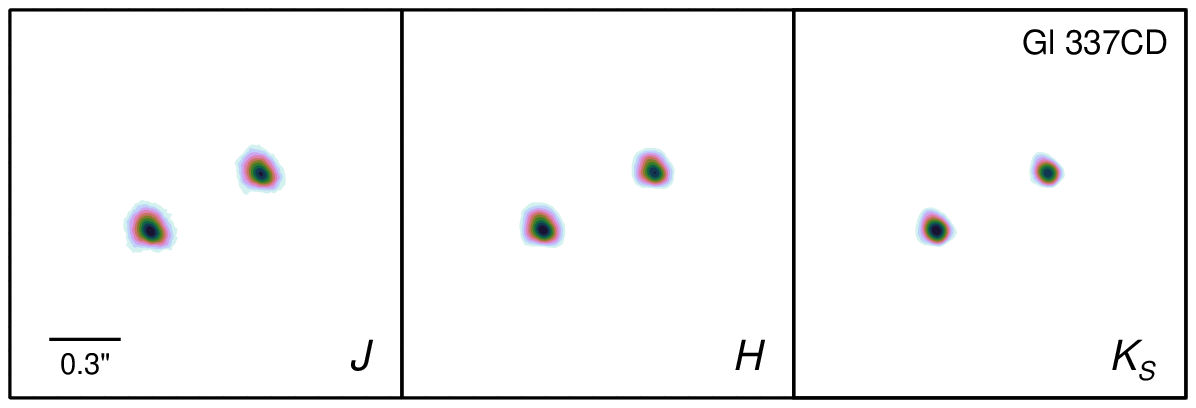}  \hskip 0.10in
            \includegraphics[width=1.6in,angle=0]{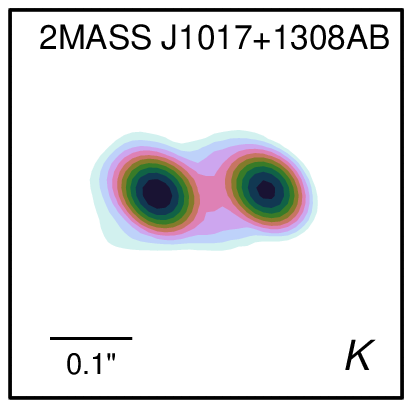}}

\caption{\normalsize Contour plots of the Keck AO images for which we
  present resolved photometry in Table~\ref{tbl:keck}.  Contours are
  in logarithmic intervals from unity to 10\% of the peak flux in each
  band.  North is always up, but the angular scale used for each
  binary varies, so scale bars are given. \label{fig:lgs1}}

\end{figure}

\begin{figure}

\centerline{\includegraphics[width=4.8in,angle=0]{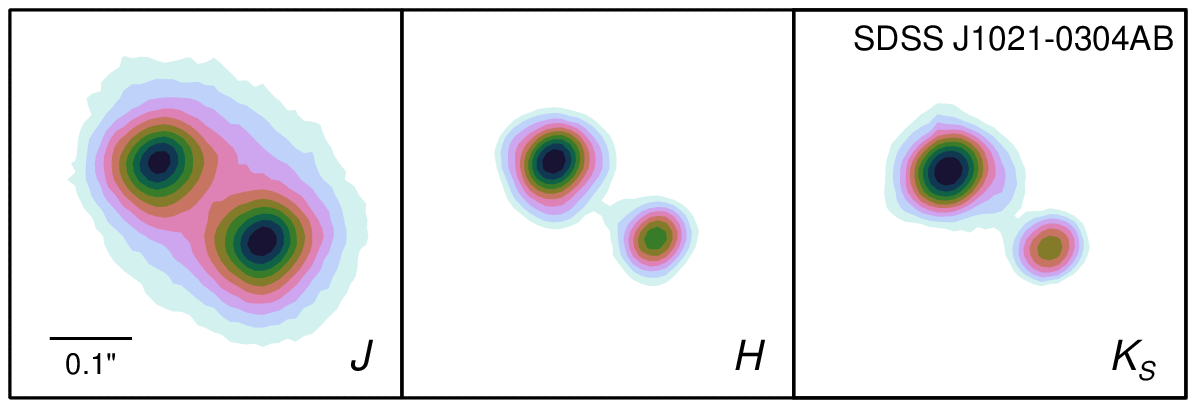}  \hskip 0.10in
            \includegraphics[width=1.6in,angle=0]{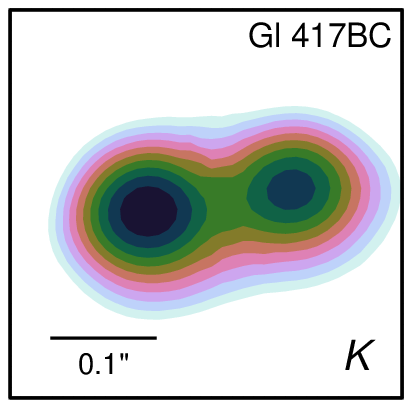}} \vskip 0.10in
\centerline{\includegraphics[width=6.4in,angle=0]{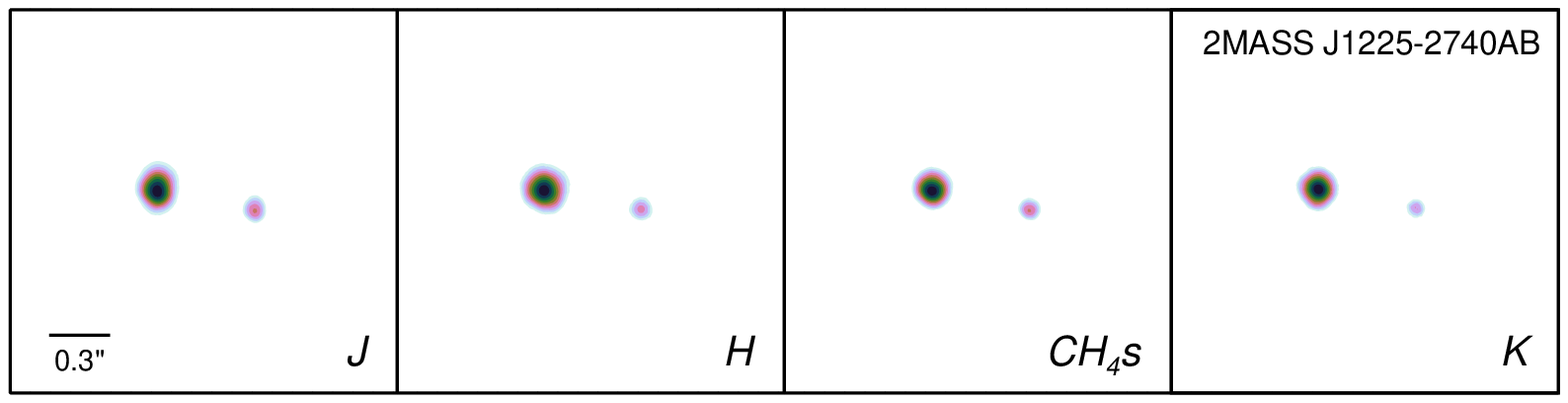}} \vskip 0.10in
\centerline{\includegraphics[width=1.6in,angle=0]{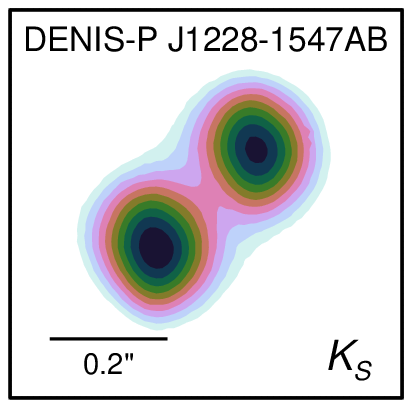}  \hskip 0.10in
            \includegraphics[width=4.8in,angle=0]{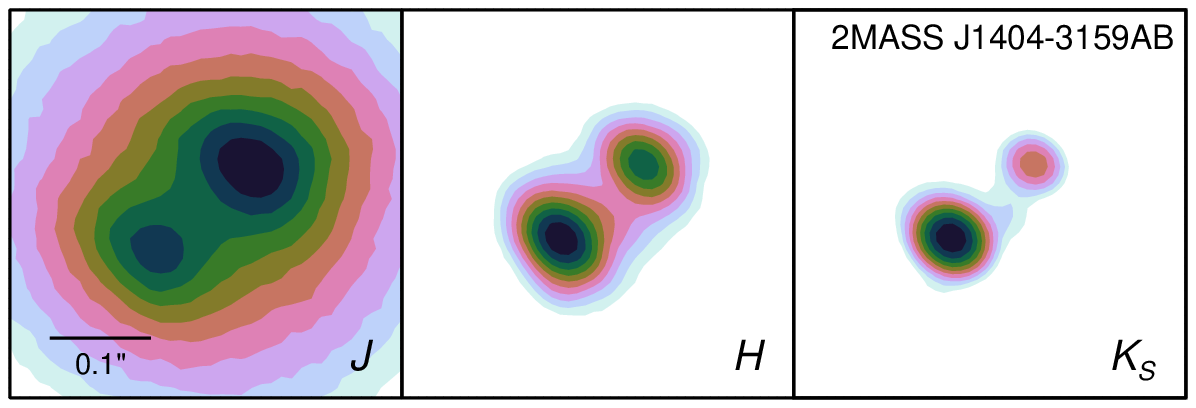}} \vskip 0.10in
\centerline{\includegraphics[width=6.4in,angle=0]{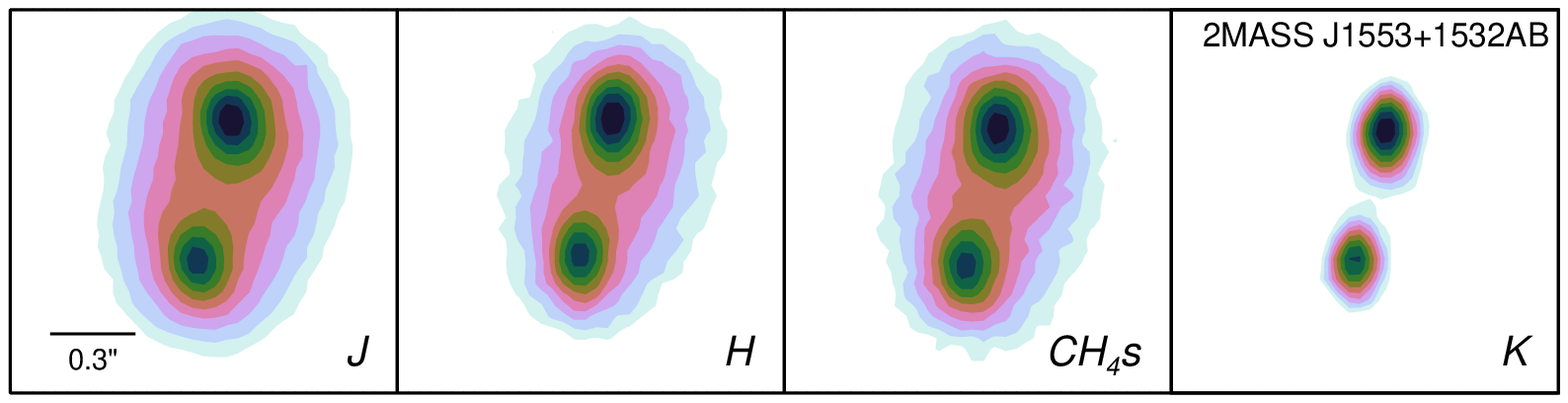}}

\caption{\normalsize Same as Figure~\ref{fig:lgs1}. \label{fig:lgs2}}

\end{figure}

\begin{figure}

\centerline{\includegraphics[width=4.8in,angle=0]{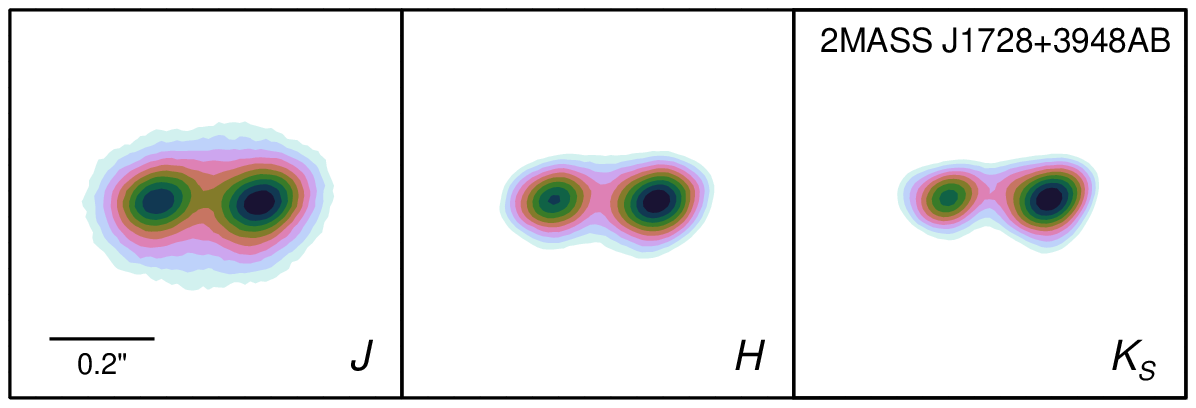}} \vskip 0.10in
\centerline{\includegraphics[width=6.4in,angle=0]{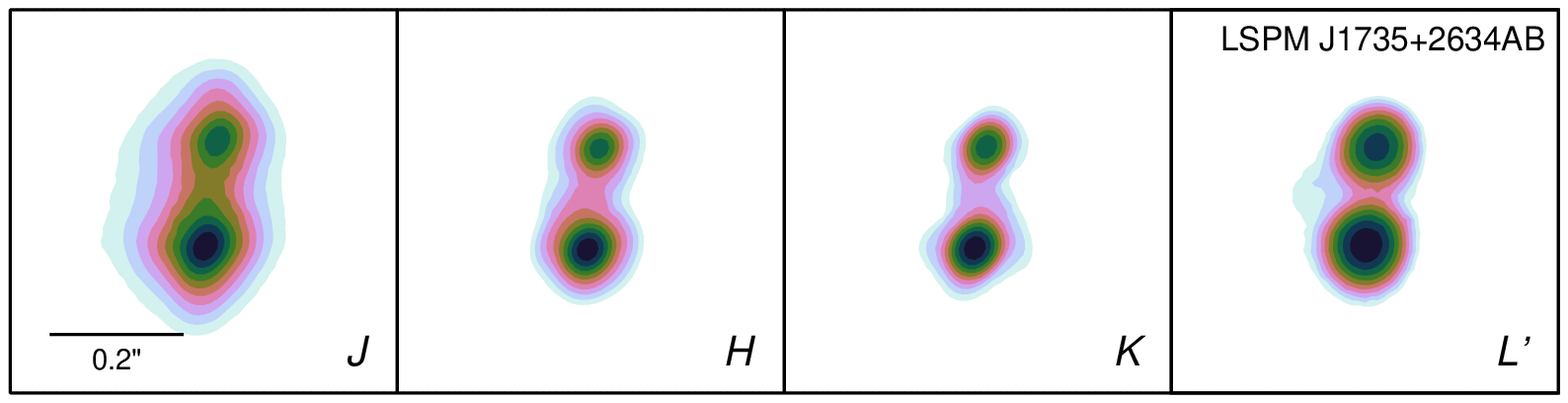}} \vskip 0.10in
\centerline{\includegraphics[width=4.8in,angle=0]{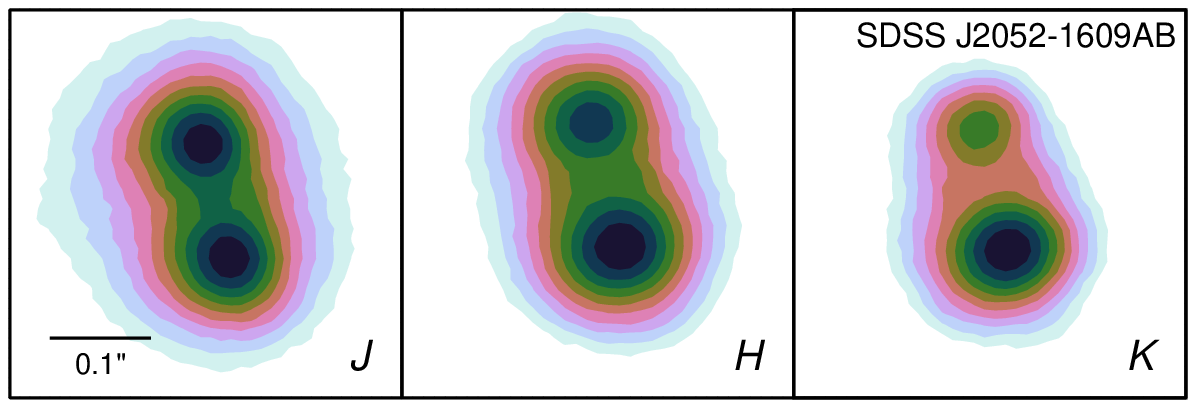}} \vskip 0.10in
\centerline{\includegraphics[width=4.8in,angle=0]{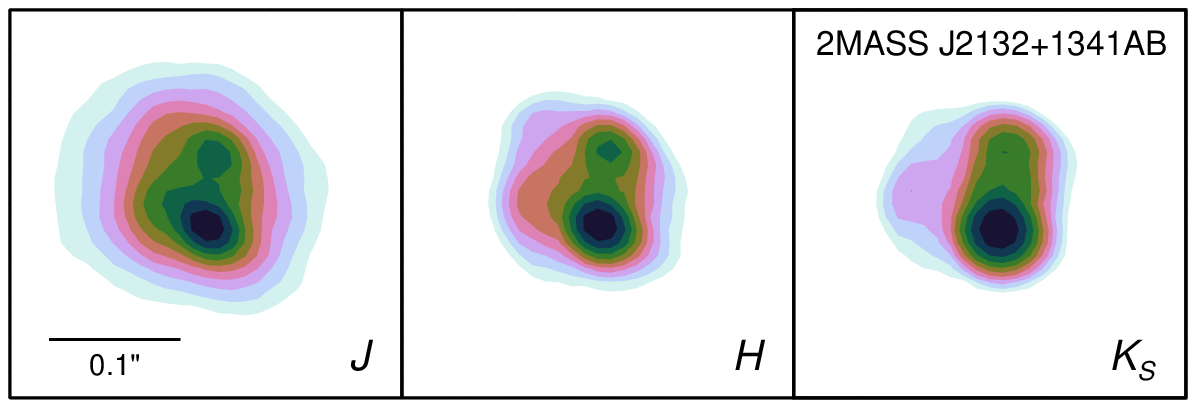}}

\caption{\normalsize Same as Figure~\ref{fig:lgs1}. \label{fig:lgs3}}

\end{figure}

\begin{figure}

\centerline{\includegraphics[width=3.1in,angle=0]{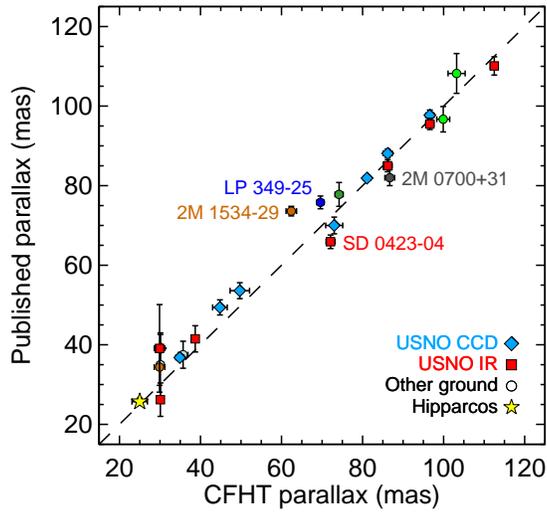}}

\caption{\normalsize Comparison of our CFHT parallaxes to published
  values.  This comparison sample consists mostly of binaries for
  which we wish to independently check and/or improve the published
  parallax, as well as some single control objects.  Note that USNO IR
  parallaxes \citep{2004AJ....127.2948V} are considered preliminary,
  which explains one of the 4 discrepant cases (labeled objects).  At
  least 2 of the other 3 cases can be explained by underestimated
  errors in the published parallaxes, as shown by Monte Carlo
  simulations in Section~\ref{sec:compare}.  For the remaining 23 of
  27 comparison cases, our parallaxes agree to within 1.8$\sigma$ of
  the published results with a reasonable $\chi^2$ of 19.9 (23 dof).
  In 19 of these 23 cases our parallax errors are less than or equal
  to published errors.  \label{fig:vs-pub}}

\end{figure}

\begin{figure}

\centerline{\includegraphics[width=2.1in,angle=0]{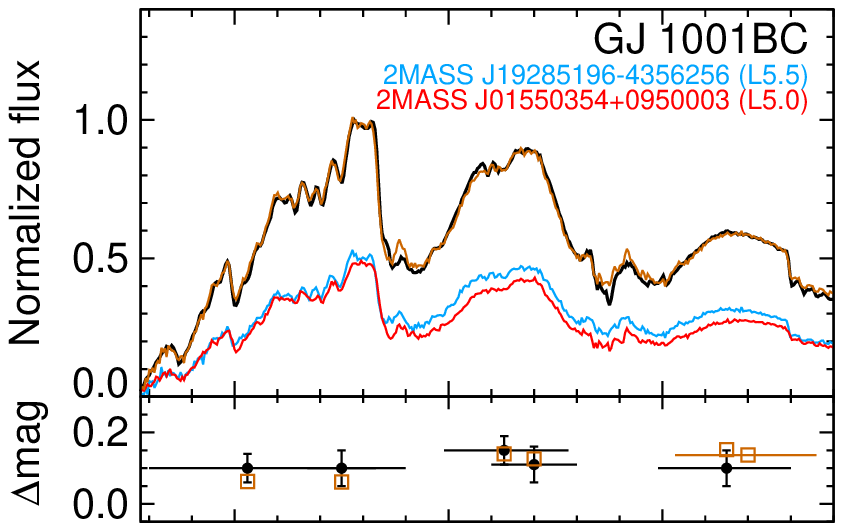} \hskip 0.05in
            \includegraphics[width=2.1in,angle=0]{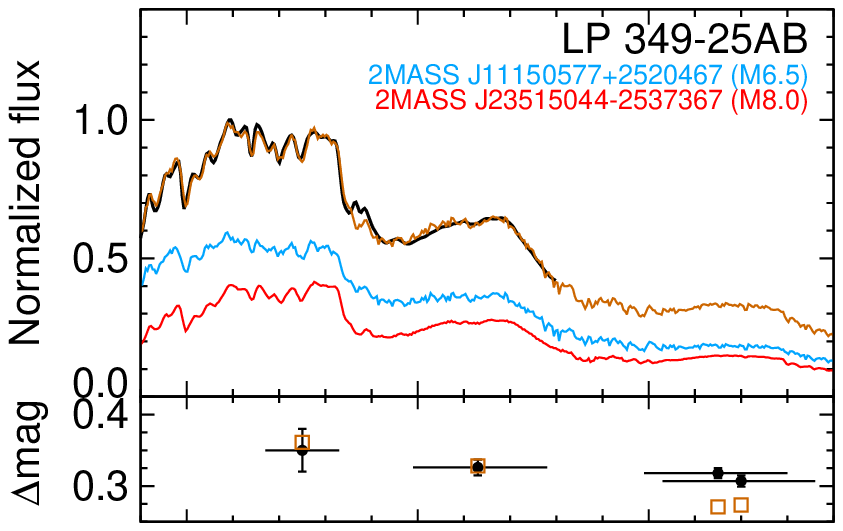} \hskip 0.05in
            \includegraphics[width=2.1in,angle=0]{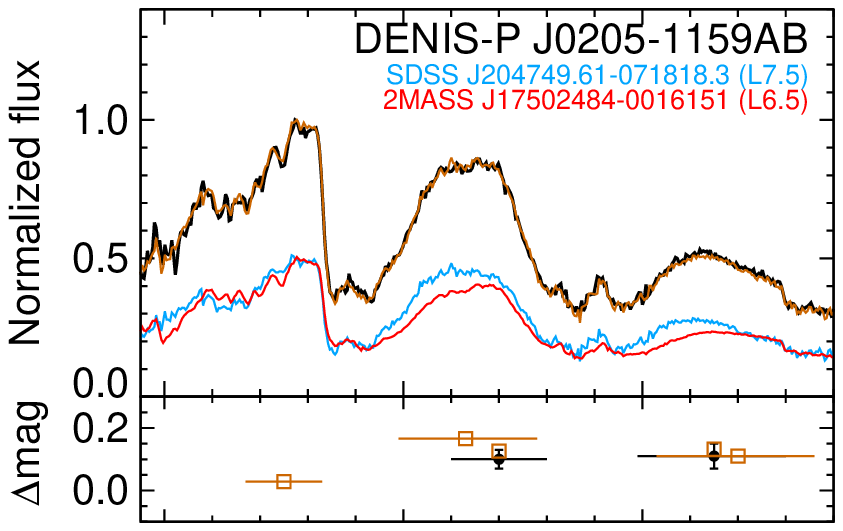}} \vskip -0.15in
\centerline{\includegraphics[width=2.1in,angle=0]{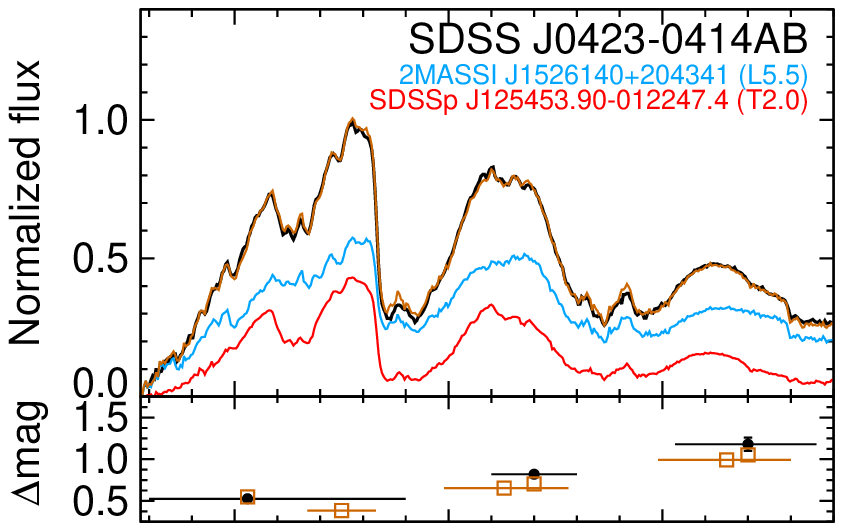} \hskip 0.05in
            \includegraphics[width=2.1in,angle=0]{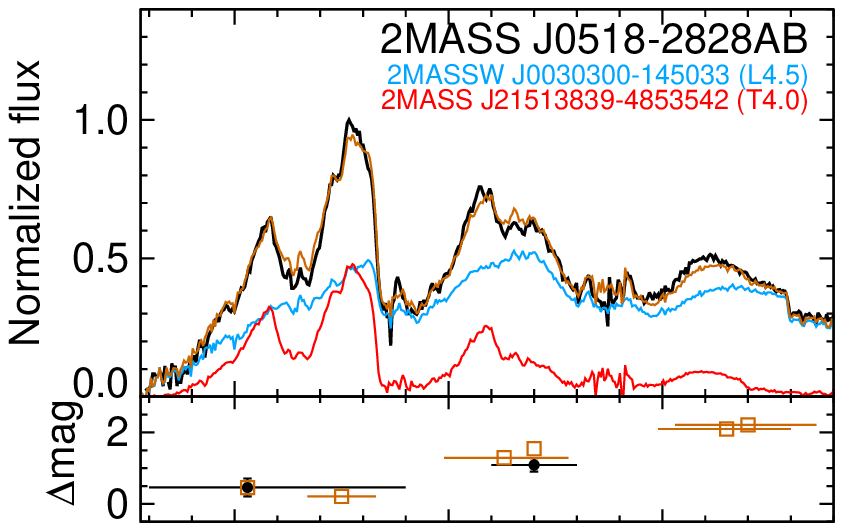} \hskip 0.05in
            \includegraphics[width=2.1in,angle=0]{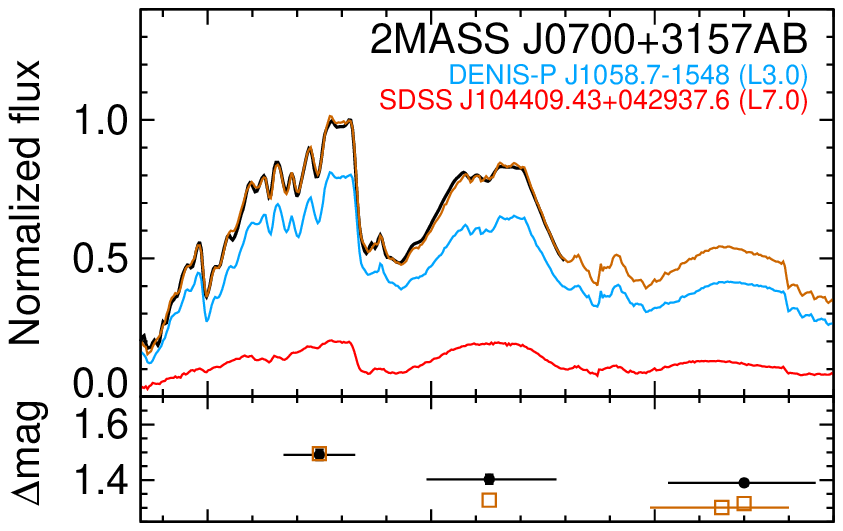}} \vskip -0.15in
\centerline{\includegraphics[width=2.1in,angle=0]{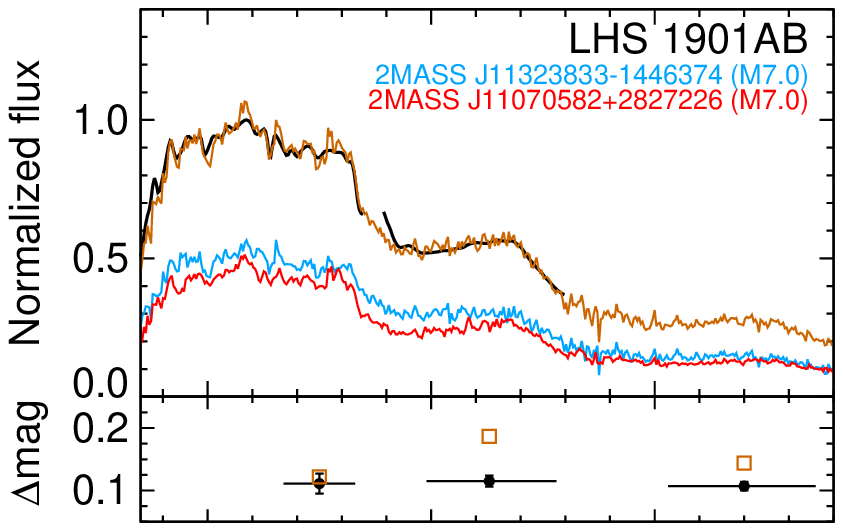} \hskip 0.05in
            \includegraphics[width=2.1in,angle=0]{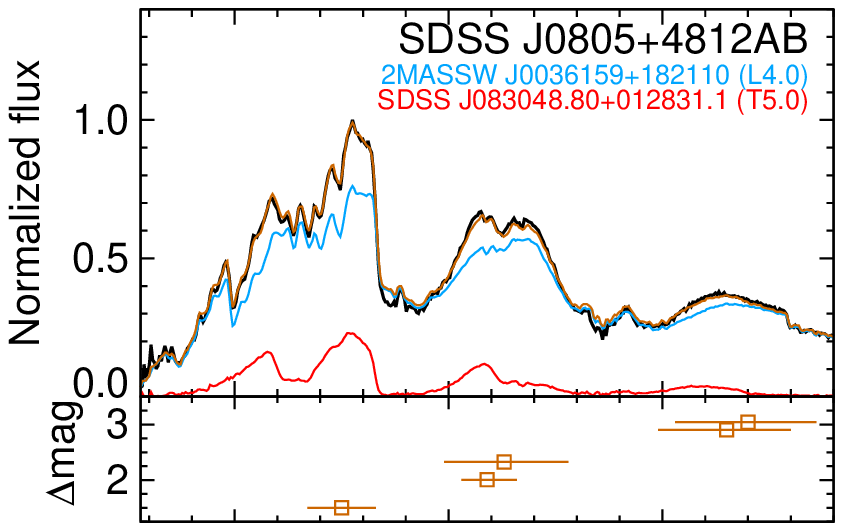} \hskip 0.05in
            \includegraphics[width=2.1in,angle=0]{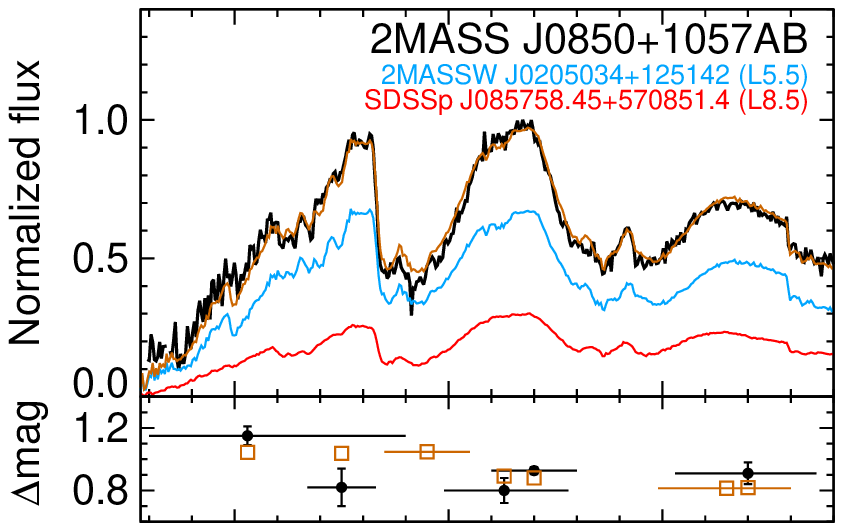}} \vskip -0.15in
\centerline{\includegraphics[width=2.1in,angle=0]{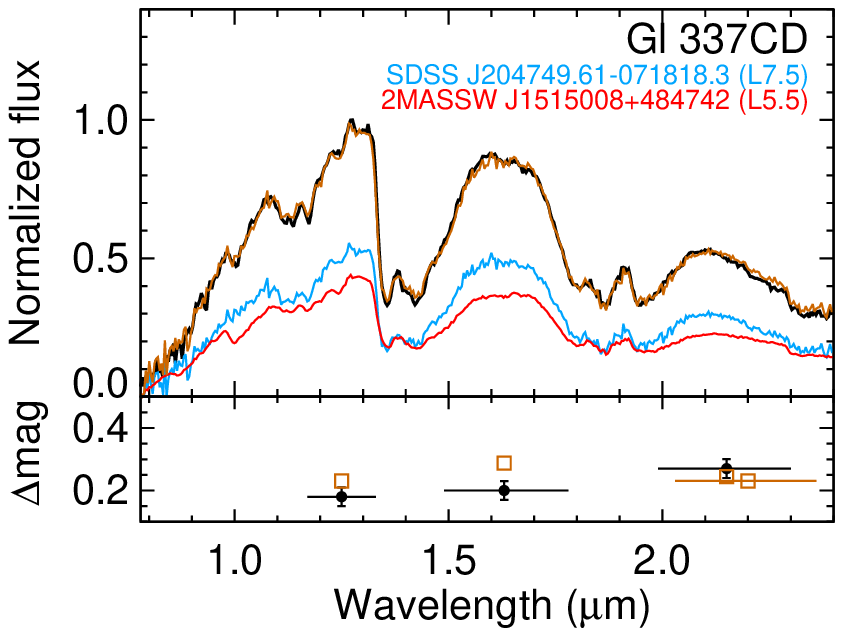} \hskip 0.05in
            \includegraphics[width=2.1in,angle=0]{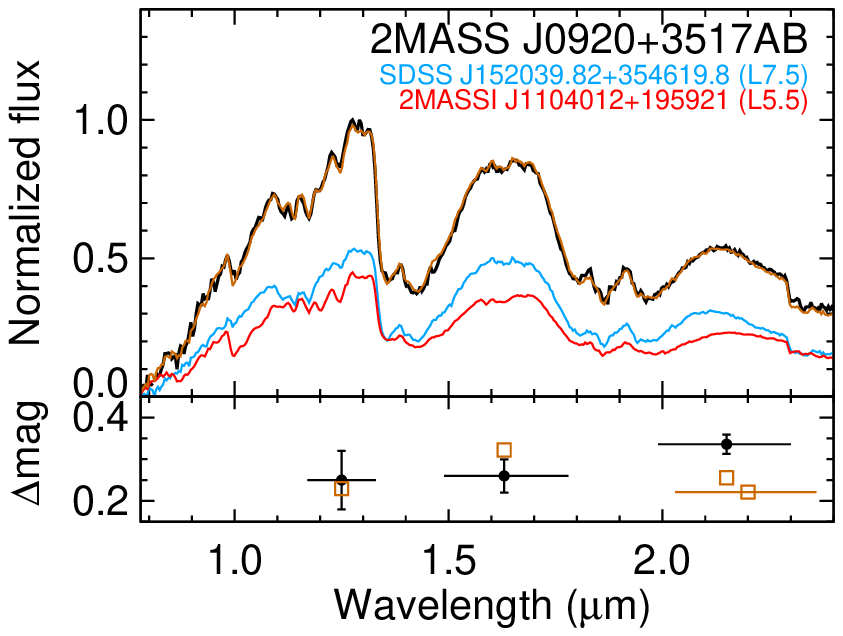} \hskip 0.05in
            \includegraphics[width=2.1in,angle=0]{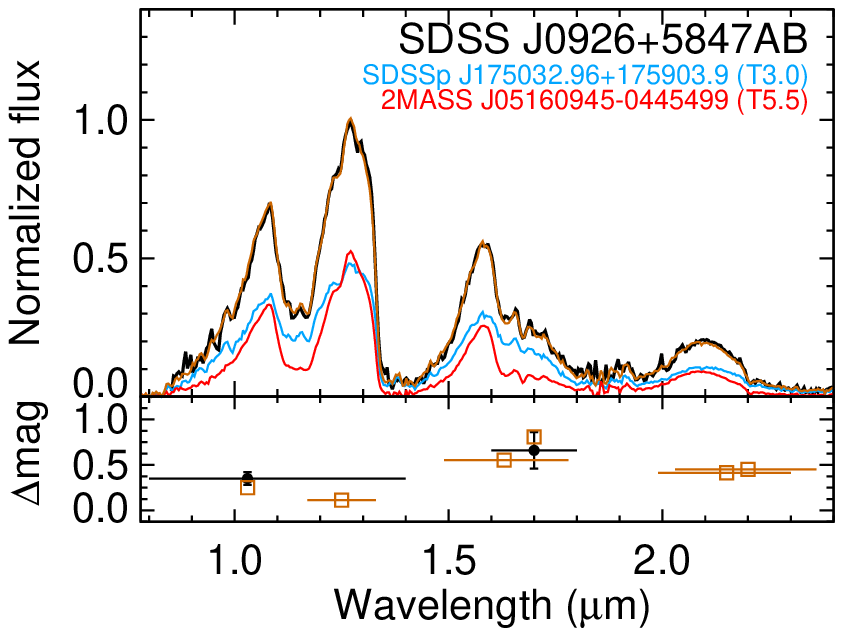}}

\caption{\normalsize Integrated-light spectra (black) and best
  matching component templates (colored lines) for binaries analyzed
  by our spectral decomposition method (see Section~\ref{sec:spt},
  Table~\ref{tbl:spt}).  The bottom subpanels show the resolved
  photometry used to constrain the decomposition (filled black circles
  with errors) and the resulting flux ratios computed from the best
  matching template pair (open brown squares).  In some cases there
  exists both \HST/NICMOS medium-band and Keck photometry (e.g.,
  2MASS~J0850+1057AB), and in one case there is no resolved photometry
  available (SDSS~J0805+4812AB).  For some binaries (e.g.,
  2MASS~J0700+3157AB) we degrade our SpeX SXD spectra ($R =
  1200$--2000) to the resolution of the prism template spectra ($R =
  120$) and exclude $K$ band from the analysis. \label{fig:spec1}}

\end{figure}

\begin{figure}

\centerline{\includegraphics[width=2.1in,angle=0]{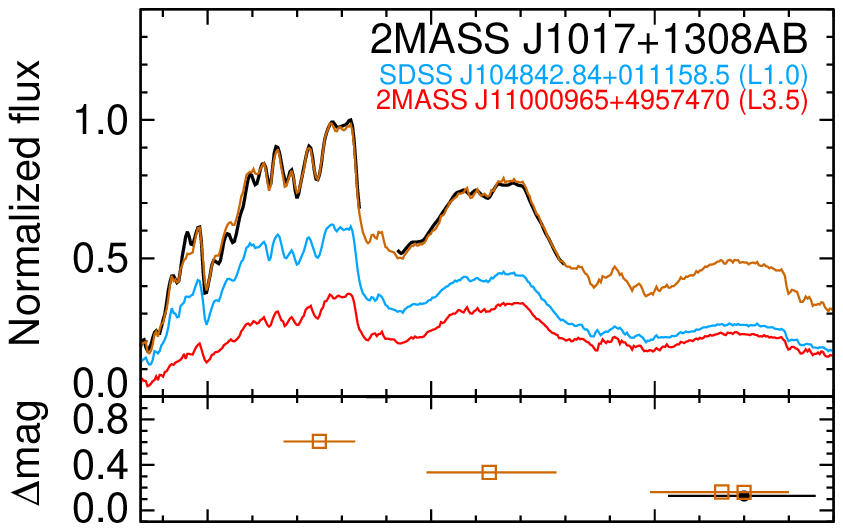}  \hskip 0.05in
            \includegraphics[width=2.1in,angle=0]{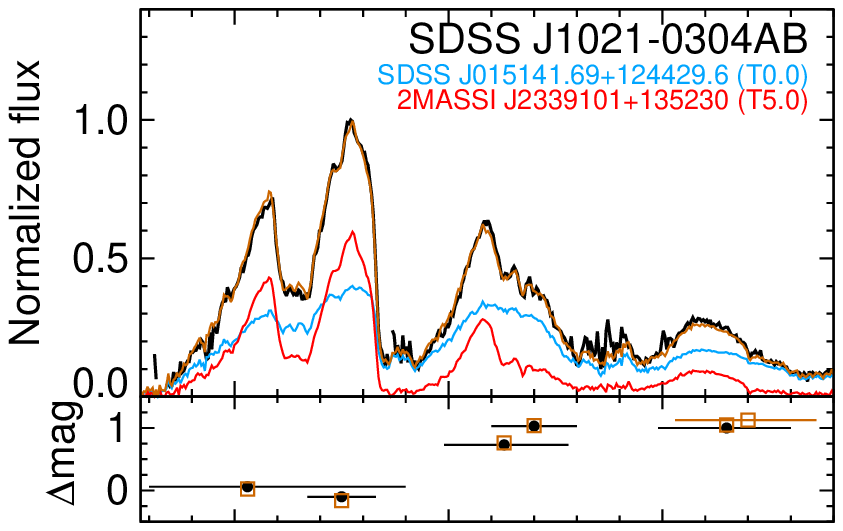}  \hskip 0.05in
            \includegraphics[width=2.1in,angle=0]{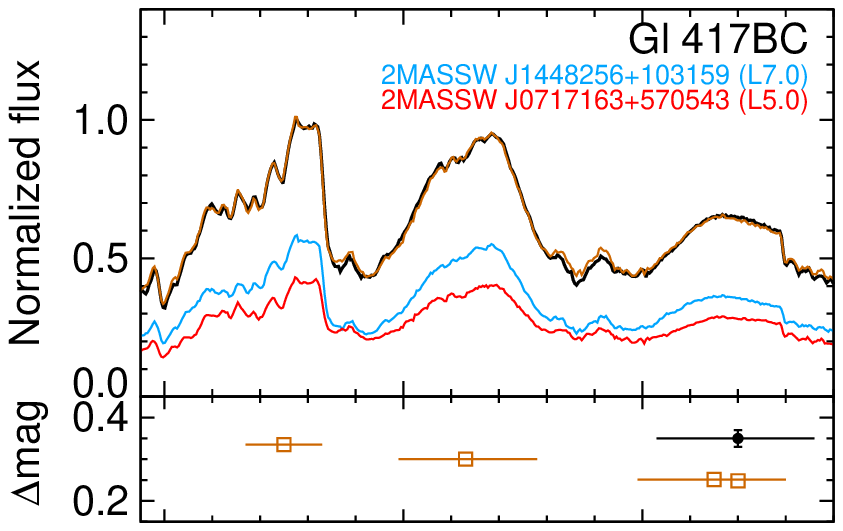}} \vskip -0.15in
\centerline{\includegraphics[width=2.1in,angle=0]{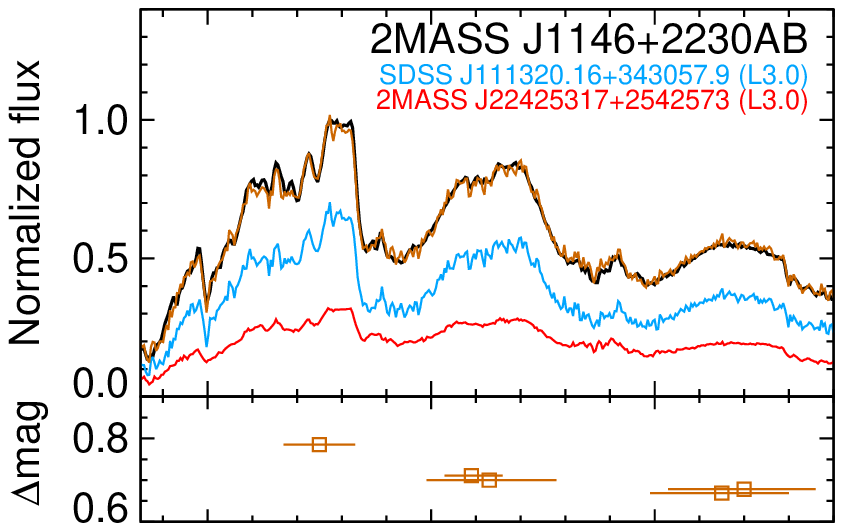}  \hskip 0.05in
            \includegraphics[width=2.1in,angle=0]{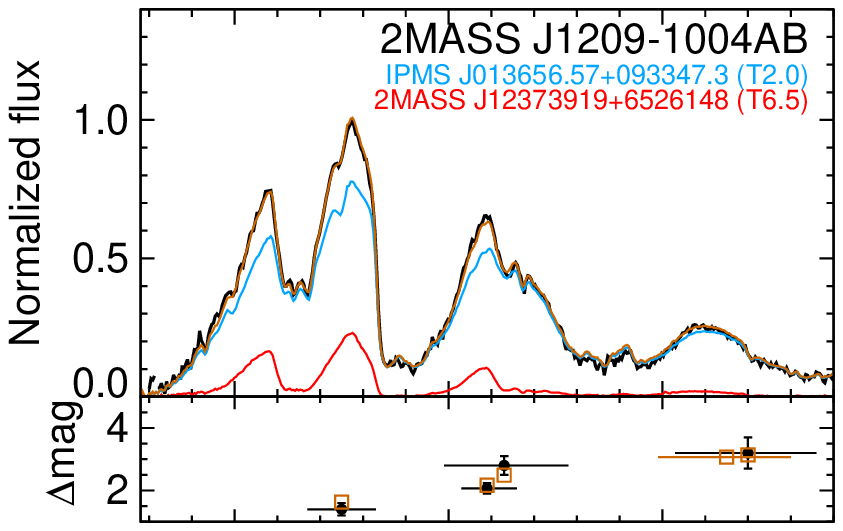}  \hskip 0.05in
            \includegraphics[width=2.1in,angle=0]{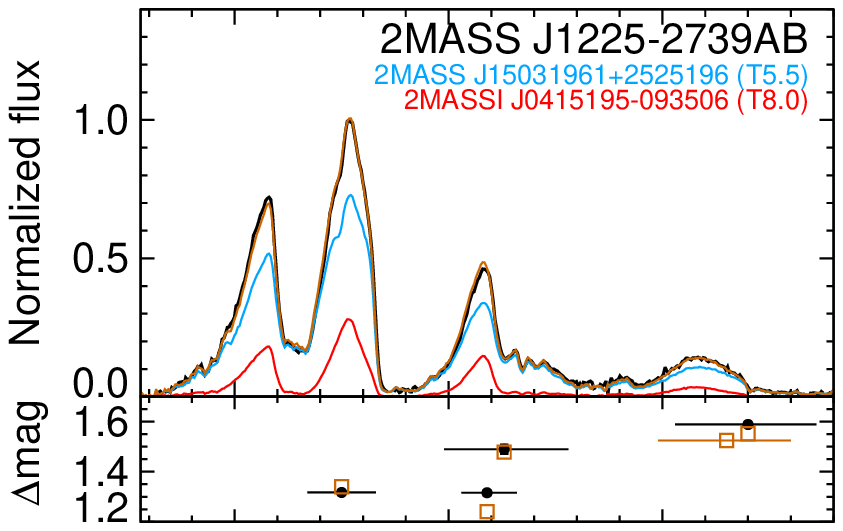}} \vskip -0.15in
\centerline{\includegraphics[width=2.1in,angle=0]{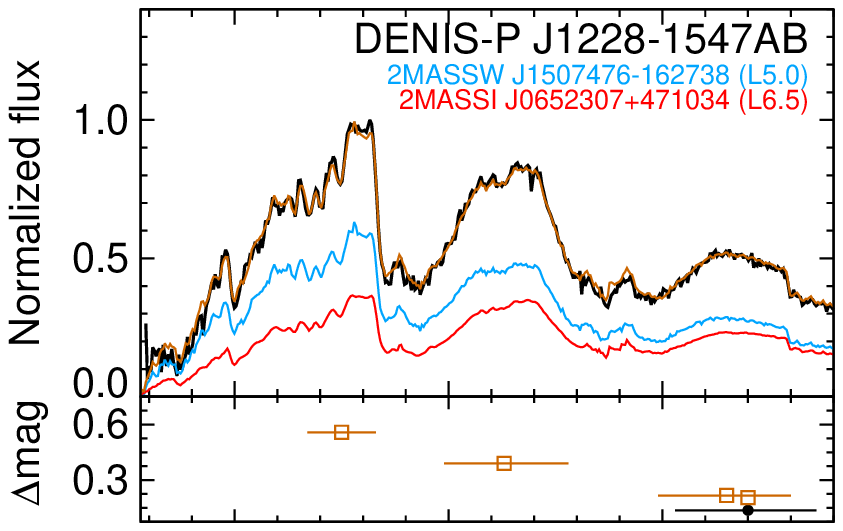}  \hskip 0.05in
            \includegraphics[width=2.1in,angle=0]{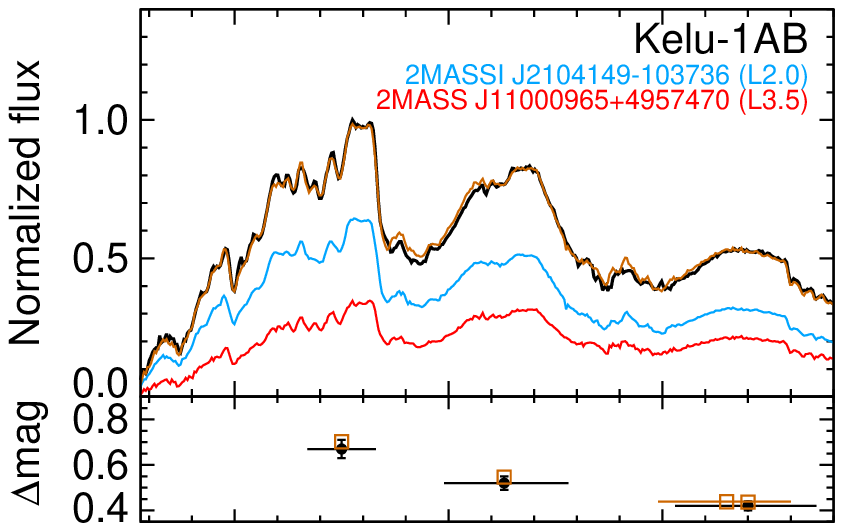}  \hskip 0.05in
            \includegraphics[width=2.1in,angle=0]{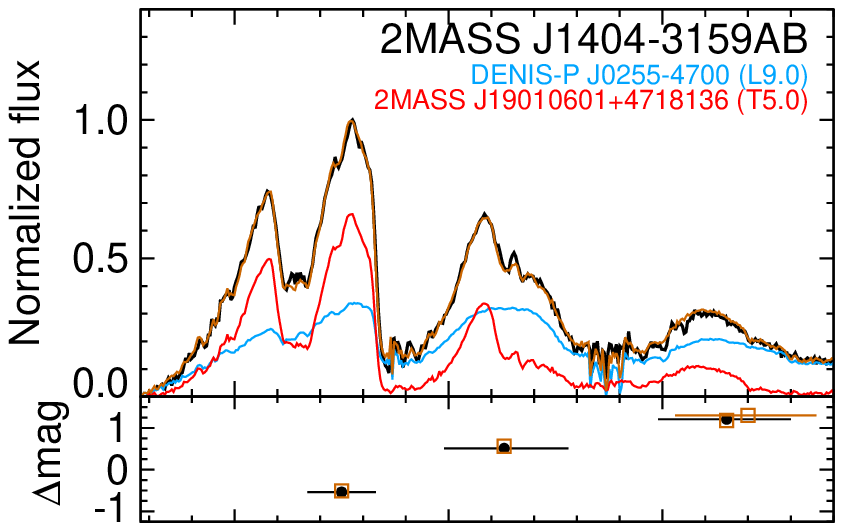}} \vskip -0.15in
\centerline{\includegraphics[width=2.1in,angle=0]{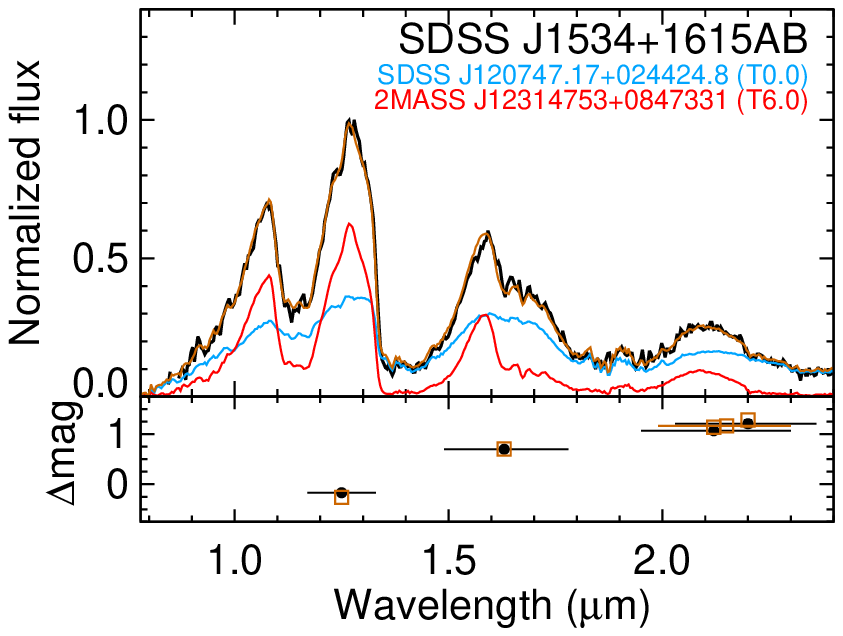}  \hskip 0.05in
            \includegraphics[width=2.1in,angle=0]{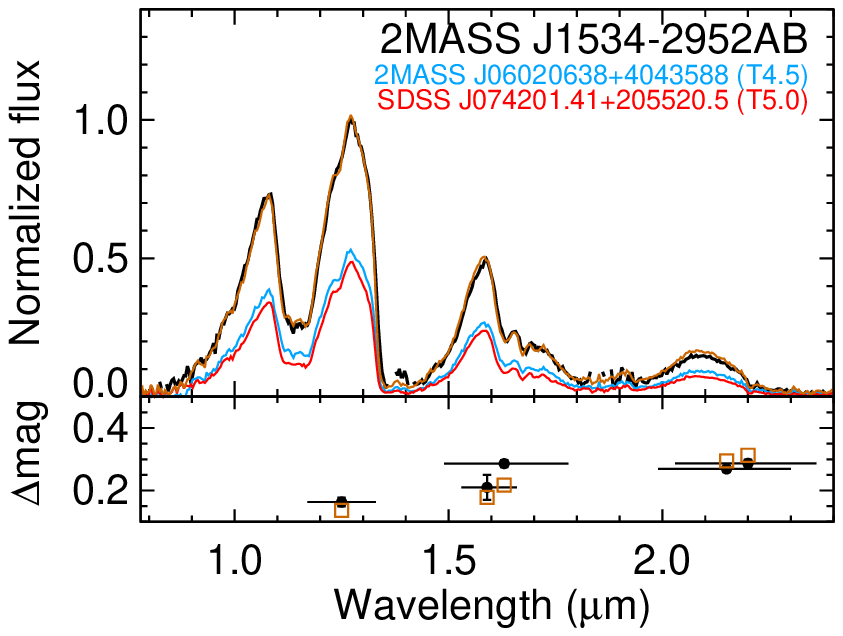}  \hskip 0.05in
            \includegraphics[width=2.1in,angle=0]{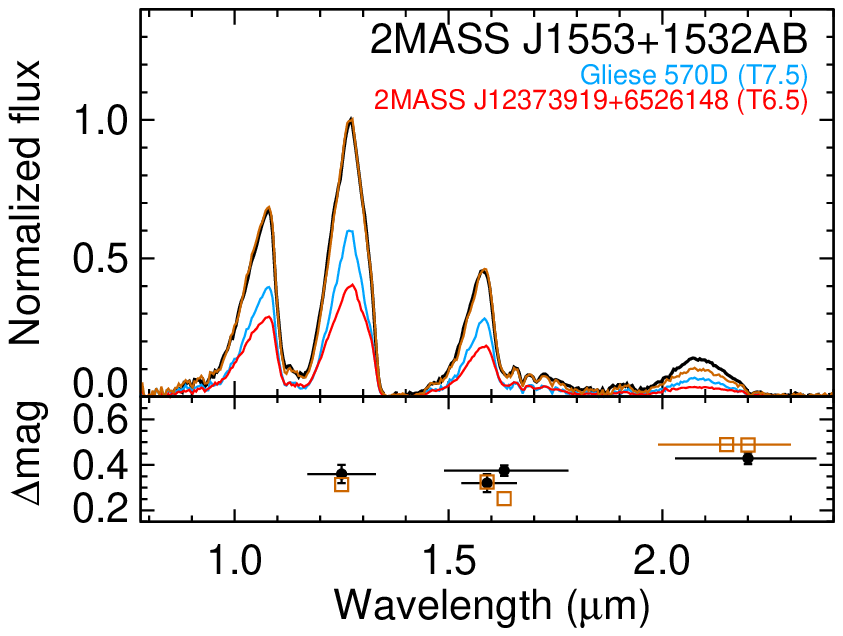}}

\caption{\normalsize Same as Figure~\ref{fig:spec1}. \label{fig:spec2}}

\end{figure}

\clearpage

\begin{figure}

\centerline{\includegraphics[width=2.1in,angle=0]{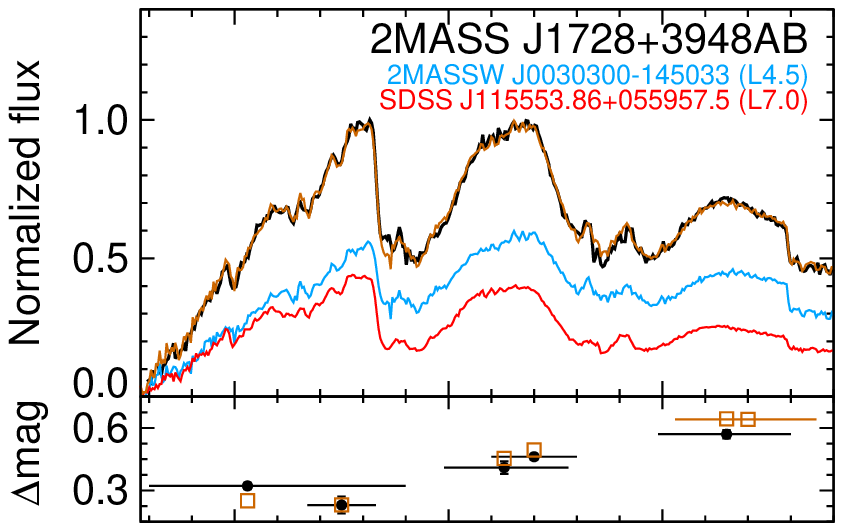}  \hskip 0.05in
            \includegraphics[width=2.1in,angle=0]{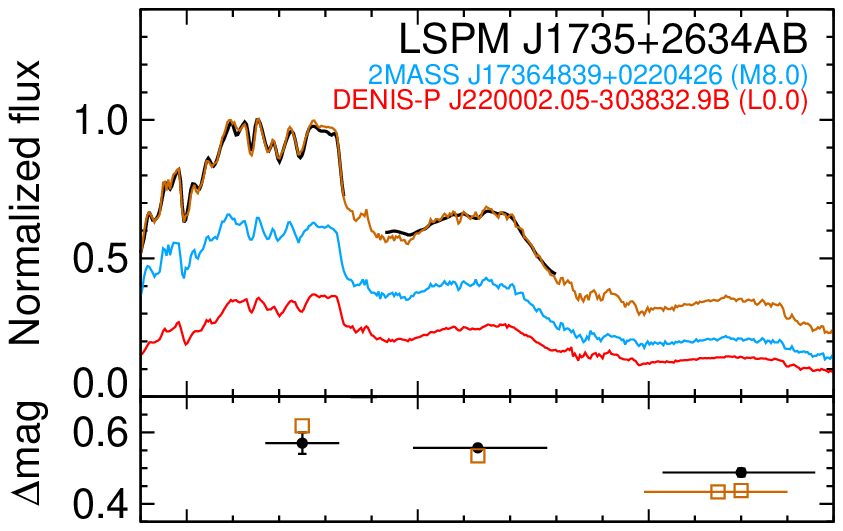}  \hskip 0.05in
            \includegraphics[width=2.1in,angle=0]{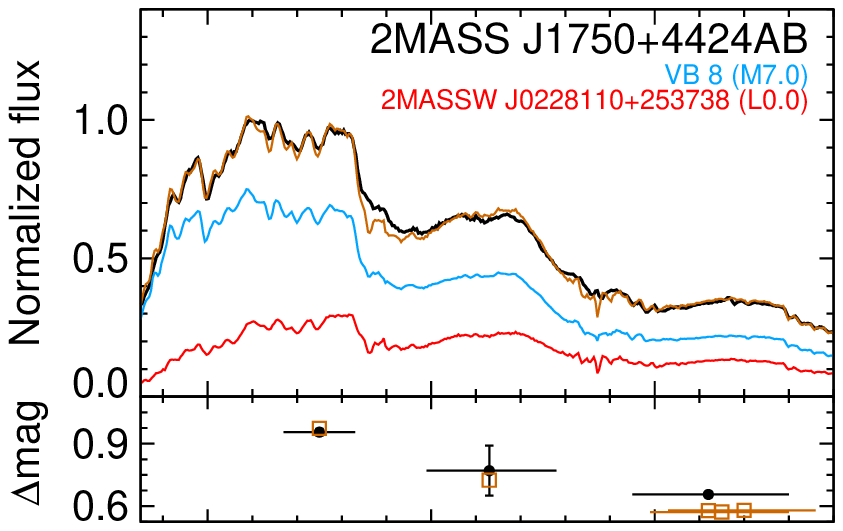}} \vskip -0.15in
\centerline{\includegraphics[width=2.1in,angle=0]{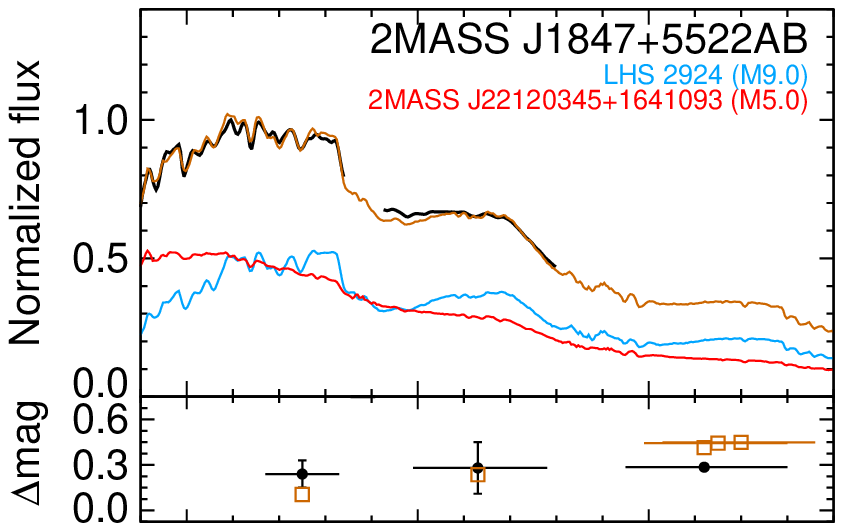}  \hskip 0.05in
            \includegraphics[width=2.1in,angle=0]{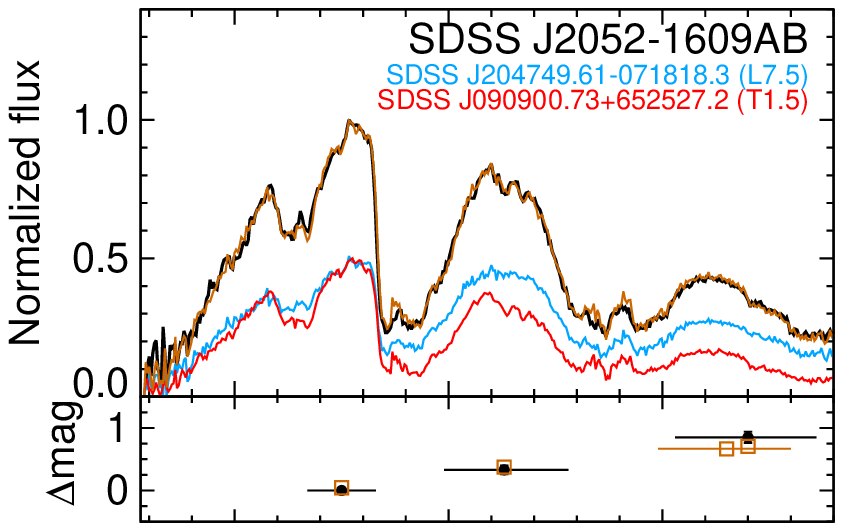}  \hskip 0.05in
            \includegraphics[width=2.1in,angle=0]{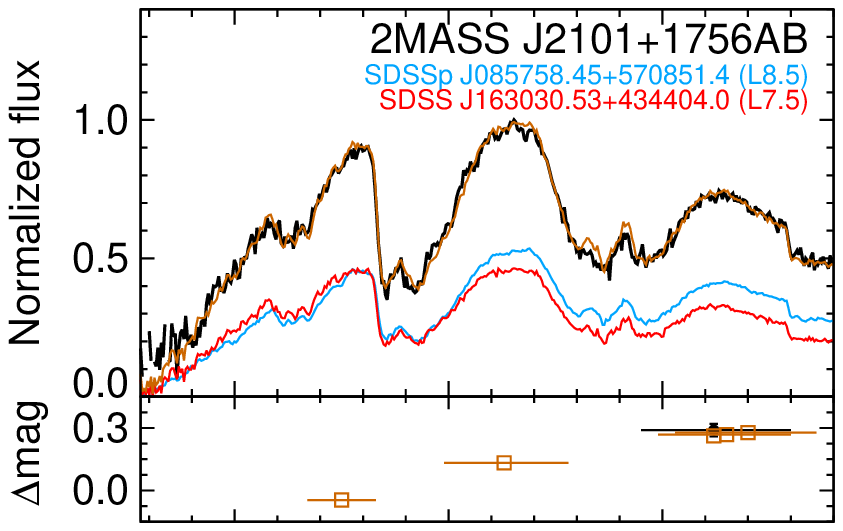}} \vskip -0.15in
\centerline{\includegraphics[width=2.1in,angle=0]{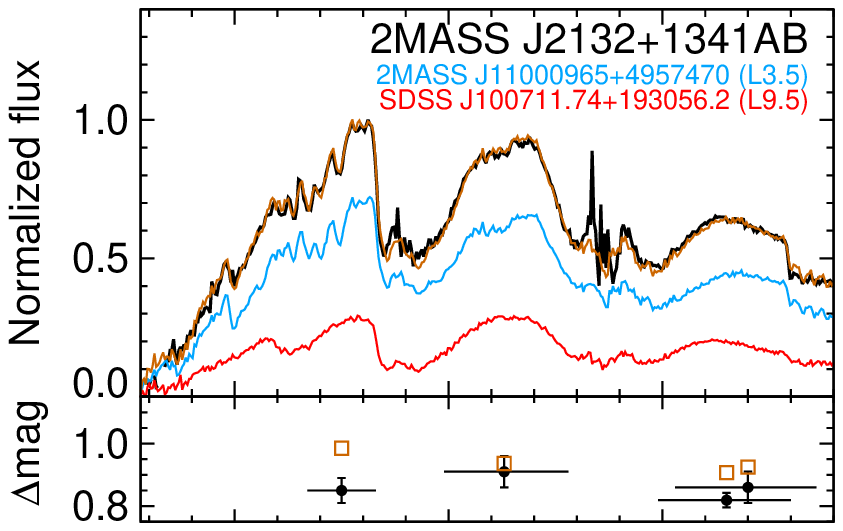}  \hskip 0.05in
            \includegraphics[width=2.1in,angle=0]{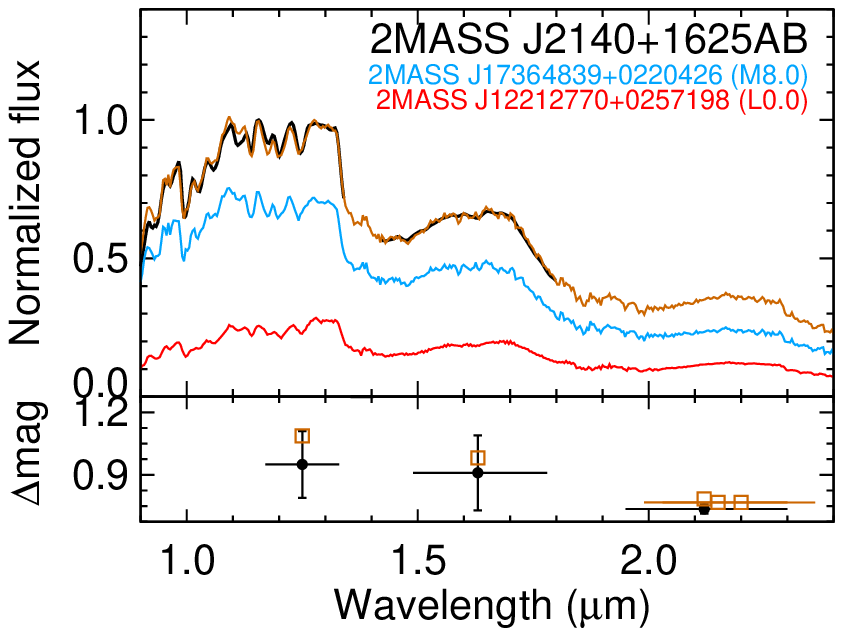}  \hskip 0.05in
            \includegraphics[width=2.1in,angle=0]{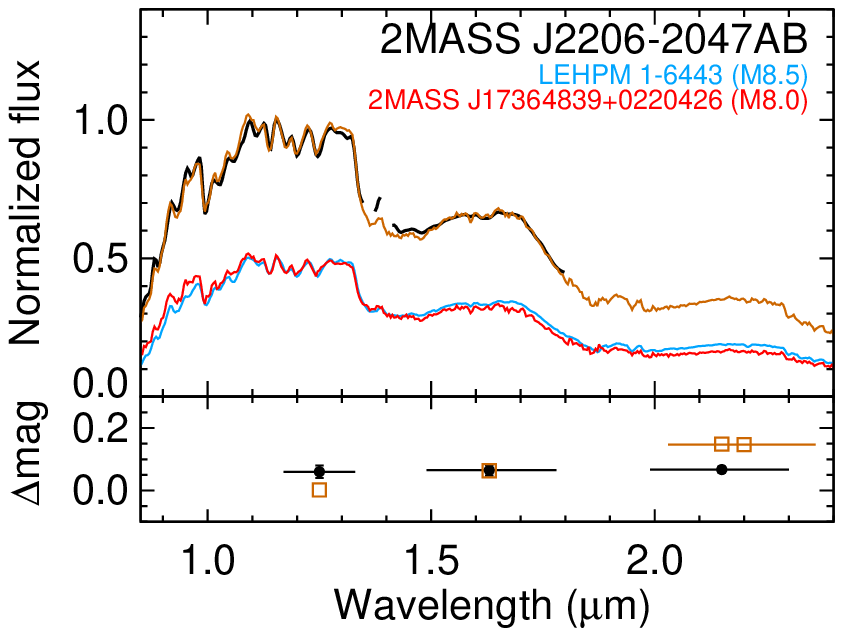}} \vskip -0.15in
            \includegraphics[width=2.1in,angle=0]{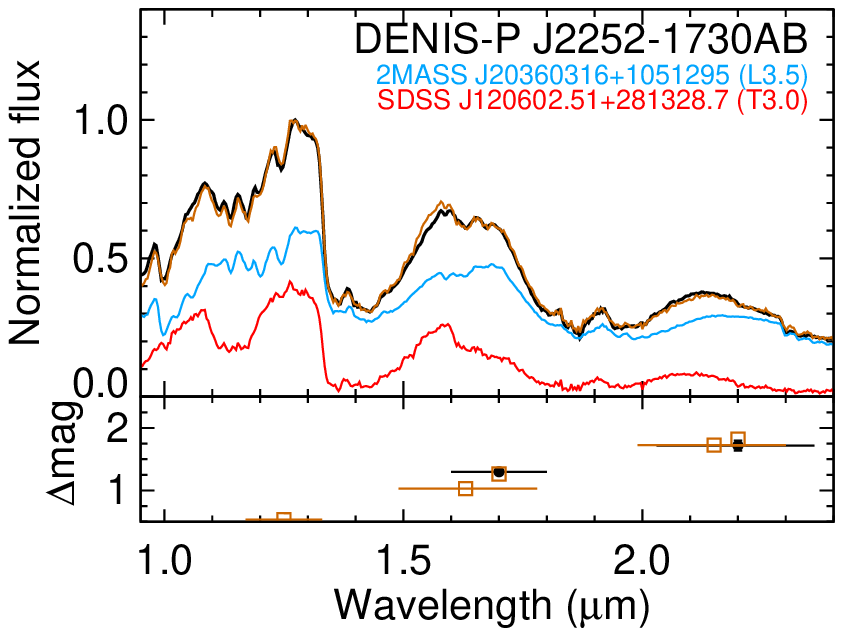}

\caption{\normalsize Same as Figure~\ref{fig:spec1}. \label{fig:spec3}}

\end{figure}

\begin{figure}

\vskip -0.70in

\centerline{\includegraphics[width=2.5in,angle=0]{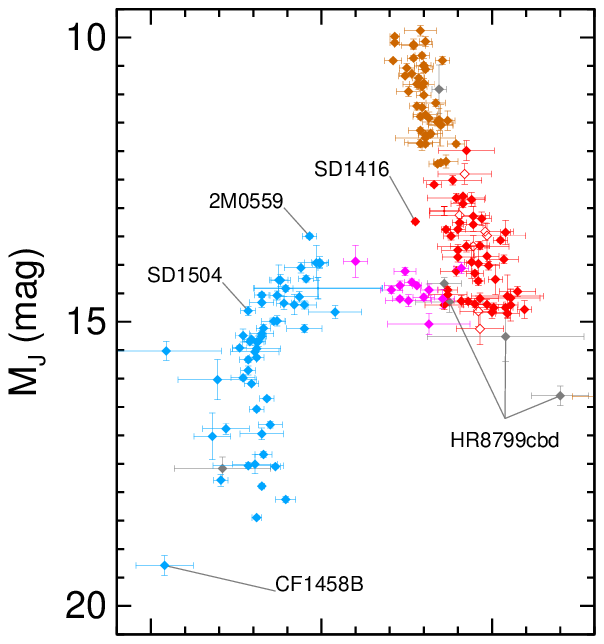} \hskip -0.60in
            \includegraphics[width=2.5in,angle=0]{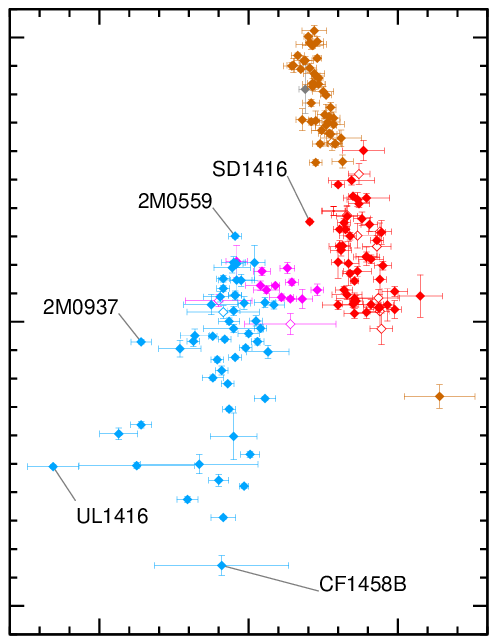} \hskip -0.60in
            \includegraphics[width=2.5in,angle=0]{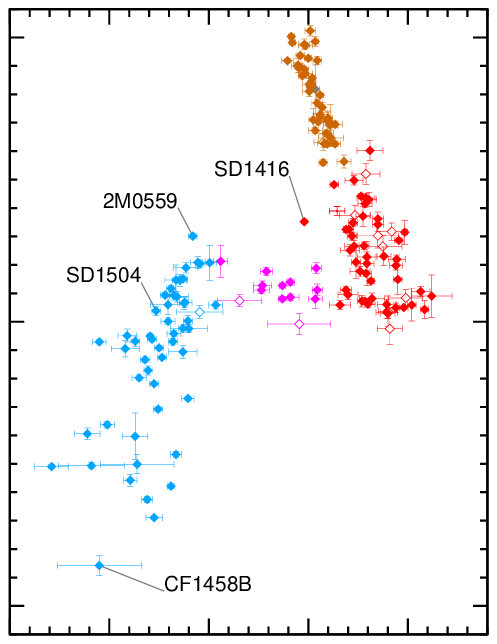}} \vskip -0.48in
\centerline{\includegraphics[width=2.5in,angle=0]{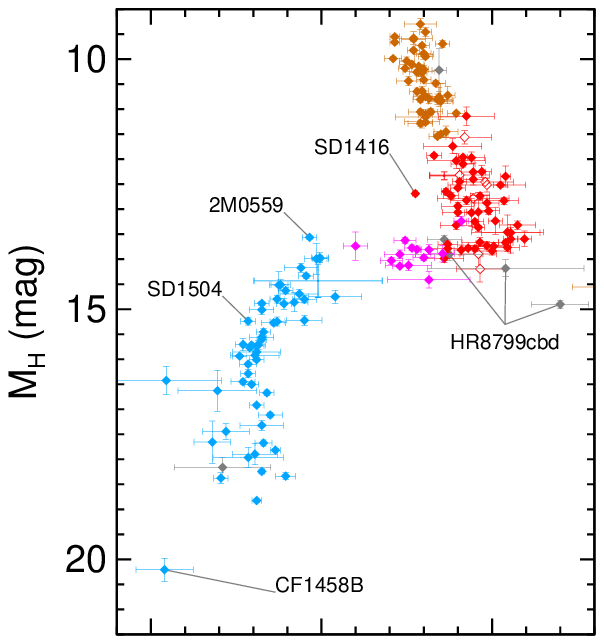} \hskip -0.60in
            \includegraphics[width=2.5in,angle=0]{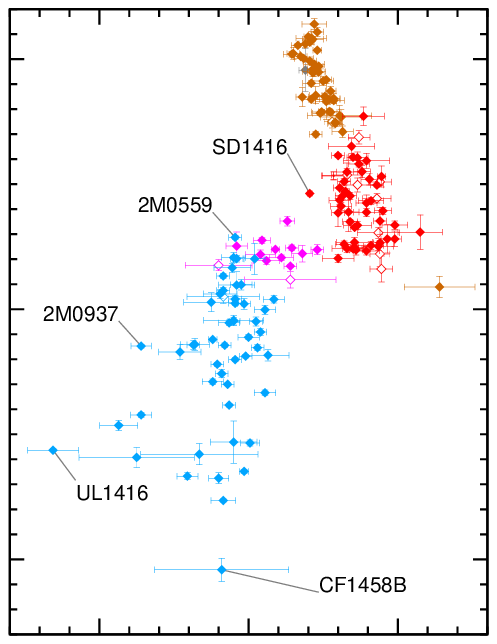} \hskip -0.60in
            \includegraphics[width=2.5in,angle=0]{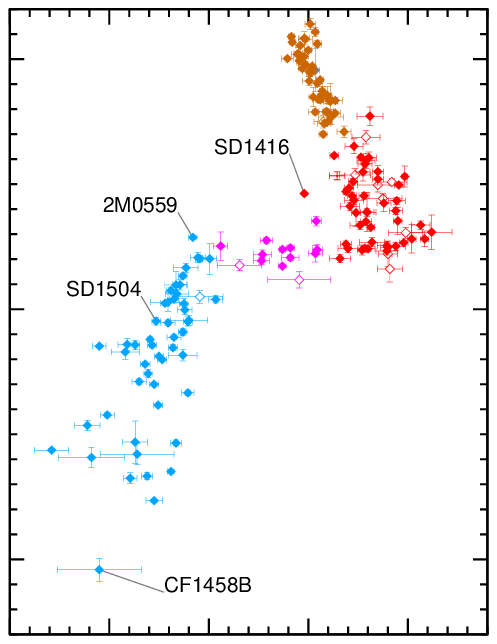}} \vskip -0.48in
\centerline{\includegraphics[width=2.5in,angle=0]{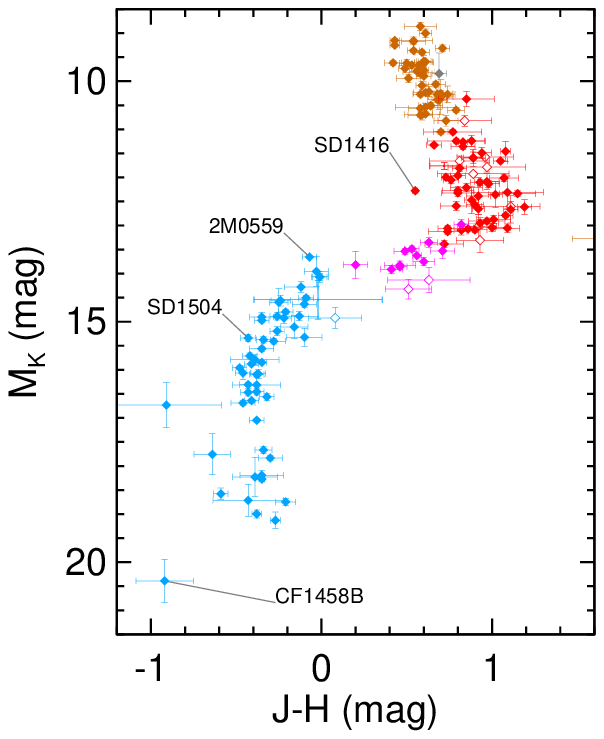} \hskip -0.60in
            \includegraphics[width=2.5in,angle=0]{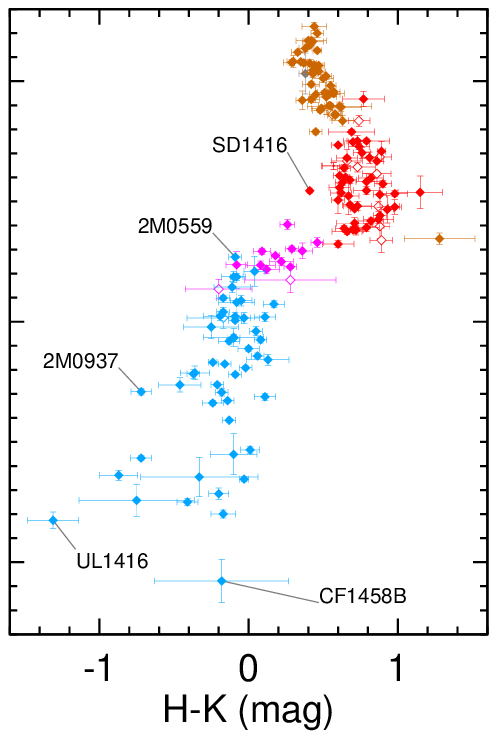} \hskip -0.60in
            \includegraphics[width=2.5in,angle=0]{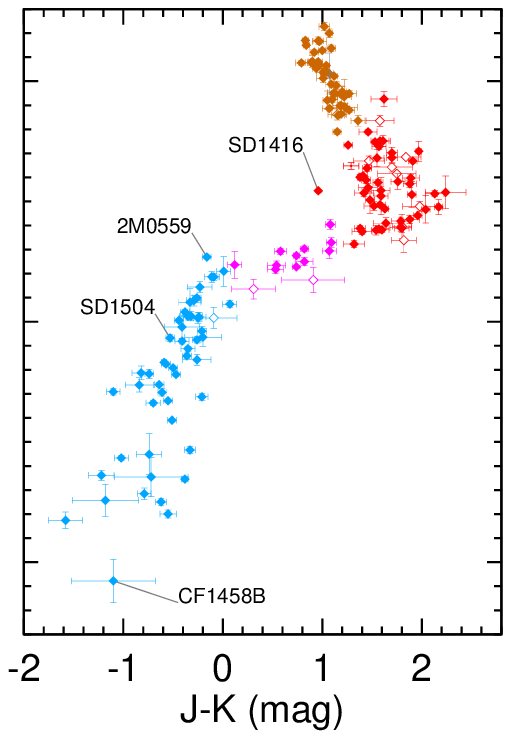}}

\vskip -0.15in

\caption{\normalsize Color--magnitude diagrams in the MKO system
  showing all ultracool dwarfs from Table~\ref{tbl:nir-mags} with
  errors $\leq$0.5~mag.  Nearly half (40\%) of the data points here
  use our new CFHT parallaxes. Symbol colors indicate spectral types:
  M6--L2 (brown), L2.5--L9 (red), L9.5--T4 (purple), $\geq$T4.5
  (blue), and unknown (gray).  Solid symbols indicate photometry that
  is measured either directly or converted, e.g., from 2MASS to MKO,
  using the object's own spectrum.  Open symbols indicate binary
  components where the flux ratio in one or more, but not all, bands
  was estimated from spectral decomposition.  For one binary
  (SDSS~J0805+4812), no symbol (i.e., error bar only) indicates that
  all flux ratios were estimated from spectral
  decomposition. \label{fig:cmd-jhk1-mko}}

\end{figure}

\begin{figure}

\vskip -0.35in

\centerline{\includegraphics[width=2.5in,angle=0]{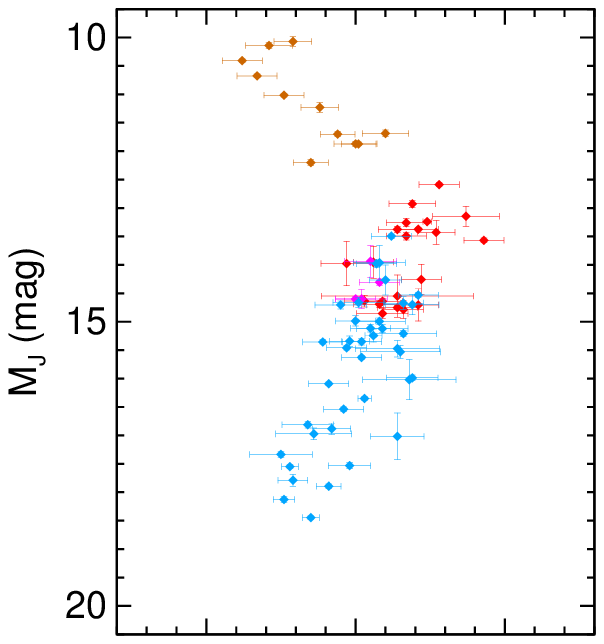} \hskip -0.60in
            \includegraphics[width=2.5in,angle=0]{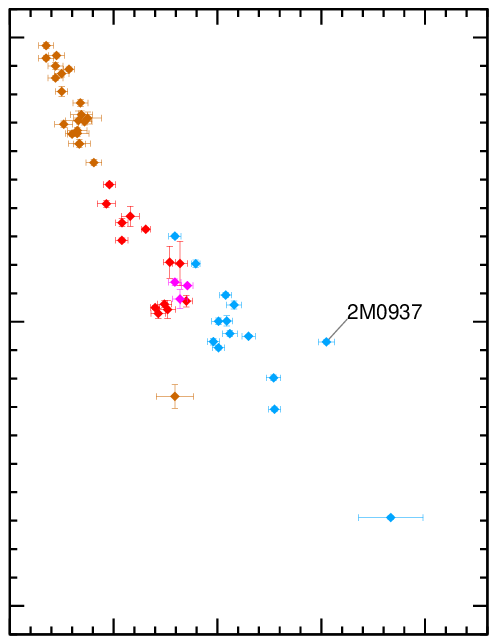}} \vskip -0.48in
\centerline{\includegraphics[width=2.5in,angle=0]{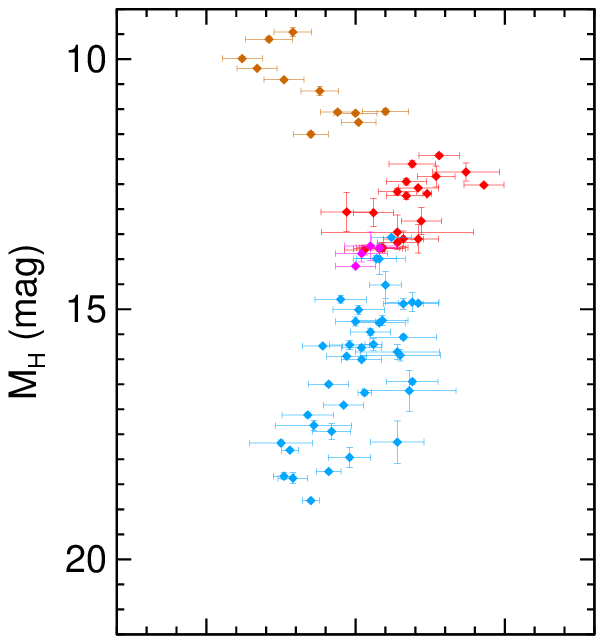} \hskip -0.60in
            \includegraphics[width=2.5in,angle=0]{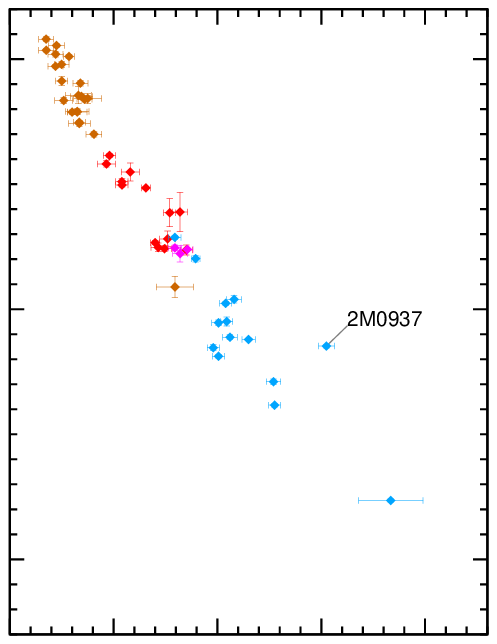}} \vskip -0.48in
\centerline{\includegraphics[width=2.5in,angle=0]{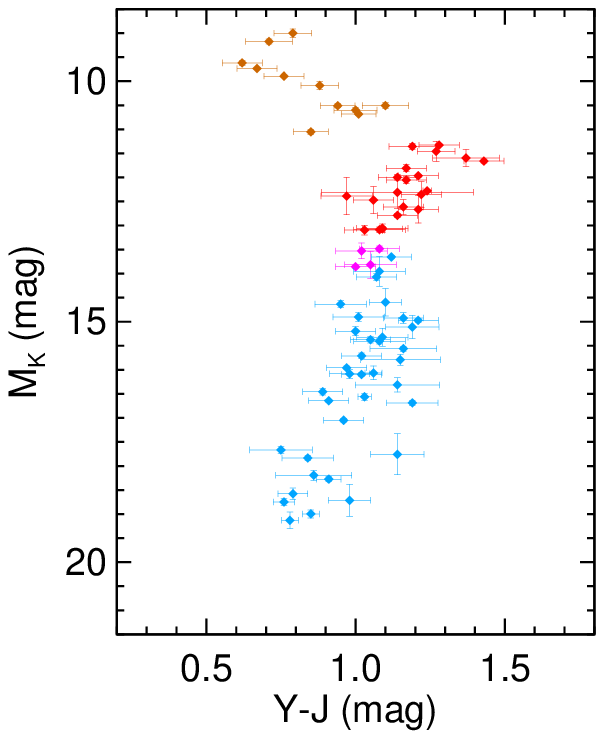} \hskip -0.60in
            \includegraphics[width=2.5in,angle=0]{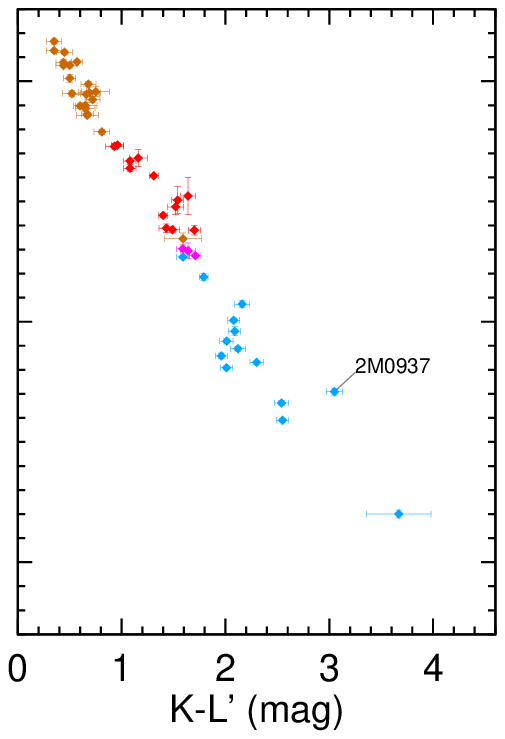}}

\caption{\normalsize Same as Figure~\ref{fig:cmd-jhk1-mko} but with
  $Y-J$ and $K-\Lp$ colors. \label{fig:cmd-jhk2-mko}}

\end{figure}

\begin{figure}

\vskip -0.35in
\centerline{\includegraphics[width=2.5in,angle=0]{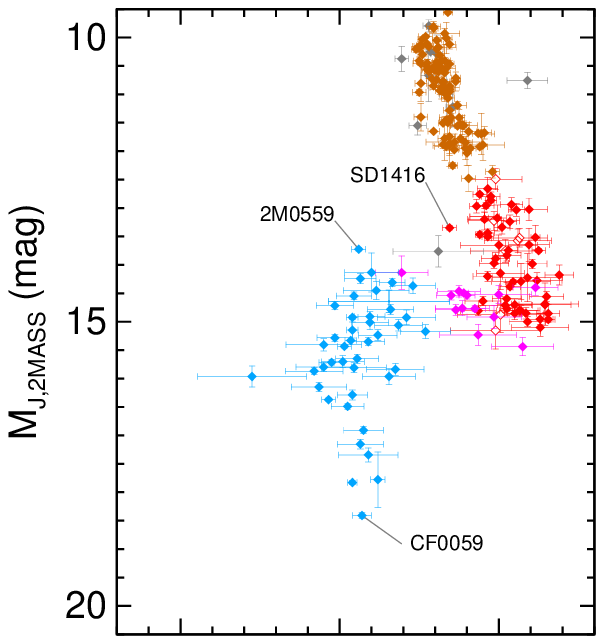} \hskip -0.60in
            \includegraphics[width=2.5in,angle=0]{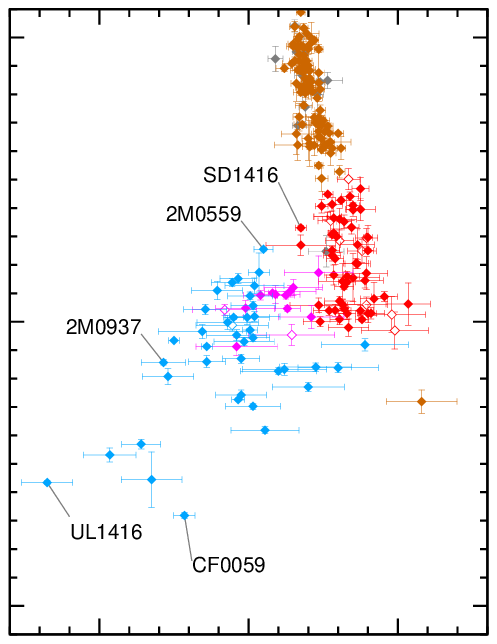} \hskip -0.60in
            \includegraphics[width=2.5in,angle=0]{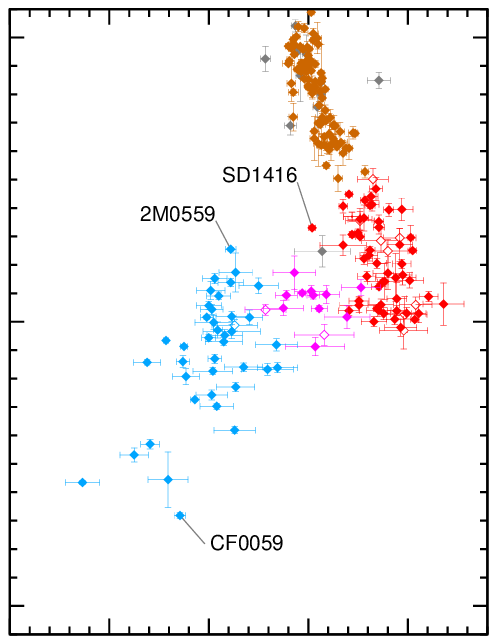}} \vskip -0.48in
\centerline{\includegraphics[width=2.5in,angle=0]{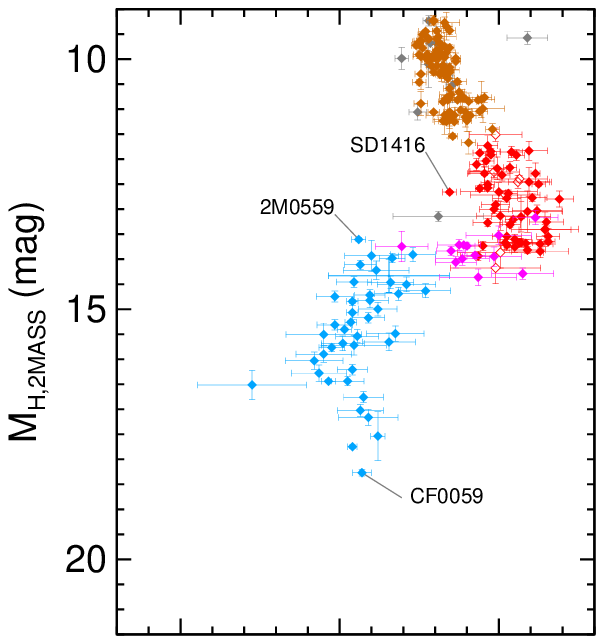} \hskip -0.60in
            \includegraphics[width=2.5in,angle=0]{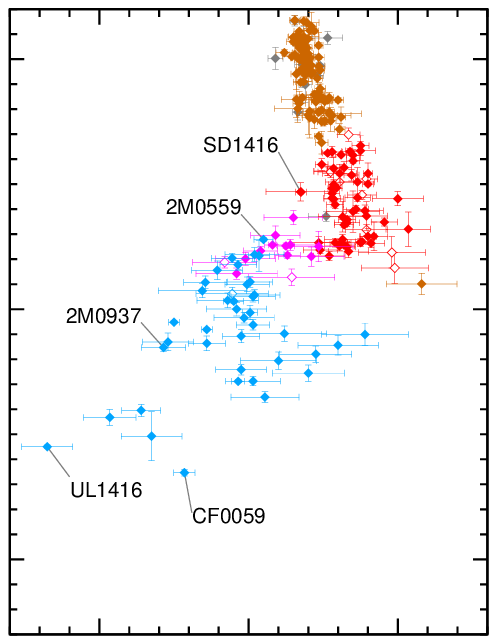} \hskip -0.60in
            \includegraphics[width=2.5in,angle=0]{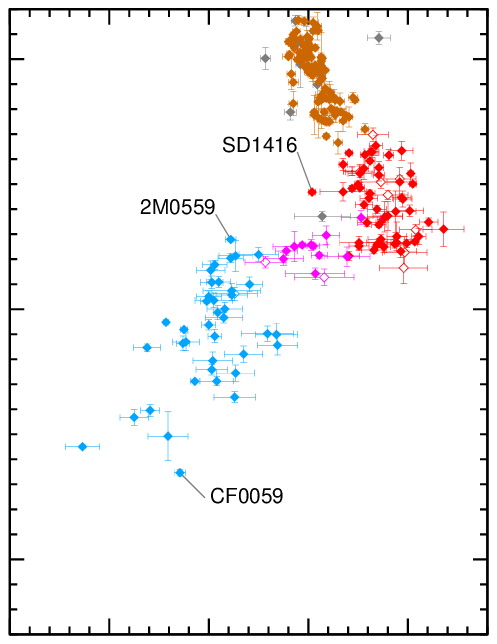}} \vskip -0.48in
\centerline{\includegraphics[width=2.5in,angle=0]{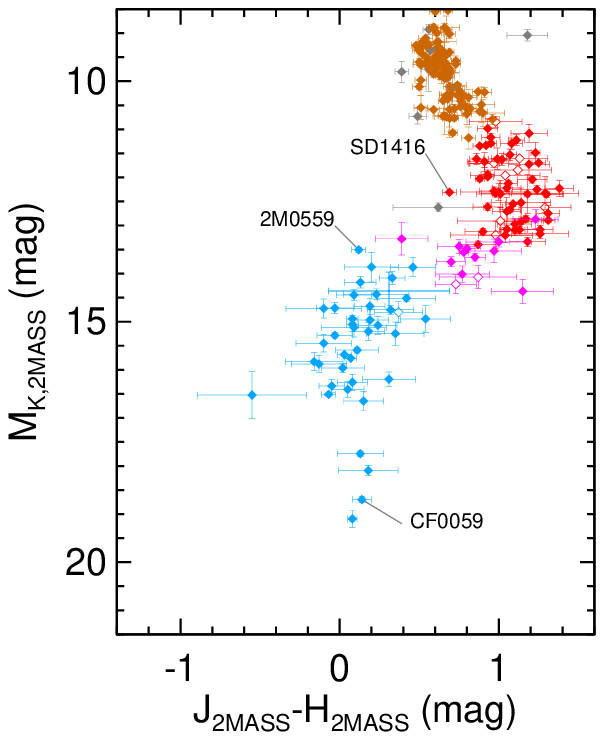} \hskip -0.60in
            \includegraphics[width=2.5in,angle=0]{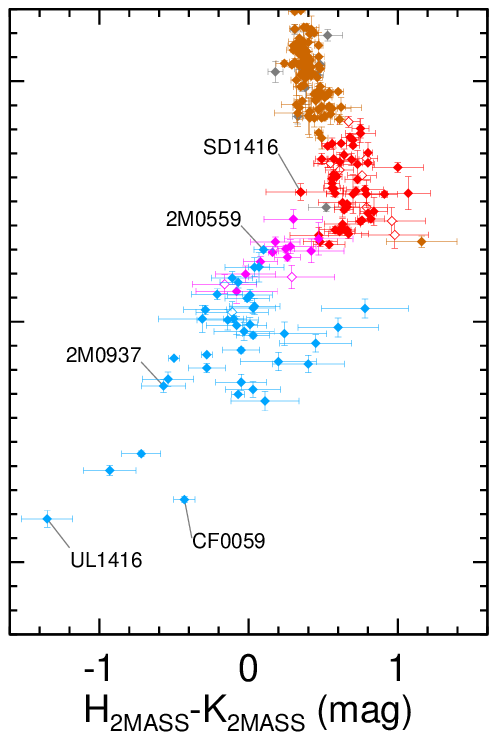} \hskip -0.60in
            \includegraphics[width=2.5in,angle=0]{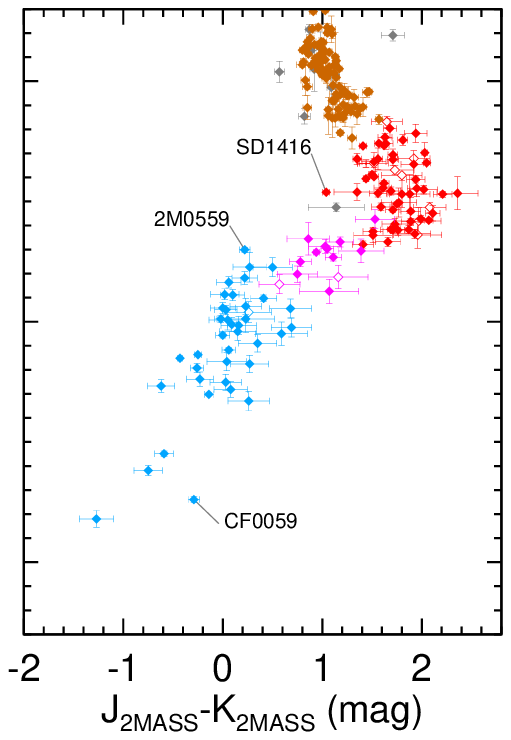}}

\caption{\normalsize Same as Figure~\ref{fig:cmd-jhk1-mko} but with
  photometry in the 2MASS system. \label{fig:cmd-jhk1-2mass}}

\end{figure}

\begin{figure}

\vskip -0.55in

\centerline{\includegraphics[width=2.5in,angle=0]{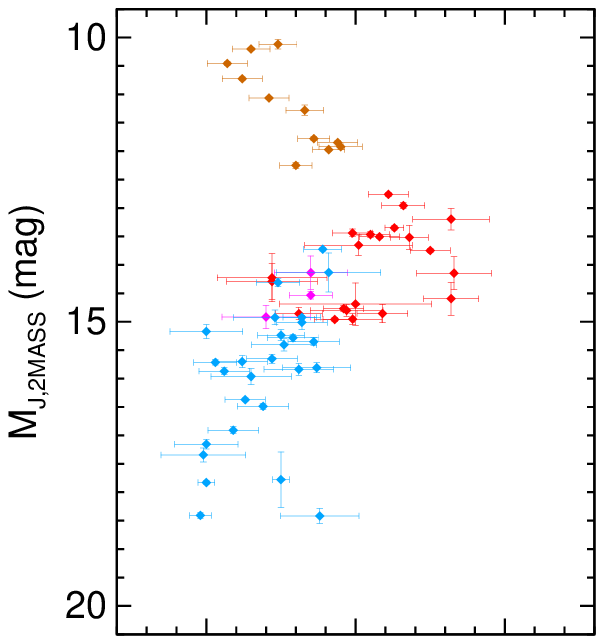} \hskip -0.60in
            \includegraphics[width=2.5in,angle=0]{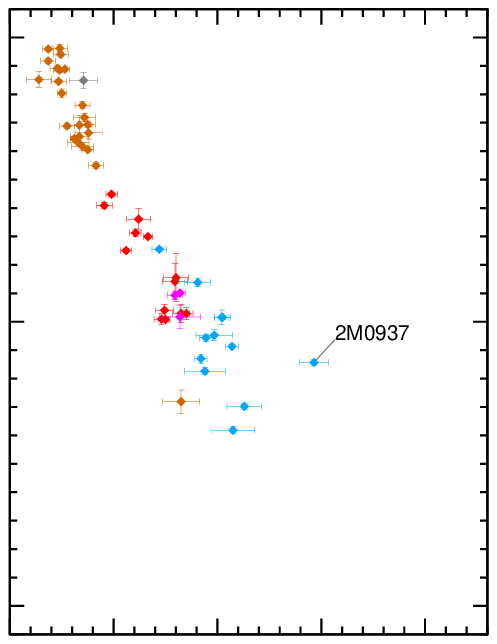}} \vskip -0.48in
\centerline{\includegraphics[width=2.5in,angle=0]{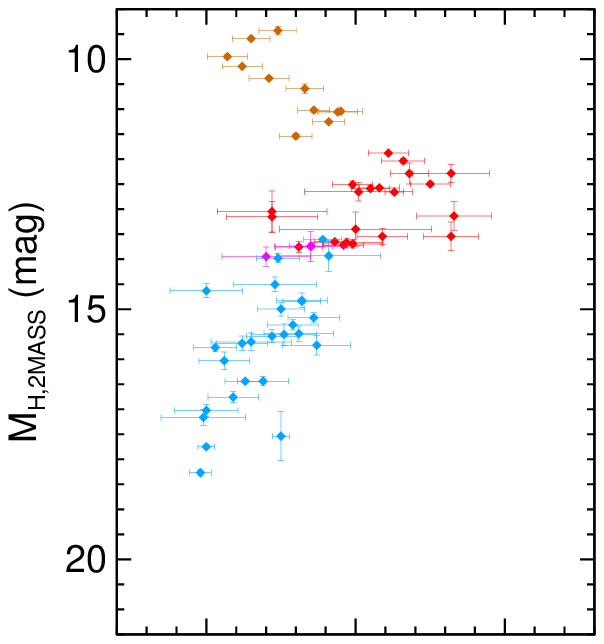} \hskip -0.60in
            \includegraphics[width=2.5in,angle=0]{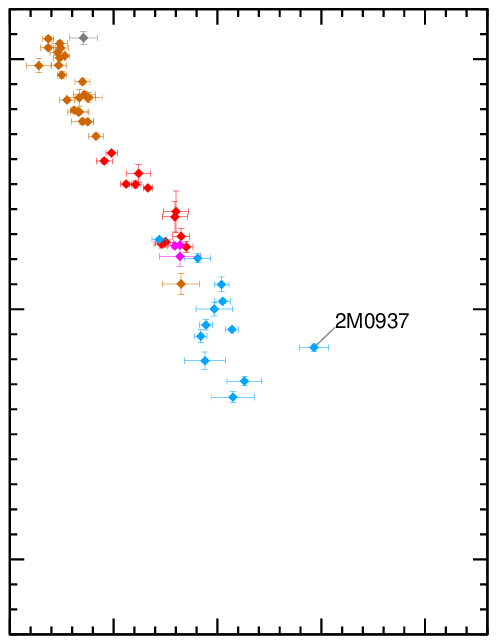}} \vskip -0.48in
\centerline{\includegraphics[width=2.5in,angle=0]{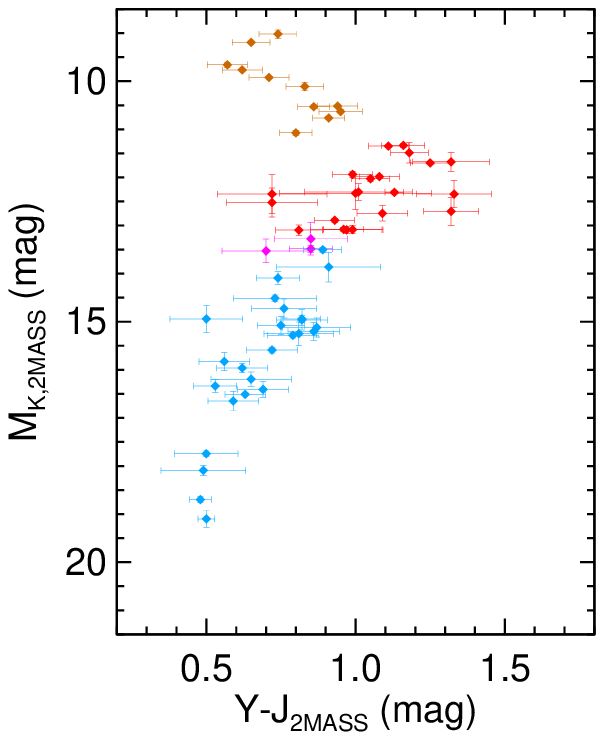} \hskip -0.60in
            \includegraphics[width=2.5in,angle=0]{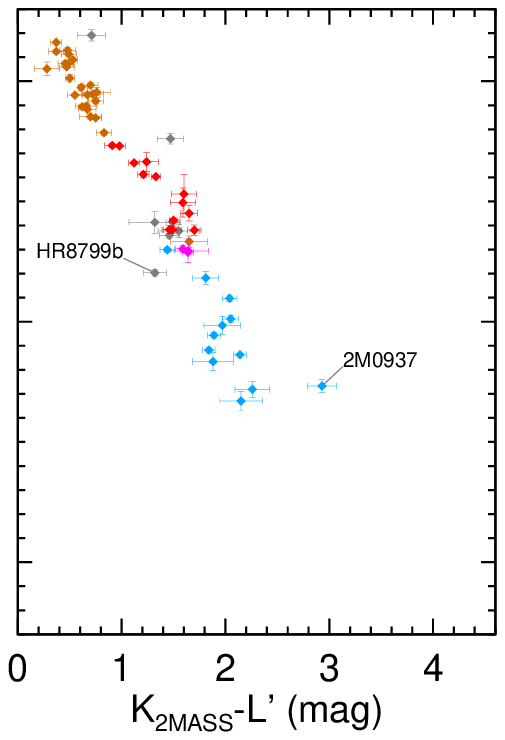}}

\caption{\normalsize Same as Figure~\ref{fig:cmd-jhk2-mko} but with
  $JHK$ photometry in the 2MASS system. \label{fig:cmd-jhk2-2mass}}

\end{figure}

\begin{landscape}

\begin{figure}

\centerline{\includegraphics[width=2.5in,angle=0]{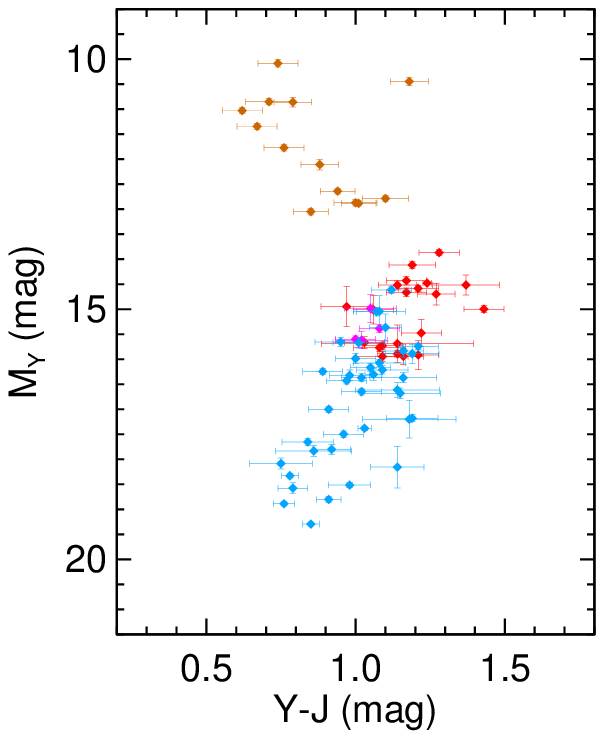} \hskip -0.60in
            \includegraphics[width=2.5in,angle=0]{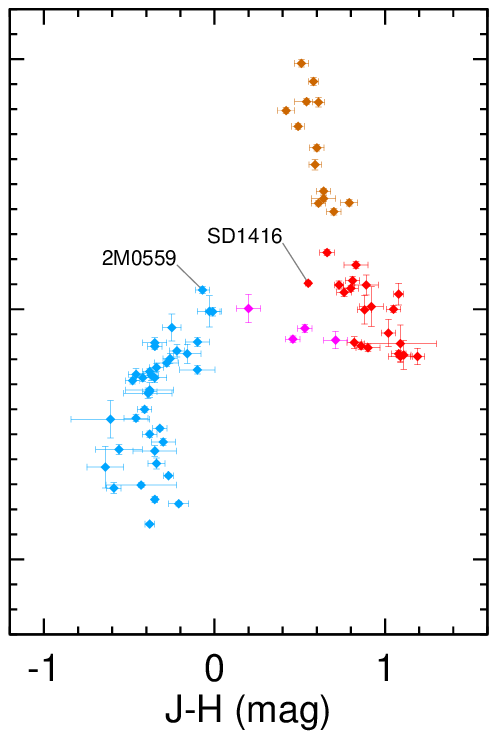} \hskip -0.60in
            \includegraphics[width=2.5in,angle=0]{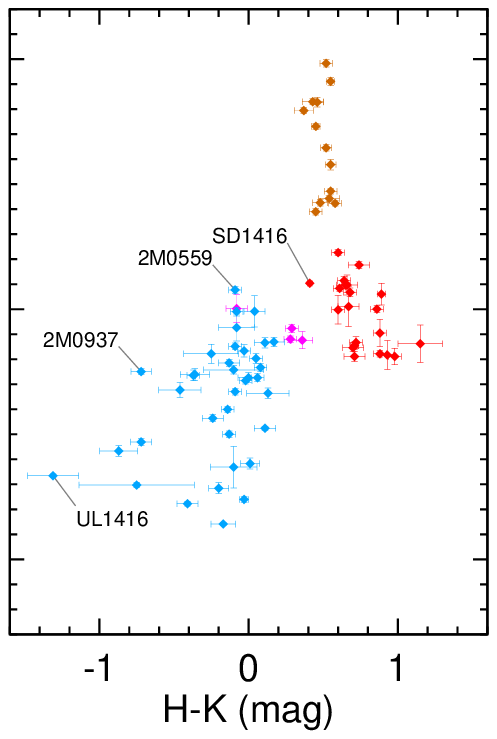} \hskip -0.60in
            \includegraphics[width=2.5in,angle=0]{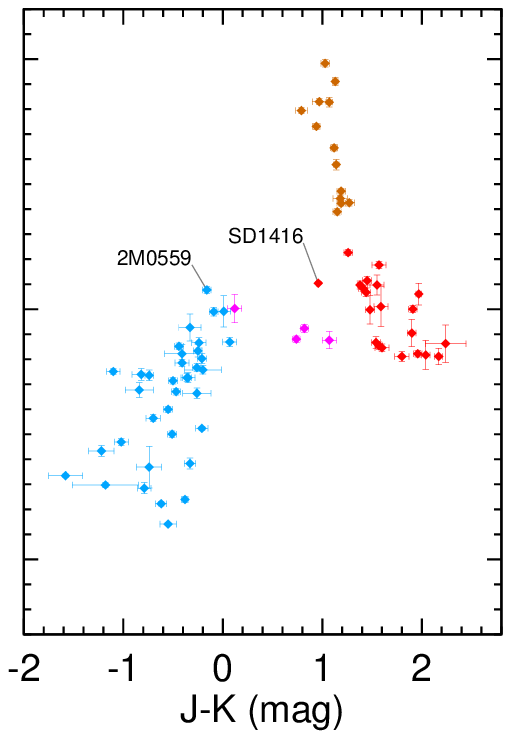}}

\caption{\normalsize Same as Figures~\ref{fig:cmd-jhk1-mko} and
  \ref{fig:cmd-jhk2-mko} but for $Y$-band absolute magnitudes
  (MKO). \label{fig:cmd-y}}

\end{figure}

\begin{figure}

\centerline{\includegraphics[width=2.5in,angle=0]{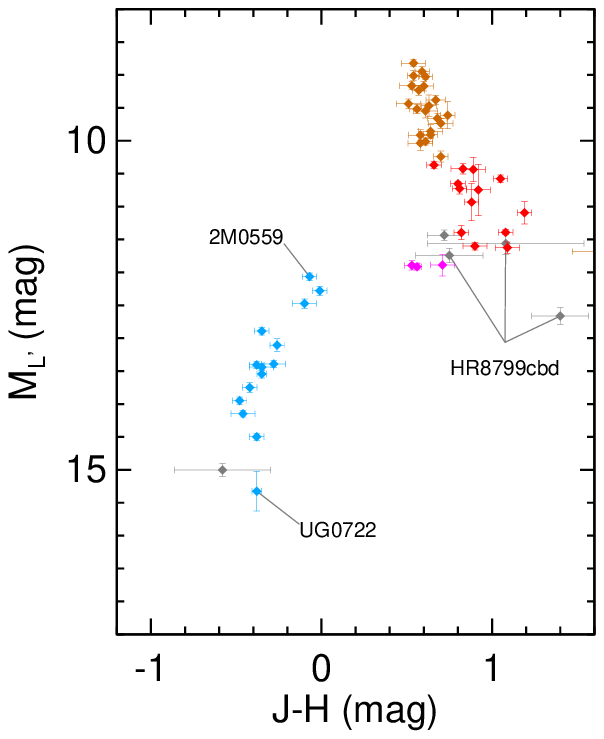} \hskip -0.60in
            \includegraphics[width=2.5in,angle=0]{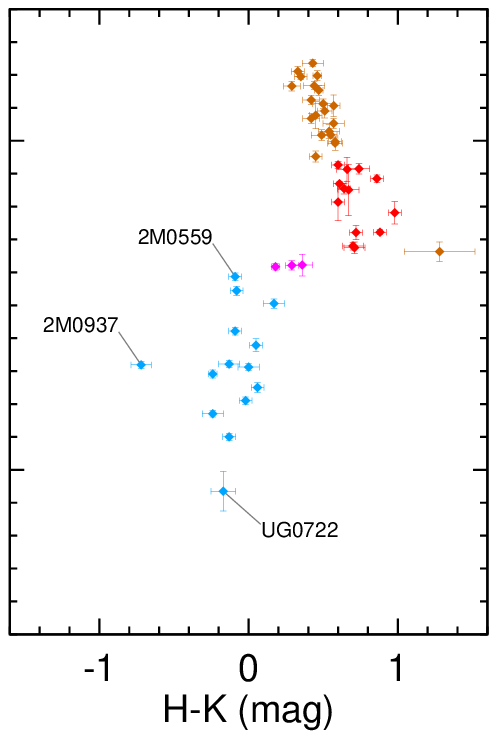} \hskip -0.60in
            \includegraphics[width=2.5in,angle=0]{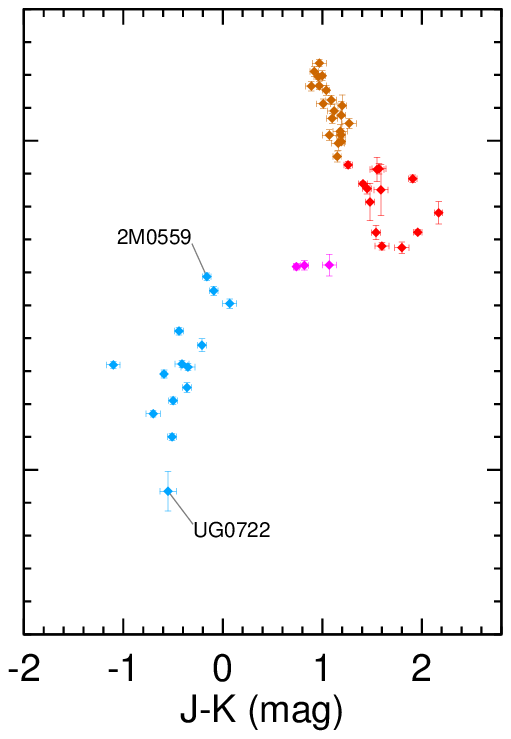} \hskip -0.60in
            \includegraphics[width=2.5in,angle=0]{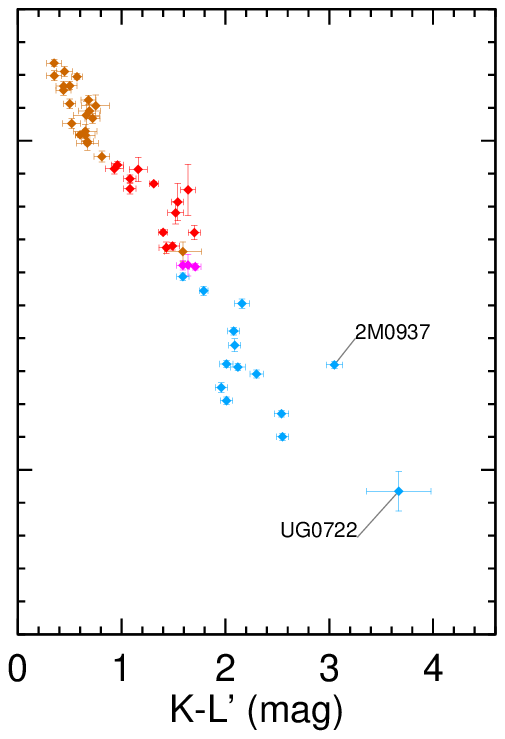}}

\caption{\normalsize Same as Figures~\ref{fig:cmd-jhk1-mko} and
  \ref{fig:cmd-jhk2-mko} but for \Lp-band absolute magnitudes
  (MKO). \label{fig:cmd-l}}

\end{figure}

\begin{figure}

\centerline{\includegraphics[width=2.5in,angle=0]{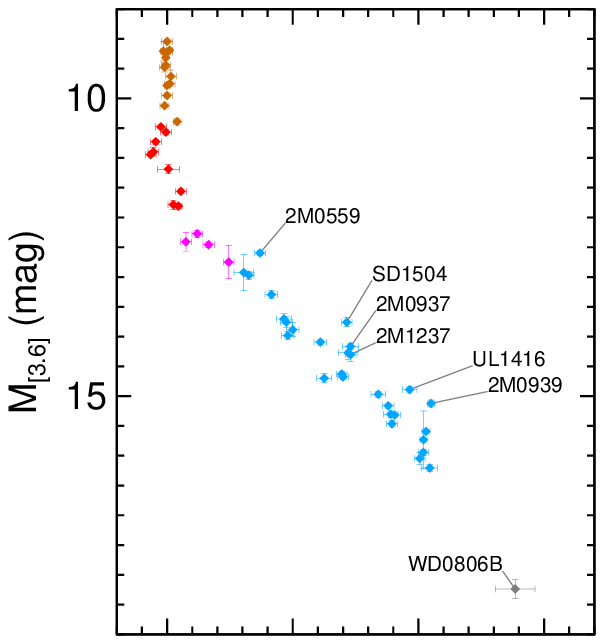} \hskip -0.60in
            \includegraphics[width=2.5in,angle=0]{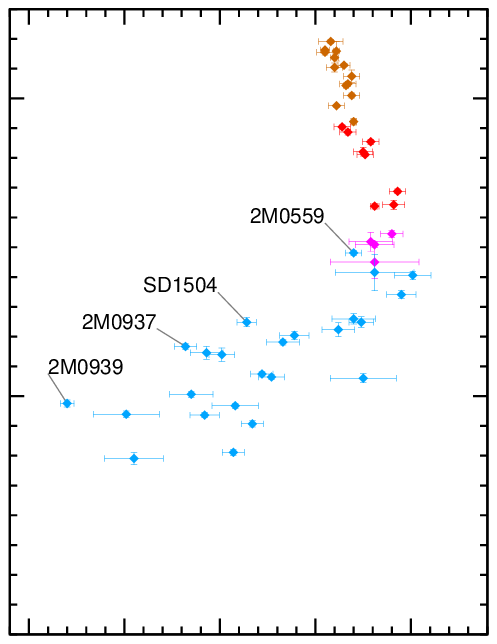} \hskip -0.60in
            \includegraphics[width=2.5in,angle=0]{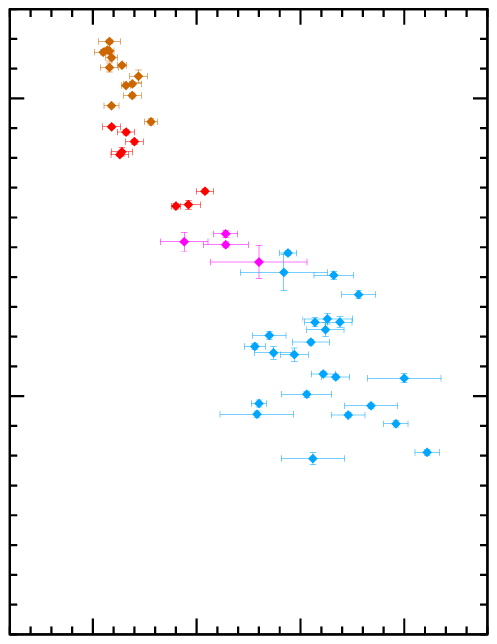} \hskip -0.60in
            \includegraphics[width=2.5in,angle=0]{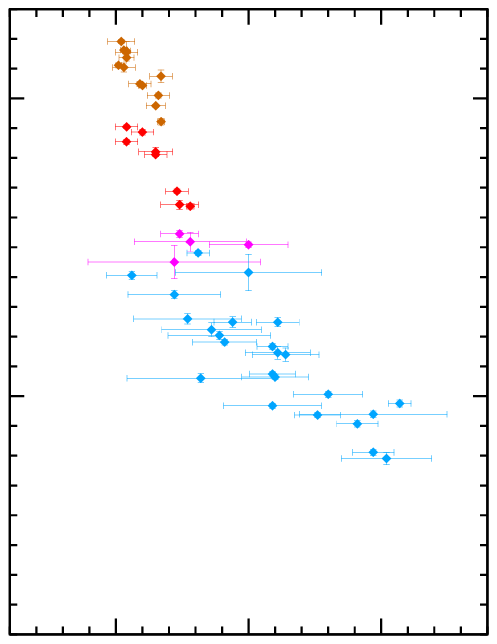}} \vskip -0.48in
\centerline{\includegraphics[width=2.5in,angle=0]{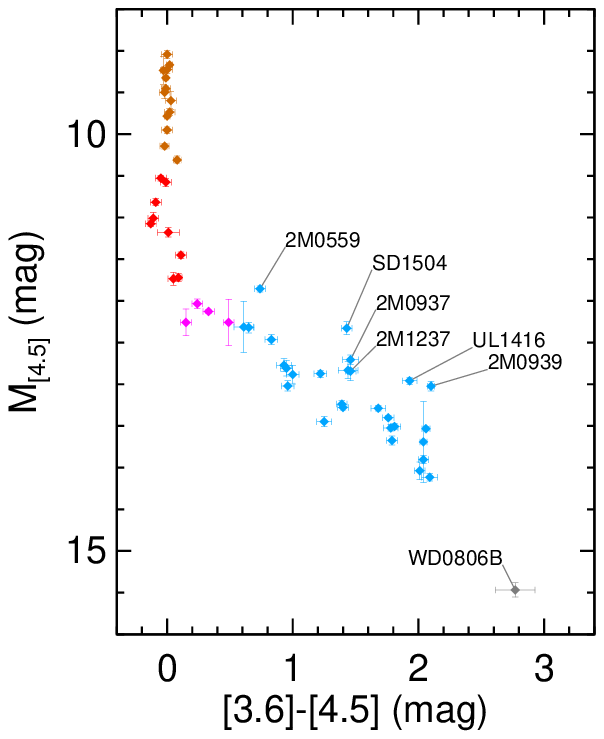} \hskip -0.60in
            \includegraphics[width=2.5in,angle=0]{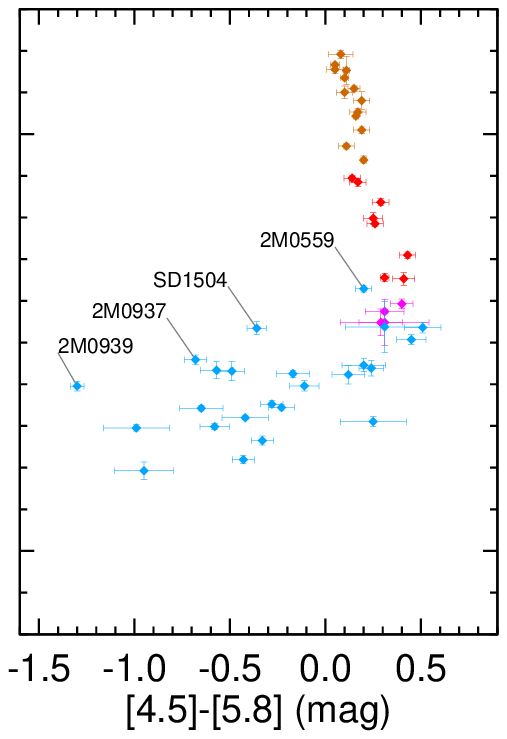} \hskip -0.60in
            \includegraphics[width=2.5in,angle=0]{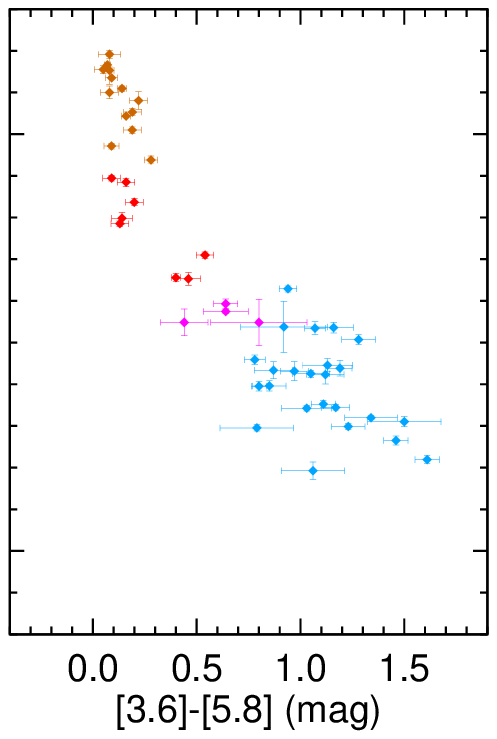} \hskip -0.60in
            \includegraphics[width=2.5in,angle=0]{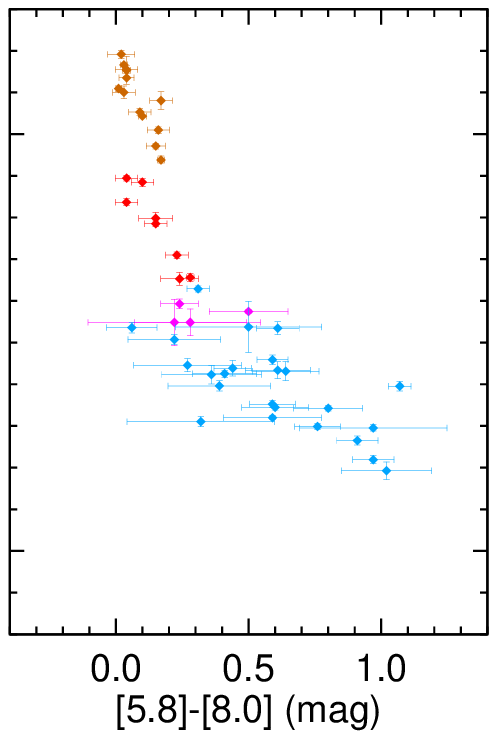}}

\caption{\normalsize Color--magnitude diagrams of \Spitzer/IRAC
  photometry showing all ultracool dwarfs from
  Table~\ref{tbl:mir-mags} with errors $\leq$0.5~mag.  Symbols are the
  same as Figure~\ref{fig:cmd-jhk1-mko}. \label{fig:cmd-irac1}}

\end{figure}

\begin{figure}

\centerline{\includegraphics[width=2.5in,angle=0]{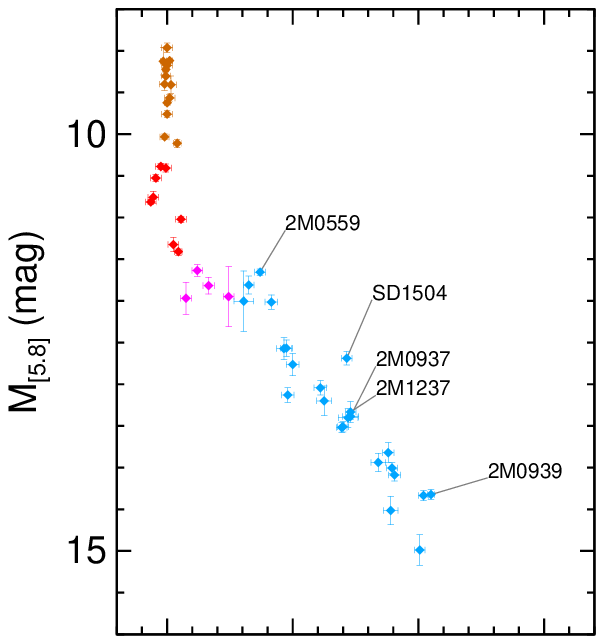} \hskip -0.60in
            \includegraphics[width=2.5in,angle=0]{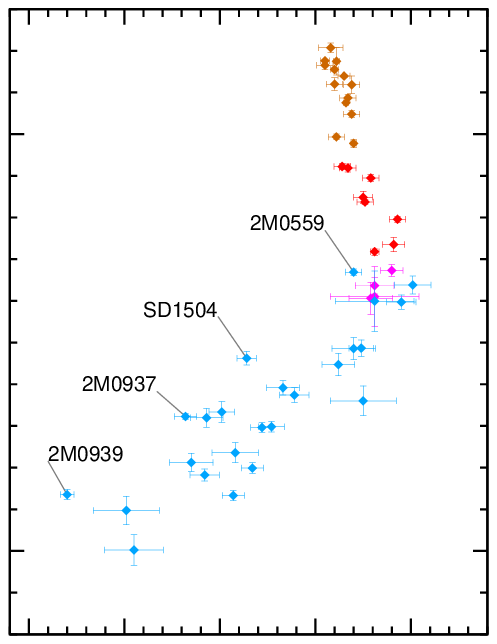} \hskip -0.60in
            \includegraphics[width=2.5in,angle=0]{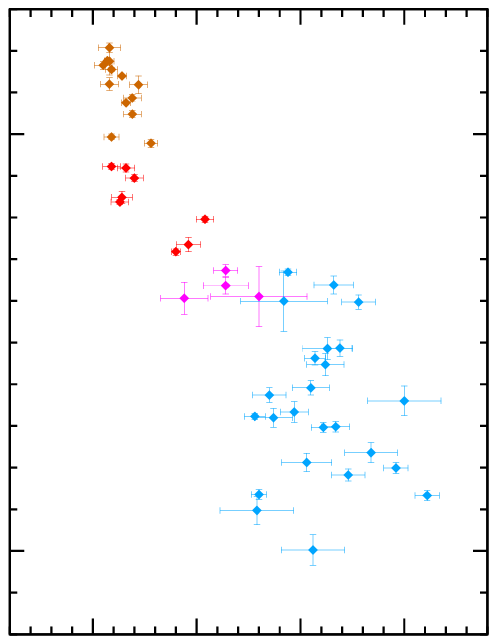} \hskip -0.60in
            \includegraphics[width=2.5in,angle=0]{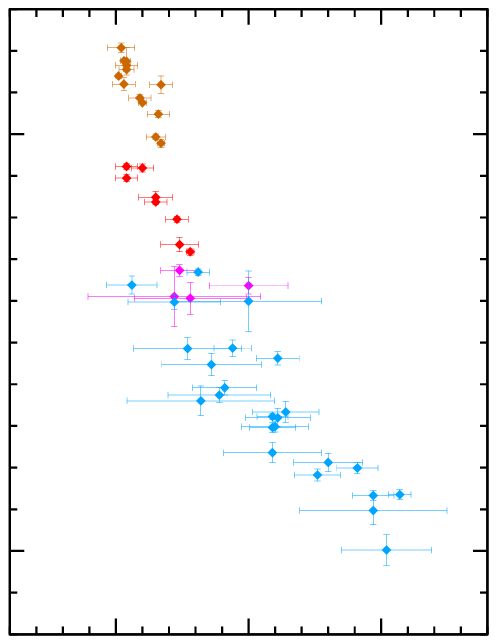}} \vskip -0.48in
\centerline{\includegraphics[width=2.5in,angle=0]{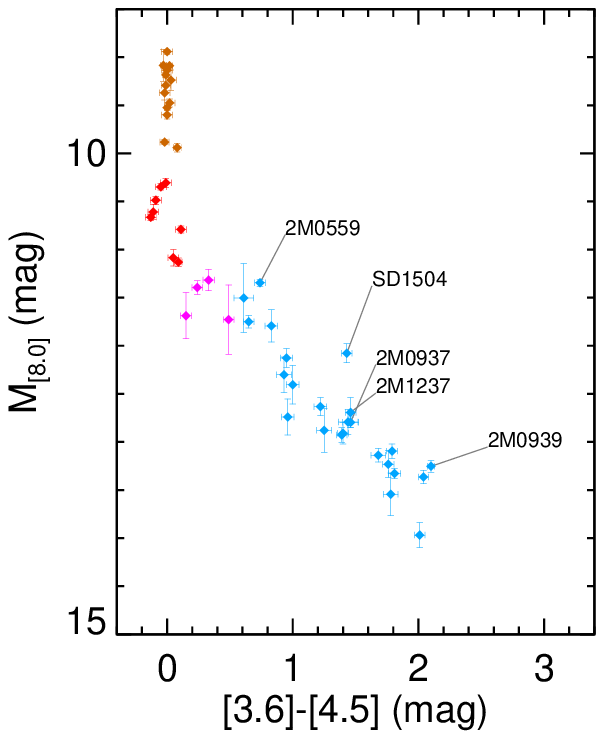} \hskip -0.60in
            \includegraphics[width=2.5in,angle=0]{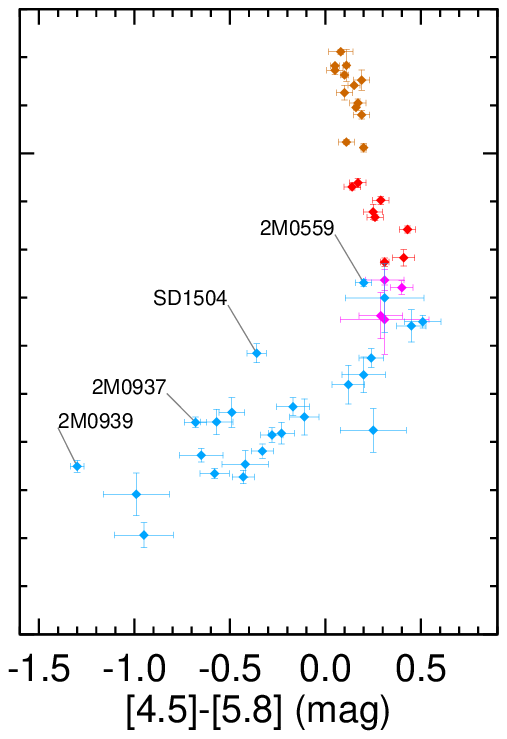} \hskip -0.60in
            \includegraphics[width=2.5in,angle=0]{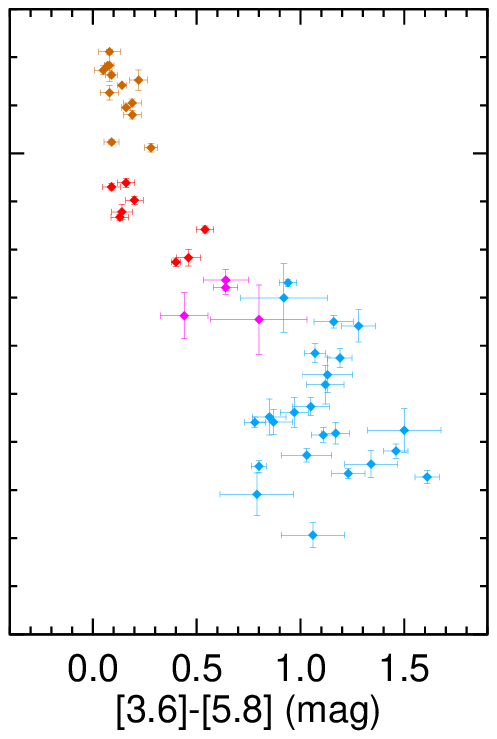} \hskip -0.60in
            \includegraphics[width=2.5in,angle=0]{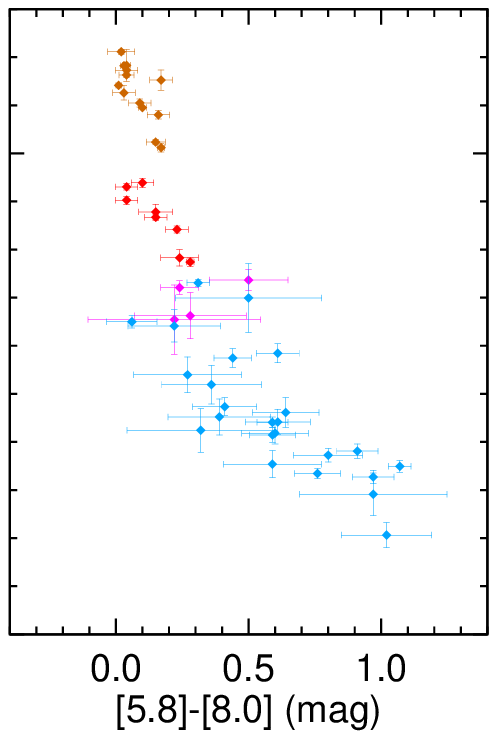}}

\caption{\normalsize Same as Figure~\ref{fig:cmd-irac1} but for the
  \Spitzer/IRAC $[5.8]$- and $[8.0]$-band absolute
  magnitudes. \label{fig:cmd-irac2}}

\end{figure}

\end{landscape}

\begin{figure}

\centerline{\includegraphics[width=2.5in,angle=0]{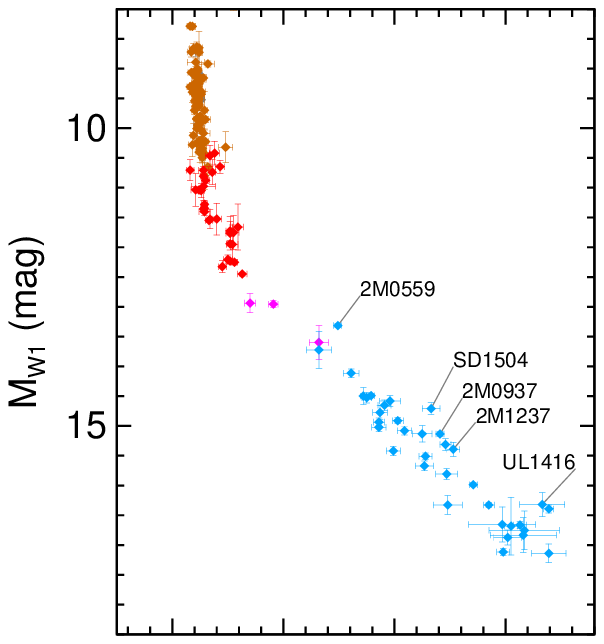} \hskip -0.60in
            \includegraphics[width=2.5in,angle=0]{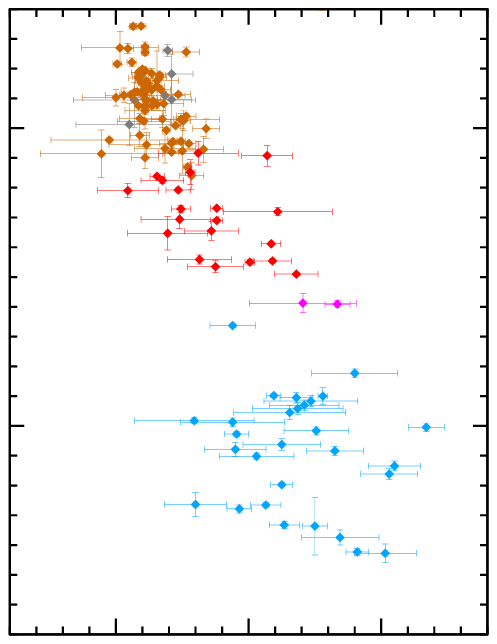} \hskip -0.60in
            \includegraphics[width=2.5in,angle=0]{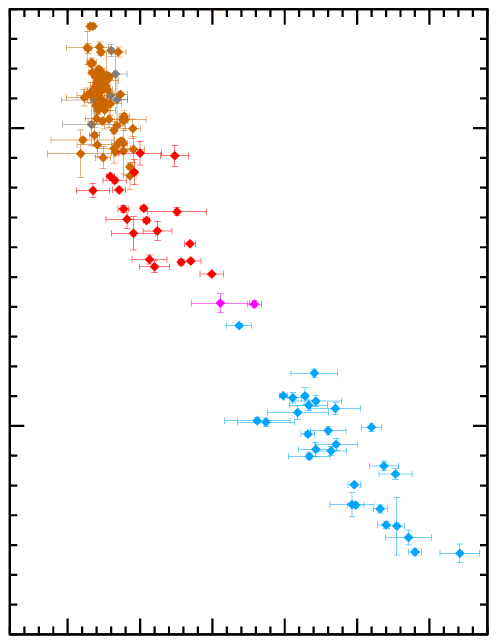}} \vskip -0.48in
\centerline{\includegraphics[width=2.5in,angle=0]{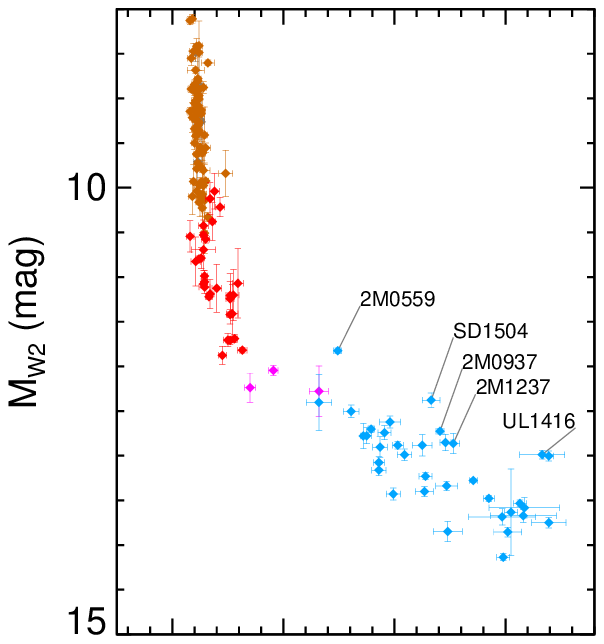} \hskip -0.60in
            \includegraphics[width=2.5in,angle=0]{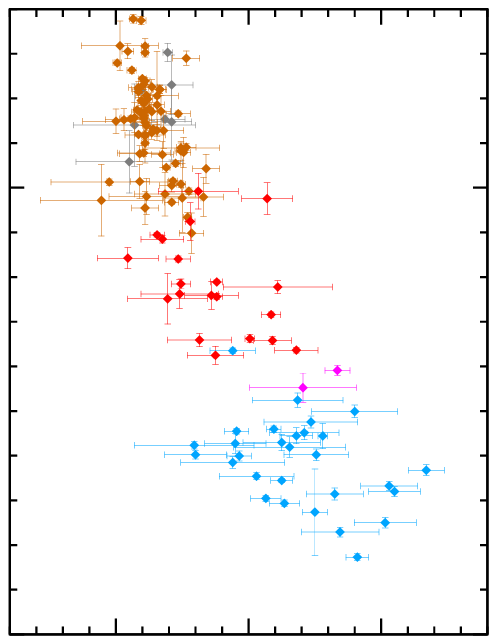} \hskip -0.60in
            \includegraphics[width=2.5in,angle=0]{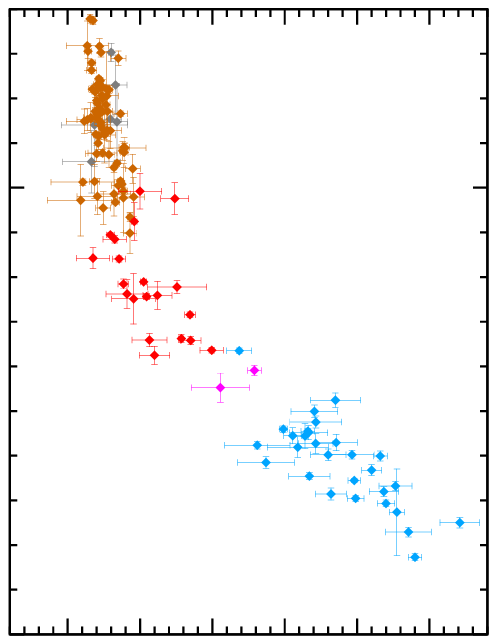}} \vskip -0.48in
\centerline{\includegraphics[width=2.5in,angle=0]{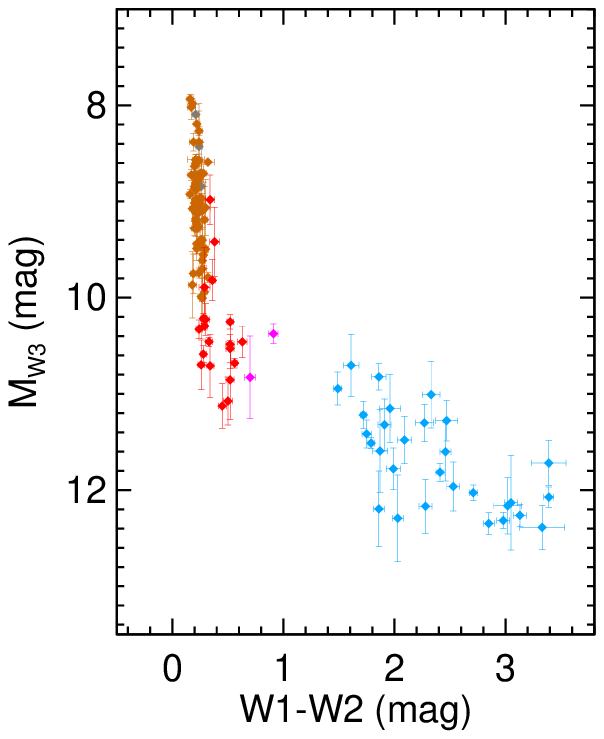} \hskip -0.60in
            \includegraphics[width=2.5in,angle=0]{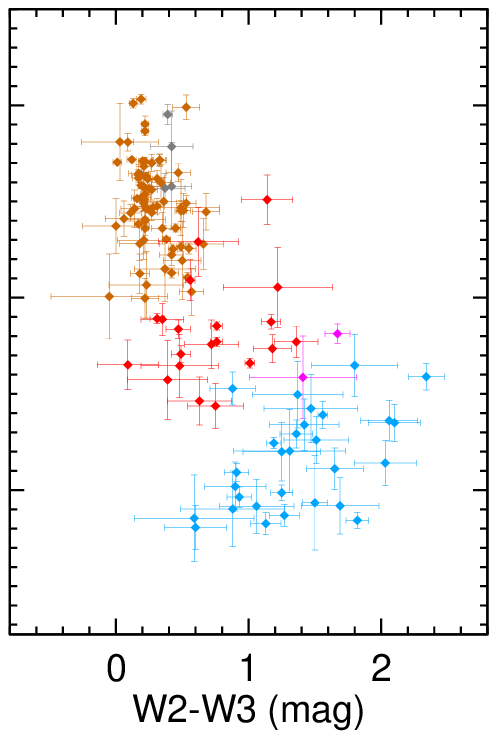} \hskip -0.60in
            \includegraphics[width=2.5in,angle=0]{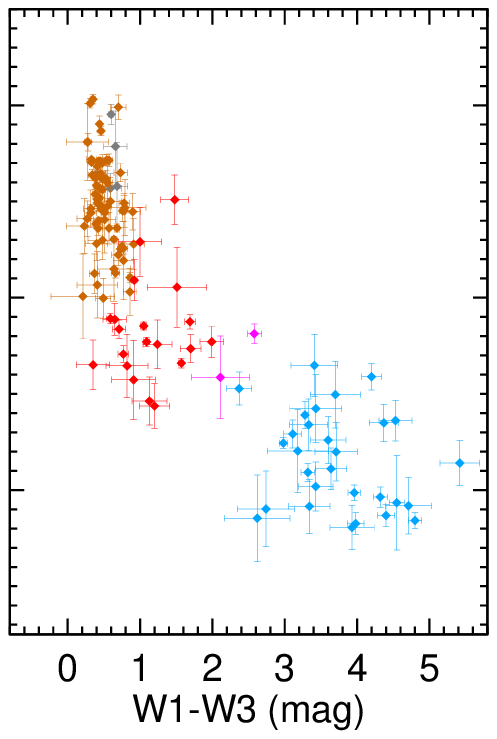}}

\caption{\normalsize Color--magnitude diagrams of \WISE\ photometry
  showing all ultracool dwarfs from Table~\ref{tbl:mir-mags} with
  errors $\leq$0.5~mag.  Symbols are the same as
  Figure~\ref{fig:cmd-jhk1-mko}. \label{fig:cmd-wise}}

\end{figure}

\begin{figure}

\vskip -0.35in
\centerline{\includegraphics[width=3.2in,angle=0]{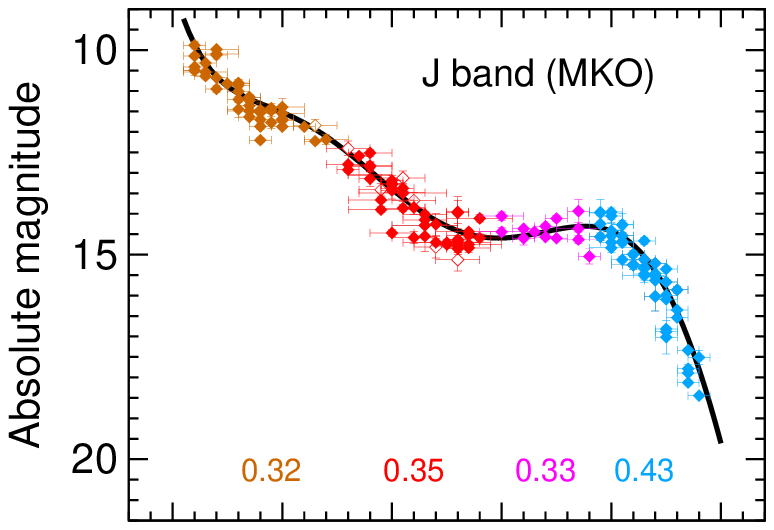} \hskip -0.60in
            \includegraphics[width=3.2in,angle=0]{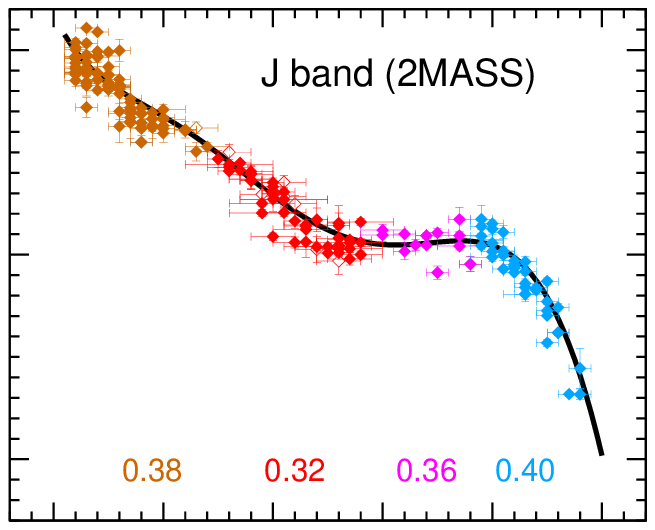}} \vskip -0.40in
\centerline{\includegraphics[width=3.2in,angle=0]{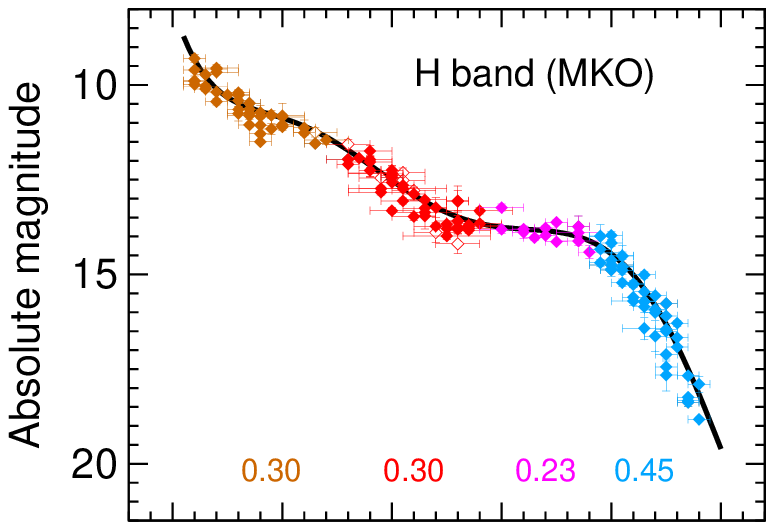} \hskip -0.60in
            \includegraphics[width=3.2in,angle=0]{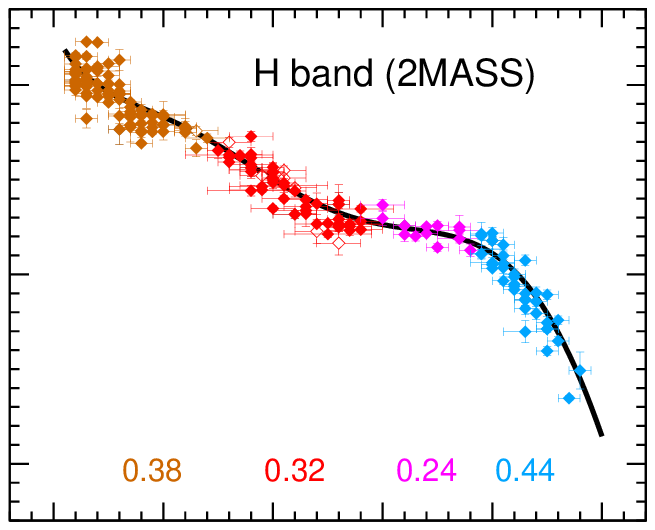}} \vskip -0.40in
\centerline{\includegraphics[width=3.2in,angle=0]{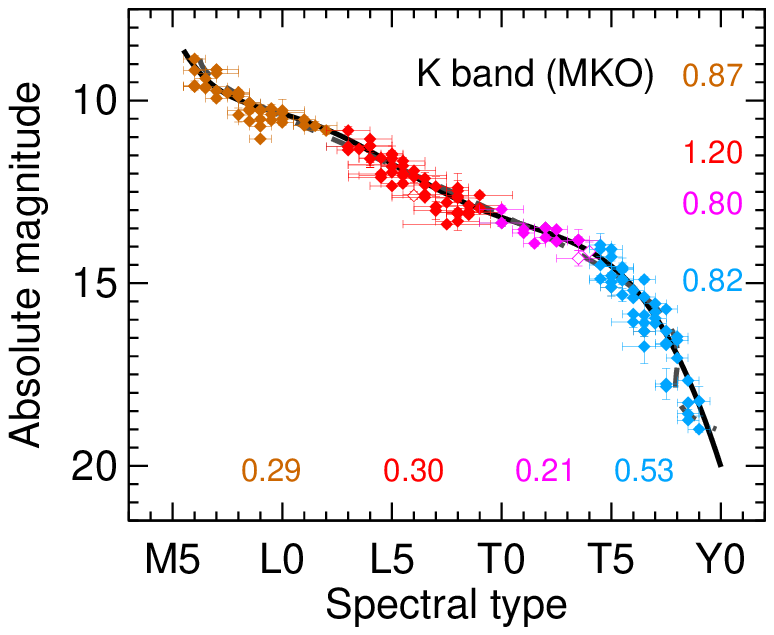} \hskip -0.60in
            \includegraphics[width=3.2in,angle=0]{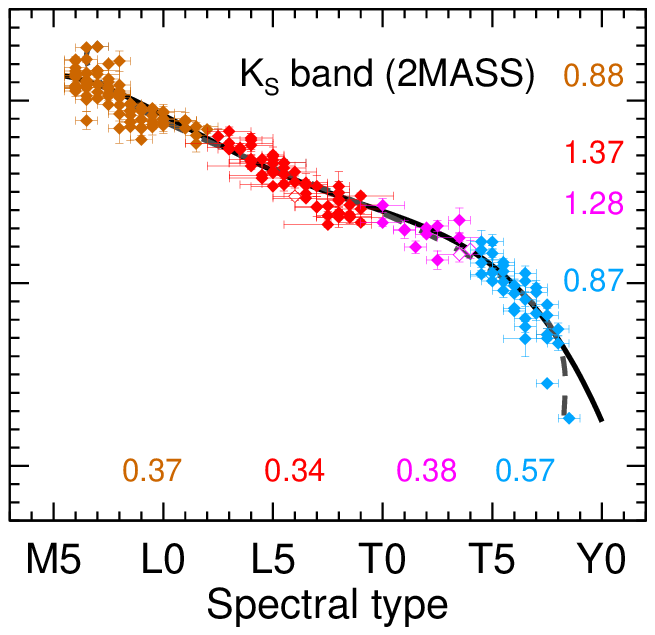}}

\caption{\normalsize Near-IR absolute magnitude as a function of
  spectral type for all ultracool field dwarfs with parallaxes
  (photometry and spectral types from Table~\ref{tbl:nir-mags};
  subdwarfs and known young or planetary-mass objects excluded from
  plot).  Thick solid lines are polynomial fits to the data
  (coefficients given in Table~\ref{tbl:poly}).  At the bottom of each
  panel the rms about the fit is given, broken down by spectral type
  range: M6--L2 (brown), L2.5--L9 (red), L9.5--T4 (purple), and
  $\geq$T4.5 (blue).  Dashed gray lines show inverted polynomial fits,
  i.e., spectral type as a function of magnitude, for bands that are
  sufficiently monotonic ($K$ and \Ks\ here).  The rms about these
  fits for the same spectral type ranges as listed above are given
  along the right side of the respective panels. We use optical
  spectral types for M and L~dwarfs when available (infrared types
  otherwise) and infrared types for T~dwarfs. \label{fig:spt-abs-jhk}}

\end{figure}

\begin{figure}

\centerline{\includegraphics[width=3.2in,angle=0]{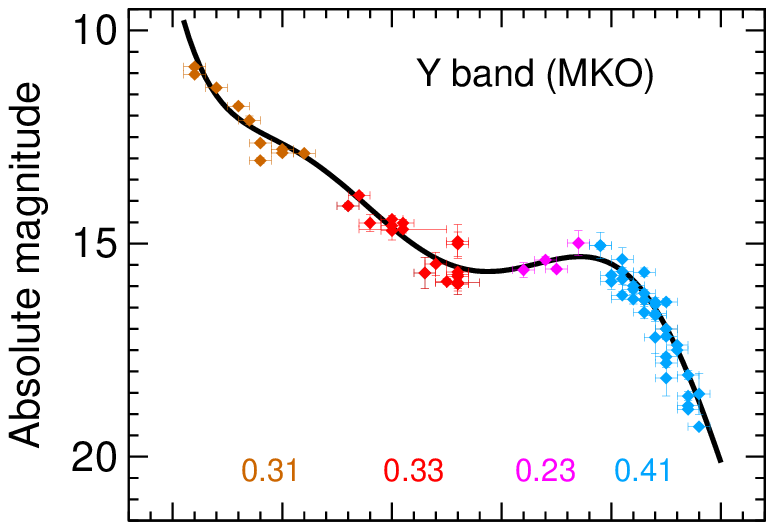}} \vskip -0.40in
\centerline{\includegraphics[width=3.2in,angle=0]{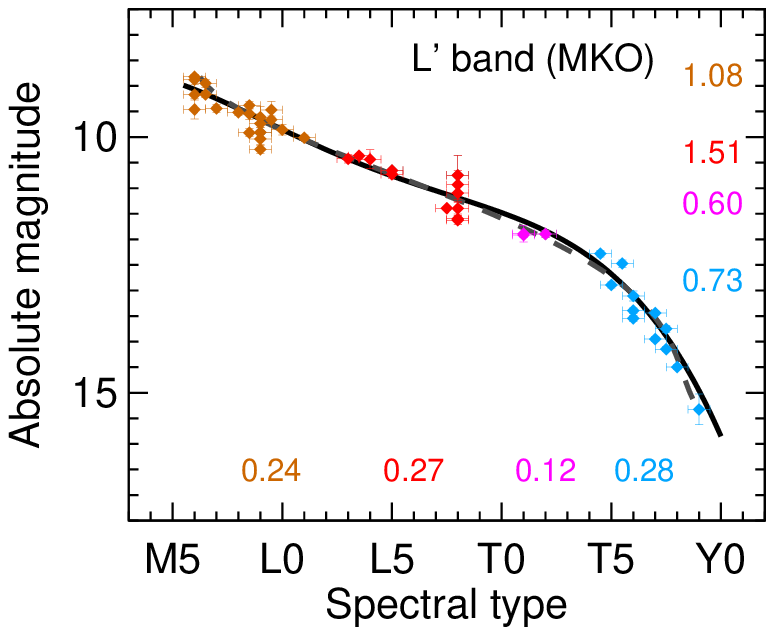}}

\caption{\normalsize Same as Figure~\ref{fig:spt-abs-jhk} but for $Y$-
  and \Lp-band absolute magnitudes. \label{fig:spt-abs-yl}}

\end{figure}

\begin{figure}

\centerline{\includegraphics[width=3.2in,angle=0]{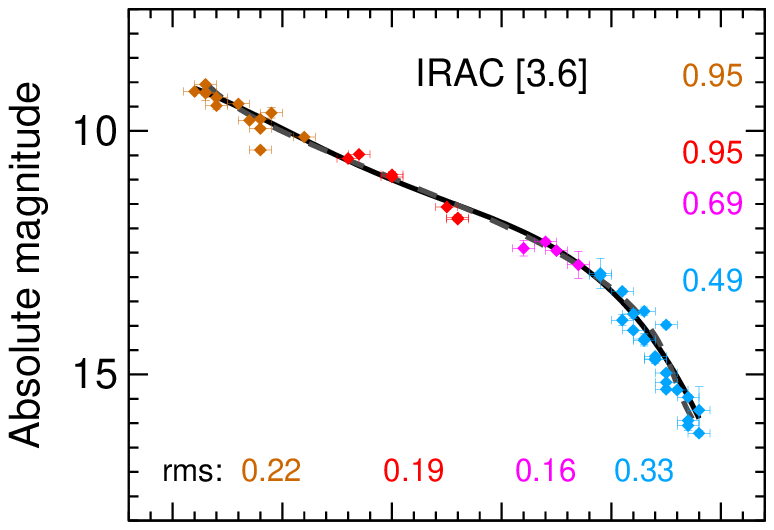} \hskip -0.60in
            \includegraphics[width=3.2in,angle=0]{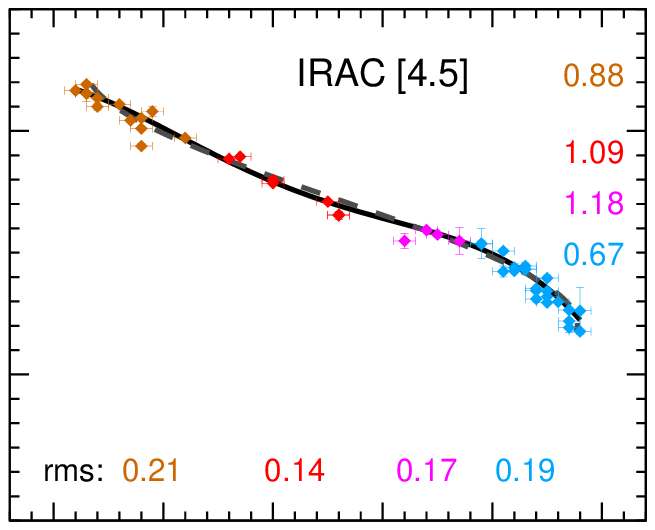}} \vskip -0.48in
\centerline{\includegraphics[width=3.2in,angle=0]{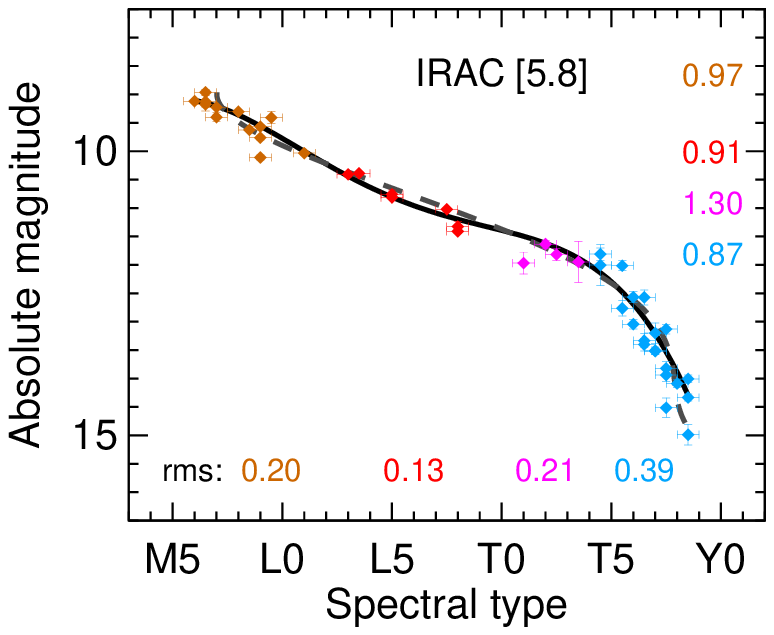} \hskip -0.60in
            \includegraphics[width=3.2in,angle=0]{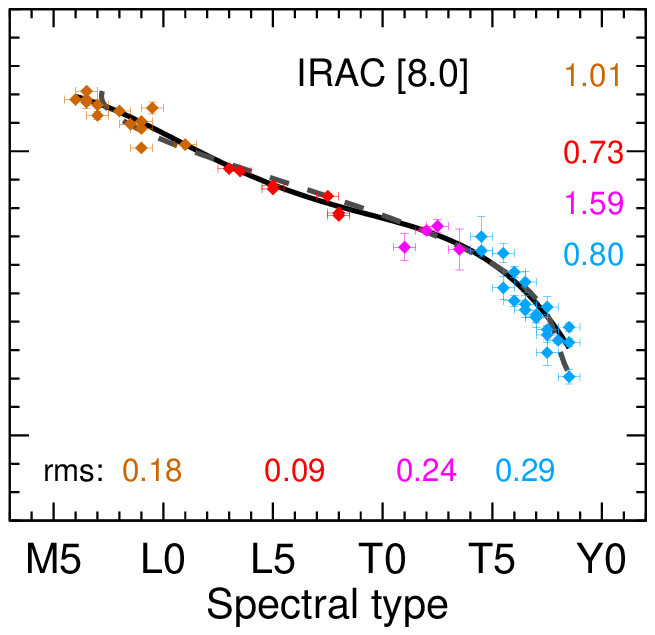}}

\caption{\normalsize Same as Figures~\ref{fig:spt-abs-jhk} and
  \ref{fig:spt-abs-yl} but for \Spitzer/IRAC
  photometry. \label{fig:spt-abs-irac}}

\end{figure}

\begin{figure}

\centerline{\includegraphics[width=3.2in,angle=0]{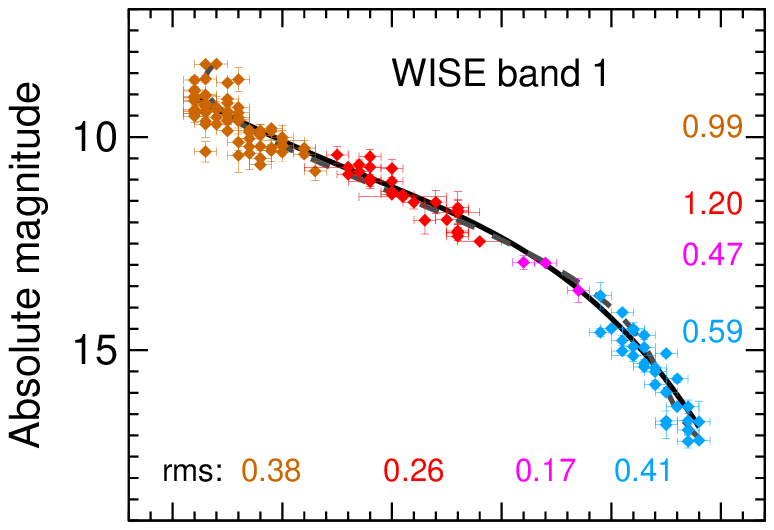} \hskip -0.60in
            \includegraphics[width=3.2in,angle=0]{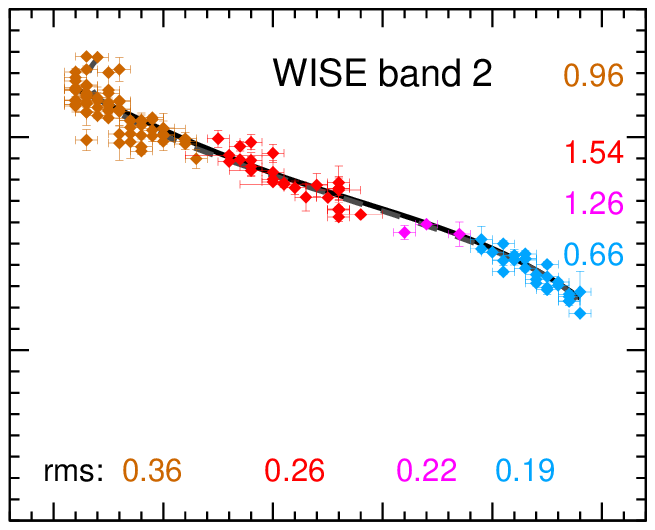}} \vskip -0.48in
\centerline{\includegraphics[width=3.2in,angle=0]{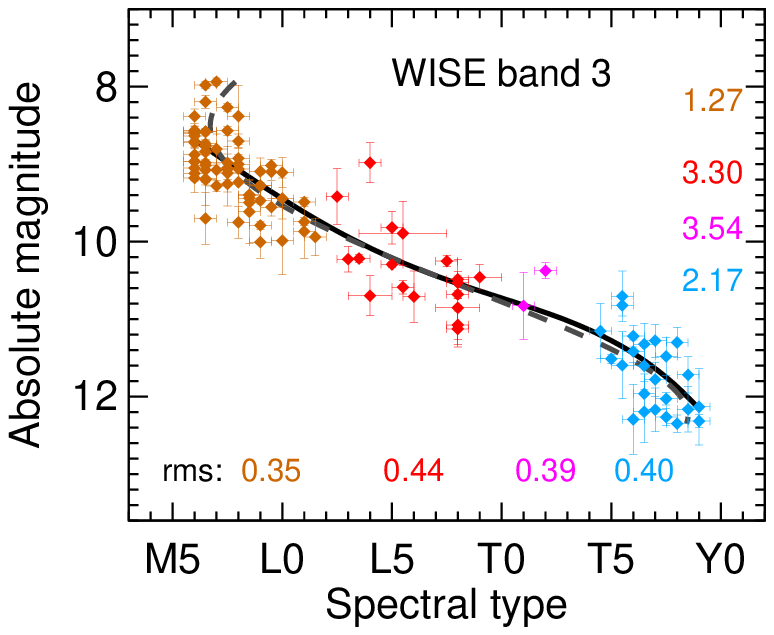} \hskip -0.60in
            \includegraphics[width=3.2in,angle=0]{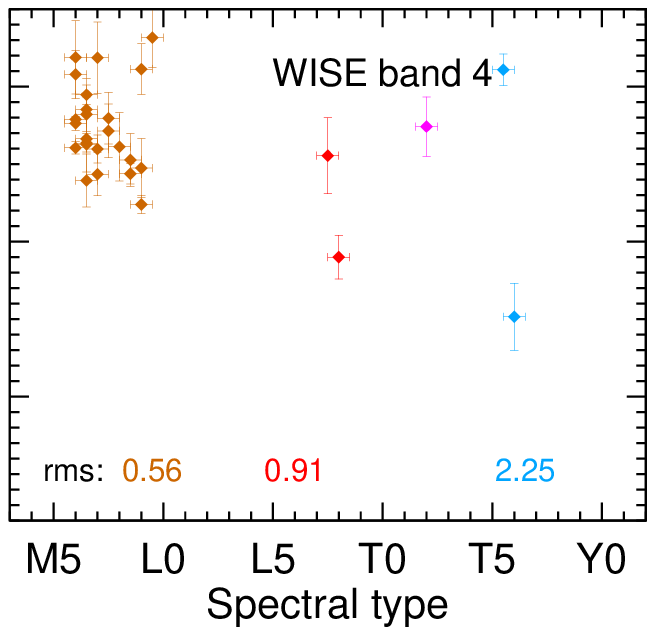}}

\caption{\normalsize Same as Figures~\ref{fig:spt-abs-jhk},
  \ref{fig:spt-abs-yl}, and \ref{fig:spt-abs-irac} but for \WISE\
  photometry. \label{fig:spt-abs-wise}}

\end{figure}

\begin{landscape}

\begin{figure}

\centerline{\includegraphics[width=2.5in,angle=0]{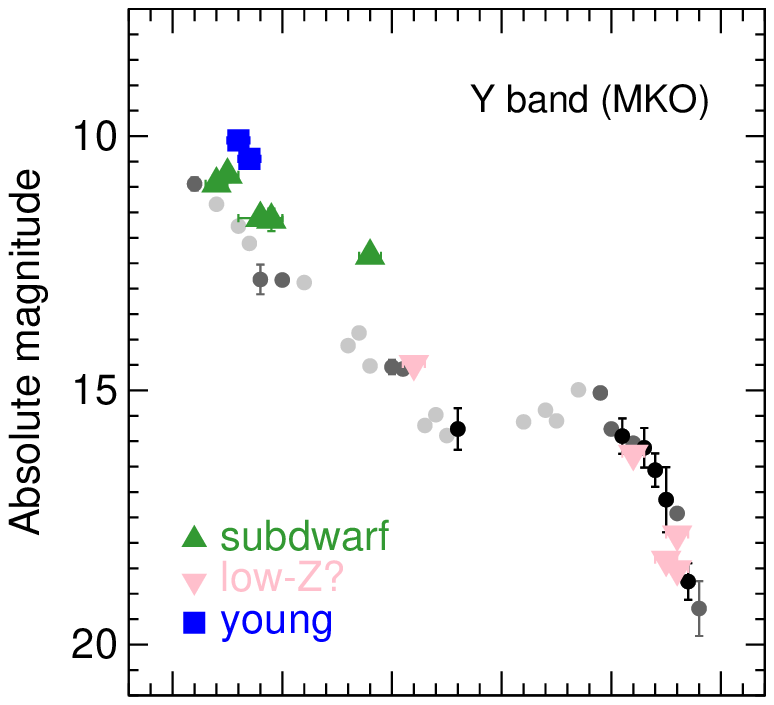} \hskip -0.50in
            \includegraphics[width=2.5in,angle=0]{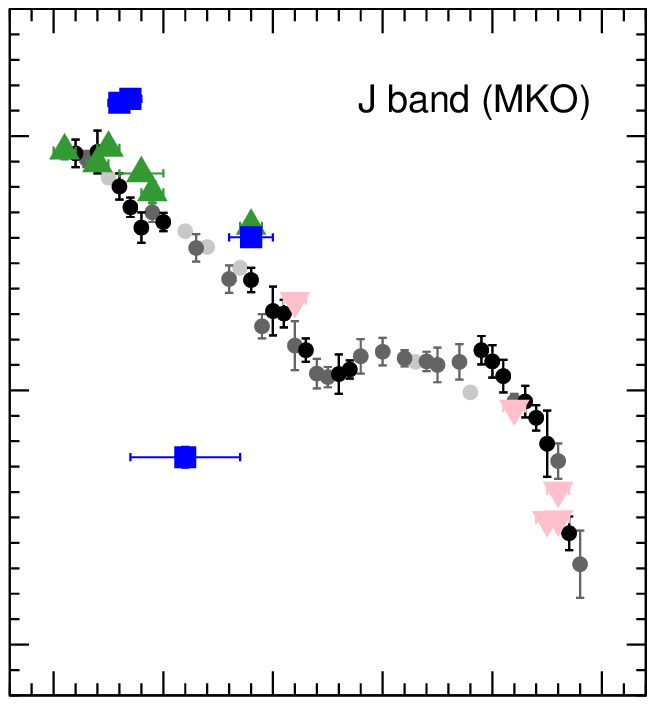} \hskip -0.50in
            \includegraphics[width=2.5in,angle=0]{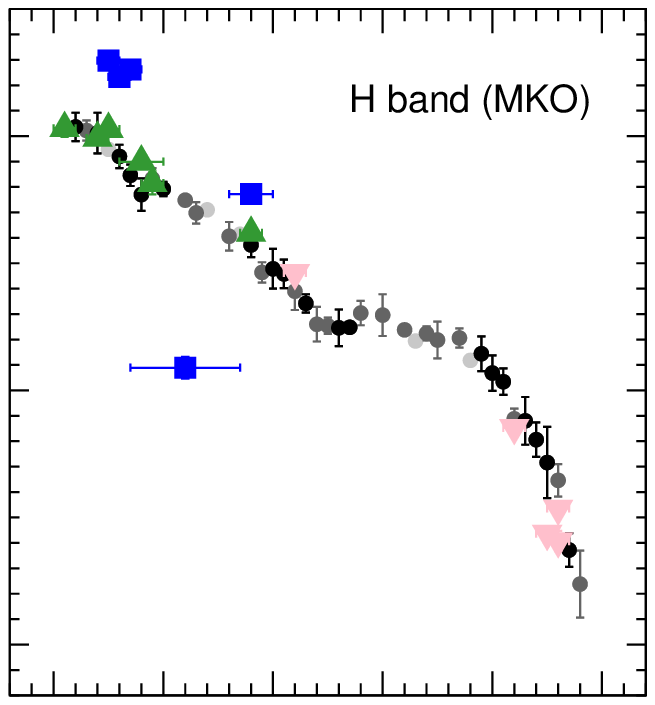} \hskip -0.50in
            \includegraphics[width=2.5in,angle=0]{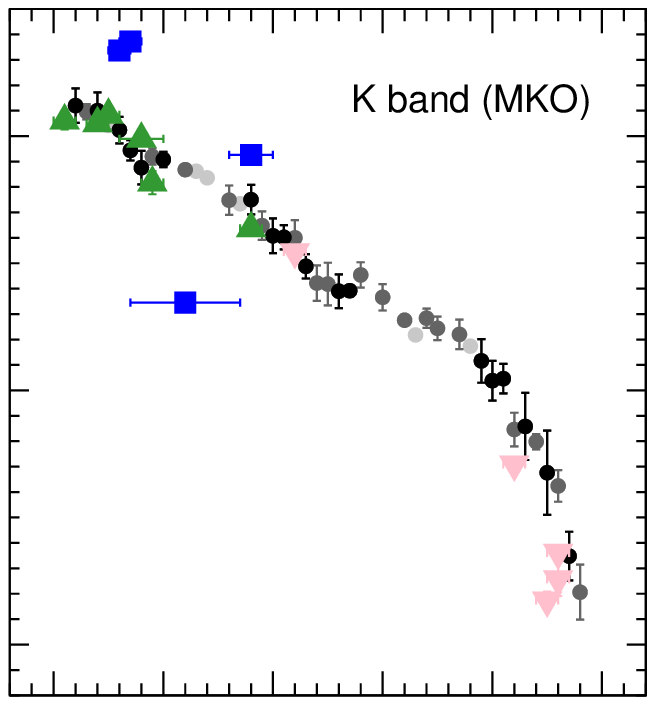}} \vskip -0.40in
\centerline{\includegraphics[width=2.5in,angle=0]{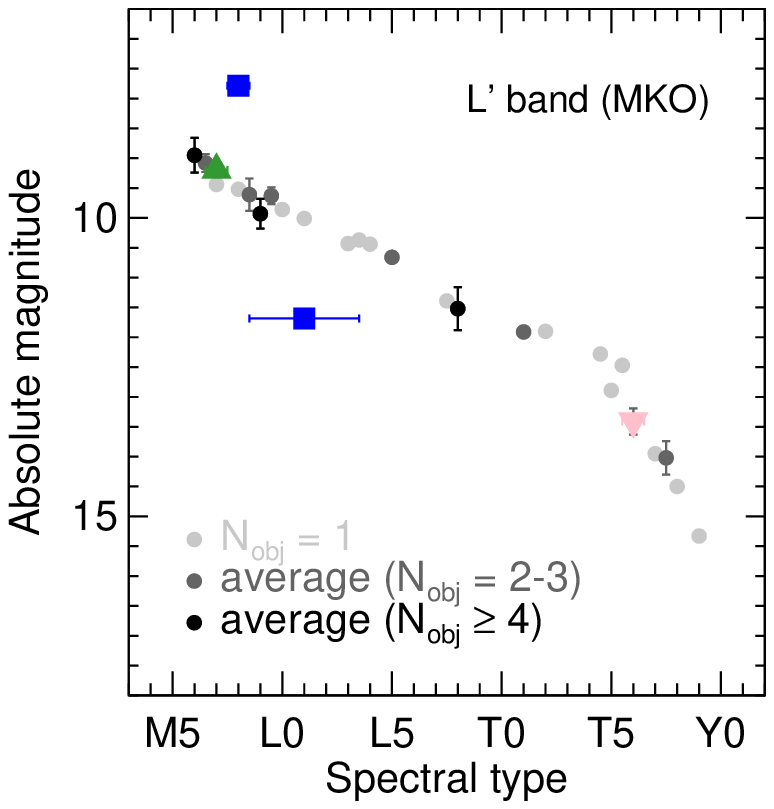} \hskip -0.50in
            \includegraphics[width=2.5in,angle=0]{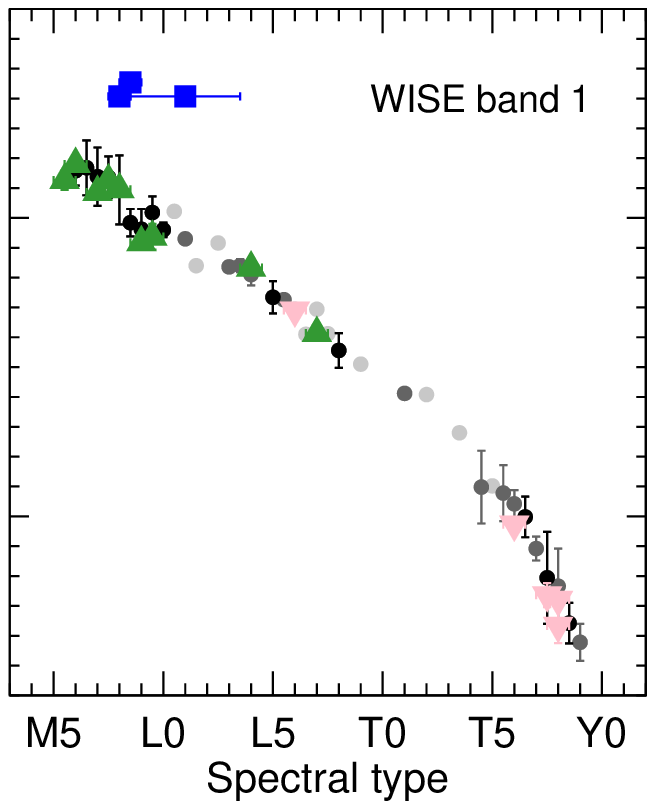} \hskip -0.50in
            \includegraphics[width=2.5in,angle=0]{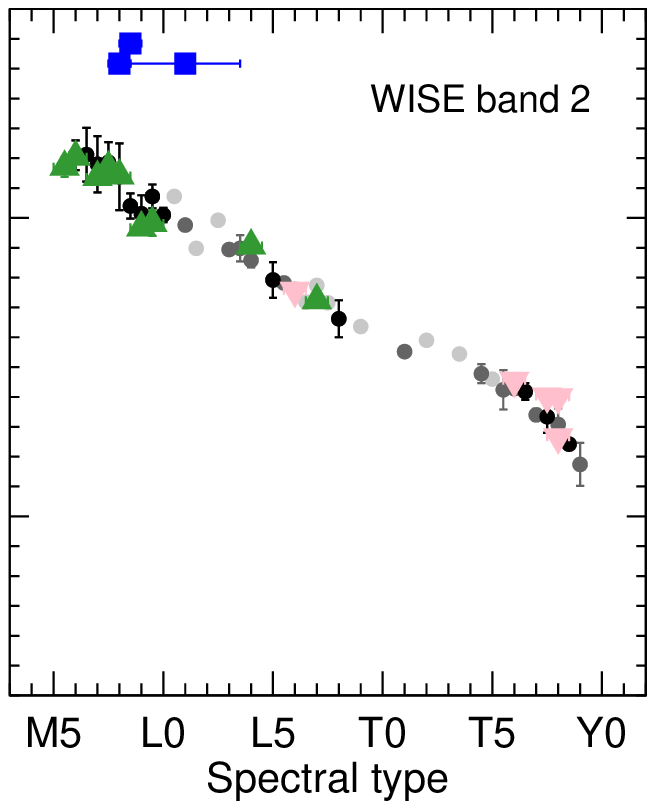} \hskip -0.50in
            \includegraphics[width=2.5in,angle=0]{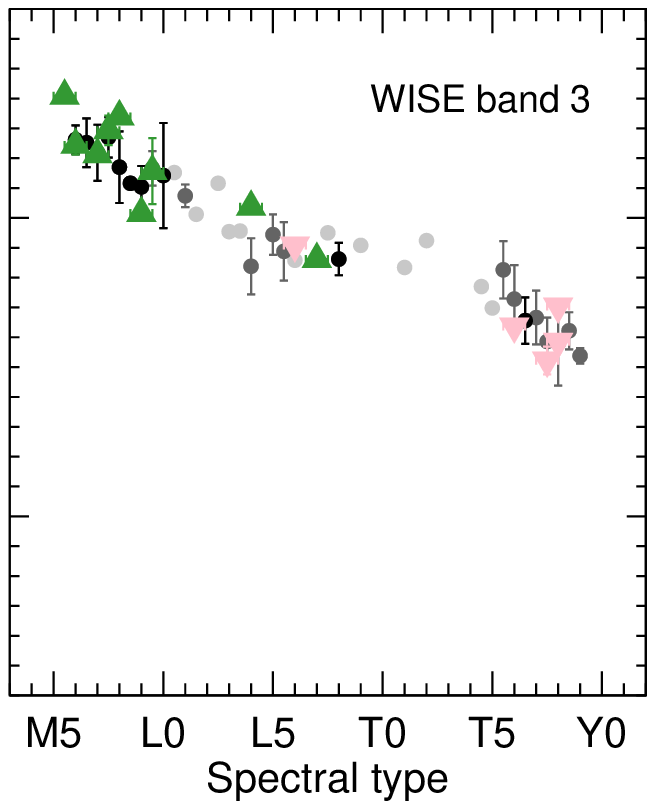}}

\caption{\normalsize Gray and black circles show the weighted averages
  of absolute magnitudes for normal field dwarfs as a function of
  spectral type with error bars showing the rms among objects of a
  given type (Tables~\ref{tbl:spt-abs-mko}--\ref{tbl:spt-abs-wise}).
  We also show individual objects that were not considered to be
  normal field dwarfs: known subdwarfs as determined from optical
  spectroscopy (green triangles); suspected low-metallicity brown
  dwarfs (pink upside-down triangles); and members of young moving
  groups, including planetary mass companions (blue squares; the 
  underluminous young object is 2M~1207b).  The
  objects plotted here are marked in the ``Note'' column in
  Table~\ref{tbl:plx-all} and are excluded from our analysis of
  absolute magnitudes as a function of spectral type, along with
  over-luminous objects suspected to be unresolved binaries (not shown
  here). \label{fig:spt-wavg}}

\end{figure}

\begin{figure}

\centerline{\includegraphics[width=3.2in,angle=0]{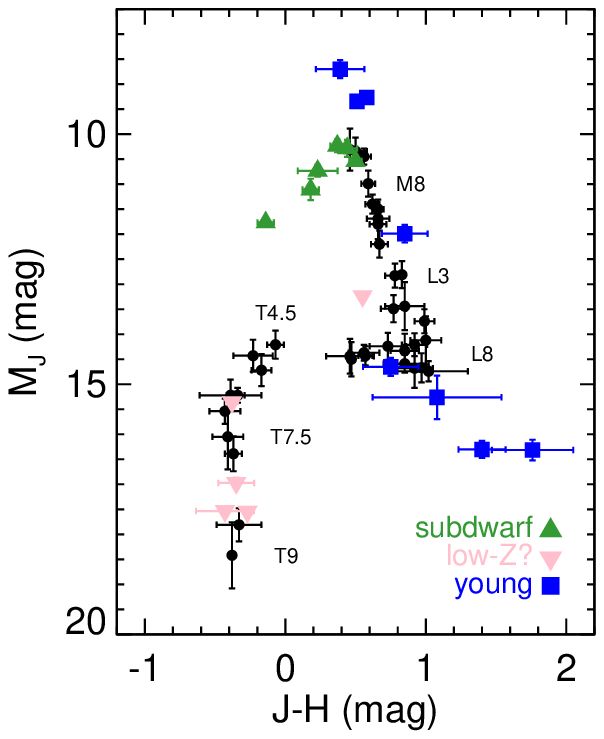} \hskip -0.70in
            \includegraphics[width=3.2in,angle=0]{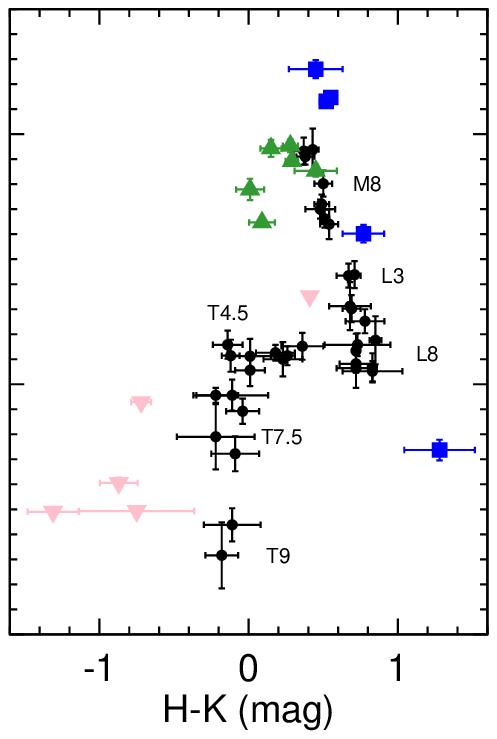} \hskip -0.70in
            \includegraphics[width=3.2in,angle=0]{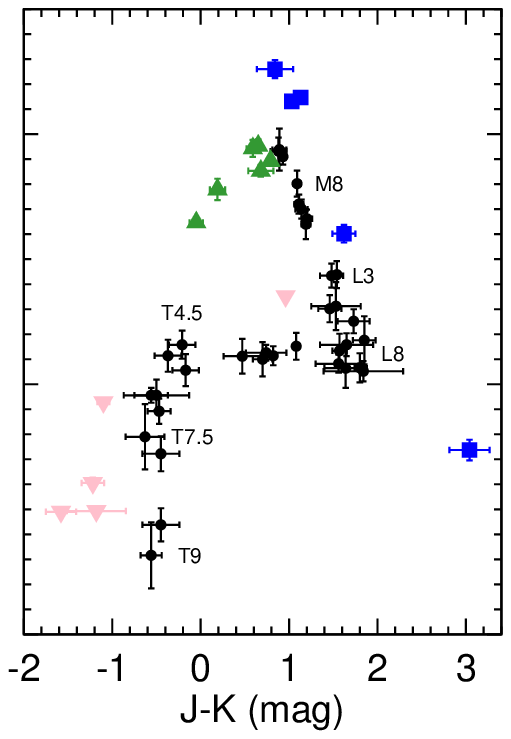}}

\caption{\normalsize Color--magnitude diagrams on the MKO system,
  where black circles show weighted averages of normal field dwarf
  magnitudes in spectral type bins from M6 to T9 (see
  Table~\ref{tbl:spt-abs-mko}; only bins with 2 or more objects are
  shown).  Error bars indicate the rms in absolute magnitude and color
  for each bin.  Individual objects that were not considered to be
  normal field dwarfs are also shown: known subdwarfs as determined
  from optical spectroscopy (green triangles); suspected
  low-metallicity brown dwarfs (pink upside-down triangles); and
  members of young moving groups, including planetary mass companions
  (blue squares).   \label{fig:cmd-wavg}}

\end{figure}

\end{landscape}

\begin{figure}

\centerline{\includegraphics[width=3.2in,angle=0]{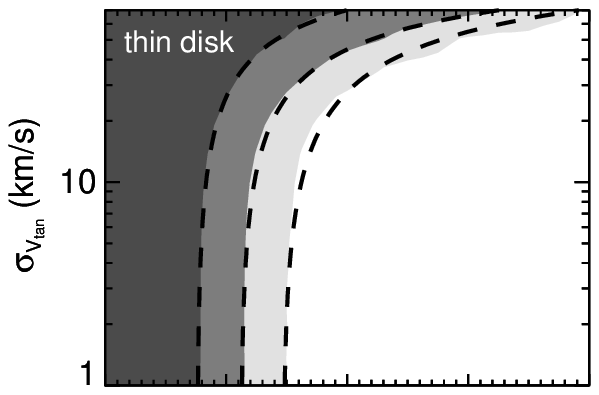}} \vskip -0.55in
\centerline{\includegraphics[width=3.2in,angle=0]{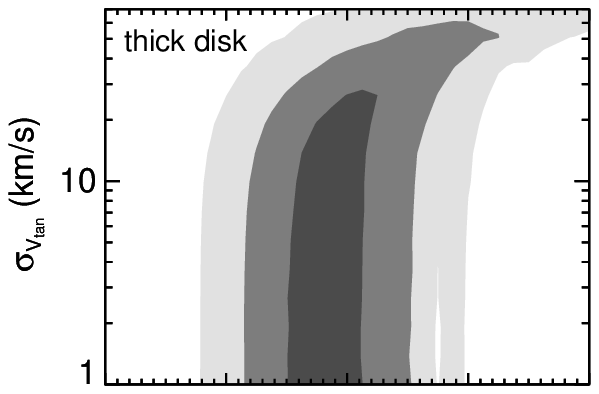}} \vskip -0.55in
\centerline{\includegraphics[width=3.2in,angle=0]{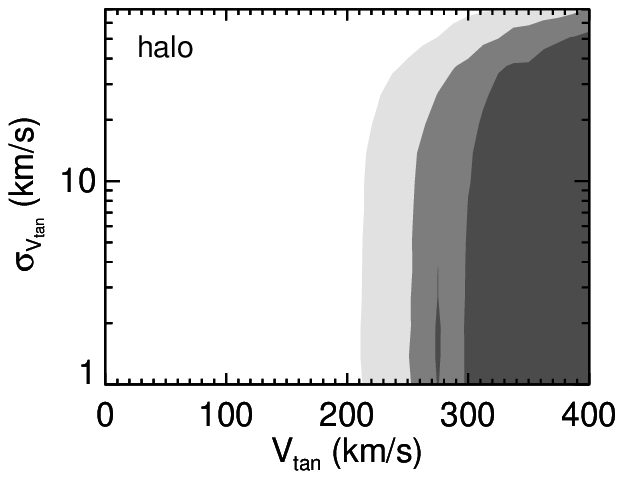}}

\caption{\normalsize Membership probability as a function of
  tangential velocity and its uncertainty for the thin disk, thick
  disk, and halo.  Probabilities are computed from the Besan\c{c}on
  Galaxy model \citep{2003A&A...409..523R} as described in
  Section~\ref{sec:vtan}, and contours are drawn at 10\%, 50\%, and
  90\%.  The dashed lines on the top panel are exponential fits to
  the contours (Equations 5, 6, and 7).  (Note that contours are not
  perfectly smooth for the thick disk and halo because of numerical
  noise in the Galaxy model.)  \label{fig:vtan}}

\end{figure}

\begin{figure}

\centerline{\includegraphics[width=3.2in,angle=0]{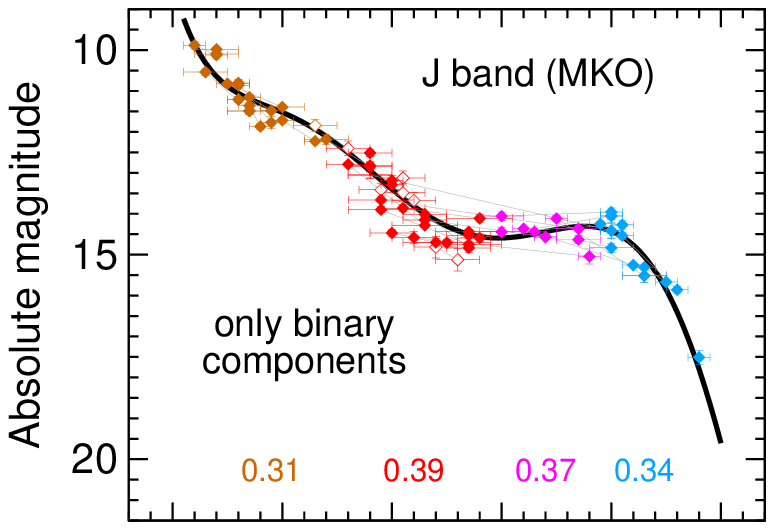} \hskip -0.60in
            \includegraphics[width=3.2in,angle=0]{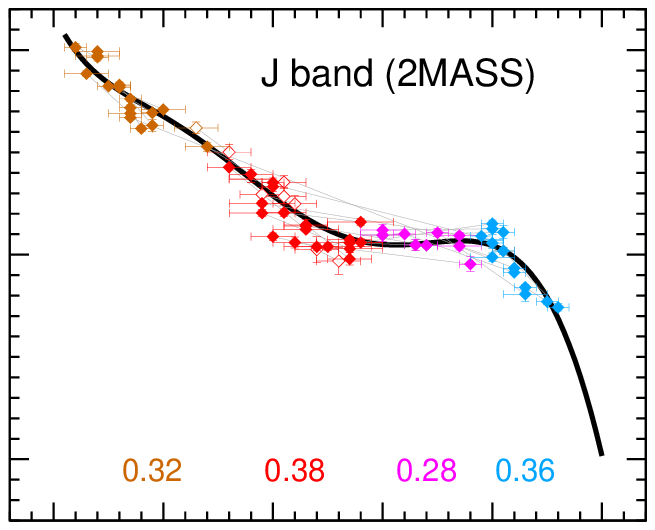}} \vskip -0.40in
\centerline{\includegraphics[width=3.2in,angle=0]{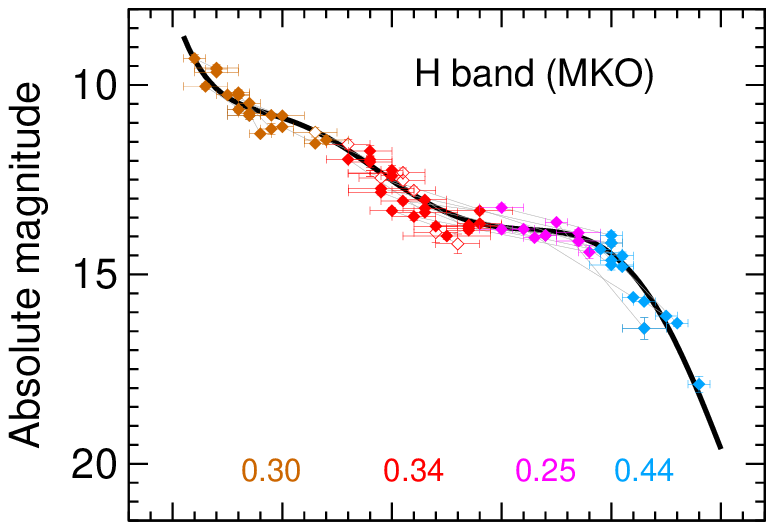} \hskip -0.60in
            \includegraphics[width=3.2in,angle=0]{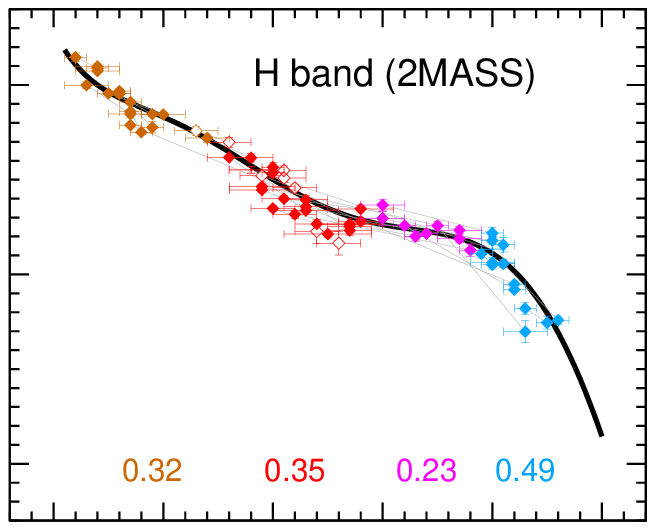}} \vskip -0.40in
\centerline{\includegraphics[width=3.2in,angle=0]{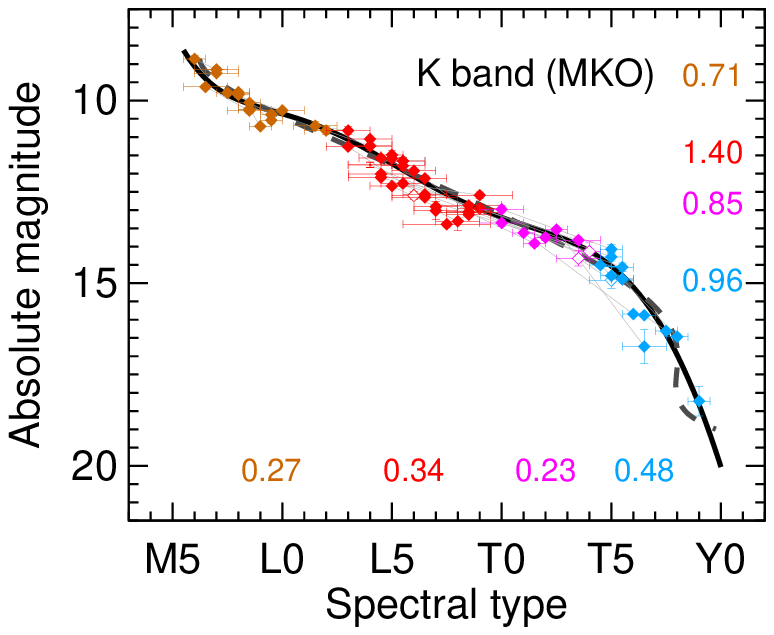} \hskip -0.60in
            \includegraphics[width=3.2in,angle=0]{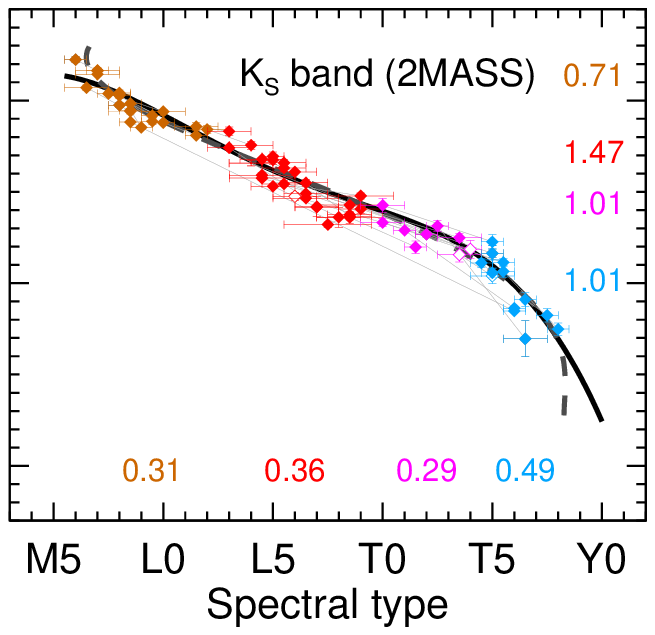}}

\caption{\normalsize Same as Figure~\ref{fig:spt-abs-jhk}, but showing
  only components of binaries, with pairs connected by thin gray
  lines.  The polynomial fits displayed are derived from the full data
  set (i.e., the data in Figure~\ref{fig:spt-abs-jhk}); they are not a
  fit to the data points plotted here. \label{fig:spt-abs-bin}}

\end{figure}

\begin{figure}

\centerline{\includegraphics[width=2.5in,angle=0]{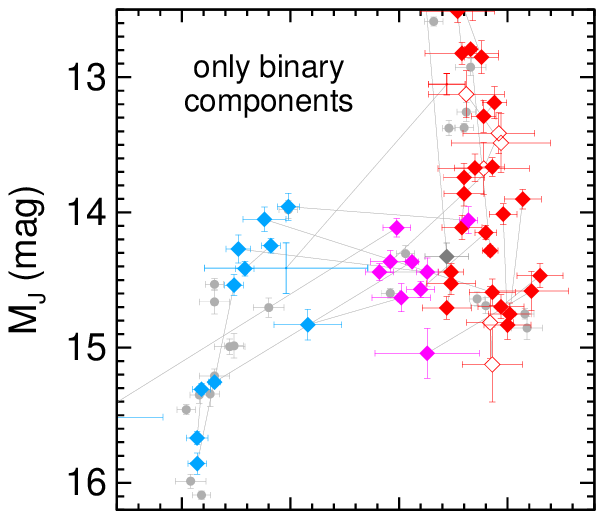} \hskip -0.60in
            \includegraphics[width=2.5in,angle=0]{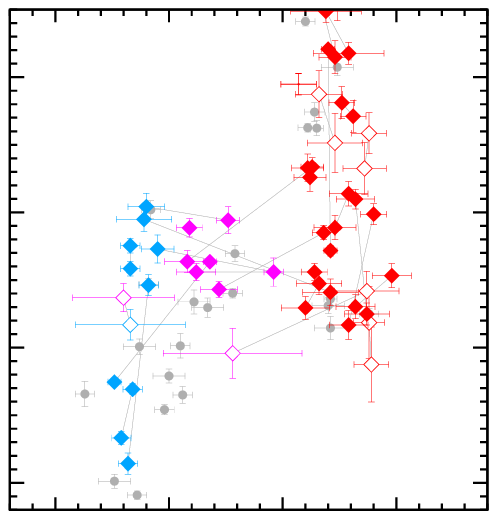} \hskip -0.60in
            \includegraphics[width=2.5in,angle=0]{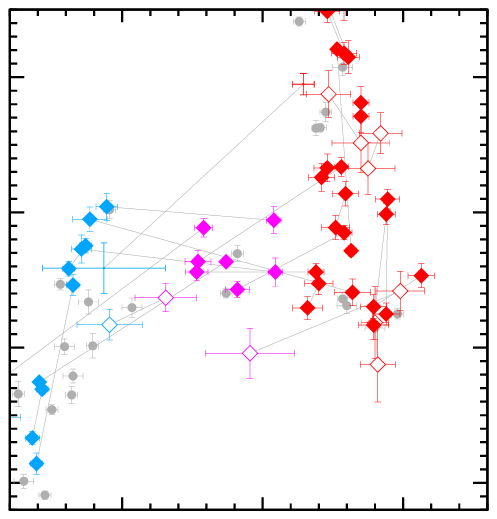}} \vskip -0.45in
\centerline{\includegraphics[width=2.5in,angle=0]{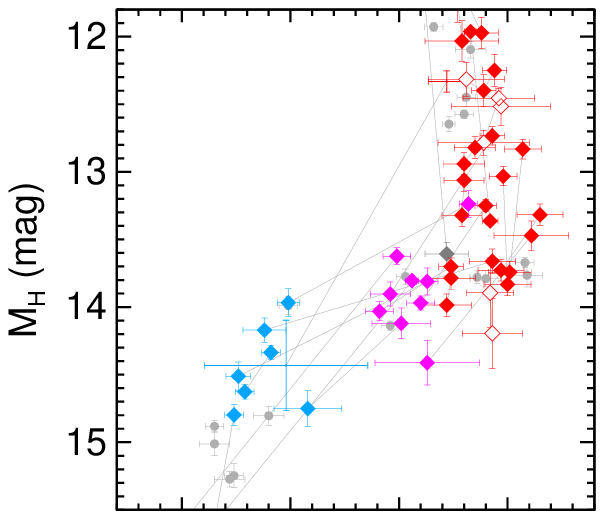} \hskip -0.60in
            \includegraphics[width=2.5in,angle=0]{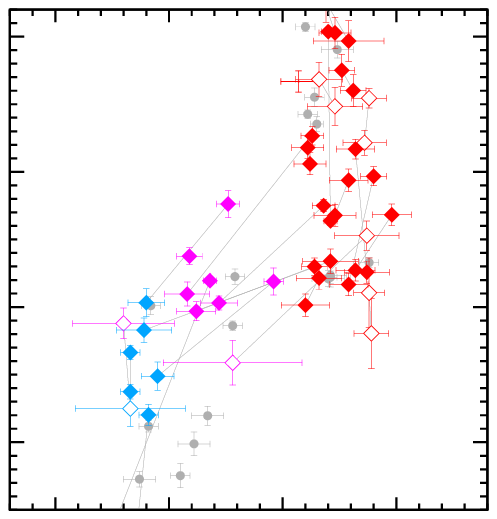} \hskip -0.60in
            \includegraphics[width=2.5in,angle=0]{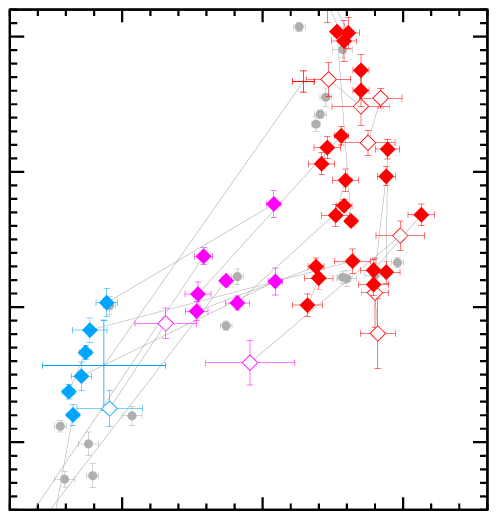}} \vskip -0.45in
\centerline{\includegraphics[width=2.5in,angle=0]{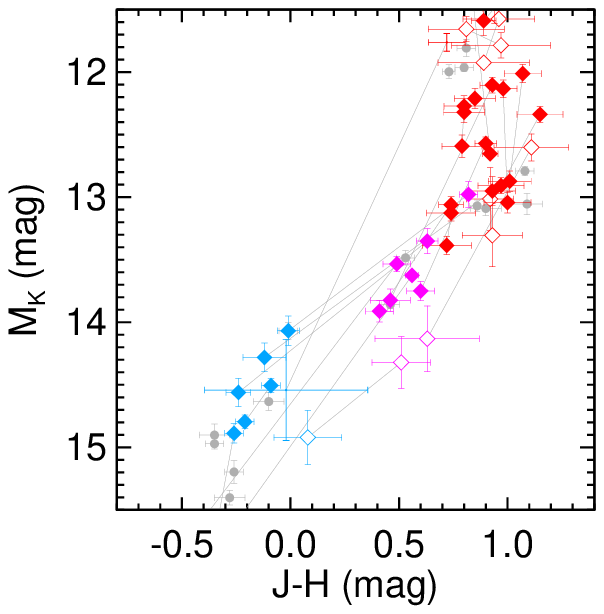} \hskip -0.60in
            \includegraphics[width=2.5in,angle=0]{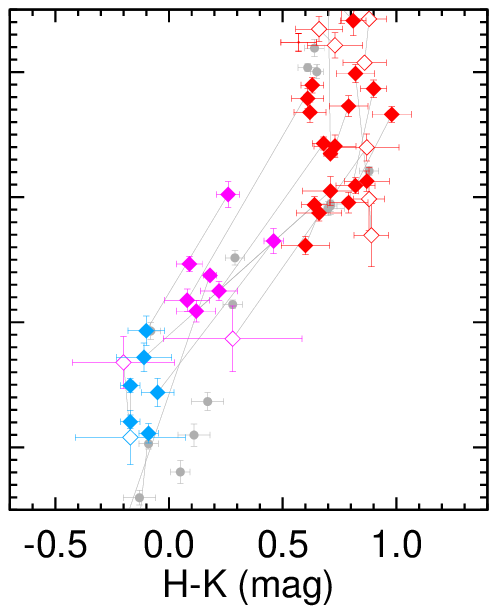} \hskip -0.60in
            \includegraphics[width=2.5in,angle=0]{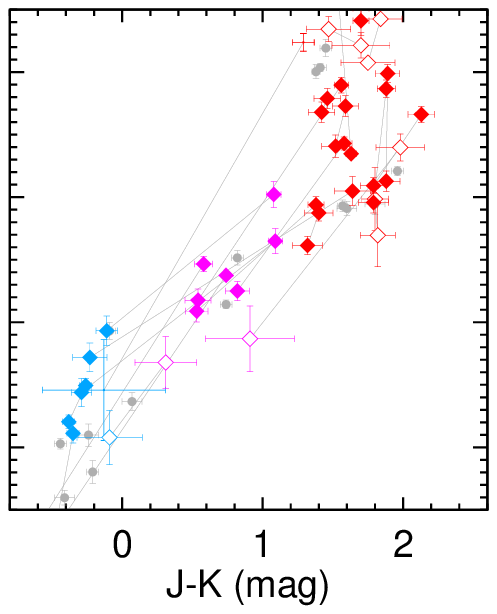}}

\caption{\normalsize Color--magnitude diagrams in the MKO system
  showing binary components (large colored symbols) zoomed in on the
  L/T transition.  The majority of data points here (72\%) use our new
  CFHT parallaxes, and even more rely on our Keck resolved photometry.
  Thin gray lines connect binary pairs, and symbols are the same as
  Figure~\ref{fig:cmd-jhk1-mko} for binary components.  Open symbols
  indicate data where a flux ratios was estimated from spectral
  decomposition constrained by flux ratios measured in other bands.
  For one binary (SDSS~J0805+4812), no symbol indicates that the flux
  ratio was not measured in any band, so resolved photometry is based
  solely on spectral decomposition.  Smaller gray points show the
  locus of likely single objects (i.e., those unresolved in \HST/AO
  imaging), only those with $<$0.10~mag errors for
  clarity.  \label{fig:cmd-mko-zoom}}

\end{figure}

\begin{figure}

\centerline{\includegraphics[width=2.5in,angle=0]{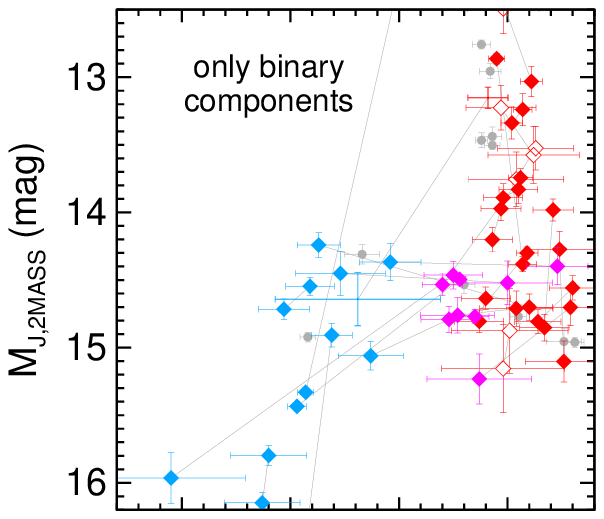} \hskip -0.60in
            \includegraphics[width=2.5in,angle=0]{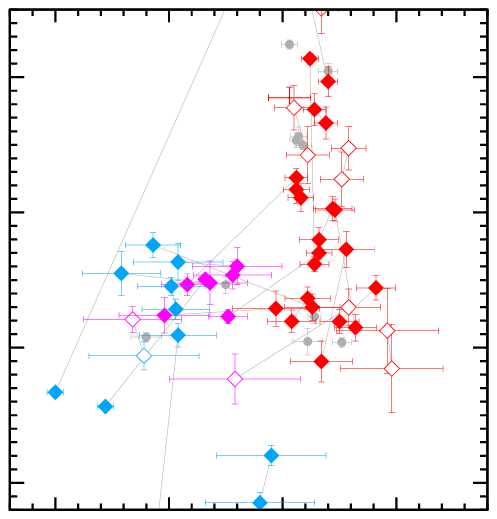} \hskip -0.60in
            \includegraphics[width=2.5in,angle=0]{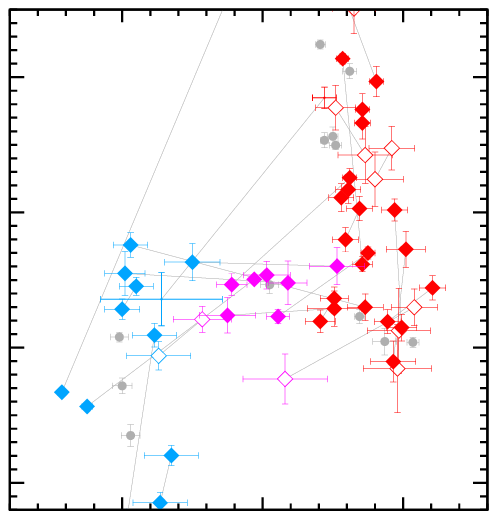}} \vskip -0.45in
\centerline{\includegraphics[width=2.5in,angle=0]{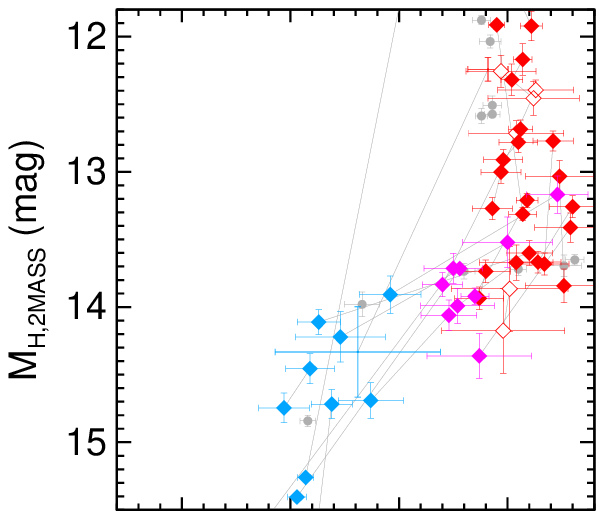} \hskip -0.60in
            \includegraphics[width=2.5in,angle=0]{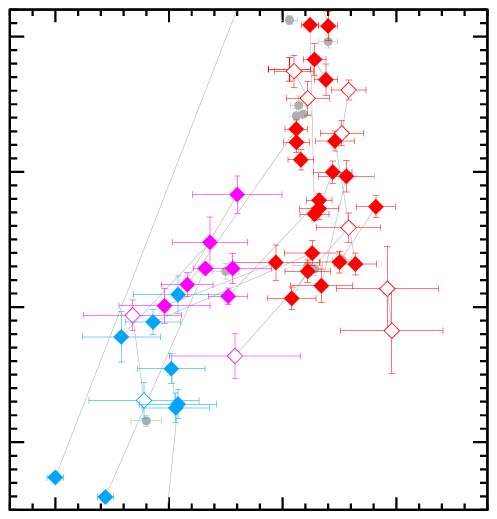} \hskip -0.60in
            \includegraphics[width=2.5in,angle=0]{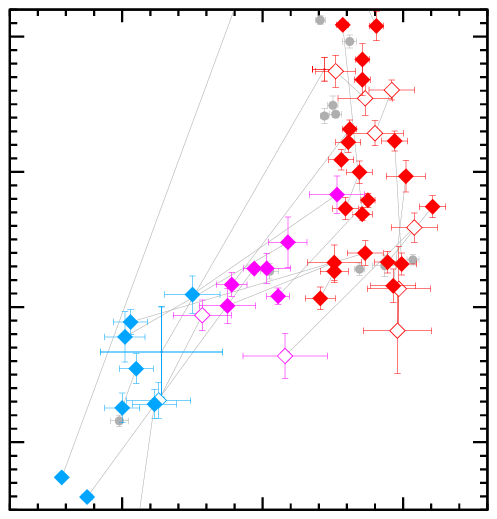}} \vskip -0.45in
\centerline{\includegraphics[width=2.5in,angle=0]{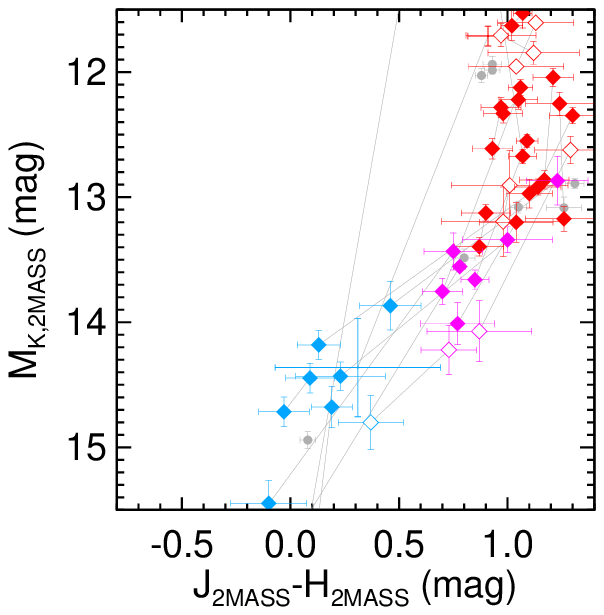} \hskip -0.60in
            \includegraphics[width=2.5in,angle=0]{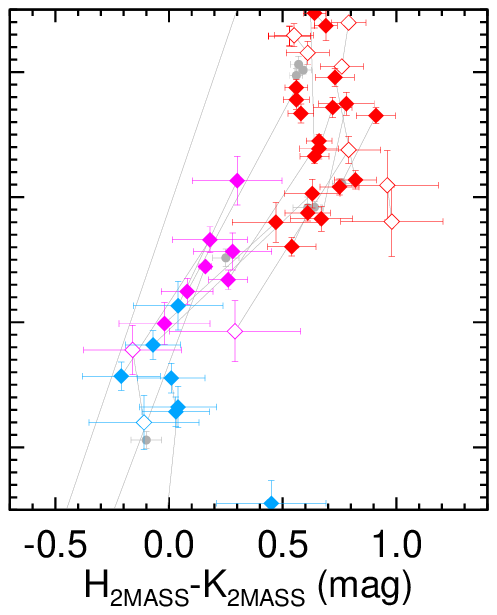} \hskip -0.60in
            \includegraphics[width=2.5in,angle=0]{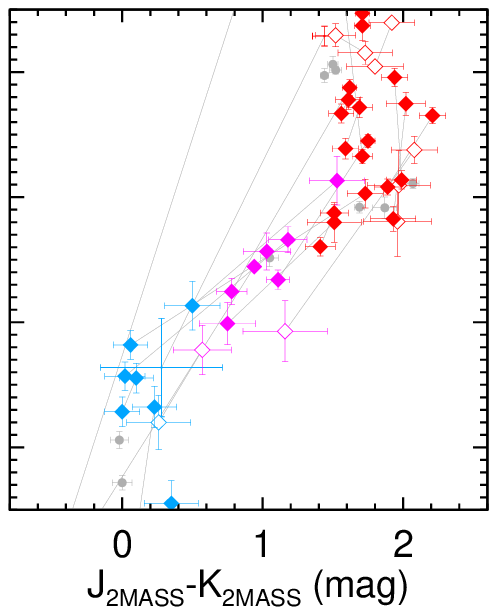}}

\caption{\normalsize Same as Figure~\ref{fig:cmd-mko-zoom} but with
  photometry in the 2MASS system. \label{fig:cmd-2mass-zoom}}

\end{figure}

\begin{figure}

\centerline{\includegraphics[width=5.5in,angle=0]{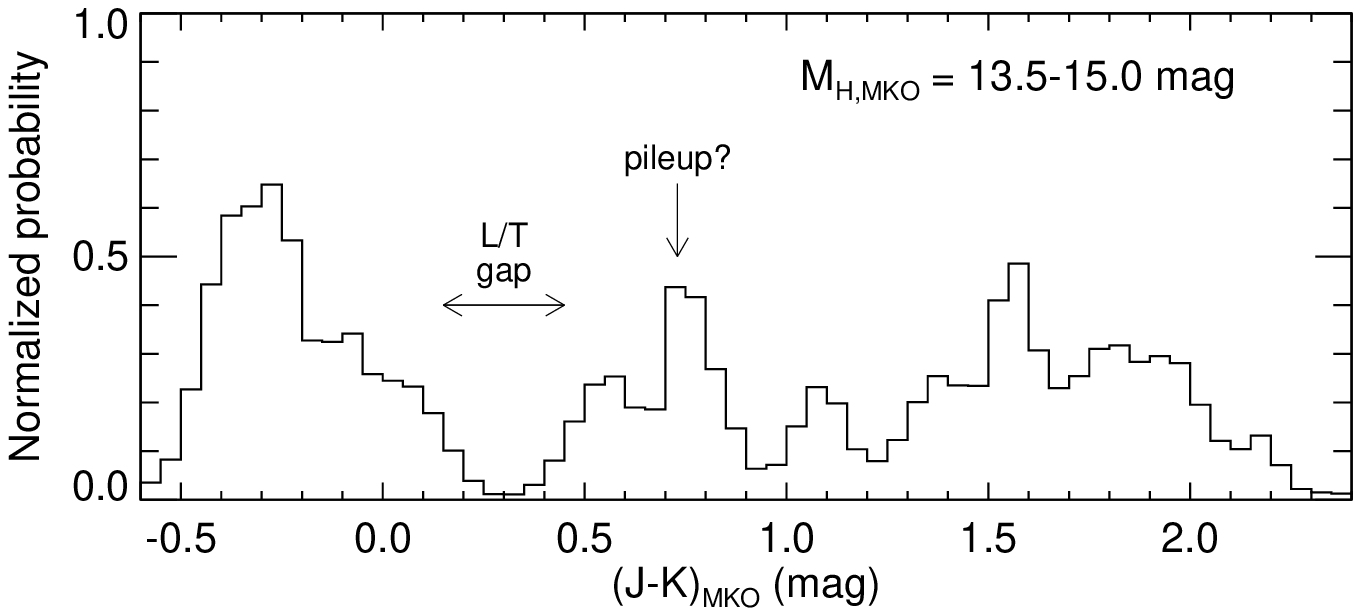}} \vskip 0.10in
\centerline{\includegraphics[width=3.0in,angle=0]{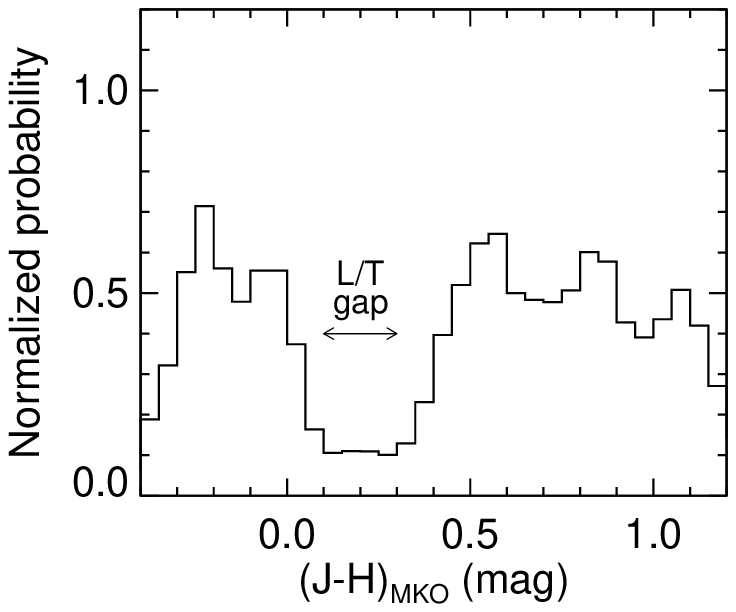} \hskip -0.50in
            \includegraphics[width=3.0in,angle=0]{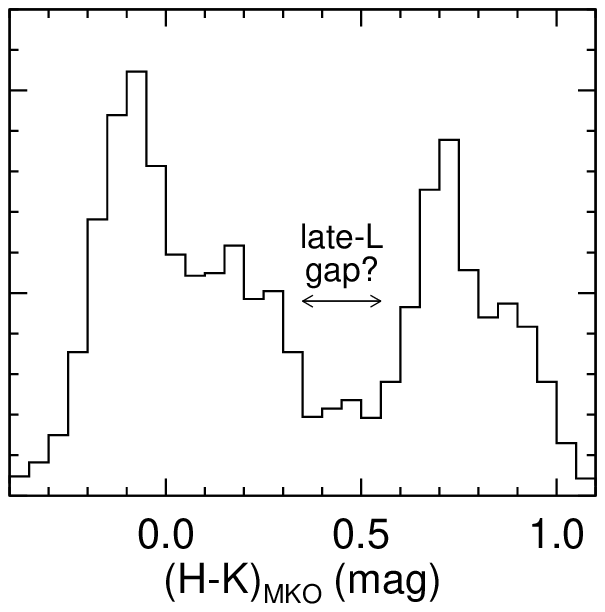}}

\caption{\normalsize Distributions of near-IR colors for objects in
  the L/T transition, as selected by absolute magnitude ($M_{H, {\rm
      MKO}} = 13.5$--15.0~mag).  Histograms were computed in a Monte
  Carlo fashion accounting for errors in colors and absolute
  magnitudes.  The most prominent feature seen in near-IR CMDs
  (Figures~\ref{fig:cmd-jhk1-mko} and \ref{fig:cmd-mko-zoom}) is a gap
  in $(J-H)_{\rm MKO}$ and $(J-K)_{\rm MKO}$ colors just blueward of
  the late-L/early-T~dwarf sequence.  This feature is clear in these
  histograms (labeled ``L/T gap'' here), and other features less
  obvious to the eye in the CMDs also appear.  There is a less
  prominent gap in $(H-K)_{\rm MKO}$ just blueward of the red L~dwarf
  peak (labeled ``late-L gap?'') and an enhanced number of objects
  with $(J-K)_{\rm MKO} = 0.5$--0.9~mag (labeled ``pileup?'').  The
  L/T gap and the pileup are qualitatively similar to the behavior of
  near-IR colors along the L/T transition in the hybrid evolutionary
  models of \citet{2008ApJ...689.1327S}.  In these models, brown dwarf
  evolution slows as a direct consequence of the removal of condensate
  cloud opacity across the L/T transition, resulting in a pileup of
  objects in $J-K$ and a gap just blueward of this (though the
  particular colors of these features differ between their models and
  the data shown here). \label{fig:col-hist}}

\end{figure}

\clearpage


\begin{deluxetable}{lcccccccrrc}
\rotate
\tabletypesize{\footnotesize}
\tablecaption{CFHT/WIRCam Parallax Observations \label{tbl:sample}}
\tablewidth{0pt}
\tablehead{
\colhead{Target} &
\colhead{Spec.\ Type} &
\colhead{CFHT} &
\colhead{FWHM} &
\colhead{max($\Delta$AM)} &
\colhead{$N_{\rm fr}$} &
\colhead{$N_{\rm ep}$} &
\colhead{$\Delta{t}$} &
\colhead{$N_{\rm ref}$} &
\colhead{$N_{\rm cal}$} &
\colhead{$\pi_{\rm abs}-\pi_{\rm rel}$} \\
\colhead{} &
\colhead{Optical/IR} &
\colhead{Filter} &
\colhead{(\arcsec)} &
\colhead{} &
\colhead{} &
\colhead{} &
\colhead{(yr)} &
\colhead{} &
\colhead{} &
\colhead{(mas)} }
\startdata

SDSS~J000013.54+255418.6            & \nodata/T4.5        & $J$ & $0.58\pm0.07$ & 0.014 & 291 & 12 & 2.43 & 124 & 114 & $1.31\pm0.11$ \\
2MASSI~J0003422$-$282241            & M7.5/\nodata        & \Kn & $0.59\pm0.14$ & 0.031 & 213 & 11 & 2.32 &  21 & 17  & $2.07\pm0.59$ \\
LP~349-25AB                         & M8/M8               & \Kn & $0.62\pm0.09$ & 0.062 & 456 & 15 & 2.96 &  33 & 30  & $1.74\pm0.31$ \\
ULAS~J003402.77$-$005206.7          & \nodata/T8.5        & $J$ & $0.57\pm0.09$ & 0.018 &  66 &  9 & 2.18 &  73 & 64  & $1.46\pm0.18$ \\
2MASS~J00501994$-$3322402           & \nodata/T7          & $J$ & $0.82\pm0.14$ & 0.023 & 137 &  7 & 2.19 &  77 & 37  & $1.56\pm0.25$ \\
CFBDS~J005910.90$-$011401.3         & \nodata/T8.5        & $J$ & $0.63\pm0.15$ & 0.021 &  71 &  8 & 2.14 &  88 & 53  & $1.37\pm0.17$ \\
2MASSI~J0415195$-$093506            & T8/T8               & $J$ & $0.70\pm0.08$ & 0.026 & 136 &  8 & 2.28 & 124 & 44  & $1.38\pm0.19$ \\
SDSSp~J042348.57$-$041403.5AB       & L7.5/T0             & $J$ & $0.72\pm0.15$ & 0.027 & 100 & 11 & 4.27 & 128 & 63  & $1.41\pm0.17$ \\
\enddata

\tablecomments{Opt./IR Spec.\ Type: For targets that are binaries, the
  integrated-light spectral type is listed.  Spectrally peculiar
  objects are denoted by ``p'' and types uncertain by $\pm$1 and
  $\pm$2 are denoted by ``:'' and ``::'', respectively. FWHM: The
  median and rms of the FWHM as measured from the science target.
  $\Delta$AM$_{\rm max}$: Maximum difference in airmass over all
  epochs.  $N_{\rm ep}$: Number of distinct observing epochs (i.e.,
  nights).  $N_{\rm fr}$: Total number of frames obtained (typically
  20--30 per epoch).  $N_{\rm ref}$: Number of reference stars used.
  $N_{\rm cal}$: Subset of reference stars used in the absolute
  astrometric calibration (i.e., those available in SDSS, 2MASS, or
  USNO-B).  $\pi_{\rm abs}-\pi_{\rm rel}$: Offset from relative to
  absolute parallax computed for each field using the Besan\c{c}on
  model of the Galaxy \citep{2003A&A...409..523R} as described in
  Section~\ref{sec:abscorr}.}

\tablenotetext{a}{Kelu-1AB and 2MASS~J1553+1532AB are extended in our
  CFHT imaging, which resulted in somewhat larger FWHM than for other
  targets observed at similar airmass.  This is consistent with the
  fact that these are both wide, $\approx$0$\farcs$3 binaries
  \citep{2005astro.ph..8082L, 2006ApJS..166..585B}.}

\end{deluxetable}

\clearpage
\begin{deluxetable}{cll}
\tabletypesize{\small}
\tablecaption{Distortion Coefficients for WIRCam Northeast Array \label{tbl:distort}}
\tablewidth{0pt}
\tablehead{
\colhead{Term} &
\colhead{$a_{ij}$} &
\colhead{$b_{ij}$}}
\startdata

$x^2$  &\phs $ 1.173\times10^{-6} $ &     $-6.409\times10^{-7} $  \\
$xy$   &     $-1.303\times10^{-6} $ &\phs $ 1.117\times10^{-6} $  \\
$y^2$  &\phs $ 5.105\times10^{-7} $ &     $-1.191\times10^{-6} $  \\
$x^3$  &     $-5.287\times10^{-10}$ &     $-1.466\times10^{-10}$  \\
$x^2y$ &     $-4.130\times10^{-10}$ &     $-4.589\times10^{-10}$  \\
$xy^2$ &     $-5.338\times10^{-10}$ &     $-3.884\times10^{-10}$  \\
$y^3$  &     $-1.353\times10^{-10}$ &     $-5.872\times10^{-10}$  \\

\enddata

\tablecomments{To apply this distortion correction, the origin must
  first be redefined as the optical axis: 
  $$x^{\prime} = x -2122.6900$$
  $$y^{\prime} = y + 81.6789$$
  where $x$ and $y$ are the pixel positions measured by SExtractor.
  Distortion-free positions may then be computed:
  $$x^{\prime\prime} = x^{\prime} + a_{20}x^{\prime 2} +
  a_{11}x^{\prime}y^{\prime} + a_{02}y^{\prime 2} + a_{30}x^{\prime 3} + ...$$
  $$y^{\prime\prime} = y^{\prime} + b_{20}x^{\prime 2} +
  b_{11}x^{\prime}y^{\prime} + b_{02}y^{\prime 2} + b_{30}x^{\prime 3} + ...$$
  This distortion correction only applies for the northeast array in
  the WIRCam mosaic.}

\end{deluxetable}

\clearpage
\begin{deluxetable}{lccclllclc}
\rotate
\tabletypesize{\scriptsize}
\tablecaption{Parallax and Proper Motion MCMC Results \label{tbl:plx}}
\tablewidth{0pt}
\tablehead{
\colhead{Target} &
\colhead{$\alpha_{J2000}$} &
\colhead{$\delta_{J2000}$} &
\colhead{Epoch} &
\colhead{$\mu_{\alpha}\cos{\delta}$} &
\colhead{$\mu_{\delta}$} &
\colhead{$\mu$} &
\colhead{P.A.} &
\colhead{$\pi_{\rm abs}$} &
\colhead{$\chi^2$/dof} \\
\colhead{} &
\colhead{(deg)} &
\colhead{(deg)} &
\colhead{(MJD)} &
\colhead{(\asyr)} &
\colhead{(\asyr)} &
\colhead{(\asyr)} &
\colhead{(deg)} &
\colhead{(\arcsec)} &
\colhead{}}
\startdata

SDSS~J000013.54+255418.6            & 000.0563857 & $+$25.9054561 & 54301.63 &   $-$0.0191(15)    &  \phs0.1267(13)    &  \phs0.1281(13)    &         $ 351.4\pm0.7 $ & 0.0708(19)    &     22.5/19     \\
2MASSI~J0003422$-$282241            & 000.9277249 & $-$28.3782531 & 55050.53 &  \phs0.2803(15)    &   $-$0.1233(17)    &  \phs0.3062(15)    &         $ 113.7\pm0.3 $ & 0.0250(19)    &     21.4/17     \\
LP~349-25AB                         & 006.9841925 & $+$22.3255463 & 54687.57 &  \phs0.4040(10)    &   $-$0.1654(15)    &  \phs0.4365(9)\phn &         $112.27\pm0.21$ & 0.0696(9)\phn &     23.2/25     \\
ULAS~J003402.77$-$005206.7          & 008.5116117 & $-$00.8687246 & 55051.60 &   $-$0.0167(10)    &   $-$0.3588(8)\phn &  \phs0.3592(8)\phn &         $182.66\pm0.16$ & 0.0687(14)    &     13.3/13     \\
2MASS~J00501994$-$3322402           & 012.5872589 & $-$33.3749337 & 55050.57 &  \phs1.1505(22)    &  \phs0.9391(21)    &  \phs1.4851(21)    &     \phn$ 50.78\pm0.08$ & 0.0946(24)    &     10.2/9 \phn \\
CFBDS~J005910.90$-$011401.3         & 014.7960832 & $-$01.2335758 & 55068.57 &  \phs0.8847(11)    &  \phs0.0440(13)    &  \phs0.8858(11)    &     \phn$ 87.15\pm0.08$ & 0.1032(21)    &     11.7/11     \\
2MASSI~J0415195$-$093506            & 063.8381022 & $-$09.5835266 & 55070.64 &  \phs2.2143(12)    &  \phs0.5359(12)    &  \phs2.2782(12)    &     \phn$ 76.39\pm0.03$ & 0.1752(17)    &     13.6/11     \\
SDSSp~J042348.57$-$041403.5AB       & 065.9516865 & $-$04.2338814 & 54341.64 &   $-$0.3276(5)\phn &  \phs0.0912(5)\phn &  \phs0.3401(5)\phn &         $285.56\pm0.09$ & 0.0721(11)    &     12.5/17     \\
\enddata

\tablecomments{This table gives all the astrometric parameters derived
  from our MCMC analysis for each target.  For parameters in units of
  arcseconds, errors are given in parentheses in units of
  $10^{-4}$~arcsec.  ($\alpha$, $\delta$, MJD): Coordinates that
  correspond to the epoch listed, which is the first epoch of our
  observations for that target.  ($\mu_{\alpha}\cos{\delta}$,
  $\mu_{\delta}$, $\mu$, P.A.): Proper motion parameters are listed
  both as the direct fitting results (i.e., in $\alpha$ and $\delta$)
  and the computed quantities of total amplitude ($\mu$) and position
  angle. $\pi_{\rm abs}$: The absolute parallax as computed by
  combining the relative parallax that comes directly from our fits
  with the relative-to-absolute corrections (see
  Section~\ref{sec:abscorr}).  Note that where applicable proper
  motion and parallax parameters contain orbital motion correction
  offsets (see Section~\ref{sec:orbcorr} and
  Table~\ref{tbl:orbcorr}). $\chi^2$/dof: The lowest $\chi^2$ in each
  set of MCMC chains along with the degrees of freedom.}

\end{deluxetable}

\clearpage
\begin{deluxetable}{lcccccccccc}
\rotate
\tabletypesize{\scriptsize}
\tablecaption{Orbital Motion Corrections to Parallax and Proper Motion \label{tbl:orbcorr}}
\tablewidth{0pt}
\tablehead{
\colhead{Target} &
\colhead{$a_{\rm phot}$} &
\colhead{$q$} &
\colhead{$\Delta{m}$} &
\colhead{$\Delta\mu_{\alpha}\cos{\delta}$} &
\colhead{$\Delta\mu_{\delta}$} &
\colhead{$\Delta\mu$} &
\colhead{$\Delta$P.A.} &
\colhead{$\Delta\pi$} &
\colhead{$\Delta\chi^2$} &
\colhead{Orbit} \\
\colhead{} &
\colhead{(mas)} &
\colhead{($M_2/M_1$)} &
\colhead{(mag)} &
\colhead{(\asyr)} &
\colhead{(\asyr)} &
\colhead{(\asyr)} &
\colhead{(deg)} &
\colhead{(\arcsec)} &
\colhead{} &
\colhead{Ref.}}
\startdata

LP~349-25AB              & \phn$ 5.0\pm1.7$\phs & $0.86\pm0.04$ & \phs$ 0.307\pm0.008$ &  \phs0.0018(6)\phn &  \phs0.0030(10)    &  \phs0.0005(2)\phn &   $-$0.46(16)    &  $-$0.00040(13)    & \phs$  0.0$ & 3 \\
LP~415-20AB              &     $10.0\pm1.1$\phs & $0.80\pm0.03$ & \phs$ 0.728\pm0.023$ &  \phs0.0026(3)\phn &  \phs0.0000(1)\phn &  \phs0.0025(3)\phn &   $-$0.32(5)\phn &  $-$0.00006(7)\phn &     $ -1.3$ & 5 \\
LHS~1901AB               & \phn$ 7.4\pm1.0$\phs & $1.00\pm0.00$ & \phs$ 0.113\pm0.016$ &   $-$0.0006(1)\phn &   $-$0.0032(5)\phn &  \phs0.0024(4)\phn &  \phs0.19(3)\phn & \phs0.00016(5)\phn & \phs$  0.0$ & 3 \\
2MASS~J0746$+$2000AB     &     $14.1\pm1.9$\phs & $0.92\pm0.02$ & \phs$ 0.356\pm0.024$ &  \phs0.0025(3)\phn &   $-$0.0018(2)\phn &   $-$0.0022(3)\phn &   $-$0.33(4)\phn &  $-$0.00008(2)\phn & \phs$  0.0$ & 6 \\
2MASS~J0920$+$3517AB     & \phn$ 3.3\pm1.1$\phs & $0.98\pm0.02$ & \phs$  0.25\pm0.07 $ &  \phs0.0017(6)\phn &  \phs0.0007(3)\phn &   $-$0.0017(6)\phn &   $-$0.15(5)\phn &  $-$0.00044(18)    &     $-11.8$ & 4 \\
LHS~2397aAB              &     $  78\pm7  $\phs & $0.77\pm0.08$ & \phs$  2.80\pm0.03 $ &  \phs0.0270(24)    &   $-$0.0104(12)    &   $-$0.0257(22)    &   $-$1.51(16)    &  $-$0.00081(16)    &     $-19.4$ & 2 \\
2MASS~J1534$-$2952AB     & \phn$ 6.3\pm2.0$\phs & $0.95\pm0.03$ & \phs$ 0.162\pm0.014$ &   $-$0.0003(2)\phn &   $-$0.0014(6)\phn &  \phs0.0012(5)\phn &  \phs0.15(7)\phn &  $-$0.00001(3)\phn &     $ -0.2$ & 7 \\
2MASS~J2132$+$1341AB     &     $11.8\pm2.0$\phs & $0.81\pm0.07$ & \phs$  0.85\pm0.04 $ &   $-$0.0067(12)    &   $-$0.0008(6)\phn &   $-$0.0005(5)\phn &  \phs3.11(57)    & \phs0.00016(9)\phn &     $ -2.2$ & 4 \\
\enddata

\tablecomments{Offsets to parallax and proper motion parameters due to
  orbital motion during our astrometric monitoring program.  This is
  computed from the relative orbit parameters (reference given in last
  column), our evolutionary model-derived mass ratio estimate ($q$),
  and the flux ratio in the observed bandpass ($\Delta{m}$).  The
  semimajor axis of the resulting photocenter motion ($a_{\rm phot}$)
  is shown for each binary.  The difference in $\chi^2$ between the
  original best-fit astrometric solution and orbit-corrected solution
  is also given ($\Delta\chi^2$).  These offsets and their errors have
  already been accounted for in values given in Table~\ref{tbl:plx}.}

\tablerefs{(1)~\citet{2009ApJ...706..328D}; (2)~\citet{2009ApJ...699..168D};
  (3)~\citet{2010ApJ...721.1725D}; (4)~\citet{me-phd}; (5)~\citet{me-ecc};
  (6)~\citet{2010ApJ...711.1087K}; (7)~\citet{2008ApJ...689..436L}.}

\end{deluxetable}

\clearpage
\begin{deluxetable}{llcccc}
\tabletypesize{\small}
\tablecaption{Keck AO Observations of Sample Binaries \label{tbl:keck}}
\tablewidth{0pt}
\tablehead{
\colhead{Target} &
\colhead{Epoch} &
\colhead{NIRC2} &
\colhead{FWHM} &
\colhead{Strehl} &
\colhead{$\Delta{m}$} \\
\colhead{} &
\colhead{(UT)} &
\colhead{filter} &
\colhead{(mas)} &
\colhead{ratio} &
\colhead{(mag)}}
\startdata

SDSS~J0423$-$0414AB    & 2007 Sep 6  & $K$ &    \nodata      &     \nodata     & $  1.18\pm0.08 $     \\

 & & & & &  \\

2MASS~J0700$+$3157AB   & 2008 Nov 3  & $J$ &    $ 65\pm5   $  & $0.026\pm0.004$ & $ 1.491\pm0.019$     \\
                       &             & $H$ &    $61.0\pm2.0$  & $0.081\pm0.008$ & $ 1.403\pm0.017$     \\
                       &             & \Ks &    $63.6\pm1.8$ & $0.194\pm0.021$ & $ 1.390\pm0.011$     \\
                       &             & \Lp &    $86.5\pm1.4$ & $ 0.61\pm0.10 $ & $  0.92\pm0.03 $     \\

 & & & & &  \\

2MASS~J0850$+$1057AB   & 2006 Dec 19 & $J$ &     $ 58\pm6 $  & $0.044\pm0.004$ & $  0.82\pm0.12 $     \\
\enddata

\tablecomments{Epochs without FWHM or Strehl ratio information
  correspond to aperture masking observations.  The errors on the FWHM
  and Strehl ratios are the rms scatter among individual dithered
  images.}


\end{deluxetable}

\clearpage
\begin{deluxetable}{llccc}
\tabletypesize{\scriptsize}
\tablecaption{Analysis of Archival Imaging for Sample Binaries \label{tbl:arch}}
\tablewidth{0pt}
\tablehead{
\colhead{Target} &
\colhead{Epoch} &
\colhead{Instrument} &
\colhead{Filter} &
\colhead{$\Delta{m}$} \\
\colhead{} &
\colhead{(UT)} &
\colhead{} &
\colhead{} &
\colhead{(mag)}}
\startdata

GJ~1001BC                       & 2004 Sep 17 &  \HST/NICMOS  &   F110W   & $  0.10\pm0.04 $     \\
                                &             &               &   F170M   & $  0.11\pm0.05 $     \\
                                & 2004 Oct  7 &    VLT/NACO   &    $J$    & $  0.10\pm0.05 $     \\
                                &             &               &    $H$    & $  0.15\pm0.04 $     \\
                                &             &               &    \Ks    & $  0.10\pm0.05 $     \\
                                                               
 & & & &  \\                                                   
                                                               
LHS~1070AB\tablenotemark{a}     & 2003 Dec 12 &    VLT/NACO   &    $J$    & $ 0.648\pm0.036$     \\
                                &             &               &    $H$    & $ 0.579\pm0.032$     \\
\enddata

\tablenotetext{a}{Triple PSF-fitting was performed on LHS~1070ABC
  using the StarFinder-based routine described in \citet{2009ApJ...692..729D}.
  The ``LHS~1070AB'' entry gives the flux ratio of B/A, while the
  ``LHS~1070BC'' entry gives C/B.}

\tablecomments{\HST\ program IDs: GO-9833 (PI Burgasser), GO-9843 (PI
  Gizis), GO-10143 (PI Reid), GO-10247 (PI Cruz), GO-11136 (PI Liu).
  VLT program IDs: 072.C-0022 (PI Leinert), 074.C-0407 (PI Minniti),
  077.C-0062 (PI Bouy).}

\end{deluxetable}

\clearpage
\begin{deluxetable}{lccccr}
\tabletypesize{\small}
\tablecaption{Comparison to Published Parallaxes \label{tbl:compare}}
\tablewidth{0pt}
\tablehead{
\colhead{Target} &
\multicolumn{2}{c}{Parallax 1}      &
\multicolumn{2}{c}{Parallax 2}      &
\colhead{} \\
\colhead{} &
\colhead{$\pi$ (mas)} &
\colhead{Ref.} &
\colhead{$\pi$ (mas)} &
\colhead{Ref.} &
\colhead{$\Delta\pi/\sigma_{\Delta\pi}$}}
\startdata

\multicolumn{6}{c}{\bf CFHT vs.\ Other Published Values} \\
\cline{1-6}

2MASSI~J0003422$-$282241        & $ 25.0\pm 1.9$     &  C & $ 25.7\pm 0.9$     & 26 & $ 0.33\sigma$ \\ 
LP~349-25AB                     & $ 69.6\pm 0.9$     &  C & $ 75.8\pm 1.6$     & 11 & $ 3.36\sigma$ \\ 
CFBDS~J005910.90$-$011401.3     & $103.2\pm 2.1$\phn &  C & $108.2\pm 5.0$\phn & 17 & $ 0.92\sigma$ \\ 
2MASSI~J0415195$-$093506        & $175.2\pm 1.7$\phn &  C & $174.3\pm 2.8$\phn & 27 & $-0.27\sigma$ \\ 
SDSSp~J042348.57$-$041403.5AB   & $ 72.1\pm 1.1$     &  C & $ 65.9\pm 1.7$     & 27 & $-3.05\sigma$ \\ 
2MASSI~J0559191$-$140448        & $ 96.6\pm 1.0$     &  C & $ 97.7\pm 1.3$     &  6 & $ 0.67\sigma$ \\ 
                                &                    &    & $ 95.5\pm 1.4$     & 27 & $-0.61\sigma$ \\
\enddata

\tablerefs{(C)~This work; 
(1)~\citet{2011AJ....141...54A};  (2)~\citet{2010ApJ...710...45B};  (3)~\citet{2008ApJ...672.1159B}; 
(4)~\citet{2005AJ....130..337C};  (5)~\citet{2006AJ....132.1234C};  (6)~\citet{2002AJ....124.1170D}; 
(7)~\citet{2008ApJ...686..548D};  (8)~\citet{2005AJ....129..409D};  (9)~\citet{2008A&A...477L...1D}; 
(10)~\citet{2011AJ....141...71F};  (11)~\citet{2009AJ....137..402G};  (12)~\citet{2007ApJ...669L..45G}; 
(13)~\citet{2006AJ....132.2360H};  (14)~\citet{2012ApJ...748...74L};  (15)~\citet{2009AJ....137.4109L}; 
(16)~\citet{2010MNRAS.408L..56L};  (17)~\citet{2010A&A...524A..38M};  (18)~\citet{1992AJ....103..638M}; 
(19)~\citet{2009A&A...493L..27S};  (20)~\citet{2010A&A...510L...8S};  (21)~\citet{2003PASP..115.1207T}; 
(22)~\citet{1995AJ....110.3014T};  (23)~\citet{1996MNRAS.281..644T};  (24)~\citet{2003AJ....126..975T}; 
(25)~\citet{1995gcts.book.....V};  (26)~\citet{2007hnrr.book.....V};  (27)~\citet{2004AJ....127.2948V}.}

\end{deluxetable}

\clearpage
\begin{deluxetable}{lllccc}
\tabletypesize{\small}
\tablecaption{Component Spectral Types of Ultracool Binaries with Parallaxes \label{tbl:spt}}
\tablewidth{0pt}
\tablehead{
\colhead{Object} &
\colhead{Primary} &
\colhead{Secondary} &
\colhead{Broadband} &
\colhead{\HST/NICMOS} &
\colhead{Phot.}  \\
\colhead{} &
\colhead{Type} &
\colhead{Type} &
\colhead{data} &
\colhead{data} &
\colhead{Ref.}}
\startdata

\multicolumn{6}{c}{\bf Derived from Our Template Matching Method} \\
\cline{1-6}

GJ~1001BC                 & L$5\pm0.5$   & L$5\pm0.5$   &   $JHK$   & 110W,170M           &    1    \\
LP~349-25AB               & M$6.5\pm1$   & M$8\pm1$     &   $JHK$   & \nodata             &    5    \\
SDSSp~J0423$-$0414AB      & L$6.5\pm1$   & T$2\pm0.5$   &     $K$   & 110W,170M           &   1,2   \\
2MASS~J0518$-$2828AB      & L$6\pm1$     & T$4\pm0.5$   &  \nodata  & 110W,170M           &    1    \\
2MASS~J0700+3157AB        & L$3\pm1$     & L$6.5\pm1.5$ &   $JHK$   & \nodata             &    1    \\
LHS~1901AB                & M$7\pm1$     & M$7\pm1$     &   $JHK$   & \nodata             &    5    \\
SDSS~J0805+4812AB         & L$4\pm1$     & T$5\pm0.5$   &  \nodata  & \nodata             & \nodata \\
\enddata

\tablecomments{We list component spectral types derived using the
  template matching method described in Section~\ref{sec:spt}
  supplemented by spectral type determinations from the literature
  based on resolved spectroscopy or resolved photometric indices.  The
  fourth and fifth columns list the MKO broad-band and \HST/NICMOS
  medium-band flux ratios used in the template matching, respectively.
  The last column gives references for the photometry used or for the
  source of the resolved spectroscopy.}

\tablerefs{(1)~This work (Tables~\ref{tbl:keck} and \ref{tbl:arch});
  (2)~\citet{2006ApJS..166..585B}; (3)~\citet{2010ApJ...710.1142B};
  (4)~\citet{2009ApJ...706..328D}; (5)~\citet{2010ApJ...721.1725D}; (6)~\citet{2010ApJ...711.1087K};
  (7)~\citet{2005astro.ph..8082L}; (8)~\citet{2006ApJ...647.1393L};
  (9)~\citet{2008ApJ...689..436L}; (10)~\citet{2010ApJ...722..311L};
  (11)~\citet{2011A&A...525A.123S}.  (A)~\citet{2000A&A...353..691L};
  (B)~\citet{2004A&A...423..341B}; (C)~\citet{2002ApJ...567L..59G};
  (D)~\citet{2001ApJ...560..390L}; (E)~\citet{2011ApJ...740..108L};
  (F)~\citet{2007A&A...471..655K}; (G)~\citet{2010A&A...510A..99K};
  (H)~\citet{2009ApJ...697..824A}.}

\end{deluxetable}

\clearpage
\begin{deluxetable}{lcccllllcccl}
\rotate
\setlength{\tabcolsep}{0.06in}
\tabletypesize{\tiny}
\tablecaption{All Ultracool Dwarfs with Parallaxes \label{tbl:plx-all}}
\tablewidth{0pt}
\tablehead{
\colhead{Object} &
\colhead{$\alpha_{\rm J2000}$} &
\colhead{$\delta_{\rm J2000}$} &
\colhead{Epoch} &
\colhead{$\pi_{\rm abs}$} &
\colhead{$\mu_{\alpha}\cos{\delta}$} &
\colhead{$\mu_{\delta}$} &
\colhead{$\mu$} &
\colhead{P.A.} &
\colhead{$V_{\rm tan}$} &
\colhead{Ref.} &
\colhead{Note} \\
\colhead{} &
\colhead{(deg)} &
\colhead{(deg)} &
\colhead{(MJD)} &
\colhead{(\arcsec)} &
\colhead{(\asyr)} &
\colhead{(\asyr)} &
\colhead{(\asyr)} &
\colhead{(deg)} &
\colhead{(km/s)} &
\colhead{} &
\colhead{}}
\startdata

SDSS~J000013.54+255418.6        & 000.0564 & $+25.9055$ & 54301.63 & 0.0708(19)   &  $-$0.0191(14)   & \phs0.1267(13)   & 0.1281(13)   & $ 351.4\pm0.6 $\phn\phn & $  8.58\pm0.24$         &  1 &                       \\
LSR~J0011+5908                  & 002.8826 & $+59.1445$ & 51492.22 & 0.1083(14)   &  $-$0.8997       &  $-$1.1654       & 1.4723       & $   218       $         & $    64       $         & 18 &                       \\
BRI~0021$-$0214                 & 006.1027 & $-01.9723$ & 51071.24 & 0.0866(40)   &  $-$0.0804(38)   & \phs0.1330(60)   & 0.1550(70)   & $ 328.8\pm0.7 $\phn\phn & $   8.5\pm0.5 $         & 26 &                       \\
LHS~1070A                       & 006.1841 & $-27.1401$ & 51542.05 & 0.1295(25)   &  $-$0.1330(50)   & \phs0.6401(31)   & 0.6537(30)   & $ 348.3\pm0.4 $\phn\phn & $  23.9\pm0.5 $\phn     &  6 &                       \\
PC~0025+0447                    & 006.9249 & $+05.0616$ & 51768.41 & 0.0138(16)   & \phs0.0105(4)    &  $-$0.0008(3)    & 0.0105(4)    & $  94.6\pm1.8 $\phn     & $   3.6\pm0.5 $         &  8 &                       \\
LP~349-25AB                     & 006.9842 & $+22.3255$ & 54687.57 & 0.0696(9)    & \phs0.4039(10)   &  $-$0.1654(15)   & 0.4365(9)    & $112.27\pm0.21$\phn\phn & $  29.7\pm0.4 $\phn     &  1 &                young? \\
2MASSW~J0030300$-$145033        & 007.6256 & $-14.8426$ & 51840.16 & 0.0374(45)   & \phs0.2450(35)   &  $-$0.0282(18)   & 0.2466(36)   & $  96.6\pm0.4 $\phn     & $    31\pm  4 $\phn     & 31 &                       \\
SDSSp~J003259.36+141036.6       & 008.2474 & $+14.1770$ & 51878.20 & 0.0300(50)   & \phs0.2730(70)   & \phs0.0391(35)   & 0.2760(70)   & $  81.8\pm0.7 $\phn     & $    43\pm  8 $\phn     & 31 &                       \\
\enddata
\end{deluxetable}

\clearpage
\begin{deluxetable}{lcccccccccccc}
\rotate
\setlength{\tabcolsep}{0.03in}
\tabletypesize{\tiny}
\tablecaption{Near-Infrared Photometry for All Ultracool Dwarfs with Parallaxes \label{tbl:nir-mags}}
\tablewidth{0pt}
\tablehead{
\colhead{}   &
\colhead{}   &
\colhead{}   &
\multicolumn{5}{c}{MKO}      &
\colhead{}   &
\multicolumn{3}{c}{2MASS}      &
\colhead{} \\
\cline{4-8}
\cline{10-12}
\colhead{Object} &
\colhead{Spec.\ Type} &
\colhead{$m-M$} &
\colhead{$Y$} &
\colhead{$J$} &
\colhead{$H$} &
\colhead{$K$} &
\colhead{\Lp} &
\colhead{}   &
\colhead{$J$} &
\colhead{$H$} &
\colhead{\Ks} &
\colhead{References} \\
\colhead{} &
\colhead{Optical/IR} &
\colhead{(mag)} &
\colhead{(mag)} &
\colhead{(mag)} &
\colhead{(mag)} &
\colhead{(mag)} &
\colhead{(mag)} &
\colhead{} &
\colhead{(mag)} &
\colhead{(mag)} &
\colhead{(mag)} &
\colhead{Plx.; SpT; Phot.}}
\startdata

SDSS~J000013.54+255418.6        & \nodata/T4.5       & $  0.75\pm0.06 $     &     $[15.80(6) ]$ &        $14.73(3)    $ &        $14.74(3)    $ &        $14.82(3)    $ &         $13.03(3) $ & &         $15.06(4)     $ &        $14.73(7)    $ &    \phn$14.84(12)   $ & 1; 14; 1, 2, 86, 95        \\  
2MASSI~J0003422$-$282241        & M7.5/\nodata       & $  2.95\pm0.08 $     &     $[13.81(6) ]$ &       $[13.02(2)   ]$ &       $[12.41(3)   ]$ &       $[11.95(3)   ]$ &           \nodata   & &         $13.07(2)     $ &        $12.38(3)    $ &        $11.97(3)    $ & 156; 30; 1, 2              \\  
GJ~1001B                        & \nodata/L5         & $  0.57\pm0.11 $     &        \nodata    &        $13.76(4)    $ &        $12.82(4)    $ &       $[12.06(4)   ]$ &           \nodata   & &        $[13.81(3)    ]$ &       $[12.74(3)   ]$ &        $12.10(4)    $ & 66; 1; 1, 2, 94            \\  
GJ~1001C                        & \nodata/L5         & $  0.57\pm0.11 $     &        \nodata    &        $13.86(4)    $ &        $12.97(4)    $ &       $[12.16(4)   ]$ &           \nodata   & &        $[13.91(4)    ]$ &       $[12.89(3)   ]$ &        $12.20(4)    $ & 66; 1; 1, 2, 94            \\  
LSR~J0011+5908                  & M6.5/\nodata       & $-0.173\pm0.028$\phs &        \nodata    &          \nodata      &          \nodata      &          \nodata      &           \nodata   & &     \phn$ 9.94(2)     $ &    \phn$ 9.39(3)    $ &    \phn$ 9.09(2)    $ & 102; 102; 2                \\  
HD~1160B                        & \nodata/\nodata    & $  5.07\pm0.10 $     &        \nodata    &          \nodata      &          \nodata      &          \nodata      &     \phn$13.41(12)$ & &     \phn$15.83(10)    $ &        $14.65(8)    $ &        $14.12(6)    $ & 156; -- ; 126              \\
\enddata
\end{deluxetable}

\clearpage
\begin{deluxetable}{lccllllcllllcc}
\setlength{\tabcolsep}{0.03in}
\rotate
\tabletypesize{\tiny}
\tablecaption{Mid-Infrared Photometry for All Ultracool Dwarfs with Parallaxes \label{tbl:mir-mags}}
\tablewidth{0pt}
\tablehead{
\colhead{}   &
\colhead{}   &
\colhead{}   &
\multicolumn{4}{c}{\Spitzer/IRAC}      &
\colhead{}   &
\multicolumn{4}{c}{\WISE}      &
\colhead{} &
\colhead{} \\
\cline{4-7}
\cline{9-12}
\colhead{Object} &
\colhead{Spec.\ Type} &
\colhead{$m-M$} &
\colhead{$[$3.6$]$} &
\colhead{$[$4.5$]$} &
\colhead{$[$5.8$]$} &
\colhead{$[$8.0$]$} &
\colhead{}   &
\colhead{W1} &
\colhead{W2} &
\colhead{W3} &
\colhead{W4} &
\colhead{} &
\colhead{References} \\
\colhead{} &
\colhead{Optical/IR}   &
\colhead{(mag)} &
\colhead{(mag)} &
\colhead{(mag)} &
\colhead{(mag)} &
\colhead{(mag)} &
\colhead{} &
\colhead{(mag)} &
\colhead{(mag)} &
\colhead{(mag)} &
\colhead{(mag)} &
\colhead{} &
\colhead{Plx.; SpT; Phot.}}
\startdata

SDSS~J000013.54+255418.6        & \nodata/T4.5    & $  0.75\pm0.06 $     &     13.72(3)  &     13.07(3)  &     12.56(9)  &     12.50(3)  & &      \nodata  &      \nodata  &      \nodata  &      \nodata  & & 1; 9; 60           \\  
2MASSI~J0003422$-$282241        & M7.5/\nodata    & $  2.95\pm0.08 $     &      \nodata  &      \nodata  &      \nodata  &      \nodata  & &     11.67(3)  &     11.50(2)  &     10.97(10) & \phn$>$9.02   & & 106; 23; 2         \\  
LSR~J0011+5908                  & M6.5/\nodata    & $-0.173\pm0.028$\phs &      \nodata  &      \nodata  &      \nodata  &      \nodata  & & \phn 8.86(2)  & \phn 8.63(2)  & \phn 8.41(3)  & \phn 7.93(12) & & 65; 65; 2          \\  
BRI~0021$-$0214                 & M9.5/M9.5       & $  0.31\pm0.10 $     & \phn 9.94(3)  & \phn 9.91(3)  & \phn 9.72(3)  & \phn 9.55(3)  & &     10.17(2)  & \phn 9.90(2)  & \phn 9.41(4)  & \phn$>$8.80   & & 101; 33, 50; 2, 80 \\  
PC~0025+0447                    & M9.5/\nodata    & $  4.30\pm0.26 $     &      \nodata  &      \nodata  &      \nodata  &      \nodata  & &     14.62(4)  &     14.14(5)  &     $>$12.24  & \phn$>$8.89   & & 25; 50; 2          \\  
2MASSW~J0030300$-$145033        & L7/\nodata      & $  2.13\pm0.27 $     &      \nodata  &      \nodata  &      \nodata  &      \nodata  & &     13.66(3)  &     13.26(3)  &     $>$12.27  & \phn$>$9.15   & & 107; 53; 2         \\  
SDSSp~J003259.36+141036.6       & \nodata/L8      & $   2.6\pm0.4  $     &      \nodata  &      \nodata  &      \nodata  &      \nodata  & &     14.26(3)  &     13.67(4)  &     $>$12.22  & \phn$>$8.98   & & 107; 33; 2         \\  
ULAS~J003402.77$-$005206.7      & \nodata/T8.5    & $  0.82\pm0.04 $     &     16.28(3)  &     14.49(3)  &     14.82(5)  &     13.91(6)  & &     17.47(29) &     14.50(8)  &     $>$12.06  & \phn$>$8.68   & & 1; 24, 108; 2, 108 \\
\enddata
\end{deluxetable}

\clearpage
\begin{deluxetable}{lcccccccccccc}
\rotate
\setlength{\tabcolsep}{0.03in}
\tabletypesize{\tiny}
\tablecaption{Near-Infrared Absolute Magnitudes for All Ultracool Dwarfs with Parallaxes \label{tbl:nir-absmag}}
\tablewidth{0pt}
\tablehead{
\colhead{}   &
\colhead{}   &
\multicolumn{5}{c}{MKO}      &
\colhead{}   &
\multicolumn{3}{c}{2MASS} &
\colhead{}   &
\colhead{} \\
\cline{3-7}
\cline{9-11}
\colhead{Object} &
\colhead{Spec.\ Type} &
\colhead{$M_Y$} &
\colhead{$M_J$} &
\colhead{$M_H$} &
\colhead{$M_K$} &
\colhead{$M_{L^{\prime}}$} &
\colhead{}   &
\colhead{$M_J$} &
\colhead{$M_H$} &
\colhead{$M_{K_S}$} &
\colhead{}   &
\colhead{\HST/AO} \\
\colhead{} &
\colhead{Optical/IR}   &
\colhead{(mag)} &
\colhead{(mag)} &
\colhead{(mag)} &
\colhead{(mag)} &
\colhead{(mag)} &
\colhead{} &
\colhead{(mag)} &
\colhead{(mag)} &
\colhead{(mag)} &
\colhead{}   &
\colhead{References}}
\startdata

L~726-8A                       & M5.5/\nodata    &        \nodata     &            \nodata       &            \nodata       &            \nodata       &           \nodata   & &      \phn$ 9.73(3)     $ &      \phn$ 9.17(4)     $ &     \phn$ 8.78(5)     $ & & \nodata                    \\  
Proxima~Cen                    & M5.5/\nodata    &        \nodata     &            \nodata       &            \nodata       &            \nodata       &           \nodata   & &      \phn$ 9.79(2)     $ &      \phn$ 9.27(6)     $ &     \phn$ 8.81(3)     $ & & 33b                        \\  
LHS~1742a                      & esdM5.5/\nodata &        \nodata     &      \phn$10.28(17)    $ &  \phn\phn$ 9.85(17)    $ &  \phn\phn$ 9.70(17)    $ &           \nodata   & &      \phn$10.28(17)    $ &  \phn\phn$ 9.86(17)    $ & \phn\phn$ 9.74(17)    $ & & \nodata                    \\  
2MASS~J22344161+4041387A       & \nodata/M6:     &        \nodata     & \phn\phn$[ 5.69(41)   ]$ & \phn\phn$[ 5.02(41)   ]$ & \phn\phn$[ 4.58(41)   ]$ & \phn\phn$ 3.60(42)$ & & \phn\phn$[ 5.74(41)   ]$ & \phn\phn$[ 4.98(41)   ]$ & \phn\phn$ 4.61(41)    $ & & 3                          \\  
2MASS~J22344161+4041387B       & \nodata/M6:     &        \nodata     & \phn\phn$[ 5.75(41)   ]$ & \phn\phn$[ 5.11(41)   ]$ & \phn\phn$[ 4.63(41)   ]$ & \phn\phn$ 3.83(42)$ & & \phn\phn$[ 5.80(41)   ]$ & \phn\phn$[ 5.08(41)   ]$ & \phn\phn$ 4.66(41)    $ & & 3                          \\  
CE~507                         & M6/\nodata      &        \nodata     &            \nodata       &            \nodata       &            \nodata       &           \nodata   & &      \phn$ 9.81(9)     $ &      \phn$ 9.22(9)     $ &     \phn$ 8.91(9)     $ & & \nodata                    \\
\enddata
\end{deluxetable}

\clearpage
\begin{deluxetable}{lcccccccccccc}
\rotate
\setlength{\tabcolsep}{0.03in}
\tabletypesize{\tiny}
\tablecaption{Mid-Infrared Absolute Magnitudes for All Ultracool Dwarfs with Parallaxes \label{tbl:mir-absmag}}
\tablewidth{0pt}
\tablehead{
\colhead{}   &
\colhead{}   &
\multicolumn{4}{c}{\Spitzer/IRAC} &
\colhead{}   &
\multicolumn{4}{c}{\WISE} &
\colhead{}   &
\colhead{}   \\
\cline{3-6}
\cline{8-11}
\colhead{Object} &
\colhead{Spec.\ Type} &
\colhead{$M_{[3.6]}$} &
\colhead{$M_{[4.5]}$} &
\colhead{$M_{[5.8]}$} &
\colhead{$M_{[8.0]}$} &
\colhead{}   &
\colhead{$M_{\rm W1}$} &
\colhead{$M_{\rm W2}$} &
\colhead{$M_{\rm W3}$} &
\colhead{$M_{\rm W4}$} &
\colhead{}   &
\colhead{\HST/AO} \\
\colhead{} &
\colhead{Optical/IR}   &
\colhead{(mag)} &
\colhead{(mag)} &
\colhead{(mag)} &
\colhead{(mag)} &
\colhead{} &
\colhead{(mag)} &
\colhead{(mag)} &
\colhead{(mag)} &
\colhead{(mag)} &
\colhead{} &
\colhead{References}}

\startdata

Proxima~Cen                     & M5.5/\nodata    &          \nodata   &              \nodata   &              \nodata   &              \nodata   & &     \phn 8.63(9)   &     \phn 8.00(3)   &     \phn 8.26(1)   &     \phn 8.09(2)   & & 30                  \\  
LHS~1742a                       & esdM5.5/\nodata &          \nodata   &              \nodata   &              \nodata   &              \nodata   & & \phn\phn 9.36(17)  & \phn\phn 9.14(17)  &          \nodata   &          \nodata   & & \nodata             \\  
CE~507                          & M6/\nodata      &          \nodata   &              \nodata   &              \nodata   &              \nodata   & &     \phn 8.66(9)   &     \phn 8.47(9)   &     \phn 8.38(9)   &          \nodata   & & \nodata             \\  
Wolf~359                        & M6/M6           &          \nodata   &              \nodata   &              \nodata   &              \nodata   & &     \phn 8.92(5)   &     \phn 8.60(3)   &     \phn 8.59(1)   &     \phn 8.42(3)   & & 39, 56              \\  
GJ~1245B                        & M6/M6           &          \nodata   &              \nodata   &              \nodata   &              \nodata   & &     \phn 8.89(7)   &     \phn 8.68(3)   &     \phn 8.56(3)   &     \phn 8.47(9)   & & 56                  \\  
LP~397-10                       & M6/\nodata      &          \nodata   &              \nodata   &              \nodata   &              \nodata   & &     \phn 9.05(5)   &     \phn 8.85(5)   &     \phn 8.64(7)   &          \nodata   & & \nodata             \\  
LSPM~J2158+6117                 & M6/\nodata      &          \nodata   &              \nodata   &              \nodata   &              \nodata   & &     \phn 9.08(8)   &     \phn 8.87(8)   &     \phn 8.60(9)   & \phn\phn 7.84(31)  & & \nodata             \\
\enddata
\end{deluxetable}

\clearpage
\begin{deluxetable}{cclllllllc}
\setlength{\tabcolsep}{0.03in}
\tabletypesize{\tiny}
\tablecaption{Coefficients of Polynomial Fits to Absolute Magnitudes \label{tbl:poly}}
\tablewidth{0pt}
\tablehead{
\colhead{$y$}   &
\colhead{$x$}   &
\colhead{$c_0$}   &
\colhead{$c_1$}   &
\colhead{$c_2$}   &
\colhead{$c_3$}   &
\colhead{$c_4$}   &
\colhead{$c_5$}   &
\colhead{$c_6$}   &
\colhead{rms}}
\startdata

$Y_{\rm MKO}$   & SpT &  $-$3.51560$\times$10$^{  1}$ & \phs1.95444$\times$10$^{  1}$ &  $-$3.26895                   & \phs2.79438$\times$10$^{ -1}$ &  $-$1.26151$\times$10$^{ -2}$ & \phs2.85027$\times$10$^{ -4}$ &  $-$2.52638$\times$10$^{ -6}$ &  0.40 \\
$J_{\rm MKO}$   & SpT &  $-$2.83129$\times$10$^{  1}$ & \phs1.63986$\times$10$^{  1}$ &  $-$2.74405                   & \phs2.32771$\times$10$^{ -1}$ &  $-$1.03332$\times$10$^{ -2}$ & \phs2.27641$\times$10$^{ -4}$ &  $-$1.94920$\times$10$^{ -6}$ &  0.39 \\
$H_{\rm MKO}$   & SpT &  $-$2.97306$\times$10$^{  1}$ & \phs1.69138$\times$10$^{  1}$ &  $-$2.85705                   & \phs2.45209$\times$10$^{ -1}$ &  $-$1.10960$\times$10$^{ -2}$ & \phs2.51601$\times$10$^{ -4}$ &  $-$2.24083$\times$10$^{ -6}$ &  0.38 \\
$K_{\rm MKO}$   & SpT &  $-$1.52200$\times$10$^{  1}$ & \phs1.01248$\times$10$^{  1}$ &  $-$1.63930                   & \phs1.35177$\times$10$^{ -1}$ &  $-$5.84342$\times$10$^{ -3}$ & \phs1.25731$\times$10$^{ -4}$ &  $-$1.04935$\times$10$^{ -6}$ &  0.40 \\
$\Lp_{\rm MKO}$ & SpT & \phs8.89928                   &  $-$1.96584$\times$10$^{ -1}$ & \phs5.30581$\times$10$^{ -2}$ &  $-$2.93191$\times$10$^{ -3}$ & \phs5.46366$\times$10$^{ -5}$ & \phs0.00000                   & \phs0.00000                   &  0.28 \\

 & & & & & & & & \\
$J_{\rm 2MASS}$ & SpT &  $-$9.67994                   & \phs8.16362                   &  $-$1.33053                   & \phs1.11715$\times$10$^{ -1}$ &  $-$4.82973$\times$10$^{ -3}$ & \phs1.00820$\times$10$^{ -4}$ &  $-$7.84614$\times$10$^{ -7}$ &  0.40 \\
$H_{\rm 2MASS}$ & SpT &  $-$1.17526$\times$10$^{  1}$ & \phs9.00279                   &  $-$1.50370                   & \phs1.29202$\times$10$^{ -1}$ &  $-$5.80847$\times$10$^{ -3}$ & \phs1.29363$\times$10$^{ -4}$ &  $-$1.11499$\times$10$^{ -6}$ &  0.40 \\
\enddata

\tablecomments{These polynomial fits are applicable from spectral
  types of M6 to T9 (inclusive), with the exceptions of $H_{\rm
    2MASS}$, $K_{\rm 2MASS}$, $[$5.8$]$, and $[$8.0$]$ (M6--T8.5).
  The coefficients are defined as:
  $$y = \sum_{i=0} c_ix^i$$
  where $y$ and $x$ are the quantities listed in the first two
  columns.  Numerical spectral types are defined such that M6~=~6 and
  T9~=~29. We use optical spectral types for M and L~dwarfs when
  available, infrared types otherwise, and infrared types for
  T~dwarfs.  (As described in the text, using infrared types for all
  objects would result in polynomial relations different by
  0.01--0.02~mag for MKO bands and 0.03--0.06~mag for 2MASS bands,
  i.e., negligible compared to the scatter about the fits.)  The
  rightmost column gives the rms about the fit over all spectral
  types.  Note that the rms may be significantly different over more
  restricted ranges of spectral type as discussed in
  Section~\ref{sec:absmag} and shown in
  Figures~\ref{fig:spt-abs-jhk}--\ref{fig:spt-abs-wise}.}

\end{deluxetable}

\clearpage
\begin{deluxetable}{ccccccccccccccccccccccccc}
\rotate
\setlength{\tabcolsep}{0.035in}
\tabletypesize{\scriptsize}
\tablecaption{Mean MKO Absolute Magnitudes as a Function of Spectral Type \label{tbl:spt-abs-mko}}
\tablewidth{0pt}
\tablehead{
\colhead{Spectral}   &
\multicolumn{4}{c}{$Y$ band} &
\colhead{} &
\multicolumn{4}{c}{$J$ band} &
\colhead{} &
\multicolumn{4}{c}{$H$ band} &
\colhead{} &
\multicolumn{4}{c}{$K$ band} &
\colhead{} &
\multicolumn{4}{c}{\Lp\ band}  \\
\cline{2-5}
\cline{7-10}
\cline{12-15}
\cline{17-20}
\cline{22-25}
\colhead{Type}   &
\colhead{mean} &
\colhead{rms} &
\colhead{$\sigma_{\rm add}$} &
\colhead{$N$} &
\colhead{} &
\colhead{mean} &
\colhead{rms} &
\colhead{$\sigma_{\rm add}$} &
\colhead{$N$} &
\colhead{} &
\colhead{mean} &
\colhead{rms} &
\colhead{$\sigma_{\rm add}$} &
\colhead{$N$} &
\colhead{} &
\colhead{mean} &
\colhead{rms} &
\colhead{$\sigma_{\rm add}$} &
\colhead{$N$} &
\colhead{} &
\colhead{mean} &
\colhead{rms} &
\colhead{$\sigma_{\rm add}$} &
\colhead{$N$}}
\startdata

M6   &     $10.94\pm0.04$ &    0.13 & \nodata &  2 &  &     $10.34\pm0.02$ &    0.27 &    0.12 &  5 &  & \phn$ 9.82\pm0.03$ &    0.28 &    0.13 &  5 &  & \phn$ 9.40\pm0.03$ &    0.34 &    0.16 &  5 &  & \phn$ 8.95\pm0.04$ &    0.29 &    0.07 &  4 \\
M6.5 &         \nodata    & \nodata & \nodata &  0 &  &     $10.45\pm0.03$ &    0.16 &    0.03 &  3 &  & \phn$ 9.89\pm0.03$ &    0.20 &    0.06 &  3 &  & \phn$ 9.52\pm0.03$ &    0.15 &    0.01 &  3 &  & \phn$ 9.08\pm0.05$ &    0.15 & \nodata &  2 \\
M7   &     $11.34\pm0.06$ & \nodata & \nodata &  1 &  &     $10.31\pm0.02$ &    0.42 &    0.33 &  5 &  & \phn$ 9.94\pm0.02$ &    0.40 &    0.31 &  5 &  & \phn$ 9.50\pm0.02$ &    0.36 &    0.29 &  5 &  & \phn$ 9.44\pm0.08$ & \nodata & \nodata &  1 \\
M7.5 &         \nodata    & \nodata & \nodata &  0 &  &     $10.82\pm0.05$ & \nodata & \nodata &  1 &  &     $10.26\pm0.05$ & \nodata & \nodata &  1 &  & \phn$ 9.79\pm0.05$ & \nodata & \nodata &  1 &  &         \nodata    & \nodata & \nodata &  0 \\
M8   &     $11.77\pm0.06$ & \nodata & \nodata &  1 &  &     $10.99\pm0.02$ &    0.26 &    0.12 &  6 &  &     $10.40\pm0.02$ &    0.23 &    0.11 &  6 &  & \phn$ 9.88\pm0.02$ &    0.26 & \nodata &  5 &  & \phn$ 9.52\pm0.07$ & \nodata & \nodata &  1 \\
M8.5 &     $12.11\pm0.11$ & \nodata & \nodata &  1 &  &     $11.40\pm0.03$ &    0.19 &    0.13 &  5 &  &     $10.77\pm0.03$ &    0.21 &    0.15 &  5 &  &     $10.28\pm0.03$ &    0.20 &    0.14 &  5 &  & \phn$ 9.61\pm0.05$ &    0.27 &    0.12 &  3 \\
M9   &     $12.82\pm0.04$ &    0.29 &    0.10 &  2 &  &     $11.80\pm0.03$ &    0.30 &    0.22 &  5 &  &     $11.15\pm0.03$ &    0.32 &    0.24 &  5 &  &     $10.62\pm0.02$ &    0.33 &    0.25 &  5 &  & \phn$ 9.93\pm0.03$ &    0.25 &    0.15 &  5 \\
\enddata

\tablecomments{For each band the weighted average and standard error
  are given (``mean'' column) as computed from all objects of a given
  spectral type with measured photometry (see
  Table~\ref{tbl:nir-absmag}).  The rms of those objects is also
  given, which includes the scatter due to both measurement errors and
  intrinsic variations at that spectral type.  The $\sigma_{\rm add}$
  column additional uncertainty needed to make $p(\chi^2) = 0.5$
  (i.e., reduced $\chi^2 \approx 1$) for the set of magnitudes.  This
  is essentially a lower limit on the intrinsic scatter because
  scatter due to measurement error can always mask intrinsic
  variations at some level and because small sample sizes at some
  spectral types may not fully capture the intrinsic variations. The
  $N$ column notes how many objects were used to derive the mean, rms,
  and $\sigma_{\rm add}$; when $N=1$ the ``mean'' is simply the
  magnitude of the one object available.  We only use ``normal'' field
  dwarfs, i.e., those not flagged as atypical in
  Table~\ref{tbl:plx-all}, to compute the weighted averages.  When
  individual object magnitude errors are large enough to explain the
  observed scatter (i.e., $p(\chi^2) > 0.5$) no value for $\sigma_{\rm
    add}$ is listed. Note that we use optical spectral types for M and
  L~dwarfs when available (infrared types otherwise) and infrared
  types for T~dwarfs.}

\end{deluxetable}

\clearpage
\begin{deluxetable}{ccccccccccccccc}
\setlength{\tabcolsep}{0.06in}
\tabletypesize{\tiny}
\tablecaption{Mean 2MASS Absolute Magnitudes as a Function of Spectral Type \label{tbl:spt-abs-2mass}}
\tablewidth{0pt}
\tablehead{
\colhead{Spectral}   &
\multicolumn{4}{c}{$J$ band} &
\colhead{} &
\multicolumn{4}{c}{$H$ band} &
\colhead{} &
\multicolumn{4}{c}{\Ks\ band} \\
\cline{2-5}
\cline{7-10}
\cline{12-15}
\colhead{Type}   &
\colhead{mean} &
\colhead{rms} &
\colhead{$\sigma_{\rm add}$} &
\colhead{$N$} &
\colhead{} &
\colhead{mean} &
\colhead{rms} &
\colhead{$\sigma_{\rm add}$} &
\colhead{$N$} &
\colhead{} &
\colhead{mean} &
\colhead{rms} &
\colhead{$\sigma_{\rm add}$} &
\colhead{$N$}}
\startdata

M6   &     $10.28\pm0.01$ &    0.27 &    0.25 & 16 &  & \phn$ 9.73\pm0.01$ &    0.27 &    0.26 & 16 &  & \phn$ 9.31\pm0.01$ &    0.28 &    0.28 & 16 \\
M6.5 &     $10.33\pm0.01$ &    0.44 &    0.44 & 16 &  & \phn$ 9.75\pm0.01$ &    0.45 &    0.44 & 16 &  & \phn$ 9.41\pm0.01$ &    0.44 &    0.43 & 16 \\
M7   &     $10.25\pm0.02$ &    0.45 &    0.48 &  8 &  & \phn$ 9.79\pm0.02$ &    0.45 &    0.47 &  8 &  & \phn$ 9.42\pm0.01$ &    0.44 &    0.46 &  8 \\
M7.5 &     $10.63\pm0.02$ &    0.30 &    0.31 &  9 &  &     $10.01\pm0.02$ &    0.32 &    0.33 &  9 &  & \phn$ 9.61\pm0.02$ &    0.33 &    0.34 &  9 \\
M8   &     $10.93\pm0.01$ &    0.46 &    0.34 & 11 &  &     $10.24\pm0.01$ &    0.46 &    0.34 & 11 &  & \phn$ 9.83\pm0.01$ &    0.44 &    0.35 & 12 \\
M8.5 &     $11.58\pm0.01$ &    0.20 &    0.20 &  9 &  &     $10.88\pm0.01$ &    0.23 &    0.23 &  9 &  &     $10.46\pm0.01$ &    0.22 &    0.23 &  9 \\
M9   &     $11.81\pm0.02$ &    0.29 &    0.26 &  7 &  &     $11.10\pm0.02$ &    0.31 &    0.29 &  7 &  &     $10.57\pm0.02$ &    0.30 &    0.30 &  8 \\
M9.5 &     $11.69\pm0.03$ &    0.14 &    0.12 &  6 &  &     $10.89\pm0.03$ &    0.20 &    0.20 &  6 &  &     $10.37\pm0.03$ &    0.22 &    0.23 &  6 \\
\enddata

\tablecomments{Same as Table~\ref{tbl:spt-abs-mko} but for 2MASS
  magnitudes.}

\end{deluxetable}

\clearpage
\begin{deluxetable}{cccccccccccccccccccc}
\setlength{\tabcolsep}{0.035in}
\tabletypesize{\tiny}
\tablecaption{Mean \Spitzer/IRAC Absolute Magnitudes as a Function of Spectral Type \label{tbl:spt-abs-irac}}
\tablewidth{0pt}
\tablehead{
\colhead{Spectral}   &
\multicolumn{4}{c}{$[3.6]$ band} &
\colhead{} &
\multicolumn{4}{c}{$[4.5]$ band} &
\colhead{} &
\multicolumn{4}{c}{$[5.8]$ band} &
\colhead{} &
\multicolumn{4}{c}{$[8.0]$ band} \\
\cline{2-5}
\cline{7-10}
\cline{12-15}
\cline{17-20}
\colhead{Type}   &
\colhead{mean} &
\colhead{rms} &
\colhead{$\sigma_{\rm add}$} &
\colhead{$N$} &
\colhead{} &
\colhead{mean} &
\colhead{rms} &
\colhead{$\sigma_{\rm add}$} &
\colhead{$N$} &
\colhead{} &
\colhead{mean} &
\colhead{rms} &
\colhead{$\sigma_{\rm add}$} &
\colhead{$N$} &
\colhead{} &
\colhead{mean} &
\colhead{rms} &
\colhead{$\sigma_{\rm add}$} &
\colhead{$N$}}
\startdata

M6   & \phn$ 9.19\pm0.01$ & \nodata & \nodata &  1 &  & \phn$ 9.17\pm0.02$ & \nodata & \nodata &  1 &  & \phn$ 9.12\pm0.01$ & \nodata & \nodata &  1 &  & \phn$ 9.09\pm0.01$ & \nodata & \nodata &  1 \\
M6.5 & \phn$ 9.10\pm0.03$ &    0.10 & \nodata &  3 &  & \phn$ 9.14\pm0.03$ &    0.11 & \nodata &  3 &  & \phn$ 9.09\pm0.03$ &    0.11 & \nodata &  3 &  & \phn$ 8.99\pm0.02$ &    0.10 &    0.00 &  3 \\
M7   & \phn$ 9.33\pm0.02$ &    0.12 & \nodata &  2 &  & \phn$ 9.33\pm0.01$ &    0.12 &    0.03 &  2 &  & \phn$ 9.24\pm0.02$ &    0.12 &    0.02 &  2 &  & \phn$ 9.20\pm0.02$ &    0.13 &    0.03 &  2 \\
M7.5 &         \nodata    & \nodata & \nodata &  0 &  &         \nodata    & \nodata & \nodata &  0 &  &         \nodata    & \nodata & \nodata &  0 &  &         \nodata    & \nodata & \nodata &  0 \\
M8   & \phn$ 9.44\pm0.02$ & \nodata & \nodata &  1 &  & \phn$ 9.45\pm0.03$ & \nodata & \nodata &  1 &  & \phn$ 9.30\pm0.01$ & \nodata & \nodata &  1 &  & \phn$ 9.29\pm0.01$ & \nodata & \nodata &  1 \\
M8.5 & \phn$ 9.78\pm0.02$ & \nodata & \nodata &  1 &  & \phn$ 9.78\pm0.01$ & \nodata & \nodata &  1 &  & \phn$ 9.62\pm0.01$ & \nodata & \nodata &  1 &  & \phn$ 9.52\pm0.01$ & \nodata & \nodata &  1 \\
M9   & \phn$ 9.97\pm0.03$ &    0.32 &    0.22 &  3 &  & \phn$ 9.98\pm0.02$ &    0.29 &    0.19 &  3 &  & \phn$ 9.79\pm0.02$ &    0.28 &    0.18 &  3 &  & \phn$ 9.65\pm0.02$ &    0.24 &    0.16 &  3 \\
M9.5 & \phn$ 9.63\pm0.11$ & \nodata & \nodata &  1 &  & \phn$ 9.60\pm0.11$ & \nodata & \nodata &  1 &  & \phn$ 9.41\pm0.11$ & \nodata & \nodata &  1 &  & \phn$ 9.24\pm0.11$ & \nodata & \nodata &  1 \\
\enddata

\tablecomments{Same as Table~\ref{tbl:spt-abs-mko} but for
  \Spitzer/IRAC magnitudes.}

\end{deluxetable}

\clearpage
\begin{deluxetable}{cccccccccccccccccccc}
\setlength{\tabcolsep}{0.035in}
\tabletypesize{\tiny}
\tablecaption{Mean \WISE\ Absolute Magnitudes as a Function of Spectral Type \label{tbl:spt-abs-wise}}
\tablewidth{0pt}
\tablehead{
\colhead{Spectral}   &
\multicolumn{4}{c}{$W1$ band} &
\colhead{} &
\multicolumn{4}{c}{$W2$ band} &
\colhead{} &
\multicolumn{4}{c}{$W3$ band} &
\colhead{} &
\multicolumn{4}{c}{$W4$ band} \\
\cline{2-5}
\cline{7-10}
\cline{12-15}
\cline{17-20}
\colhead{Type}   &
\colhead{mean} &
\colhead{rms} &
\colhead{$\sigma_{\rm add}$} &
\colhead{$N$} &
\colhead{} &
\colhead{mean} &
\colhead{rms} &
\colhead{$\sigma_{\rm add}$} &
\colhead{$N$} &
\colhead{} &
\colhead{mean} &
\colhead{rms} &
\colhead{$\sigma_{\rm add}$} &
\colhead{$N$} &
\colhead{} &
\colhead{mean} &
\colhead{rms} &
\colhead{$\sigma_{\rm add}$} &
\colhead{$N$}}
\startdata

M6   & \phn$ 9.21\pm0.01$ &    0.25 &    0.25 & 14 &  & \phn$ 8.95\pm0.01$ &    0.25 &    0.24 & 14 &  & \phn$ 8.69\pm0.01$ &    0.24 &    0.24 & 14 &  & \phn$ 8.46\pm0.03$ &    0.48 &    0.11 &  5 \\
M6.5 & \phn$ 9.16\pm0.01$ &    0.46 &    0.46 & 15 &  & \phn$ 8.94\pm0.01$ &    0.45 &    0.44 & 15 &  & \phn$ 8.74\pm0.01$ &    0.41 &    0.39 & 15 &  & \phn$ 8.59\pm0.05$ &    0.67 &    0.25 &  8 \\
M7   & \phn$ 9.31\pm0.02$ &    0.49 &    0.54 &  6 &  & \phn$ 9.10\pm0.02$ &    0.47 &    0.51 &  6 &  & \phn$ 8.91\pm0.02$ &    0.47 &    0.51 &  6 &  & \phn$ 8.78\pm0.14$ &    0.79 &    0.37 &  3 \\
M7.5 & \phn$ 9.32\pm0.02$ &    0.35 &    0.37 &  8 &  & \phn$ 9.07\pm0.02$ &    0.34 &    0.36 &  8 &  & \phn$ 8.65\pm0.03$ &    0.34 &    0.39 &  8 &  & \phn$ 8.43\pm0.20$ &    0.13 & \nodata &  4 \\
M8   & \phn$ 9.53\pm0.01$ &    0.58 &    0.52 &  9 &  & \phn$ 9.31\pm0.01$ &    0.56 &    0.50 &  9 &  & \phn$ 9.15\pm0.02$ &    0.60 &    0.46 &  9 &  & \phn$ 8.58\pm0.34$ &    0.35 & \nodata &  2 \\
M8.5 &     $10.08\pm0.01$ &    0.23 &    0.19 &  5 &  & \phn$ 9.80\pm0.01$ &    0.21 &    0.17 &  5 &  & \phn$ 9.42\pm0.02$ &    0.09 &    0.03 &  5 &  & \phn$ 9.10\pm0.12$ &    0.13 & \nodata &  2 \\
M9   &     $10.19\pm0.02$ &    0.34 &    0.34 &  6 &  & \phn$ 9.93\pm0.02$ &    0.31 &    0.31 &  6 &  & \phn$ 9.48\pm0.02$ &    0.35 &    0.33 &  6 &  & \phn$ 9.30\pm0.10$ &    0.82 &    0.68 &  4 \\
M9.5 & \phn$ 9.91\pm0.03$ &    0.27 &    0.31 &  4 &  & \phn$ 9.64\pm0.03$ &    0.20 &    0.23 &  4 &  & \phn$ 9.17\pm0.06$ &    0.29 &    0.25 &  3 &  &         \nodata    & \nodata & \nodata &  0 \\
\enddata

\tablecomments{Same as Table~\ref{tbl:spt-abs-mko} but for \WISE\
  magnitudes.}

\end{deluxetable}

\end{document}